%% file: main.tex
\documentclass{these-dbl}

\hypersetup{
   pdfauthor   = {Pierre Fernandez},
   pdftitle    = {Watermarking Across Modalities for Content Tracing and Generative AI},
   pdfsubject  = {Watermarking Across Modalities for Content Tracing and Generative AI},
   pdfkeywords = {watermarking, generative, ai, tracing, moderation},
}

\usepackage{amsmath, bm, amsthm, amsfonts, amssymb} %
\usepackage{dsfont} %
\usepackage{enumitem} %
\usepackage{wrapfig}
\usepackage{pifont} %
\usepackage{array}
\usepackage{tcolorbox} %

\usepackage{multirow} %
\usepackage{booktabs} %
\usepackage{tabularx}
\usepackage{tablefootnote}
\usepackage{colortbl} %
\usepackage{makecell}
\usepackage{float}
\usepackage{stmaryrd} %
\usepackage[export]{adjustbox} %

\usepackage{xurl} %

\usepackage[scientific-notation=true]{siunitx} %

\usepackage{algorithm} %
\usepackage{algpseudocode}
\usepackage{subcaption} %
\usepackage{graphicx} %
\usepackage{comment} %
\usepackage{annotate} %

\input{macros}

\usepackage[toc, sort=def]{glossaries} %

\input{./0-conclusion/glossary}

\begin{document}
\selectlanguage{english}

\begin{titlepage}
    \begin{center}
        \vspace*{1cm}

        \Huge
        \textbf{Watermarking across Modalities for Content Tracing and Generative AI} 

        \vspace{1cm}
        \large
        \textit{by} \\
        \Large
        Pierre Fernandez

        \vspace*{3cm}

        \raisebox{-0.5\height}{\includegraphics[width=0.3\textwidth]{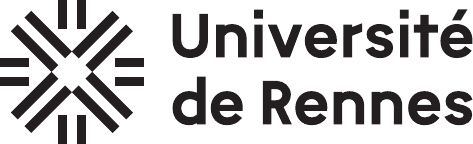}}
        \hspace{1.5cm}
        \raisebox{-0.4\height}{\includegraphics[width=0.3\textwidth]{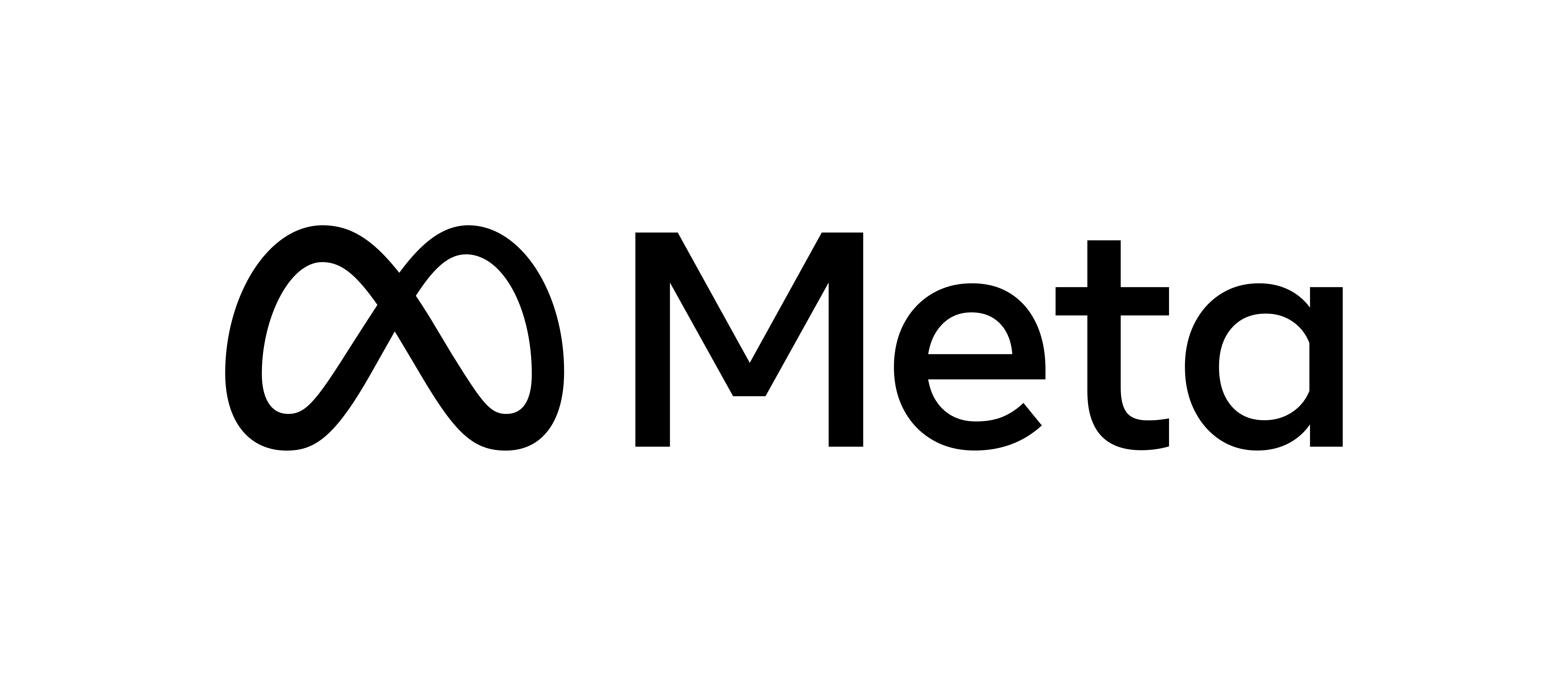}}

        \vspace*{0.8cm}

        \Large
        Centre Inria de l'Université de Rennes  \\
        Meta, Fundamental AI Research Lab (FAIR), Paris 

        \vspace{3cm}

        \large
        A thesis submitted for the degree of \\
        Doctor of Philosophy 

        \vspace{0.8cm}
        \large
        January 2025

    \end{center}
\end{titlepage}

\clearpage
\newgeometry{top=3cm, left=2.5cm, right=2.5cm, bottom=3cm}
\renewcommand{\contentsname}{Table of Contents}
\setcounter{tocdepth}{1}
\tableofcontents %
\restoregeometry 

\clearpage \input{./0-introduction/acknowledgement}

\clearpage \part{Introduction}\label{part:introduction}

\clearpage \input{./0-introduction/introduction}
\clearpage \input{./0-introduction/related-all}

\clearpage \input{./0-introduction/technical-background}

\mainmatter

\clearpage \part{Content Moderation}\label{part:content-moderation}

\clearpage \input{./chapter-1/main}

\clearpage \input{./chapter-2/main}

\clearpage \part{Tracing AI-Generated Content}\label{part:genai-tracing}

\clearpage \input{./chapter-3/main}

\clearpage \input{./chapter-4/main}

\clearpage \input{./chapter-5/main}

\clearpage \part{Monitoring AI Models}\label{part:model-monitoring}

\clearpage \input{./chapter-6/main}

\clearpage \input{./chapter-7/main}

\clearpage \part{Conclusion and Appendices}\label{part:conclusion}
\clearpage \input{./0-conclusion/conclusion}

\backmatter

\appendix
\clearpage
\chapter{Proofs}

\input{./0-conclusion/proofs}

\clearpage 
\chapter{List of Publications}
\input{./0-conclusion/publications}

\clearpage
\renewcommand{\glsnamefont}[1]{\makefirstuc{#1}}
\printnoidxglossaries

\clearpage 
\addcontentsline{toc}{chapter}{Bibliography}
{\small
\bibliographystyle{plainnat}
\bibliography{biblio/consolidated}
}

\end{document}

%% file: macros.tex
\newcommand{\real}{\mathbb{R}}
\newcommand{\R}{\mathbb{R}}

\newcommand{\abs}[1]{\lvert{#1}\rvert}
\newcommand{\norm}[1]{\lVert{#1}\rVert}
\newcommand{\T}{^\intercal}

\renewcommand{\vec}[1]{\mathbf{#1}}
\newcommand{\temperature}{\theta}

\def\A{\mathcal A}
\def\B{\mathcal B}

\def\V{\mathcal{V}}

\def\G{\mathcal{G}}
\def\H{\mathcal{H}}
\def\Prob{\mathds{P}}
\def\E{\mathds{E}}

\newcommand{\etal}{{et al.}\@ }
\newcommand{\eg}{{e.g.}}
\newcommand{\ie}{{i.e.}}
\newcommand{\aka}{{a.k.a.}}

\newcommand{\Io}{I_o}
\newcommand{\Iw}{I_w}
\newcommand{\Ispace}{\mathcal{I}}
\newcommand{\Fspace}{\mathcal{F}}
\newcommand{\Tr}{\mathrm{Tr}}
\def\payload{k}
\def\im{x}

\renewcommand{\L}{\mathcal{L}}
\newcommand{\Lw}{ \mathcal{L}_{\mathrm{w}} }
\newcommand{\Li}{ \mathcal{L}_{\mathrm{i}} }
\newcommand{\Lf}{ \mathcal{L}_{\mathrm{f}} }

\newcommand{\soutx}[1]{}

\newcommand{\acti}{\star}

\newcommand{\cmark}{\textcolor{blue!90}{\ding{51}}\xspace}%
\newcommand{\cmarkg}{\ding{51}\xspace}%
\newcommand{\xmark}{\textcolor{red!90}{\ding{55}}\xspace}%
\newcommand{\xmarkg}{\ding{55}\xspace}%
\newcommand{\amark}{\bm{$\sim$}} %

\definecolor{Apricot}{gray}{0.8}
\definecolor{apricot}{gray}{0.8}

\newcommand{\fullcite}[1]{\bibentry{#1}}

\newcommand{\fullref}[1]{\hyperref[#1]{\autoref*{#1}: \nameref*{#1}}}

\addto\extrasenglish{

}

\setlist[itemize]{itemsep=0.5em, leftmargin=3em, label={$\bullet$}}
\setlist[enumerate]{itemsep=0.5em}
\newcommand\blfootnote[1]{%
  \begingroup
  \renewcommand\thefootnote{}\footnote{#1}%
  \addtocounter{footnote}{-1}%
  \endgroup
}

\newcommand{\muap}{$\mu$AP\@ }
\newcommand{\rone}{$R$@$1$\@ }

\newcommand{\aux}[1]{{\scriptsize{\textcolor{gray}{#1}}}}
\newcommand{\ploss}{TF-Loudness\xspace} 

\newcommand{\wm}{\delta} %

\newcommand{\sk}{\mathsf{s}}
\newcommand{\logit}{\boldsymbol{\ell}}
\newcommand{\logpval}{\log_{10}(p)}
\newcommand{\pval}{p\textrm{-value}}

\definecolor{metablue}{HTML}{0064E0}
\definecolor{metafg}{HTML}{1C2B33}
\definecolor{metabg}{HTML}{F1F4F7}

\definecolor{Gray}{gray}{0.95}
\newcommand{\colorcell}{\cellcolor{Gray}}

\newlength\savewidth\newcommand\shline{\noalign{\global\savewidth\arrayrulewidth
\global\arrayrulewidth 1pt}\hline\noalign{\global\arrayrulewidth\savewidth}}

\newtheorem{definition}{Definition}

%% file: 0-conclusion/glossary.tex
\makenoidxglossaries

\newglossaryentry{watermarking}
{
    name={watermarking},
    description={The practice of concealing information into digital media to protect copyright, verify authenticity, or convey other metadata}
}

\newglossaryentry{steganography}
{
    name={steganography},
    description={The practice of concealing information into digital media in such a way that the presence of the information is hidden}
}

\newglossaryentry{zero-bit watermarking}
{
    name={zero-bit watermarking},
    description={A type of watermarking that does not embed any binary message, but alters the cover medium in a way that can be detected}
}

\newglossaryentry{multi-bit watermarking}
{
    name={multi-bit watermarking},
    description={A type of watermarking that embeds a binary message into the cover medium}
}

\newglossaryentry{cover}
{
    name={cover},
    description={The original media content where a watermark is embedded}
}

\newglossaryentry{robustness}
{
    name={robustness},
    description={The ability of a watermarking scheme (or any other system) to withstand various transformations of the cover medium}
}

\newglossaryentry{payload}
{
    name={payload},
    description={The amount of information that can be embedded into a cover medium, usually measured in bits}
}

\newglossaryentry{security}
{
    name={security},
    description={The ability of a watermarking scheme to prevent unauthorized parties from intentionally detecting, removing, forging or altering the watermark}
}

\newglossaryentry{fingerprinting}
{
    name={fingerprinting},
    description={A technique used to uniquely identify digital content by generating a unique identifier or hash based on the content's features}
}

\newglossaryentry{copy detection}
{
    name={copy detection},
    description={The process of identifying duplicate or near-duplicate content in a large dataset}
}

\newglossaryentry{hash}
{
    name={hash},
    description={(or fingerprint) A fixed-size vector representation of data}
}

\newglossaryentry{indexing}
{
    name={indexing},
    description={The process of organizing and storing data in a way that makes it easy to search and retrieve}
}

\newglossaryentry{FAISS}
{
    name={FAISS},
    description={(Facebook AI Similarity Search) A library for efficient similarity search and clustering of dense vectors}
}

\newglossaryentry{digital forensics}
{
    name={digital forensics},
    description={The process of analyzing and collecting digital evidence to determine the authenticity and provenance of synthetic or manipulated content}
}

\newglossaryentry{C2PA}
{
    name={C2PA},
    description={(Coalition for Content Provenance and Authenticity) An initiative that aims to address the prevalence of misleading information online by developing technical standards for certifying the source and history of media content}
}

\newglossaryentry{RSA}
{
    name={RSA},
    description={A public-key cryptosystem that is widely used for secure data transmission. Stands for Rivest-Shamir-Adleman, the inventors of the technique}
}

\newglossaryentry{GAN}
{
    name={GAN},
    description={(Generative Adversarial Network) A class of machine learning frameworks designed by opposing two neural networks against each other to generate new, synthetic instances of data that can pass for real data}
}

\newglossaryentry{diffusion model}
{
    name={diffusion model},
    description={A type of generative model that generates data by starting from a random noise and gradually reducing the noise through a learned iterative process}
}

\newglossaryentry{LDM}
{
    name={LDM},
    description={(Latent Diffusion Model) A diffusion model that operates in the latent space of a variational autoencoder. Stable Diffusion is a LDM}
}

\newglossaryentry{LLM}
{
    name={LLM},
    description={(Large Language Model) A type of deep learning model designed to understand and generate human language. These models are trained on vast amounts of text data and can perform a variety of language-based tasks. Examples include GPT-3 and BERT}
}

\newglossaryentry{token}
{
    name={token},
    description={A basic unit of a language model's input, usually a word or subword}
}

\newglossaryentry{TTS}
{
    name={TTS},
    description={(Text-To-Speech) A type of technology that converts written text into spoken voice output}
}

\newglossaryentry{p-value}
{
    name={p-value},
    description={The probability of obtaining a test statistic at least as extreme as the one that was actually observed, assuming that the null hypothesis is true. Used to determine the statistical significance of the watermark detection}
}

\newglossaryentry{test statistic}
{
    name={test statistic},
    description={The quantity upon which the statistical test relies, often referred to as watermark score in this thesis}
}

\newglossaryentry{FPR}
{
    name={FPR},
    description={(False Positive Rate) The proportion of non-watermarked instances that are incorrectly classified as watermarked}
}

\newglossaryentry{TPR}
{
    name={TPR},
    description={(True Positive Rate) The proportion of watermarked instances that are correctly classified as watermarked}
}

\newglossaryentry{ROC}
{
    name={ROC},
    description={(Receiver Operating Characteristic), or ROC curve, a graphical plot of the TPR against the FPR, used to evaluate the performance of the watermark detection. The area under the ROC curve (AUC) assesses the overall performance with a single value}
}

\newglossaryentry{HiDDeN}
{
    name={HiDDeN},
    description={A deep learning-based method for embedding and extracting invisible watermarks in images using an encoder-decoder network architecture}
}

\newglossaryentry{white-box}
{
    name={white-box},
    description={A type of attack where the attacker has full access to the model's architecture and parameters}
}

\newglossaryentry{black-box}
{
    name={black-box},
    description={A type of attack where the attacker has no access to the model's architecture or parameters}
}

%% file: 0-introduction/acknowledgement.tex
\chapter*{Executive Summary}
\addcontentsline{toc}{chapter}{Executive Summary}
\chaptermark{Executive Summary}

Watermarking embeds information into digital content like images, audio, or text, imperceptible to humans but robustly detectable by specific algorithms.
This technology has important applications in many challenges of the industry such as content moderation, tracing AI-generated content, and monitoring the usage of AI models.
The contributions of this thesis include the development of new watermarking techniques for images, audio, and text. 
We first introduce methods for active moderation of images on social platforms. 
We then develop specific techniques for AI-generated content.
We specifically demonstrate methods to adapt latent generative models to embed watermarks in all generated content, identify watermarked sections in speech, and improve watermarking in large language models with tests that ensure low false positive rates.
Furthermore, we explore the use of digital watermarking to detect model misuse, including the detection of watermarks in language models fine-tuned on watermarked text, and introduce training-free watermarks for the weights of large transformers.
Through these contributions, the thesis provides effective solutions for the challenges posed by the increasing use of generative AI models and the need for model monitoring and content moderation.
It finally examines the challenges and limitations of watermarking techniques and discuss potential future directions for research in this area.

\chapter*{Acknowledgement}
\addcontentsline{toc}{chapter}{Acknowledgement}
\chaptermark{Acknowledgement}

Acknowledgments are the least meaningful part of the thesis, yet funnily enough, they are the most meaningful to the student (me). 
After spending years on this project, they are clearly the part I was most eager to write.
I am deeply grateful to everyone I have met in my life, especially those with whom I have shared my PhD journey. 
I hope these paragraphs will pay tribute to what we have shared, and that the next hundred pages will pay tribute to the work you have inspired me to do.

\begin{quote}
    \centering
    \textit{``Bon, alors comme ça... tu veux faire du tatouage?''}
    --  Teddy, 2021. 
\end{quote}

First and foremost, I extend my most sincere thanks to my advisors. 
I hope to continue learning from you in the years to come and aspire to be like you.
Teddy, thank you for being the embodiment of both humor and seriousness since our first encounter. 
Your technical guidance has been exceptional, and I will sincerely miss the maths you wrote on the Inria's whiteboard -- which is probably still there today.
Matthijs, our bike rides, your down-to-earth views and steady presence have been the foundation of my PhD.
I am immensely grateful for the trust and responsibilities you have placed in me.
`We don't live out of thin air' is a lesson I will never forget.
Hervé, although you went for new adventures midway through my PhD, your momentum has never faded. 
Your humbleness and your ability to identify research directions that shape the field and society have been an immense source of inspiration. 
To all my advisors, thank you for your frankness and for being such remarkable mentors. 
I owe you everything.

To my fellow PhDs 22' at FAIR and to our stressful foosball matches, drunken afterworks, and countless coffee breaks, you all made my PhD life funny and unforgettable.
Thanks:\blfootnote{$^\star$: Order determined by the start date at FAIR.}
Théo$^\star$ and Timothée$^\star$, \aka, the Dinomates, for the citations, the funny memes on Rats@ and your unfailing presence at parties;
Quentin$^\star$, for never having said no to a break;
Badr$^\star$, for your unbelievable kindness;
Wassim$^\star$, for your daily ``check'';
Robin$^\star$, for running so fast and having shared the love of watermarking for a while;
Mathurin$^\star$, for the late discussions when getting home after the parties.
To all of you, I  am so proud to have shared this journey with you and to be friends with such talented and kind-hearted people.
You really were the best people I could have hoped to share this experience with.

I wish to thank all the FAIR PhDs who walked this path before me and to those who are walking it now. 
Sharing the office with you was a great privilege.
A special thanks to Alexandre S. for patiently onboarding me during my internship, introducing me to watermarking and giving me the tools to succeed in my PhD.
I am thankful to Pierre S., Hugo, and Alaa; being able benefit from your exceptional views has truly humbled me.
I also owe a lot to Lina and Virginie, whose advice allowed me to be part of the best PhD program in the world.
Guillaume C., I am so happy to have collaborated with someone as smart and humble as you, you serve as a great example of how people should work and live their lives.
Megi, sharing moments in Milan, Modena, Capri and the office was great, thank you for your authenticity and your energy.
Fabian, for initiating the board game sessions and for our lively discussions on information theory.
Simon for our deep philosophical discussions late at night.
Tariq, for giving me all of your insights on image generation and segmentation at breakfast.
Kruno, Jos\'ephine, Jo\~ao, Belen, Pierre C., thank you for taking over and bringing so much fun in the office.
Last but not least, Tom, I am so grateful for our friendship and for having shared so many moments with you.
I am also very proud of the trust you placed in me and what we have achieved together. 
I look forward to our future collaborations!

I want to express my gratitude to the folks at Inria Rennes. 
I haven't been able to spend as much time with you as I would have liked, especially towards the end, but I have always felt very welcome.
I would like to especially thank Benoît and Thibault, my co-bureaus for a time.
I was lucky to have shared my first conference with you, as well as the following years of Rennes' office life.
Thank you Antoine, Eva and Vivien for our deep discussions on watermarking at the office and at the pub.
Thank you Laurent and Aur\'elie for including me in the Linkmedia team.
Thank you Duc, Karim, Hugo, Guillaume, Gautier, Victor and Morgane for the coffee breaks and the foosball training which have helped me in my games against the FAIR people.

During my PhD, I have had the chance to discuss and collaborate with many colleagues at FAIR, all very talented people who have made my journey more enjoyable and fruitful: 
Federico, Marc, Francisco, Patrick, Mathilde, Vasil, Zoe, Adrien, Marl\`ene, Jakob, Camille, Olivier, Michal, Tuan, Nikita, Shalini, Tal, Diego, Damien, Adeel, and many others!
Hady, I am amazed by your good mood and your talent. 
I loved working on AudioSeal with you and Robin, and I hope our future collaborations will be as successful.
Armand and Piotr, allowing me to be part of such an impactful project as DINO has been a huge honor. 
You both have an extreme clarity of thought and I thoroughly enjoyed working with you.
Zeki and Dong, initiating a production effort for watermarking was a great experience, thank you for your trust and your guidance.
Jeremy, our climbing sessions and drinks were a great time for me.
Thank you Thomas S., for introducing me to Large Language Models and allowing me to alpha test the short-lived Galactica. 
Pierre-Louis, thanks for making the CIFRE what it is today. 
Naila, Mary, Pierre-Emmanuel and Alex M., I greatly appreciated your insightful discussions about career planning post-PhD. 
Paul-Ambroise, Charlotte, and Matthieu R., your guidance during the interview process has been invaluable.
Lastly, Martin and Victoire, in addition to being super nice human beings, your perspectives on AI policy and business have greatly enriched my views and my personal growth.

To my friends, thank you for your support, being with you makes life funnier!
Hugo, Juan and Arthur thanks for the many moments we have shared during the last few years, that I enjoyed even more than the ``nouilles Chong-Qing piment\'ees 4'' of TTZ -- this is saying something -- as well as Thibault, Alex, Benjamin, although these moments were fewer.
Thanks Kenza, Meryem, Louis and Nissim, our scarce encounters were always a good time to share our experience as PhDs.
Lastly, thanks Thomas B., Astrid and Ambroise, for our late night discussion on LLMs and our tennis games under the Parisian cold, and MH, Alice, Pierre and Thomas D., for still being in my life after all this time.

To my wonderful family, I am profoundly grateful to have been born into such an environment. 
Aunt Sandrine and Jo, I deeply appreciated your generous hospitality during my visits to Rennes.
Those times were very precious to me.
Grandma, thank you for your wisdom as well as for letting me talk to you about my research -- although you may have never understood a word of it (this last point now applies to rest of this paragraph).
Ang\'eline, thank you for being the best godmother on earth.
To my dear parents and beloved sisters, words cannot express how much your love and presence mean to me. 
Mom and Dad, I want to let you know how proud of both of you I am. 
You have been my North stars (mes ``\'Etoiles du Nord'') during my PhD, as you have always been in my life, and as you always will be.
Diana and Ines, thank you for sharing so many moments together. 
Most importantly, thank you for being my best friends in life.
I have the feeling that our brotherhood will always be there, and I am so lucky for it.

Completing this PhD has been a transformative experience, and there are many others who I have not cited but who have contributed to its success.
Each of you has left an (in)visible and robust mark on life. Thank you.

%% file: 0-introduction/introduction.tex
\chapter{Introduction}\label{chapter:introduction}

\section{A brief history of watermarking}

\Gls*{watermarking}, \ie, the process of hiding information about an object within the object itself, is a technique that has been used for centuries if not millenia.

\paragraph*{Origins.}

\begin{figure}[b]
    \centering
    \begin{minipage}{0.59\textwidth}
        \centering
        \includegraphics[width=1.0\textwidth]{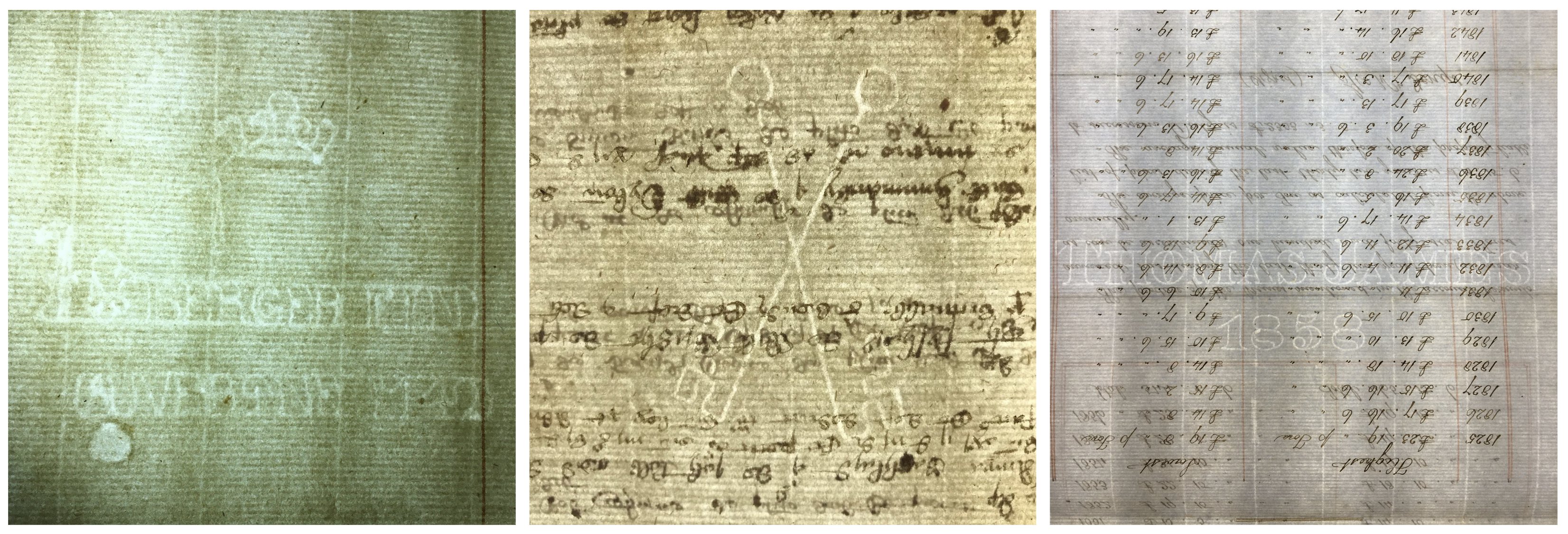}
        \caption{
            Pieces of paper with imprinted watermarks that appear when light is shining from behind because the paper is thinner in the places where a wire was. Now used by historians to date the paper~\citep{endersby2023patterns}.
        }\label{intro/fig:history}
    \end{minipage}
    \hfill
    \begin{minipage}{0.37\textwidth}
        \centering
        \includegraphics[width=0.93\textwidth]{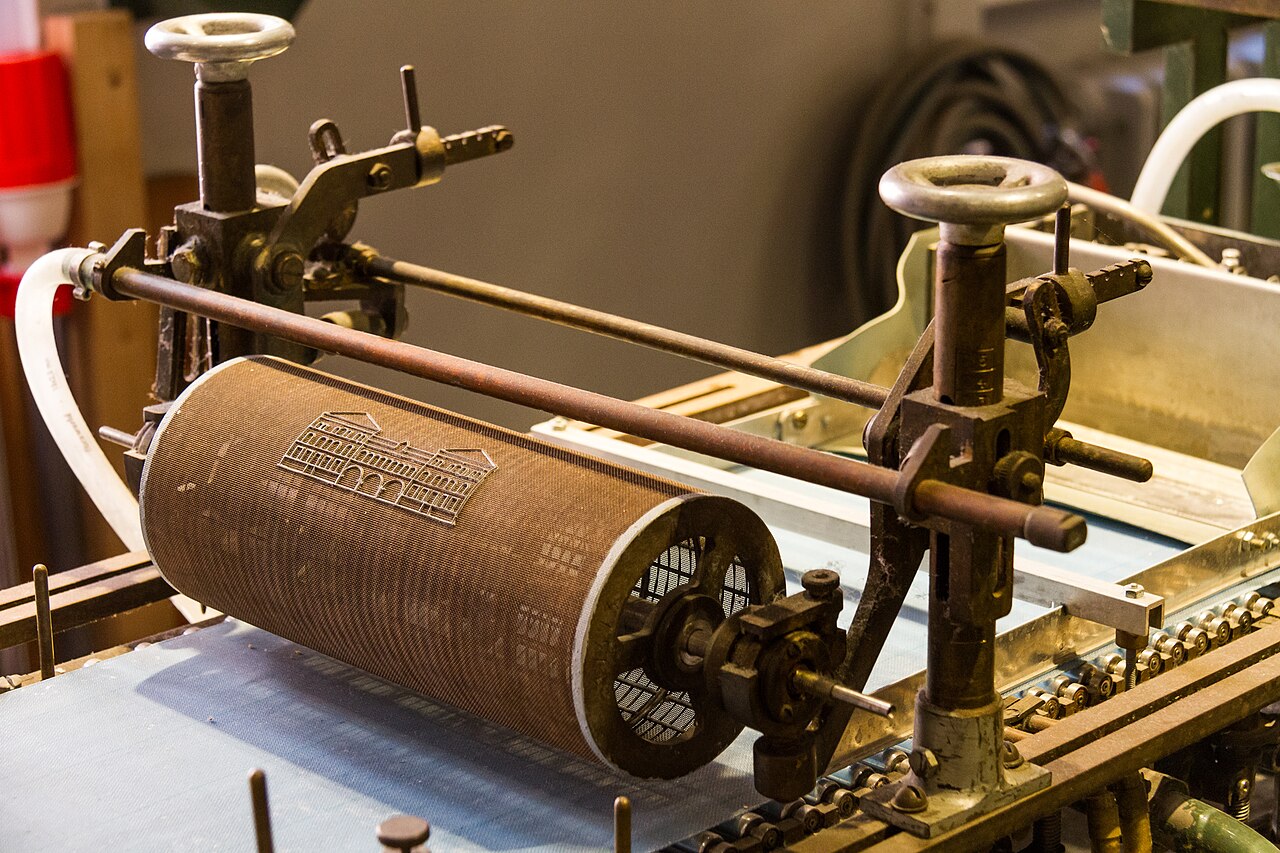}
        \caption{
            A dandy roll machine with a watermark used to create paper. \href{https://commons.wikimedia.org/wiki/File:Dandy_roll_2.jpg}{Wikimedia Commons}.
        }\label{intro/fig:dandy_roll}
    \end{minipage}
\end{figure}

The term ``watermark'' has its roots in the 13\textsuperscript{th} century in Italy, and comes from the papermaking process.
Until the 19\textsuperscript{th} century, paper was produced by shredding linen rags and other fibers and mashing them up in water to create a liquid paste of fiber.
A wire mesh mould was lowered into this paper pulp, creating a thin layer of fibers on the mould.
The mesh would create distinctive marks by altering the thickness of the paper during its production, while it was still damp, hence the term ``water-mark''.
These marks were visible when the paper was held up to the light, as shown in Fig.~\ref{intro/fig:history}.

Papermakers quickly began using watermarks as a form of logo or trademark to distinguish their products from others. 
This was a simple yet effective way to ensure the authenticity of the paper and to protect the reputation of the papermaker.
By the 14\textsuperscript{th} to 16\textsuperscript{th} centuries, watermarking had become a common practice across Europe. 
The watermarks later evolved to include symbols, logos, and dates, providing more specific information about the origin of the paper.
The invention of the dandy roll -- a wire-covered cylinder that removes excess water from the wet paper pulp -- in the early 19\textsuperscript{th} century allowed for more complex and intricate watermarks to be created (see Fig.~\ref{intro/fig:dandy_roll}).

\paragraph*{Birth of digital watermarking.}

\begin{figure}[b!]
    \centering
    \includegraphics[width=1.0\textwidth]{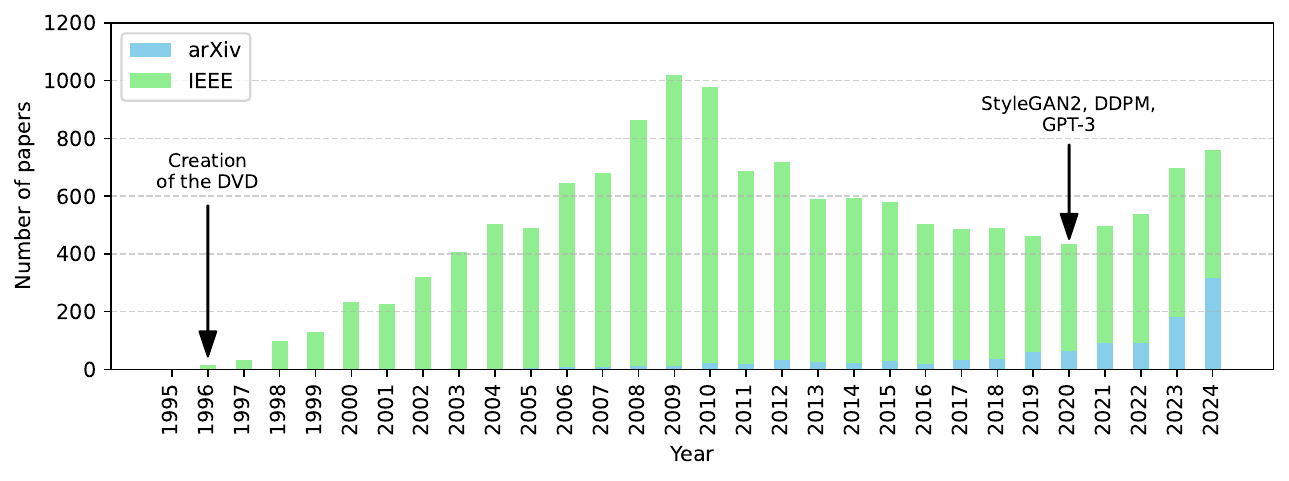}
    \caption{
        Annual number of papers on watermarking, from IEEE Xplore and arXiv (number for 2024 is extrapolated). 
        We can identify the rise of watermarking in the 1990s -- with the popularization of DVDs -- its stabilization and slight decrease in the 2010s, and its resurgence in the 2020s -- which can be attributed to the popularization of generative models, \eg, image models~\citep{karras2020analyzing,ho2020denoising} and LLMs like GPT-3~\citep{brown2020language}.
    }
    \label{intro/fig:num_papers}
\end{figure}

Watermarking evolved from a physical process in papermaking to a digital process as new types of media became available, making the original etymology somewhat obsolete. 
The arrival of digital watermarking in the early 1990s was primarily driven by the need to protect digital media from unauthorized copying and distribution in the rapidly expanding internet. 
This new challenge sparked significant interest in the research community for at least two decades (see Fig.~\ref{intro/fig:num_papers}).

Watermarking techniques were developed for numerous types of content, including audio, images, video, 3D, text, etc.
There is even an intriguing application in DNA sequences~\citep{shimanovsky2002hiding, heider2007dna}, used to protect the intellectual property of genetically modified organisms or to trace infectious agents!
More known applications of digital watermarking are exposed in \textit{Digital Watermarking and Steganograph}y, by~\cite{cox2007digital}.
A notable one is in Digital Rights Management (DRM) systems~\citep{barni2004data}, which protect digital content from unauthorized access and distribution. 
For instance, Hollywood studios use video watermarking for DVD copy protection.
It enables them to trace leaked content back to the specific distributed copy, identifying the person or entity responsible for unauthorized distribution.
Watermarking has become ubiquitous in our everyday lives, even though it often goes unnoticed because its first goal is to be imperceptible.
For example, photos in the news industry are watermarked to identify the source photo agency, and audio and video content in Video On Demand portals are watermarked to combat piracy. 
Other applications include audio or video watermarks that are used for TV broadcasts monitoring~\citep{depovere1999viva}, or in movie theaters to prevent illegal recordings and retrace the specific theater where the recording was made~\citep{nakashima2009watermarked} -- and even the seat and time of the recording!
The decline in watermarking research around the 2010s can be attributed to several factors. 
As the field matured, focus shifted towards industry applications and evolved alongside media distribution methods, particularly with the rise of streaming services. 
Additionally, funding priorities and research interests shifted due to technological advancements in areas like cloud computing and AI.

\paragraph*{Artificial Intelligence.}

Watermarking has started to become popular again since the start of the decade.
This is driven by the renewed interest in deep learning~\citep{lecun2015deep}, or more broadly, Artificial Intelligence (AI).
One possible reason is that researchers are exploring deep learning based methods to improve over traditional ones, with better accessibility, adaptability and robustness. 
However, this alone cannot fully explain the increasing interest in the field.
The factors are more likely (1) the development and popularization of image generative models such as StyleGAN2~\citep{karras2020analyzing}, DDPM~\citep{ho2020denoising} and of Large Language Models (LLM) like GPT-3~\citep{brown2020language} and (2) the productization of deep learning models and the need to protect these million-dollars assets.
This has contributed to bringing new funding and research opportunities to the watermarking community, which is now more attractive than ever.

At the moment of writing this manuscript, watermarking is therefore in the middle of the GenAI revolution given the role it could play for \emph{content authenticity}.
AI-generated content is used for swaying public opinion~\citep{shen2019fake, goldstein2023generative}, fraud, or impersonation at a higher scale and more convincingly than even authentic content~\citep{spitale2023ai}.
Governments are getting hold of the issue through new regulations that impose watermarking as a technical means for transparency and traceability~\citep{USAIAnnouncement, ChineseAIGovernance, EuropeanAIAct, ca_ab3211_2024}.
Several key players, like
\href{https://deepmind.google/discover/blog/watermarking-ai-generated-text-and-video-with-synthid/}{Google}, 
\href{https://about.fb.com/news/2024/04/metas-approach-to-labeling-ai-generated-content-and-manipulated-media/}{Meta},
and
\href{https://openai.com/index/understanding-the-source-of-what-we-see-and-hear-online/}{OpenAI},
are now paving the way for the re-birth of watermarking.

\section{New applications in the era of massive data}

The beginning of the decade, from a digital point of view, is characterized by the rapid growth of the internet (social media, search engines, cloud) and of hardware (processors, mobile devices, compute clusters).
This has led to a significant increase in the amount of data being generated and shared online, as well as new opportunities for using this data to training Machine Learning (ML) models.
In this section, we will explore new applications of watermarking that have emerged in this massive data era.

\subsection*{Content moderation}

\begin{quote}
    \textit{``The internet is the first thing that humanity has built that humanity doesn't understand, the largest experiment in anarchy that we have ever had.''}
    \begin{flushright} -- Eric Schmidt, CEO of Google, 2010 \end{flushright}
\end{quote}

Mankind now relies on internet platforms for a variety of needs: getting news, information from close circles, learning new materials, communicating with professional communities, etc.
In 2020 alone, 3.2 billion images and 720,000 hours of video were shared online daily\footnote{\scriptsize\url{https://www.qut.edu.au/insights/business/3.2-billion-images-and-720000-hours-of-video-are-shared-online-daily.-can-you-sort-real-from-fake}}, often copied and altered as they spread among users.
This explosion of digitization and data sharing presents new significant challenges: complying with intellectual property laws and moderating social platforms have all become untractable. 
Unlike TV or radio, where content is produced by a few controlled sources, everyone can create and share content online. 

The internet, much like other shared spaces, arguably needs governance to ensure it remains safe for users and content creators. 
Content moderation plays a crucial role in this governance by applying a set of rules to user-generated content, preventing the spread of harmful material (hate speech, misinformation, etc.), and ensuring it complies with ethical standards.
This moderation is required by regulations like the European Union's Digital Services Act (DSA) and the United States' Section 230 of the Communications Decency Act (CDA), which hold platforms accountable for the content they host.

Content moderation has evolved from manual verification to ML-based automated systems given the volume of images shared daily.
A key component of these systems is copy detection, which automatically identifies near-duplicate content. 
For instance, during the COVID-19 pandemic, Meta platforms used it to detect misinformation and exploitative content~\citep{sumbaly2020using, pizzi2022self}. 
By identifying and flagging near-exact copies of known misinformation, it allowed to flag millions of pieces of content and enabled human fact-checkers to focus on catching new instances of misinformation rather than near-identical variations of content they had already seen.
Similar approaches, called  NeuralHash~\citep{apple2021csamdetection} and PhotoDNA~\citep{photodna} are used to detect child sexual abuse material respectively in iCloud and on the internet, as well as other harmful content like violent terrorist imagery for the latter.

In this context, where content is transmitted on platforms, watermarking may actively trace its origin by embedding a unique identifier -- specific markers that can be used to verify the source -- in each piece of content as soon as it enters the platform.
It has the potential to increase accuracy and catch instances of misinformation or harmful content that may not be flagged by existing copy detection systems, thereby providing a more robust solution for content protection compared to traditional copy detection methods.

\subsection*{Tracing AI-generated content}

\begin{figure}[b!]
    \centering
    \begin{minipage}[b]{0.25\textwidth}
        \includegraphics[width=\textwidth]{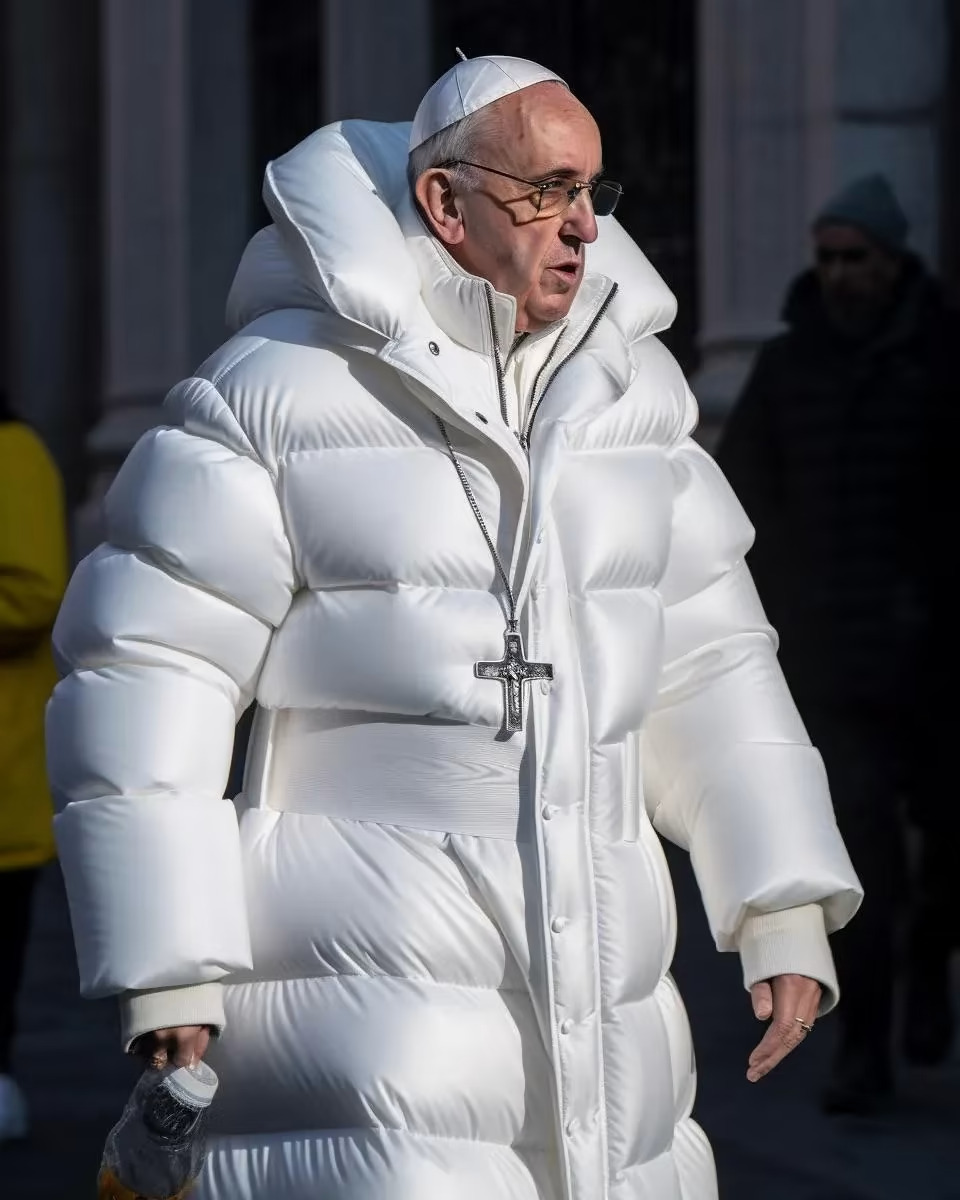}
    \end{minipage}
    \hfill
    \begin{minipage}[b]{0.72\textwidth}
        \centering
        \includegraphics[width=0.32\textwidth]{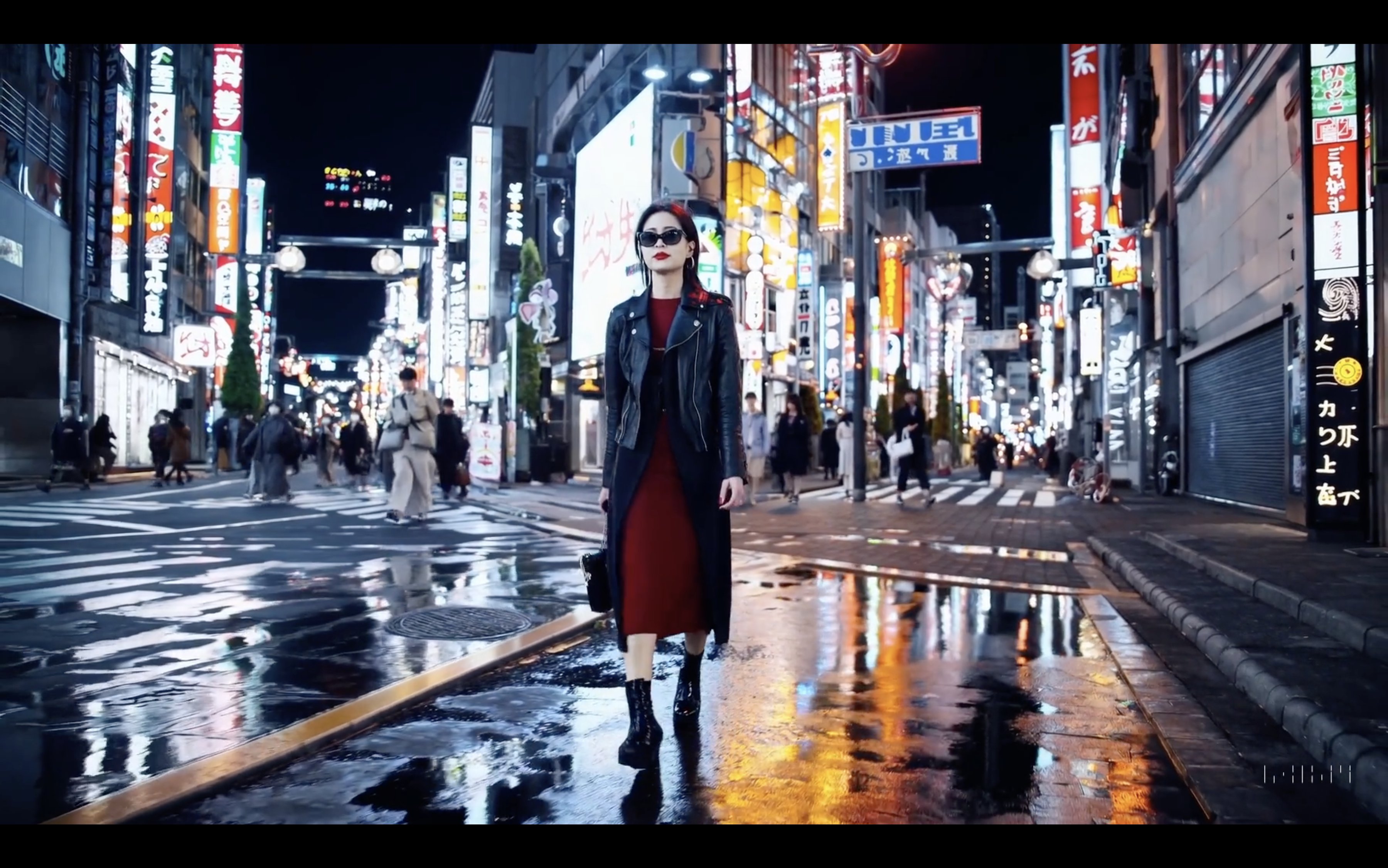}\hfill
        \includegraphics[width=0.32\textwidth]{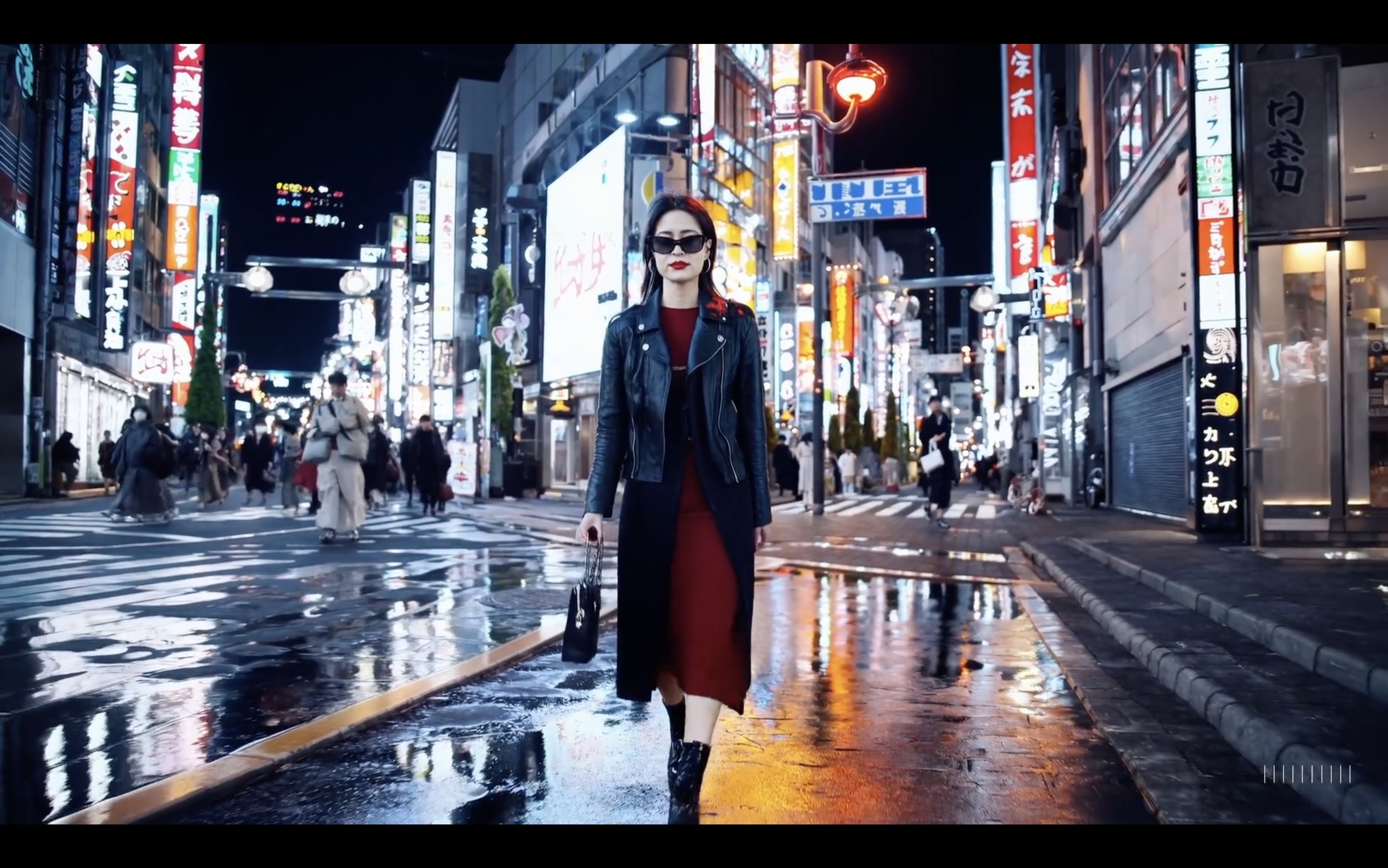}\hfill
        \includegraphics[width=0.32\textwidth]{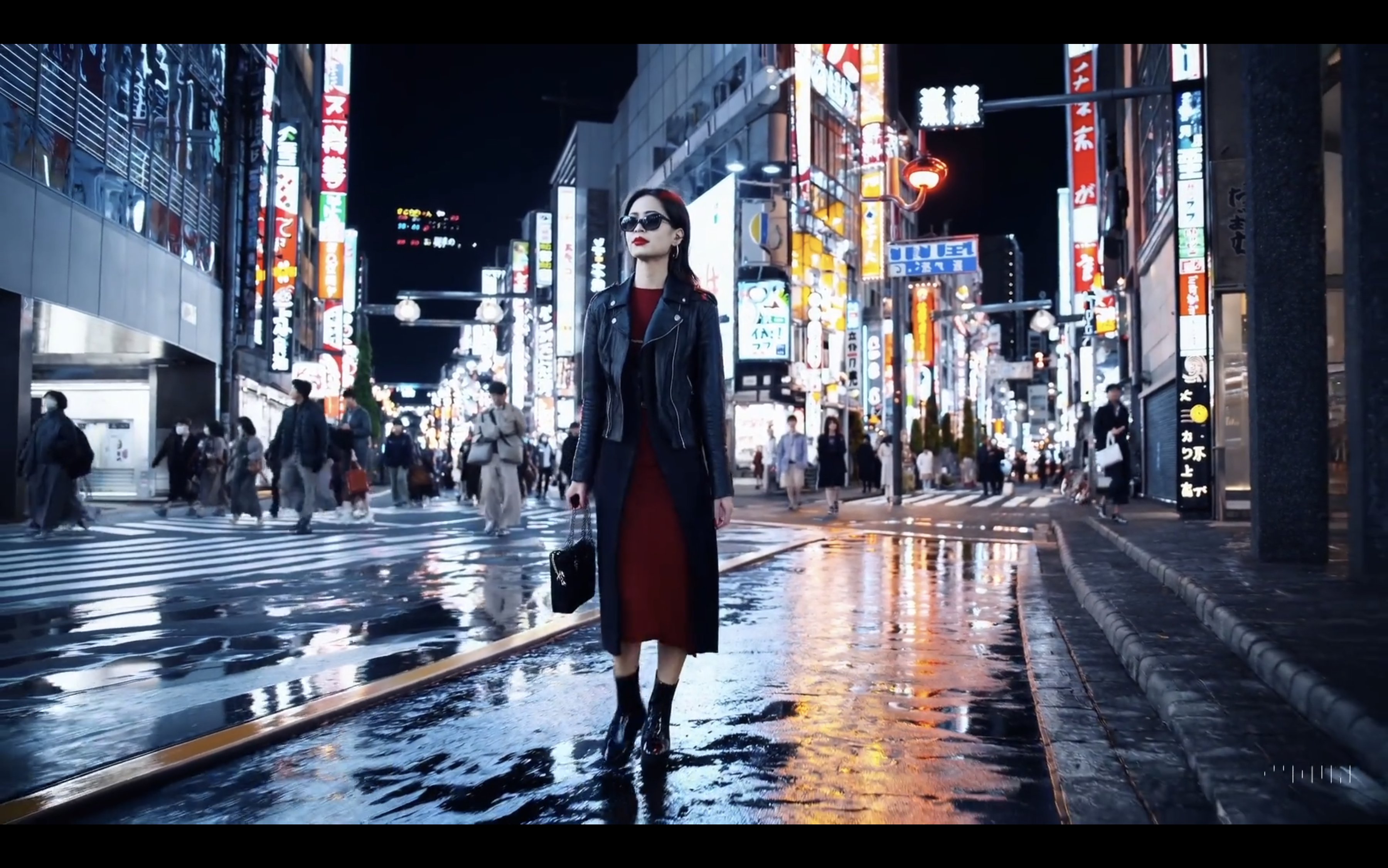} \\[10pt]
        \includegraphics[width=0.32\textwidth]{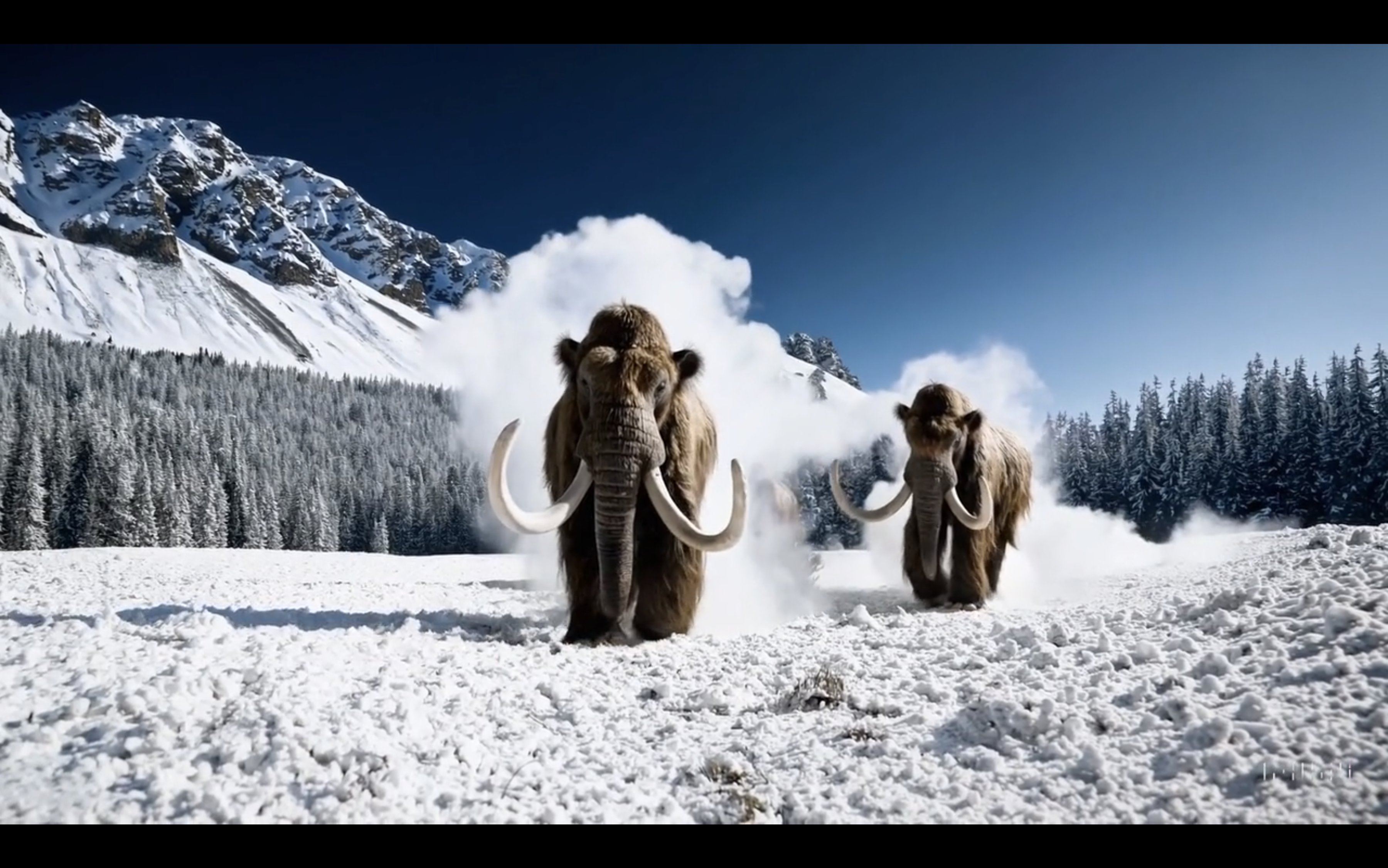}\hfill
        \includegraphics[width=0.32\textwidth]{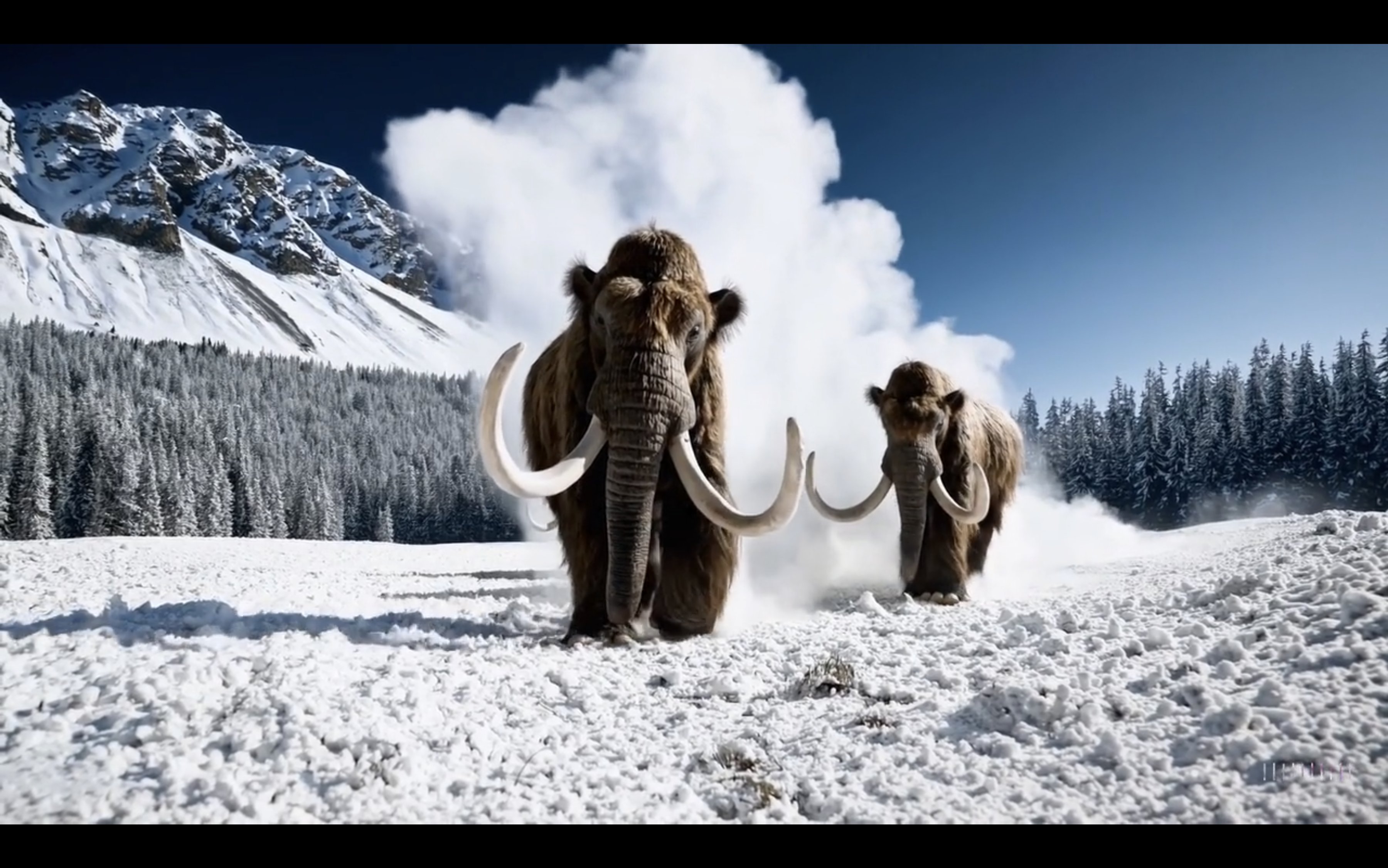}\hfill
        \includegraphics[width=0.32\textwidth]{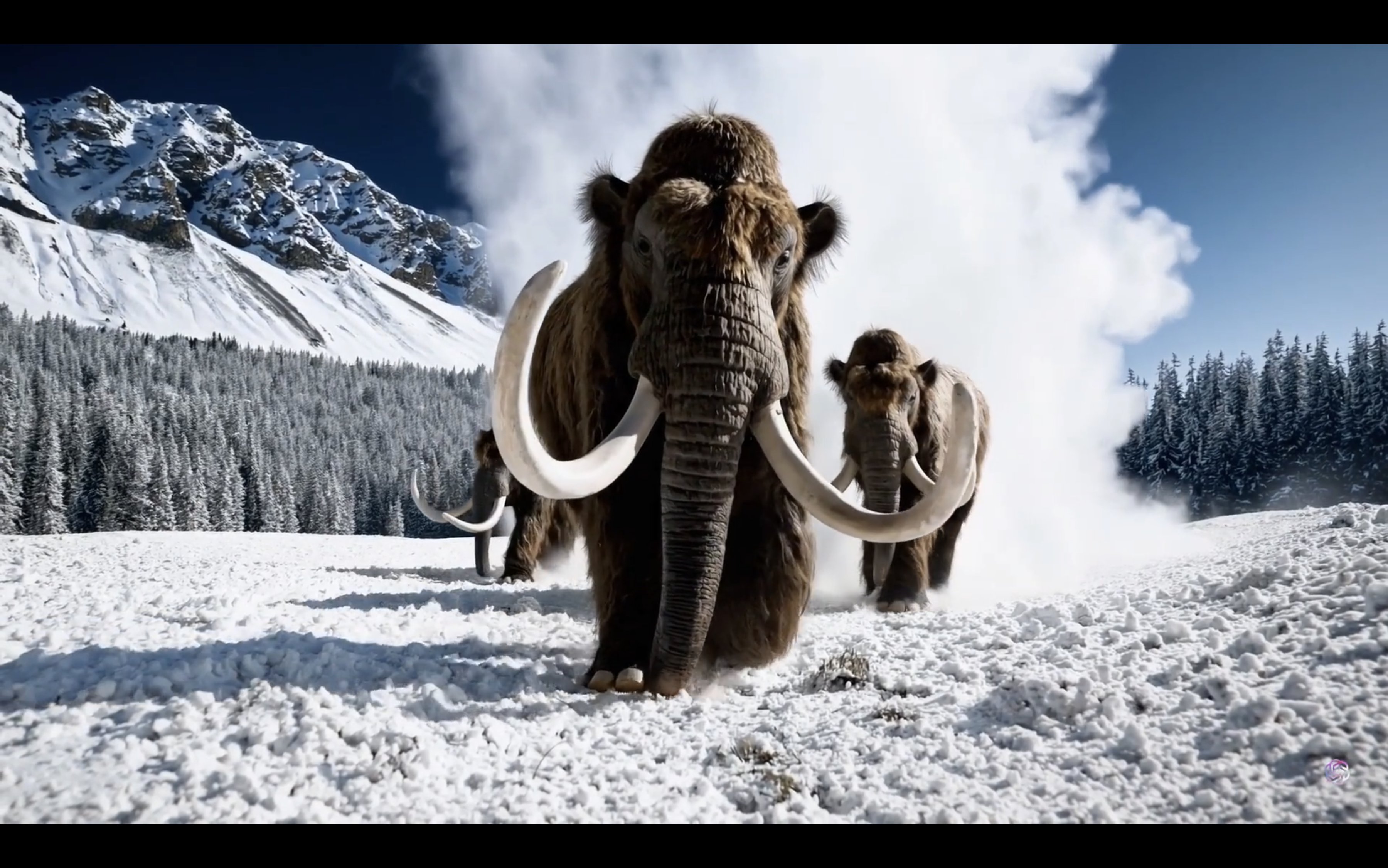}
    \end{minipage}
    \caption{
        (Left) A viral image of Pope Francis generated from text by Midjourney, which became famous as deepfake example.
        (Right) Frames from video clips generated by OpenAI's Sora.
        The fact that AI-generated content is visually indistinguishable from human-generated content highlights the need for more robust detection schemes like watermarking.
    }\label{intro/fig:pope}
\end{figure}

The above quote by Eric Schmidt is even more relevant today as the internet has become the playground of generative AI models.
These models are developed and adopted at an unprecedented pace.
In text generation, \href{https://chatgpt.com/}{ChatGPT} reached 100 million users in just two months, and has had a significant impact on the way people interact with AI since then.
It has counterparts in image generation with DALL$\cdot$E~\citep{ramesh2021zero} and \href{https://www.midjourney.com/}{Midjourney}; video with \href{https://openai.com/index/sora/}{Sora} and \href{https://kling.kuaishou.com/en}{Kling}; and audio with \href{https://suno.com/}{Suno} and \href{https://www.udio.com/}{Udio}.
Closed models are generally followed by open-weights alternative of the same quality in less than a year (\eg, Llama or Stable Diffusion).
Furthermore, they create content that is often indistinguishable from human-generated content (see in Fig.~\ref{intro/fig:pope}).
A study by \cite{nightingale2022ai} notably found that human participants exhibited a relatively low accuracy rate in distinguishing between authentic and artificially generated facial images, with an average accuracy of 48.2\% (so AI-generated faces were deemed most trustworthy).
The same finding applies to text~\citep{spitale2023ai}.

Generative models are used to create content at scale and are beginning to replace traditional content creation methods. 
By late 2023, they had already produced as many images as photographers had taken in 150 years of photography\footnote{\scriptsize\url{https://journal.everypixel.com/ai-image-statistics}}.
It raises concerns from Hollywood and the creative industry~\citep{coyle2023chatgpt} and from governments given the risks for misinformation, fraud, and impersonation.
It would be too long to cite all malicious uses of generative AI: they range from scam books sold on Amazon~\citep{knibbs2024scammy} to influence campaigns~\citep{goldstein2023generative} or the creation of deepfakes and impersonation of public figures~\citep{harris2018deepfakes, shen2019fake}.
Ironically, GPT-4o made it possible to build dynamic AI-generated websites self-sufficient by ads\footnote{\scriptsize\url{https://batchmon.com/blog/ai-cheaper-than-ads/}}.

These risks linked to AI-generated content are heightened because it is difficult to detect and attribute such content to models that generated them.
This makes it hard to hold anyone accountable and does not encourage companies to implement effective safety measures. 
Watermarking is an ideal technique for tracing content from generative models since providers have control over the model's outputs.
In this context, it is put forward by most of the regulations on AI~\citep{USAIAnnouncement, ChineseAIGovernance, EuropeanAIAct} to allow for a better transparency.
The idea is to watermark the content during or after the generation to embed a proof that it is AI-generated, or an identifier of the specific model that generated it.
Notably, traditional watermarking methods could also be employed for post-hoc watermarking of AI-generated content. However, as we will discuss later, it presents specific challenges that we address in this thesis.
For instance, in open-weights scenarios, where copy detection or post-hoc watermarking would be impossible, the model provider could even watermark the generative model prior to distribution.

\subsection*{Monitoring AI models}

Tracing AI-generated outputs serves not only regulatory purposes but more broadly economical ones with intellectual property protection and monitoring of the model usage.
Data is said to be the ``fuel of AI,''\footnote{\scriptsize Ilya Sutskever, at NeurIPS 2024.} but in contrast with motors that consume the fuel, ML models memorize and extract value from it.
Models and their training data have therefore become extremely valuable, often representing significant investments. 
How can the data being used to train models be traced and the data providers be properly credited?
How can the models themselves be traced and their use be ensured according to the licensing agreements?
The complexity and lack of transparency in AI models make it challenging to answer these questions, even more when these models can be fine-tuned, distilled or used as black-boxes in other APIs.

Watermarking may help at several levels:
\begin{itemize}
    \item \emph{Watermarking models} allows organizations to trace the origin of models and detect leaks or unauthorized use, protecting intellectual property rights. 
    It also prevents the spread of potentially harmful or insecure AI applications.
    \item \emph{Watermarking training data} enables providers to detect and track the usage of their data, to ensure that it is being used in compliance with licensing agreements.
    It should offer a proof that the said data was used and in which proportion, so that it can be attributed to its owner.
\end{itemize}

\section{Outline and contributions}

This thesis introduces watermarking techniques for each of the aforementioned applications.
\autoref{chapter:related-work} first presents related works.
\autoref{chapter:technical-background} then presents the detailed technical background on watermarking, necessary to understand the following chapters.
The contributions are structured around their applications:

\noindent
\fullref{part:content-moderation} 

\newcommand{\basedon}{$>$ Based on: }
\newcommand{\codeat}{$>$ Code at: }
\begin{itemize}
    \item \fullref{chapter:ssl-watermarking}.
    This chapter explores how to leverage the intrinsic robustness of self-supervised neural networks, often used for copy-detection, for image watermarking instead.
    It does so by hiding information in their latent representations through an iterative image optimization.
    It enables watermarking images with varying resolutions, adjustable payload, and a customizable tradeoff between robustness and quality.
    \\ \basedon
    \textit{Fernandez et al., Watermarking images in self-supervised latent spaces, ICASSP 2022}.
    \\ \codeat
    \url{github.com/facebookresearch/ssl_watermarking}
    \item \fullref{chapter:active-indexing}.
    This chapter introduces a technique that combines copy detection and watermarking.
    The goal is to greatly improve the robustness of copy detection systems by imperceptibly modifying images before their release, with similar optimization schemes as in the previous chapter.
    \\ \basedon
    \textit{Fernandez et al., Active image indexing, ICLR 2023}.
    \\ \codeat
    \url{github.com/facebookresearch/active_indexing}
\end{itemize}

\noindent
\fullref{part:genai-tracing} 

\begin{itemize}
    \item \fullref{chapter:stable-signature}.
        This chapter presents a technique that fine-tunes latent generative models such that all images they produce hide an invisible signature. 
        These signatures can be used to detect and track the origin of images generated by latent diffusion models.
        Stable Signature offers protection in open-weights scenarios, contrary to methods that would apply the watermark post-generation, like the ones of the previous part.
        \\ \basedon
        \textit{Fernandez et al., The stable signature: Rooting watermarks in latent diffusion models, ICCV 2023}.
        \\ \codeat
        \url{github.com/facebookresearch/stable_signature}
    \item \fullref{chapter:audioseal}.
        This chapter introduces a watermarking solution for the detection of AI-generated speech.
        We proactively watermark the speech signal, which can later be detected to verify the authenticity of the content. 
        Contrary to traditional methods, the watermark detection is localized and predicts for each time step (1/16k of a second) if the watermark is present or not, which makes it fast and able to pinpoint small segments of watermarked audio in longer ones.
        \\ \basedon
        \textit{San Roman et al., Proactive detection of voice cloning with localized watermarking, ICML 2024}.
        \\ \codeat
        \url{github.com/facebookresearch/audioseal}
    \item \fullref{chapter:three-bricks}.
        This chapter brings three improvements to state-of-the-art watermarking methods for large language models (LLM):
        (1) theoretically grounded and empirically validated statistical tests that guarantee false positive rates,
        (2) evaluation on classical LLM benchmarks,
        (3) extension to scalable multi-bit watermarking.
        \\ \basedon
        \textit{Fernandez et al., Three bricks to consolidate watermarks for large language models, WIFS 2023}.
        \\ \codeat
        \url{github.com/facebookresearch/three_bricks}
\end{itemize}

\noindent
\fullref{part:model-monitoring} 

\begin{itemize}
    \item \fullref{chapter:radioactive}.
        This chapter examines whether a watermark can be detected in a language model fine-tuned on watermarked text. 
        We explore scenarios based on model and data access, showing that the watermark can in some cases be identified with high significance ($p<10^{-6}$) even when only 5\% of the fine-tuning data is watermarked. 
        Thus, LLM watermarking, originally designed to detect AI-generated text, can also indicate if an LLM's outputs were used to fine-tune another model. 
        This is a pressing question for model providers concerned about the theft or misuse of their models.
        \\ \basedon
        \textit{Sander et al., Watermarking makes language models radioactive, NeurIPS 2024}.
        \\ \codeat
        \url{github.com/facebookresearch/radioactive-watermark}
    \item \fullref{chapter:invariants}.
        This chapter introduces a training-free watermark for the weights of large transformers.
        It leverages the model's invariance through operations like dimension permutations or scaling/unscaling to generate different functionally equivalent copies with different weights. 
        The use case is to distribute different weights of the exact same model to different clients, and to detect leaks or unauthorized use.
        \\ \basedon
        \textit{Fernandez et al., Functional invariants to watermark large transformers, ICASSP 2024}.
\end{itemize}

\autoref{chapter:conclusion} ends the thesis with a summary of the contributions, a discussion on the perspectives and open questions for watermarking.

%% file: 0-introduction/related-all.tex
\chapter{Related Work}
\label{chapter:related-work}

Here we provide an overview of the literature on generative modeling, content provenance, and watermarking, which intersect in many chapters of the thesis.
We provide additional related work in each chapter where it is relevant.

\section{Generative models}

Generative models are models that learn to create data samples from a given distribution.
They have benefited from the increase in computational power, the amount of data shared on the internet, and the development of algorithms able to handle these large datasets.
This section provides an overview of the most popular models in different domains, which are particularly relevant to \autoref{part:genai-tracing} which focuses on tracing AI-generated content and to \autoref{part:model-monitoring} which focuses on how to trace the usage of these models.

\paragraph{Text generation.}
Probabilistic language models are statistical models that learn the probability distribution of sequences of tokens -- \glspl*{token} being words, subwords or characters, created by tokenization algorithms that learn the optimal way to split a text into units~\citep{sennrich2015neural, kudo2018subword, dagan2024getting}.
This was originally done with $n$-grams models that count the frequency of each word given the $n-1$ previous tokens.
Given the exponential growth of the number of possible sequences with the sequence length, $n$-grams models are limited to short sequences, which explains why they have been replaced by neural models~\citep{bengio2000neural}.
RNN~\citep{mikolov2010recurrent, sutskever2011generating, jozefowicz2016exploring} and LSTM~\citep{hochreiter1997long} models were the first to generate a significant amount of interesting and high-level linguistic structures.

They were quickly outperformed by the Transformer~\citep{vaswani2017attention}.
It is a sequence-to-sequence architecture based on the self-attention mechanism, which increases the expressiveness of the model by allowing it to build the new token taking into account all the context and previously generated tokens.
Most importantly, it enables the parallelization and scaling of the training process, which is important for text given the vast amount of data available on the internet.
This is central for generative pre-trained transformers (GPT 1, 2 and 3)~\citep{Radford2018ImprovingLU, radford2019language,brown2020language} that demonstrate impressive generation quality, simply by training on the next-token prediction task on a large corpus of text.
Given some examples written in plain text of tasks they have not been trained on (few-shot demonstrations), they are able to perform these tasks competitively with models fine-tuned specifically for them (summarization, translation, etc.)~\citep{brown2020language, rae2021scaling, chowdhery2023palm}; and even better with effective prompting strategies~\citep{wei2022chain}.
What is more, this performance scales with the number of parameters~\citep{kaplan2020scaling}, and with the size of the training data~\citep{hoffmann2022training}, hence the name of Large Language Models (\Glspl*{LLM}).
Building on this, studies have found that fine-tuning on instructions significantly improves their few-shot abilities on unseen tasks~\citep{wei2021finetuned}. 
Combining this approach with Reinforcement Learning with Human Feedback (RLHF) yields even better results, enabling the resulting model, InstructGPT, to follow the human's intent more closely~\citep{ouyang2022training}.

This has led to a proliferation of generative pre-trained models (and companies built around them).
The key differentiators are: access to greater computational resources, larger and higher-quality datasets, and more advanced training algorithms (sometimes referred to as ``compute multipliers'').
There is, among many other examples, ChatGPT~\citep{openai2024chatgpt}, the first model to catch the public's attention, and its successor GPT-4~\citep{openai2023gpt}, the contender of Google DeepMind Gemini~\citep{team2023gemini} or Llama 1, 2 and 3~\citep{touvron2023llama,touvron2023llama2,dubey2024llama}, Mistral~\citep{jiang2023mistral7b}, Gemma~\citep{team2024gemma}, which are popular because openly available.

LLMs are rapidly evolving.
They integrate multi-modal capabilities to understand and generate images~\citep{alayrac2022flamingo, laurenccon2024obelics, liu2024visual, team2024chameleon}, speech~\citep{rubenstein2023audiopalm, nguyen2024spirit}, or both~\citep{lu2022unified, openai2024gpt4o}. 
The use of Mixture of Experts (MoE)~\citep{shazeer2017outrageously, jiang2024mixtral}, increases the parameter count and the expressiveness of the model, while keeping a similar inference cost. 
Models with multiple decoding heads~\citep{cai2024medusa, gloeckle2024better} increase the inference speed and performance on reasoning tasks.
State-space models aim to address the computational inefficiency on long sequences~\citep{de2024griffin, gu2023mamba}.
Models that integrate mathematical reasoning~\citep{frieder2024mathematical} or agents~\citep{park2023generative, dubois2024alpacafarm} are able to perform more complex tasks.
At the end of the day, text generation is about much more than just producing text; it involves understanding, reasoning about the world, and interacting with it, since most of our interactions are conducted through language.

\paragraph{Image generation.}
Generating images has long been dominated by Generative Adversarial Networks (\Glspl*{GAN})~\citep{goodfellow2014generative, zhu2017unpaired, karras2020training, karras2019style, karras2020analyzing, sauer2022stylegan, walton2022stylenat}, which train a generator and a discriminator to produce realistic images.
The generator learns to create images from random noise that fool the discriminator, while the latter simultaneously learns to distinguish between real and generated images.

\Glspl*{diffusion model} are a different class of models that learn to generate images by iteratively denoising a noised vector, which is a more stable training process than GANs~\citep{dhariwal2021diffusion, ho2020denoising, nichol2021improved, song2020denoising}.
They are a huge improvement in text-conditional image generation and the reason for the rise in popularity of text-to-image models.
Resulting models synthesize high-resolution photo-realistic images for a wide variety of text prompts~\citep{balaji2022ediffi, ho2022imagenvideo, ramesh2022hierarchical, saharia2022photorealistic}. 
Notably, Latent Diffusion Models (\Glspl*{LDM})~\citep{rombach2022high} are a class of diffusion models that operate in the latent space of a variational autoencoder, and Stable Diffusion is a large instance of LDM trained on the LAION dataset~\citep{schuhmann2021laion, schuhmann2022laion}.
This is the main class of model that we experiment with in Chap.~\ref{chapter:stable-signature}.
These models can also perform conditional image generation -- like inpainting or text-guided image editing -- by fine-tuning the diffusion model with additional conditioning, \eg, masked input image, segmentation map, etc.~\citep{lugmayr2022repaint, saharia2022palette}. 
Because of their iterative denoising algorithm, diffusion models can also be adapted for image editing in a zero-shot fashion by guiding the generative process~\citep{couairon2022diffedit, hertz2022prompt, kawar2022imagic, mokady2022null, valevski2022unitune,  wu2022unifying}.
These methods are also behind the video generation models \href{https://openai.com/index/sora/}{Sora} and \href{https://kling.kuaishou.com/en}{Kling}.

Another class of models treats the generation process as a sequence-to-sequence problem and leverages the advances in language models.
They are based on image tokenizers (\eg, VQVAE~\citep{van2017neural}, VQGAN~\citep{esser2021taming}, ViT-VQGAN~\citep{yu2021vector}), which are models that convert images into discrete tokens, and vice versa, and on transformers that can generate sequences of tokens.
DALL$\cdot$E~\citep{ramesh2021zero} is one of the pioneering models to generate images in this way.
Other known models include Parti~\citep{yu2022scaling}, Chameleon~\citep{team2024chameleon} which is able to generate both text from images and images from text, and many others~\citep{chen2020generative, ding2021cogview, esser2021taming, gafni2022make, singer2022makeavideo}.

\paragraph{Audio generation.}
Early models are autoregressive like WaveNet~\citep{wavnet, wang2017tacotron, shen2018natural}, and use waveform or Mel spectrogram reconstruction as objective. 
None of these objectives are deemed ideal for audio quality, leading to the adoption of adversarial models~\citep{hifigan, melgan}, and then to diffusion or auto-regressive models similar to those used in image generation.

The most popular task in audio generation is perhaps speech generation.
One key example is zero-shot text-to-speech (\Gls*{TTS}) models, \aka, voice cloning, which tries to imitate or preserve vocal style using only a small amount of data. 
VALL-E~\citep{wang2023neural} or SPEAR-TTS~\citep{kharitonov2023speak} follows the token-based autoregressive language modeling approach, which converts text or audio into tokens~\citep{defossez2022high,soundstream}, and then generates speech from these tokens, with optional conditioning (speech, text, language, etc.). 
In the context of speech machine translation, SeamlessExpressive~\citep{seamless2023} follows a similar approach to translate speech, but also retain the speaker's unique vocal style and emotional inflections.
These models are able synthesize high-quality personalized speech with only a 3-second recording. 
Other zero-shot TTS models, like NaturalSpeech2~\citep{shen_naturalspeech2}, Voicebox~\citep{le2023voicebox}, A$^{3}$T~\citep{BaiZCML022_A3T} and Audiobox~\citep{hsu2023audiobox} are non-autoregressive and based on diffusion~\citep{ho2020denoising} or flow-matching~\citep{lipman2022flow}.
They iteratively denoise the audio signal to generate speech, which is represented as spectrograms or learned latents (from HiFi-GAN or EnCodec).
They are particularly suited to perform tasks such as text-guided speech infilling, where the goal is to generate masked speech given its surrounding audio and text transcript. 
It makes them a powerful tool for speech manipulation.

Generating environmental sounds and music is another popular task, which follows the same principles.
For instance, Mo\^usai~\citep{schneider2023mo}, Make-an-Audio~\citep{huang2023make}, AudioLDM~\citep{liu2023audioldm} and AudioLDM~2~\citep{liu2024audioldm}, etc. are latent diffusion models that generate audio in the latent space of a perceptual autoencoder.
On the other hand, models like AudioGen~\citep{kreuk2022audiogen}, AudioLM~\citep{agostinelli2023musiclm} or MusicGen are autoregressive and generate audio samples as a sequence of tokens.
All these models are now able to create high-quality audio samples in different styles, from text, other musics, or even from images~\citep{girdhar2023imagebind}, and to perform inpainting or editing in a zero-shot fashion.
Note that speech and sound generation are getting unified, with models like AudioLDM 2 or Audiobox that can jointly generate both speech and music, and that start to appear commercially, \eg, in \href{https://www.suno.ai/}{Suno}.

\section{Content tracing and watermarking}

Tracing the origin of digital content is a problem that is traditionally approached passively, by copy detection or digital forensics.
On the other hand, this thesis focuses on active tracing, by embedding invisible watermarks in the content, as presented in Parts \ref{part:genai-tracing} and \ref{part:model-monitoring}.
This section provides an overview of the literature on these two approaches, and across image, audio and text modalities, with a particular focus on AI-generated content.

\subsection{Fingerprinting and copy detection}

\Gls*{fingerprinting} involves creating a unique identifier for a piece of digital content, called \gls*{hash} or \emph{fingerprint} in reference to the uniqueness of human fingerprints.
This fingerprint can then be used to identify the content even if it has been modified or compressed, but does not allow to reconstruct the original content.
\Gls*{copy detection}, on the other hand, involves comparing two pieces of digital content to determine if they are identical or similar. 
This is often done using \emph{\gls*{indexing}} algorithms that store the fingerprints of all the content in a database, and then compare the hash of the queried content to the hashes in the database to determine if it is a copy.

These hashes are vector representations that can be binary or real-valued ($\in \{0,1\}^k $ or $\mathbb{R}^k$).
They were traditionally hand-crafted with color histograms, GIST descriptors, constellation maps, etc.~\citep{swain1991color, oliva2001modeling, wang2003industrial, perronnin2010large}, but are now usually generated from self-supervised feature extractors~\citep{chen2020simclr,oquab2023dinov2, devlin2018bert, hsu2021hubert, raffel2020exploring}.
The feature extractors are not perfectly robust to content modifications.
In other words, the hashes are not perfectly invariant to transformations, \eg, an audio and its $\times$1.25 speed-up version may have different ones.
Besides, storing the hashes is cumbersome and reverse search, \ie, finding the content that has a given hash, must be approximate to be tractable at scale.
Therefore, the hashes are often stored using methods like locality-sensitive hashing (LSH)~\citep{charikar2002similarity, datar2004locality} or product quantization (PQ)~\citep{jegou2010pq}.
These indexing structures have a dual role of compressing the hashes and enabling fast approximate search.
See \Gls*{FAISS}~\citep{douze2024faiss} for a review and efficient implementations of these algorithms.
The two above factors result in errors especially in an adversarial setting~\citep{douze20212021, papakipos2022results, wang2022benchmark}.
\autoref{chapter:active-indexing} aims to reduce these errors by actively modifying images before their release, in a similar way to watermarking.
Another downside is the need of storing the hashes in a database, which makes it harder to share and impossible for open content moderation systems.

\subsection{Digital forensics}

More specific to the context of AI-generated content, \gls*{digital forensics} methods aim to detect if a piece of content has been generated or altered by an AI model.
Most methods spot imperceptible hidden traces of generated content, such as variation in words probabilities~\citep{mitchell2023detectgpt}, odd frequencies in images~\citep{corvi2023detection} or voice synthesizer artifacts~\citep{le2023voicebox}.
Relying on these traces makes the detectors very brittle to shifts in the distribution of content, and makes them fall short in effectiveness compared to watermarking techniques~\citep{sadasivan2023can, saberi2023robustness}.
As a key example, one state-of-the-art detection method~\citep{wang2023dire} is fooled to random chance simply by compressing generated images with JPEG~\citep{grommelt2024fake}, because all natural images in their training dataset were in the JPEG format.
Besides, these detectors are likely to get worse as generative models get better and as their artifacts disappear. 

\paragraph{Image.}
Detection of synthetic/manipulated images has a long history~\citep{farid2009image, barni2023information}.
It is now very active in the context of deep-fake detection~\citep{guarnera2020deepfake, zhao2021multi}. 
Many works focus on the detection of \Gls*{GAN}-generated images~\citep{chai2020makes, gragnaniello2021gan, wang2020cnn, zhang2019detecting}.
One approach is to detect inconsistencies in generated images via lights~\citep{farid2022lighting}, perspective~\citep{farid2022perspective, sarkar2024shadows}, physical objects~\citep{ma2022totems} or faces~\citep{li2018exposing, wang2019detecting, bohavcek2023geometric}.
These approaches are restricted to photo-realistic images or faces, artworks not intended to be physically correct are not covered.
Other approaches track traces left by the generators in the spatial~\citep{marra2019gans, yu2019attributing} or frequency~\citep{frank2020leveraging, zhang2019detecting} domains.
There are extensions to \glspl*{diffusion model} in recent works~\citep{corvi2022detection, sha2022fake, epstein2023online} that show encouraging results.

\paragraph{Speech.}
Detection of synthetic speech is traditionally done in the forensics community by building features and exploiting statistical differences between fake and real.
These features can be hand-crafted from the analysis of waveforms, spectrograms or formants~\citep{sahidullah2015comparison, janicki2015spoofing, albadawy2019detecting, borrelli2021synthetic, cuccovillo2024audio} and/or learned~\citep{muller2022does, barrington2023single}.
The approach of most audio generation papers~\citep{Borsos2022AudioLMAL, Kharitonov2023SpeakRA, borsos2023soundstorm, le2023voicebox} is to train end-to-end deep learning classifiers on what their models generate, similarly as \citet{zhang2017investigation, tak2021end, tak2021end2, jung2022aasist}.
These networks primarily focus on non-vocal spectrogram regions~\citep{salvi2023towards, salvi2024listening}, which explains why they are sensitive to the addition or removal of audio artifacts (see Sec.~\ref{chap4/sec:active-passive}).

\paragraph{Text.}
Detection of LLM-generated text is a relatively new field.
As for images and audio, it either relies on hand-crafted textual features or on models trained for detection.
However, contrary to images and audio, the former approach is more popular since training and running LLMs is computationally expensive and cumbersome.
The features are for instance based on $n$-gram analysis~\citep{yang2023dna}, on the probability and rank of the observed tokens~\citep{gehrmann2019gltr, ippolito2019automatic}, on the perplexity of the text observed by the LLM under scrutiny or a surrogate model~\citep{vasilatos2023howkgpt, wang2023m4}, on several of them~\citep{hans2024spotting}, or on its curvature~\citep{mitchell2023detectgpt}.
The other class of methods trains classifiers~\citep{bhattacharjee2024eagle} or other language models, often by fine-tuning the model to detect itself~\citep{solaiman2019release, zellers2019defending}.
Similarly to images and audios, these detection methods are often brittle to shifts in the text distribution and not very reliable~\citep{sadasivan2023can}.

\subsection{Cryptographic metadata}

In the context of origin tracing, cryptographic metadata is digital information associated with a piece of content (e.g., images, videos, documents) to provide evidence of its authenticity and/or provenance.
The Coalition for Content Provenance and Authenticity (\Gls*{C2PA}) and the International Press Telecommunications Council (IPTC) have recently proposed two standards.
The upside is that forging fake cryptographic signatures is extremely hard, however the metadata are often removed during re-uploads or screenshots. 
For instance, a study by~\citet{ImatagStudy} shows that only 3\% of images on the internet come with copyright metadata.
It is therefore particularly suited for authenticating real content, for which the creators want the content to be traced back to them, but less for tracing origin of AI-generated content in the wild.
Besides, they are more a subject of standardization bodies than research, because all actors of the content production chain must adhere to the same protocol for it to be effective.

This metadata includes various types of cryptographic information, such as timestamps, used to record the date and time when the content was created or modified, or provenance information about the origin of the content like its creator.
All this information is encrypted with a private key, using algorithms like \Gls*{RSA}~\citep{rivest1978method} or ECDSA~\citep{johnson2001elliptic}, which makes it impossible to forge without the private key but possible to verify with the public key.
It can also include content bindings, used to check the authenticity of the content and ensure that it has not been tampered with.
They are encrypted hash values representing the content which are attached to it as a digital signature.
The bindings are categorized into two types. 
Hard bindings (\aka, cryptographic bindings), are computed directly from the raw bits of the content and can be used to ensure that the manifest belongs with the asset and that the asset has not been modified.
Soft bindings, on the other hand, are computed from the digital content of an asset (as in fingerprinting) and can be used to identify derived assets~\citep{c2pa}.

\subsection{Visible \gls*{watermarking}}

\begin{figure}[h!]
    \centering
    \includegraphics[width=0.99\textwidth]{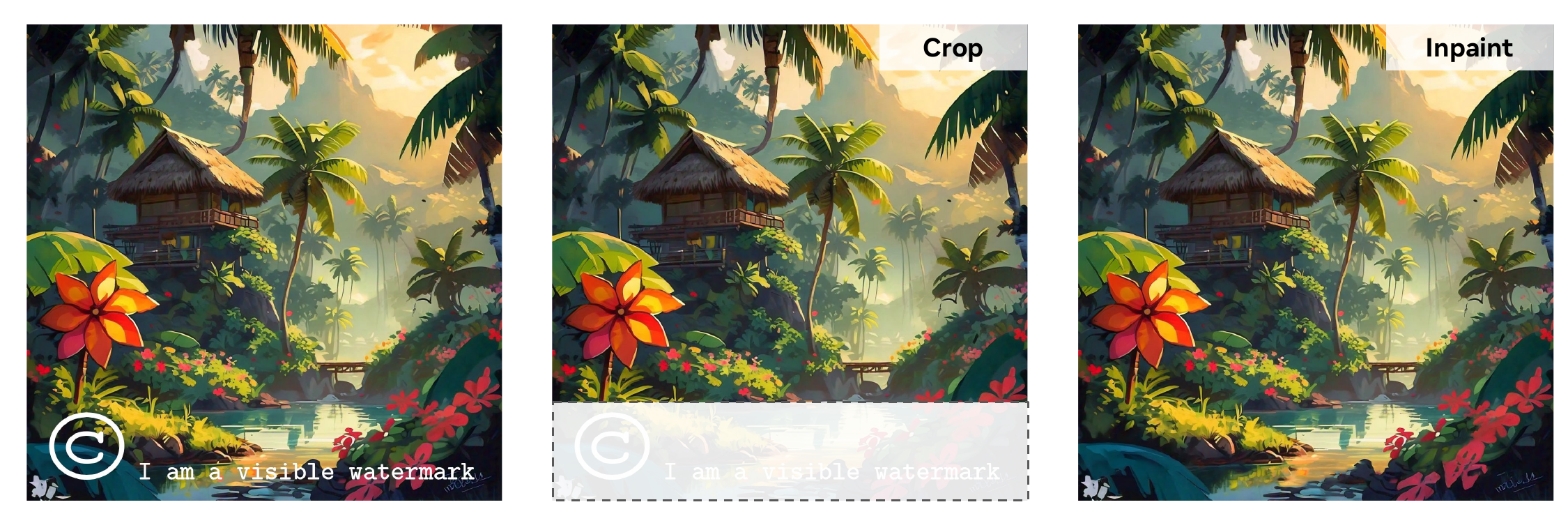}
    \caption{Example of a visible watermark, and how it can be removed in less than 10 seconds using free software (Adobe Firefly in this example).}
    \label{chap0/fig:visible-watermarking}
\end{figure}

\emph{Visible watermarks} are straightforward and widely recognized. 
However, in addition to degrading the quality of the content, they are also easy to remove or tamper with, making them less reliable.
For instance, a visible watermark on the left side of an image can be removed by cropping the image or through inpainting techniques~\citep{dekel2017effectiveness}, as illustrated in Fig.~\ref{chap0/fig:visible-watermarking}.
\emph{Invisible watermarks} (the focus of the thesis) are imperceptibility embedded within the content itself.
It makes them riskier to remove because there is no certainty that it has been removed without access to the watermark detector or extractor.

\section{Invisible \gls*{watermarking} techniques}\label{chap0/sec:watermarking}

This section provides an overview of watermarking techniques developed across the image, audio and text domains.
See \autoref{chapter:technical-background} for a more detailed description of properties\footnote{
    We notably describe in Sec.~\ref{chap0/sec:properties} the notions of robustness, imperceptibility, payload, zero-bit and multi-bit watermarking, that are used in the following paragraphs.
} and of state-of-the-art techniques, and Chapters \ref{chapter:radioactive} and \ref{chapter:invariants} for more details on model watermarking that aims to protect the model itself, instead of content.

\subsection{Image watermarking}

Traditional image watermarking methods are usually classified in two categories depending on the space on which the watermark is embedded. 
In \textit{spatial domain}, the watermark is encoded by directly modifying pixels, such as flipping low-order bits of selected pixels~\citep{van1994digital}. 
For example, \cite{nikolaidis1998robust} slightly modify the intensity of randomly selected image pixels while taking into account properties of the human visual system, robustly to JPEG compression and lowpass filtering.
\cite{bas2002geometrically} create content descriptors defined by salient points and embeds the watermark by adding a pattern on triangles formed by the tessellation of these points.
\cite{ni2006reversible} use the zero or the minimum points of the histogram of an image and slightly modifies the pixel grayscale values to embed data into the image. 
The second category is \textit{frequency domain} watermarking, which usually spreads a pseudorandom noise sequence across the entire frequency spectrum of the host signal, and provides better robustness~\citep{cox1997secure}. 
The first step is a transformation that computes the frequency coefficients.
The watermark is then added to these coefficients taking into account the human visual system.
The coefficients are mapped back onto the original pixel space through the inverse transformation to generate the watermarked image. 
The transform domains include Discrete Fourier Transform (DFT)~\citep{urvoy2014perceptual}, 
Quaternion Fourier Transform (QFT)~\citep{bas2003color, ouyang2015qdft}, 
Discrete Cosine Transform (DCT)~\citep{bors1996image, piva1997dct, barni1998dct, li2011new},
Discrete Wavelet Transform (DWT)~\citep{xia1998wavelet, barni2001improved, furon2008broken}, 
both DWT and DCT~\citep{feng2010dwt, zear2018proposed}, etc.

Deep learning-based methods have recently emerged as alternatives to traditional ones.
They are often built as encoder/decoder networks: the encoder embeds the watermark in the image and the decoder tries to extract it (see Sec.~\ref{chap0/sec:deep learning-watermarking}).
They are trained end-to-end to invisibly encode information while being resilient to transformations applied during training. 
This makes it easier to build robust systems and avoids algorithms hand-crafted for specific transformations.
\Gls*{HiDDeN}~\citep{zhu2018hidden} is the best example of this approach, and has been extended in several ways.
\cite{luo2020distortion} add adversarial training in the attack simulation, to bring robustness to unknown transformations.
\cite{zhang2019robust, zhang2020robust, yu2020attention} use an attention filter further improving imperceptibility.
\cite{ahmadi2020redmark} adds a circular convolutional layer that helps diffusing the watermark signal over the image.
\cite{wen2019romark} use robust optimization with worst-case attack as if an adversary were trying to remove the mark.
Another line of works focus on \gls*{steganography}~\citep{baluja2017hiding, wengrowski2019light, zhang2019steganogan, tancik2020stegastamp, jing2021hinet, ma2022towards}, where the goal is to hide a message in the image without being detected, rather than to robustly extract it (\eg, against crops).
Many other approaches focused on improving robustness, imperceptibility, speed, etc.~\citep{jia2021mbrs, bui2023rosteals, bui2023trustmark, huang2023arwgan, evennou2024swift, pan2024jigmark}.
In parallel to the encoder/decoder architectures, \cite{vukotic2018deep,vukotic2020classification}, followed by~\cite{kishore2021fixed} introduce an approach that is closer to traditional watermarking methods and uses neural networks as a fixed transform into a latent space.
Since there is no inverse transform, the embedding is done iteratively by gradient descent over the pixels.
This is the approach followed in Chapters \ref{chapter:ssl-watermarking} and \ref{chapter:active-indexing}.

\subsection{Audio watermarking}

Given the similar nature of the signals, audio watermarking techniques are very similar to image watermarking ones (although they lag a bit behind). 
Traditional methods relied on embedding watermarks either in the time or frequency domains~\citep{trad_wm_LieC06,trad_wm_KalantariAAA09,trad_wm_NatgunanathanXRZG12,trad_wm_freq_XiangNPHL18,trad_wm_freq_SuZYCJ018,trad_wm_freq_LiuHH19, tai2019audio}, usually including domain specific features to design the watermark and its corresponding decoding function. 
To accurately extract audio watermarks, synchronization between the encoder and decoder is crucial. 
However, this can be disrupted by desynchronization attacks such as time and pitch scaling. 
To address this issue, various techniques have been developed. 
One approach is block repetition, which repeats the watermark signal along both the time and frequency domains~\citep{blockrep2-KirovskiM03, blockrep1-Kirovski2003}. 
Another method involves implanting synchronization bits into the watermarked signal~\citep{patchwork-sync-XiangNGZN14}. During decoding, these synchronization bits serve to improve synchronization and mitigate the effects of de-synchronization attacks.

Most deep learning-based audio watermarking methods follow a \Gls*{HiDDeN}-like encoder/decoder framework~\citep{qu2023audioqr, pavlovic2022robust, DEAR_Liu0FMZY23, ren2023speaking, chen2023wavmark, o2024maskmark}. 
The approach presented in \autoref{chapter:audioseal} is similar, but is zero-bit and allows for detection at the time-step level.
Similar to the approach of \cite{vukotic2018deep} in the image domain, \cite{wu2023adversarial, kong2020adversarial} embed the watermark by iteratively modifying the audio such that its representation lies in a certain region of the feature space.

\subsection{Text watermarking}
Watermarking text is commonly thought as more challenging than images or audio: its discrete nature makes it harder to modify without altering its meaning.

The earliest works address watermarking for documents by altering text characteristics such as characters or spacing~\citep{brassil1995electronic}, which is not very robust since this may be changed directly on a text editor.
Text watermarking methods traditionally modify the grammatical or syntactical structure of the text with pre-established rules that embed watermarks without significantly altering its meaning~\citep{topkara2005natural}.
For instance, \cite{topkara2006hiding} embed information through synonym substitution, while~\cite{topkara2006words, topkara2006natural, meral2009natural} use word reordering through passivization, preposing, topicalization, etc.
Steganography methods have also been developed for text, working on the same principles~\citep{winstein1998lexical, chapman2001practical, bolshakov2004method, shirali2008new, chang2014practical, xiang2017novel}.
These edit-based systems usually suffer from low robustness to text modifications, and low payload, \eg, 1 or 2 bits per sentence as in CoverTweet~\citep{wilson2016avoiding}.
Similar to other media, deep learning-based methods have been developed more recently.
These methods either use pre-trained masked language models~\citep{ueoka2021frustratingly} or end-to-end encoder/decoder networks~\citep{abdelnabi2021adversarial}.

\subsection{Generation-time watermarking}\label{chap0/sec:generation-watermarking}

There is a growing need to track AI-generated content for transparency and for filtering such results when building new models.
In this context, some methods embed watermarks during the generation process such that it is directly in the generated content.
We call them \emph{generation-time}, as opposed to \emph{post-generation} (or \emph{post-hoc}) where the content is watermarked after it has been generated.
In generation-time watermarking, embedding the watermark does not incur additional runtime (particularly relevant for video), is easier (particularly for text), and is more robust and secure than post-hoc. 
We can broadly categorize generation-time watermarking into two categories:
\begin{itemize}[noitemsep]
    \item \emph{in-model}; embeds the watermark in the weights of the model,
    \item \emph{out-of-model}; embeds the watermark by altering the generation process.
\end{itemize}
Particularly, in-model watermarking allows for open-sourcing the model without revealing the watermark, conversely to out-of-model watermarking.
The latter is nevertheless easier to implement since it does not require training or fine-tuning the model

\paragraph*{Image.} 
The earliest methods~\citep{wu2020watermarking, yu2021artificial, zhao2023recipe} watermark the training set, which is computationally expensive and not scalable.
Alternatively, some methods~\citep{fei2022supervised, fei2024wide} train Generative Adversarial Networks (\Glspl*{GAN}) with additional watermarking losses such that generated images contain the watermark. 
Similarly, our Stable Signature work (Chap.~\ref{chapter:stable-signature}) focuses on Latent Diffusion Models (\Glspl*{LDM}) and fine-tunes the latent decoder to embed the watermark, while \cite{feng2024aqualora} fine-tune the U-Net that predicts the latent diffusion noise instead.
To eliminate the need to train or fine-tune the model for every user, \cite{yu2022responsible, fei2023robust} predict modifications to apply to GANs using a hyper-network and \cite{kim2024wouaf} apply a similar approach upon Stable Signature. 
\cite{ci2024wmadapter} and \cite{rezaei2024lawa} propose other variants that use an adapter to the decoder which takes the secret message as input.
The last two methods thus operate out-of-model in contrast to previously mentioned techniques that are in-model.

A different class of out-of-model methods, specific to diffusion models, embed the watermark by adding patterns to the initial noise. 
For example, Tree-Ring~\citep{wen2023tree} adds tree-ring-shaped (the name of the method is well chosen!) patterns to the initial noise and inverts the diffusion process to extract the watermark. 
As follow-up works, \cite{hong2024exact} improve the inversion of the diffusion process, \cite{ci2024ringid} extend the method to multi-bit watermarking and \cite{lei2024diffusetrace} use an encoder/decoder framework to embed and extract watermarks in the initial noise.

\paragraph*{Audio.} 
Most of the aforementioned literature could be extended to audio generation-based watermarking, although at the time of writing, this literature is scarce and only includes the following works.
\cite{san2024latent} watermark the training data robustly to the audio tokenization.
\cite{zhou2024traceablespeech} watermark in the latent space of a speech synthesis model.
\cite{juvela2023collaborative} use a GAN-like approach to watermark a HiFi-GAN model: it optimizes a speech generator and a collaborator, which tries to detect the watermark on the generated speech while not detecting it on the real speech.

\paragraph*{Text.} 
Literature on text generation-based watermarking is the most prominent. 
It appeared as soon as 2011 with the work of \cite{venugopal2011watermarking}, which was presumably used in Google Translate to filter out automated translations from future training data.
For text generated by Large Language Models (\Glspl*{LLM}), two concurrent methods were developed soon after the release of ChatGPT.
Both methods alter the token generation based on a secret key and a hash of some previous tokens.
\cite{kirchenbauer2023watermark} modify the probability distribution by adding a bias to the scores of one part of the vocabulary, while \cite{aaronson2023watermarking} modify the sampling method with the Gumbel trick.
These methods are detailed in Sec.~\ref{chap0/sec:llm_wm}.

They have been followed by many other works.
For instance, the work described in Chap.~\ref{chapter:three-bricks} consolidates these methods with better statistical tests and an extension to multi-bit watermarking. 
\cite{yoo2023advancing, yoo2023robust, qu2024provably} introduce further methods for multi-bit watermarking.
\cite{christ2023undetectable, kuditipudi2023robust} build methods based on the same sampling as \cite{aaronson2023watermarking}. 
They manage the pseudorandomness differently, by using a seed that is a function of the token's position in the text.
\cite{christ2023undetectable, huang2023optimal} show that detectability depends on the entropy of the generated text and suggest to embed watermarks in entropic passages, particularly for code~\citep{lee2023wrote}.
Other approaches focus on ``semantic'' watermarks, which rely on the semantic representation of the past text~\citep{liu2023semantic, liu2024adaptive, fu2024watermarking, hou2023semstamp, hou2024k}, enhancing the watermark's robustness to text alterations.
\cite{giboulot2024watermax} fix the watermark detection algorithm and choose one chunk of text among multiple ones to maximize the detection.
\cite{piet2023mark, pan2024markllm} introduce toolkits to easily benchmark watermarking methods and analyze their robustness.

All the aforementioned methods are out-of-model, since a user with access to the model can choose to generate text with or without the watermark.
There are few in-model methods in comparison.
\cite{gu2023learnability} distill the previous watermarking methods in the model itself via fine-tuning.
\cite{xu2024learning} embed the watermark with reinforcement learning methods popular for instruction fine-tuning~\citep{ouyang2022training}, by using a paired language model detector as reward model.

Watermarking generated content is therefore also very close to coverless (or generative) \gls*{steganography}, which generates stego content without the need for a cover.
It is very popular for language models~\citep{fang2017generating, yang2018rnn, ziegler2019neural, yang2020vae, cao2022generative}, and operates in a very similar way to LLM watermarking but with a focus on secrecy rather than robustness.
Some works on images also address the topic~\citep{hu2018novel, volkhonskiy2020steganographic, liu2022image, zhou2023generative}.

%% file: 0-introduction/technical-background.tex
\chapter{Technical Background}
\label{chapter:technical-background}

This section first exposes key properties and definitions which will be helpful for the rest of the thesis.
It then presents the most common deep learning based watermarking techniques for images, audio and LLM-generated text.

\section{Properties and definitions}\label{chap0/sec:properties}

\subsection{Watermarking as a communication system}

\begin{figure}[b]
    \centering
    \includegraphics[width=0.9\textwidth]{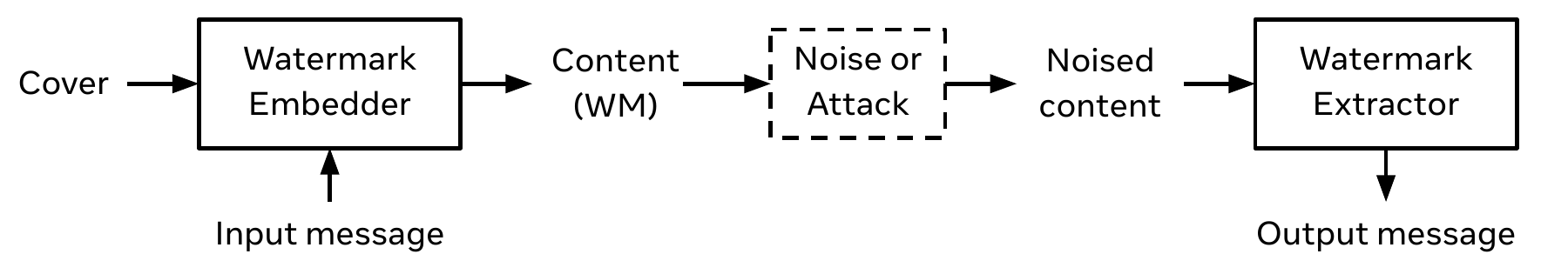}
    \caption{
        Schema of a watermarking system. 
        The embedder hides a message in a host content called cover (image, audio, etc.). 
        The host is then transmitted through a noisy channel.
        The extractor retrieves the message from the noised host.
    }\label{intro/fig:schema}
\end{figure}

Watermarking involves embedding a hidden message or signal (called watermark) into a \textit{\gls*{cover}} medium (\eg, image, audio, model's weights) without significantly altering its perceived quality. 
The process can be understood as a communication system~\citep{cover1999elements} where the cover serves as channel (see Fig.~\ref{intro/fig:schema}).
It is divided into two main stages:
The \emph{embedding} takes the original cover medium and the message as inputs and generates a watermarked version of the cover medium. 
The \emph{extraction} receives the host medium and attempts to detect and/or extract the embedded watermark. 
The extractor should be able to recover the original watermark with high accuracy, even if the host has undergone various transformations or attacks.
As a disclaimer, this is a simplified view and many variations exist depending on the application and the desired properties of the watermarking system.

\subsection{Zero-bit, multi-bit}

Watermarking techniques are broadly classified into two categories.
\begin{itemize}
    \item 
    (1) In \emph{\gls*{multi-bit watermarking}},
        the watermark is a binary message (a sequence of bits) that is embedded in the cover medium.
        The goal is to \emph{extract the embedded message} from the host medium.
        Multi-bit watermarking is used in applications where the watermark is intended to carry meaningful information that can be extracted and decoded by authorized parties, such as who owns the content.
    \item
    (2) In \emph{\gls*{zero-bit watermarking}}, 
        the watermark is not a message but a unique identifier or signature that is embedded in the cover medium. 
        The goal is to \emph{detect the presence} of the watermark in the host medium.
        Zero-bit watermarking is often used in applications where the watermark is not intended to convey any additional information but to provide a means of verifying the origin or integrity of the content.
\end{itemize}

A multi-bit watermarking system can be converted into a zero-bit system by embedding a unique binary message in the cover medium. 
The detection is then performed by comparing the extracted message with the expected one and performing a binomial test on the number of correct bits (see \hyperref[chap0/sec:test]{\nameref*{chap0/sec:test}}).

\subsection{Three (plus one) criteria}

\begin{figure}[b]
    \centering
    \includegraphics[width=0.37\linewidth, clip, trim=3cm 0.5cm 4cm 0cm]{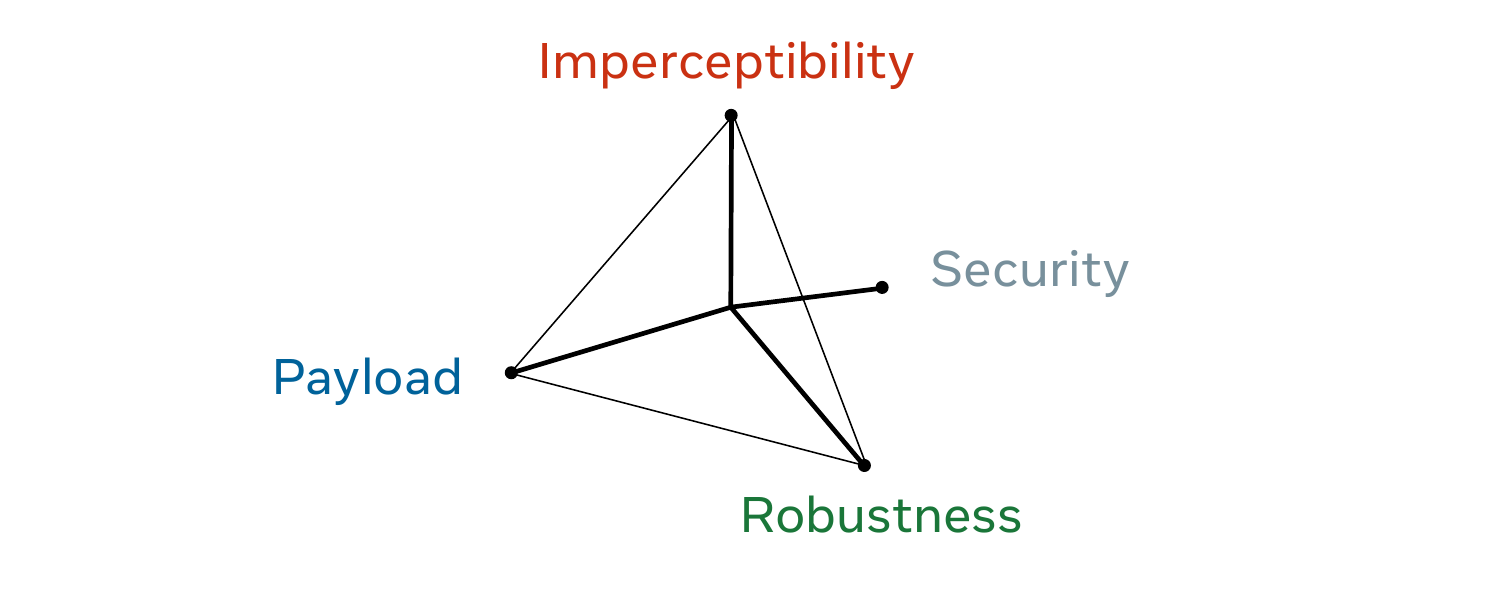}
    \caption{
        Watermarking must meet three criteria: imperceptibility, payload, and robustness.
        We can also add security, which is related to robustness but considers intentional attacks.
    }
    \label{intro/fig:criteria}
\end{figure}

Watermarking methods should meet three intertwined criteria, often represented by a triangle as in Fig.~\ref{intro/fig:criteria}:
\begin{itemize}
    \item \emph{\Gls*{robustness}}: 
    The ability of the watermark to withstand various transformations on the host without being significantly degraded or removed. 
    For example, the watermark on an image should be retrieved even after it has been compressed (JPEG, WebP), cropped, resized, etc.
    \item \emph{\Gls*{payload}}: 
    The amount of information that can be embedded into the cover. 
    For multi-bit watermarking, the payload is typically measured in bits, while for zero-bit watermarking, it is linked to the \pval\ of the detection test.
    \item \emph{Imperceptibility}: 
    The degree to which the embedded watermark remains undetectable to human perception\footnote{
        In the context of deep neural networks (DNN) watermarking, the imperceptibility is sometimes renamed \emph{unobtrusiveness}, \ie, the capacity to embed without affecting the performance of the model.
        For instance, \citet{tondi2024robust} show that unobtrusiveness is weakly linked to the other requirements, unless an additional constraint regarding the secrecy of the watermark is considered.
        }.
    For example, in an audio watermarking system, the watermark should introduce the least amount of distortion to the audio signal so that it is not audible to the human ear.
\end{itemize}
These properties are interdependent, and enhancing one property may impact others. 
Very often, decreasing the imperceptibility of a watermark (increasing its strength) will increase its robustness and payload.
Some researchers would join robustness and payload as one single criterion, and add a fourth one:
\begin{itemize}
    \item \emph{\Gls*{security}}: 
    The level of protection against intentional attacks, such as unauthorized removal, embedding, modification, or detection of the watermark.
    A secure watermarking system might use cryptographic techniques to encode the watermark so that it can only be extracted by parties possessing the correct decryption key. 
    However, it is important to note that cryptographic tools alone are often not sufficient to guarantee the security of the watermark, as shown in media watermarking.
\end{itemize}

\subsection{Other considerations}\label{chap0/sec:other-considerations}
Although the literature often focuses on the above criteria, there are several other considerations that can impact the design and implementation of a watermarking system.

\paragraph*{Speed.}
The computational cost of the watermark embedding and extraction is often left aside while of utmost importance in many applications.
Notably, for content tracing at the scale of social networks, the detection of the watermark should be able to process thousands if not millions pieces of content per second.

\paragraph*{Blindness.}
A blind watermarking system is one in which the extractor only requires access to the host medium in order to extract the watermark.
In contrast, non-blind watermarking also requires access to the original cover medium in order to extract the watermark. 
This can enhance the accuracy of the watermark extraction but at the cost of needing to securely store and retrieve the original content.
Most of the methods presented in this thesis are blind, except for the one presented in \autoref{chapter:invariants}.

\paragraph*{Private vs. public.}
Private watermarking restricts the extraction and detection of the watermark to authorized parties who possess specific tools or keys.
This is useful for ensuring that only intended recipients can embed and retrieve the information. 
Private watermarking is often more secure than public watermarking because it relies on a secret key.
In the context of deep learning based watermarking (see Sec.~\ref{chap0/sec:deep learning-watermarking}), the extractor weights can be seen as the secret key, making it private.

In contrast, a public watermarking system is one in which the watermark can be extracted by anyone who has access to the host medium. 
This would be the ideal scenario for detection of AI-generated content, as it would allow anyone to verify the origin of the content.
However, public watermarking methods that are secure and robust to attacks are very scarce.

\paragraph*{Fragile.}
A fragile watermarking system is one in which the watermark is destroyed as soon as the watermarked content is modified or tampered with.
Fragile watermarking is often used for integrity verification and tamper detection, as it provides a way to detect whether an image or audio has been altered.
Robust watermarks, as previously discussed, are on the contrary intended to withstand various types of processing and attacks without being removed or degraded.
In this thesis, we only consider robust watermarking.

\paragraph*{Reversibility.}
A reversible watermarking system is one in which the original medium can be recovered from the host medium. 
This means that the watermark can be removed from the host medium, leaving the original cover medium intact. 
Reversible watermarking is often used in applications where it is important to preserve the original cover medium, such as in medical imaging or forensic analysis.

\subsection{Detection and statistical hypothesis testing}\label{chap0/sec:test}

The detection of the watermark in a zero-bit setup is often formulated as a statistical hypothesis testing problem.
We hereby provide a brief overview of the key concepts and definitions related to hypothesis testing (FPR, TPR, \pval, etc.).

\paragraph*{Detection test.}
The \emph{null hypothesis} $\H_0$ is that the host medium does not contain the watermark, while the \emph{alternative hypothesis} $\H_1$ is that the watermark is present.
The detection test outputs a \emph{\gls*{test statistic}} (or watermark score), which is then compared to a threshold $\tau$ to decide whether to accept or reject $\H_0$.

\paragraph*{Evaluation metrics.}
The global detection performance of a watermarking system is evaluated based on the \emph{False Positive Rate} (\Gls*{FPR}) -- or type I error -- and the \emph{True Positive Rate} (\Gls*{TPR}) -- inversely proportional to the type II error -- of the detection test.
The FPR is the probability of falsely detecting the watermark when it is not present, while the TPR is the probability of correctly detecting the watermark when it is present.

There is a trade-off between the FPR and TPR, which is controlled by the threshold $\tau$. 
Large-scale applications usually operate at very low FPR to avoid human verification.
For example, at the scale of Facebook, where billions of images are uploaded every day, a FPR of $10^{-3}$ would result in millions of false detections per day, which is unacceptable.
The \emph{Receiver Operating Characteristic (\Gls*{ROC}) curve} is a graphical representation of the trade-off between the FPR and TPR, and is obtained by varying the threshold $\tau$.

\paragraph*{\pval, significance level and FPR.}
Given a test statistic, the \emph{\gls*{p-value}} is the probability of observing under $\H_0$ a test statistic as extreme as the one observed.
The \emph{significance level} $\alpha$ is the threshold for the \pval\ below which $\H_0$ is rejected -- it is therefore also the probability of falsely rejecting $\H_0$, \ie, the FPR.
\begin{itemize}
    \item The \pval\ converts the watermark score associated to a piece of content into an interpretable quantity: the probability of non-watermarked content to have a watermark score as extreme as the one observed.
    It can also be interpreted as the FPR we would need to flag this content as watermarked.
    \item The \pval\ is a local notion and concerns a specific piece of content, while the FPR is a global notion and concerns the detector.
    For instance, if the \pval\ is $10^{-6}$, we would flag this content while wrongly flagging a non-watermarked content only once every million times.
\end{itemize}

\paragraph*{Example: Binomial test.}
Let us consider a scenario where the test statistic under the null hypothesis $\H_0$ follows a binomial distribution. This is a common assumption when the watermark detection mechanism is based on counting a number of mismatches that occur randomly, which happens in many related works and in Chapters \ref{chapter:stable-signature}, \ref{chapter:audioseal}, \ref{chapter:three-bricks}, \ref{chapter:radioactive} and \ref{chapter:invariants}.
Notably, it is used when turning a multi-bit watermarking scheme into a zero-bit one, by hiding a binary message and counting the number of matching bits.

The hypotheses are:
($\H_0$) The test statistic $S$ follows a binomial distribution $B(n, p_0)$, where $n$ is the number of trials and $p_0$ is the probability of success under $\H_0$.
In the case of multi-bit watermarking, $n$ is the number of bits in the binary message and $p_0 = 0.5$ (the probability of a random bit being correct).
($\H_1$) The test statistic $S$ follows a different distribution, typically one with a higher probability of success $p_1$ ($p_1 > p_0$).

The \pval\ of the test is the probability of observing a bigger score than the observed value $s$, under the null hypothesis $\H_0$. 
Mathematically, it is given by:
\begin{align}
    \text{\pval}(s) = P(S \geq s \mid \H_0) = 1 - F_{B(n, p_0)}(s-1) = I_{p_0}(s, n-s+1),
\end{align}
where $F_{B(n, p_0)}$ is the cumulative distribution function (c.d.f.) of the binomial distribution under $\H_0$, and $I_{p_0}$ is the \href{https://en.wikipedia.org/wiki/Beta_function#Incomplete_beta_function}{regularized incomplete beta function} (the last equality is a direct consequence of the definition of the beta function).

This \pval\ helps in deciding whether to reject $\H_0$ in favor of $\H_1$ based on the significance level $\alpha$ (or FPR) chosen for the test. 
If $\text{\pval} \leq \alpha$, we reject $\H_0$; otherwise, we do not.
This is equivalent to comparing the test statistic $s$ to a threshold $\tau$ such that $\text{\pval}(\tau) = \alpha$.
In practice, the threshold $\tau$ is chosen based on the desired FPR, and $\H_0$ is rejected if $s \geq \tau$.
For more information and a practical example, see Chap.~\ref{chapter:stable-signature}, Sec.~\ref{chap3/subsec:statistical-test} and Chap.~\ref{chapter:audioseal}, Sec.~\ref{chap4/app:fpr}.

\section{Deep learning based image/audio watermarking}\label{chap0/sec:deep learning-watermarking}

\begin{figure}[b!]
    \centering
    \includegraphics[width=\textwidth, clip, trim=0 3cm 0 0]{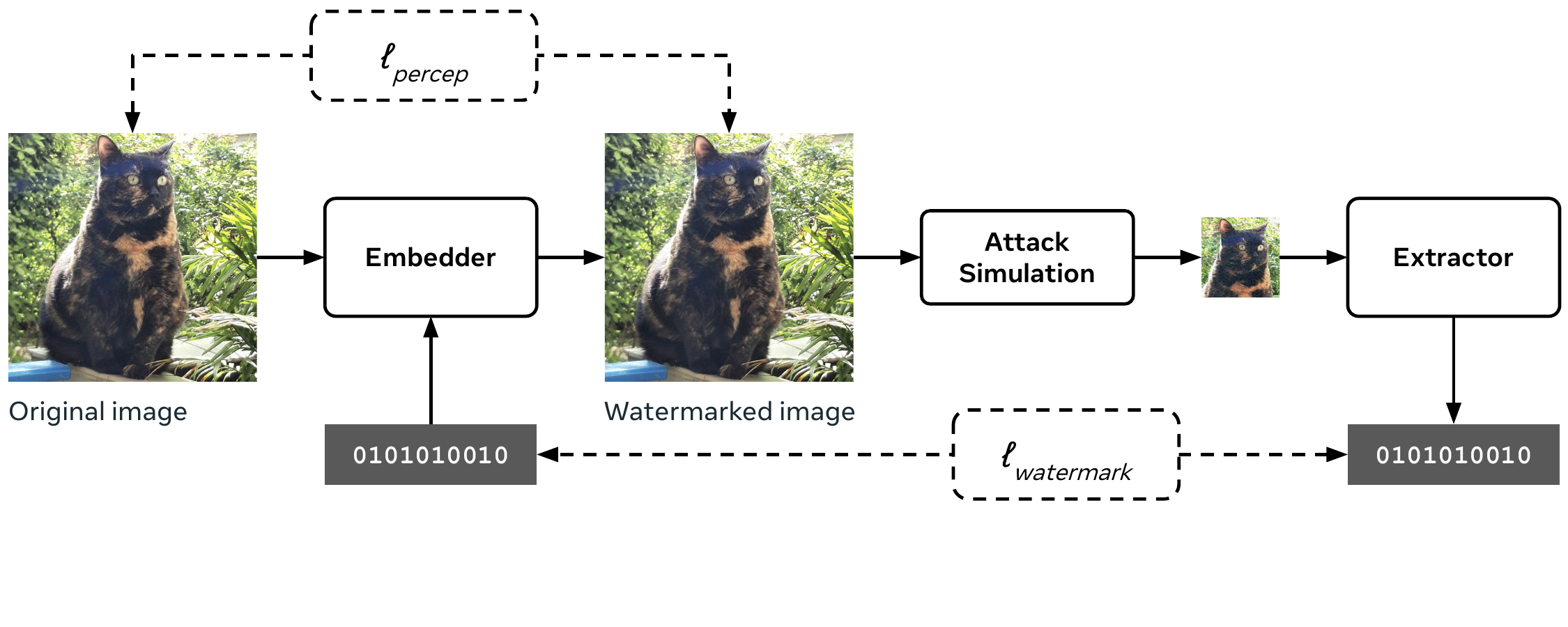}
    \caption{
        Overview of encoder/decoder based image watermarking. 
        The embedder (or encoder) takes an image and a binary message and outputs a new watermarked image that slightly differs from the original.
        The extractor (or decoder) takes an augmented version of the image and outputs a binary message.
        The training is done by minimizing 2 losses: a perceptual loss $\ell_{percep}$ between the watermarked and original, which controls the imperceptibility and a watermark loss $\ell_{watermark}$ which makes the output message similar to the original.
    }
    \label{chap0/fig:hidden}
\end{figure}

State-of-the-art watermarking methods for continuous domains, \eg, image, audio, etc. are predominantly based on \emph{encoder/decoder} deep neural networks~\citep{ahmadi2020redmark, DEAR_Liu0FMZY23, chen2023wavmark}, \Gls*{HiDDeN}~\citep{zhu2018hidden} being one of the most known examples.
Although they appear in different forms, they share a common structure and training procedure illustrated in Fig.~\ref{chap0/fig:hidden}.
These approaches serve as baselines and have inspired the techniques developed in Chapters \ref{chapter:ssl-watermarking}, \ref{chapter:active-indexing}, \ref{chapter:stable-signature}, and \ref{chapter:audioseal}.
We describe it bellow in the context of image watermarking\footnote{
    This chapter does not cover classical media watermarking as they are not used in the thesis, except as baselines.
    We refer the reader to Sec.~\ref{chap0/sec:watermarking} for a description and a discussion on these methods.
}.

\paragraph*{Watermark embedding.}
The \emph{encoder} (or embedder) network $E$ takes as input a cover image $\im_o \in \mathbb{R}^{W\times H\times 3}$ and a binary message $m\in\{0,1\}^\payload$.
The input message is fed into the network through a \emph{message embedding layer} that transforms it into a vector representation that can be processed by the network.
In Chap.~\ref{chapter:stable-signature}, the message is converted to $\pm 1$ values and concatenated pixel-wise to a feature map inside the network, while in Chap.~\ref{chapter:audioseal}, the message is used to select representations from a learned embedding layer, which are then aggregated and concatenated to the intermediate activations of the network.

The embedder outputs either directly the watermarked image $x_w$, or a distortion map $\delta \in \mathbb{R}^{W\times H\times 3}$.
$\delta$ is added to the cover image to produce the watermarked image $\im_w = \im_o + \alpha \cdot \delta$ with a scaling factor $\alpha$ that controls the imperceptibility of the watermark.
There are two common strategies to operate on various resolutions.
The first one is to use architectures that can handle images of any size, such as fully convolutional ones (as long as the distribution of the image is covered by the training set).
The second  one -- more practical because more computationally efficient and generalizable -- is to resize the image to a fixed resolution before feeding it to the network, then resize the output to the original resolution: 
\begin{equation}
    \im_w = \im_o + \alpha \cdot \text{resize}(\delta), \quad \delta = E(\text{resize}(\im_o)).
\end{equation}

\paragraph*{Attack simulation (augmentations).}
During training, the watermarked image is augmented with various transformations to simulate attacks and ensure the robustness of the watermark.
At each optimization step we randomly sample an augmentation $T$ to create a transformed image $\im_t = T(\im_w)$.
Among the most common augmentations are: JPEG compression, rotation, resizing, cropping, brightness or contrast changes, etc.

Many times, the augmentation is not differentiable with respect to the image, which makes the training more challenging since the gradients cannot be back-propagated through the augmentation.
The most common solutions are to either use a differentiable approximation of the augmentation, such as differentiable JPEG compression~\citep{zhu2018hidden}, or to use a straight-through estimator that approximates the gradient of the non-differentiable operation with the identity function~\citep{bengio2013estimating}.
The second option is often preferred as it often gives similar or better results than the differentiable approximation and is extremely simple to implement:
\begin{equation}
    \im_t = \im_w + \mathrm{nograd} \left( T(\im_w) - \im_w \right),
\end{equation}
where $\mathrm{no grad}$ does not propagate gradients, such that $\nabla_\theta\,\im_t = \nabla_\theta\,\im_w$ (where $\theta$ are the parameters of the network).

\paragraph*{Message extraction.}
The \emph{decoder} (or extractor) network $D$ takes as input the transformed image $\im_t$ and outputs a ``soft'' message $m' = D(\im_t) \in \mathbb{R}^\payload$ (soft because continuous).
At inference time, the decoded message is obtained by thresholding the soft message, for instance with $m' > 0$.
In the same way as for the embedding, the extraction can be done at a fixed resolution or with a network that can handle any resolution.

\paragraph*{Perceptual loss.}
The \emph{perceptual loss} $\ell_{\text{percep}}$ ensures that the watermarked image is visually similar to the original.
It is computed between the watermarked image and the original image and varies a lot depending on the papers.
It can be the mean squared error (MSE) between the pixel values, the LPIPS~\citep{zhang2018unreasonable} between the feature maps of a pre-trained network, discriminative losses based on GANs~\citep{goodfellow2014generative}, etc. or weighted combinations of these losses.

\paragraph*{Extraction loss.}
The \emph{message loss} $\ell_{\text{watermark}}$ ensures that the extracted message is close to the original message.
Most of the time it is the MSE or the Binary Cross Entropy (BCE) between $m$ and the sigmoid $\sigma (m')$:
\begin{align}
    \ell_{\text{watermark}} &= \frac{1}{\payload} \sum_{i=1}^\payload \left( m_i - \sigma (m'_i) \right)^2 
    \quad \text{or} \nonumber\\
    & =- \sum_{i=1}^\payload m_i \cdot \log \sigma (m'_i) + (1-m_i) \cdot \log ( 1 - \sigma (m'_i)).
\end{align}

\section{Watermarking for large language models}\label{chap0/sec:llm_wm}

\begin{figure}[b!]
    \centering
    \includegraphics[width=1.0\textwidth]{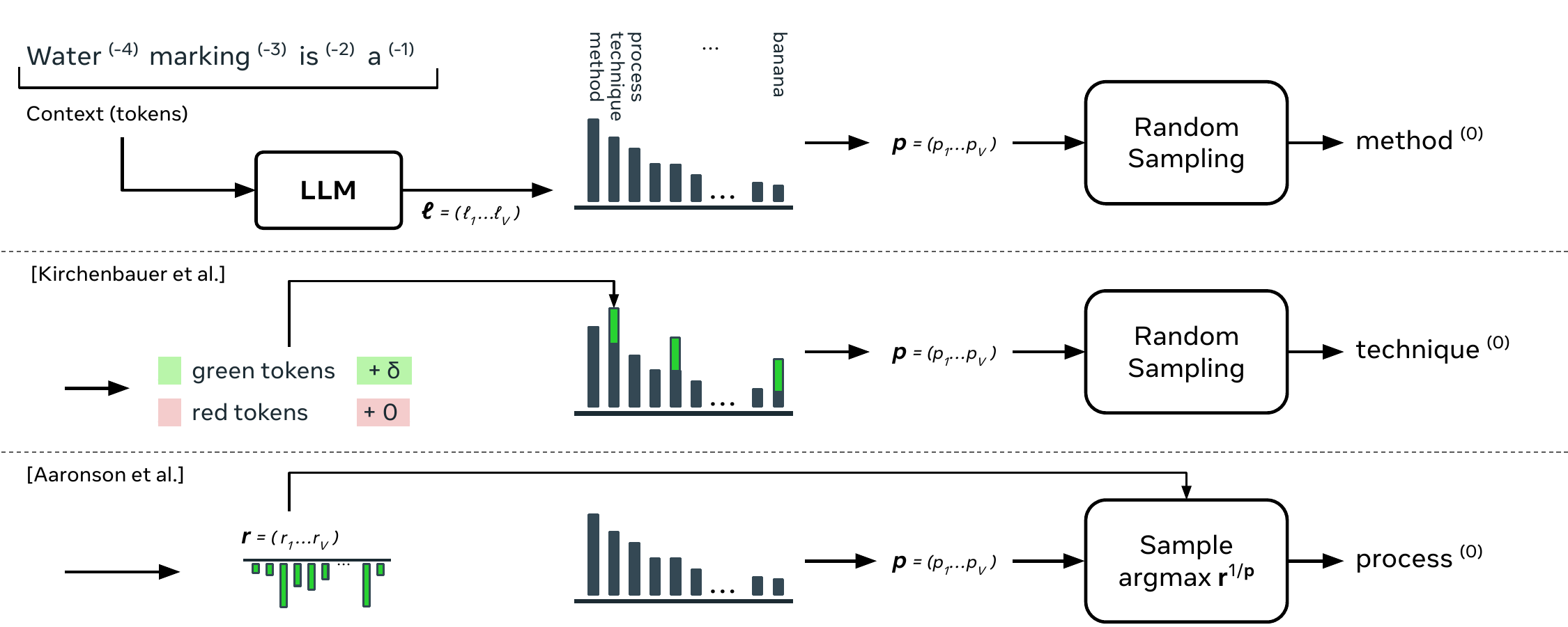}
    \caption{
        Text generation with and without watermarking.
        (Top) The LLM generates logits based on the context, then samples the next token according to a distribution $\vec{p}$.
        The watermarking process alters the generation of the next token by 
        (Middle) partitioning the vocabulary into green and red tokens and adding a bias to green tokens~\citep{kirchenbauer2023watermark},
        (Bottom) sampling an uniform vector $\vec{r}$ and choosing the token as $\arg \max \vec{r}^{1/\vec{p}}$~\citep{aaronson2023watermarking}.
        The partitioning $\G \cup \bar{\G} = \V$ or the vector $\vec{r}$ are determined by a hash value computed from the watermark window consisting of the $k$ previous tokens.
    }
    \label{chap0/fig:llm-watermarking}
\end{figure}

We hereby provide a detailed overview of two watermarking techniques for large language models (\Glspl*{LLM}) that are used in Chapters \ref{chapter:three-bricks} and \ref{chapter:radioactive}.
They have been developed concurrently and are based on approximately the same principles~\citep{aaronson2023watermarking,kirchenbauer2023watermark}.
The watermarking process is illustrated in Fig.~\ref{chap0/fig:llm-watermarking}.

\paragraph*{Text generation with LLMs.} 
LLMs generate text by computing the likelihood of generating a sequence of tokens given a context.
The thesis focuses on decoder-only models, \aka, auto-regressive LLMs.
The \glspl*{token} are pieces of words or characters from a vocabulary $\V$, that are represented as integers between 0 and $|\V|-1$.
From a context $x^{(-C)}, ..., x^{(-1)}$, the model estimates the probability of each token of $\V$ being the next.
It computes a vector $\vec{\boldsymbol\ell}\in \R^{|\V|}$ of logits, transformed into:
\begin{equation}
    \vec{p} = \text{softmax}(\vec{\boldsymbol\ell};\temperature) \approx \left(\Prob \left( X^{(0)}= v \mid x^{(-C)},\dots, x^{(-1)} \right)\right)_{v\in\V},
    \label{chap0/eq:Distrib}
\end{equation}
where $\temperature$ is a temperature and $X^{(0)}$ is the random variable representing the next token.
The generation of a sequence from the context samples a token from this distribution, then appends it to the context and iterates the process. 
Various sampling schemes exist: 
greedy search selects the token with the highest probability, 
nucleus-sampling (top-p)~\citep{holtzman2019curious} randomly select a token in the smallest subset whose cumulative probability exceeds $p$,
top-k sampling~\citep{fan2018hierarchical}, beam search, etc.

\paragraph*{Hashing.}
Both methods alter the generation of the next token $x^{(0)}$ in a way that is influenced by the window formed by the $k$ previous tokens in the context: $x^{(-k)}, ..., x^{(-1)}$. 
We call it the \emph{watermark window} in all the thesis.

A hash function, which also depends on a secret key $\sk$, maps these $k$ integers to a hash value.
It varies according to implementations. 
In our work, the hash function operates according to the equation: 
$$h_{n+1} = (h_{n} \cdot \sk + x^{(n)}) \mod (2^{64} - 1),$$ 
for $n \in -k, .., -1$, and $h_{-k} = 1$.
Here, $h_{n+1}$ is the updated hash value, $h_{n}$ is the current hash value, $\sk$ is a constant, $x^{(n)}$ is the current input token, $\mod$ is the modulus operator, and $2^{64} - 1$ is the modulus value. 
The hash value is the final one ($h_0$).

\paragraph*{Watermark embedding.} 
The hash value is used as a seed that initializes a random number generator (RNG).
RNG is then used to influence the choice of the next token $x^{(0)}$.
\begin{itemize}
    \item \citet{kirchenbauer2023watermark} use the RNG to partition the vocabulary $\V$ into a greenlist $\G$ and a redlist $\bar{\G}$, where $\G$ contains a proportion $\gamma \in [0,1]$ of the vocabulary.
    The logit of each token in the greenlist is incremented by a value $\delta>0$, and the sampling proceeds as usual.
    Intuitively, this encourages the generation of greenlist tokens by increasing their probability.

    The imperceptibility of the watermark is mainly controlled by the parameter $\delta$.
    When it is high, the sampling only selects greenlist tokens, which  degrades the text quality but increases the robustness of the watermark, and vice versa (if $\delta=0$, the generation is not altered).

    \item \citet{aaronson2023watermarking} use the RNG to sample a vector $\vec{r}\in[0,1]^{|\V|}$.
    The next token is chosen as:
    \begin{equation}
        x^{(0)} = \arg \max_{v \in \V } \vec{r}_v^{1/\vec{p}_v}
    \end{equation}
    This encourages the generation of tokens that have a high $\vec{r}$ value.
    It also presents the interesting property that $\forall v\in \V$, $\Prob_{\vec{R}\,\sim\,\mathcal{U}([0,1]^{|\V|})} (X^{(0)}=v) = \vec{p}_v$ (proof is available in App.~\ref{chap5/app:aaronson_prob}).
    In other words, the probability of generating a token is not altered on expectation over the secret hash.

    The imperceptibility of the watermark is controlled by the temperature $\theta$ of the softmax.
    For very high values of $\theta$, the softmax outputs an almost uniform probability vector $\vec{p}$, so the choice of the next token is determined entirely by $\vec{r}$ (the token with highest $\vec{r}$ value is chosen) -- whereas for very low $\theta$, distribution $\vec{p}$ is very peaky so $\vec{r}$ has little influence.
\end{itemize}

\paragraph*{Watermark scoring.} 
We first tokenize a text into a sequence of tokens $x^{(1)}, \ldots, x^{(T)}$.
We go through every token $x^{(t)}$, replay the hashing process with the secret key $\sk$ and the $k$ previous tokens $x^{(t-k)}, \ldots, x^{(t-1)}$ and seed the RNG with the hash value.
We then re-create either the greenlist or the key vector, and score the token $x^{(t)}$ based on this information.
The score function on a current token $x^{(0)}$ may therefore be summed up as $W_{\textrm{score}}$ that also takes as input the watermark window $(x^{(-k)},\dots, x^{(-1)} )$, and implicitly depends on the secret-key $\sk$:
\begin{figure}[h!]
   \vspace*{1em}
   \begin{equation}
   \label{chap6/eq:watermark_score}
   \eqnmarkbox[Plum]{token}{x^{(0)}} ;\, 
   \eqnmarkbox[Emerald]{window}{\big(x^{(-k)},\dots, x^{(-1)} \big)} 
   \mapsto 
   \eqnmarkbox[BurntOrange]{Wscore}{W_{\textrm{score}}} 
   \left(  
      \eqnmarkbox[Plum]{token2}{x^{(0)}}  ;\,  
      \eqnmarkbox[Emerald]{window2}{\big(x^{(-k)},\dots, x^{(-1)} \big)} 
   \right) \in \R.
   \end{equation}
   \annotate[yshift=-0.4em]{below,right}{token}{Current token being scored}
   \annotate[yshift=0.4em]{above,right}{window}{Watermark window ($k$ previous tokens)}
   \annotate[yshift=-0.4em]{below,right}{Wscore}{Scoring function (\eg $1$ if green token, $0$ otherwise)}
\end{figure}\\
The total score $s_T$ is finally the sum of the scores of all tokens:
\begin{equation}
    s_T = 
        \sum_{t=1}^T W_{\textrm{score}} \left( 
                x^{(t)} ;\, 
                \big(x^{(t-k)},\dots, x^{(t-1)} \big)
            \right).
\end{equation}\\
The score functions are defined as follows:
\begin{itemize}
    \item For \citet{kirchenbauer2023watermark}:
    \begin{equation}
        W_{\textrm{score}} \left( x^{(t)} ;\, \G^{(t)} \right) = \mathds{1} \left(x^{(t)}\in\G^{(t)}\right),
    \end{equation}        
    where $x^{(t)}$ and $\G^{(t)}$ are the $t^{\textrm{th}}$ token and its associated partition, \ie, $1$ if the token is in the greenlist and $0$ otherwise.
    The total score is therefore the number of greenlist tokens in the text.
    Intuitively, if the text is not watermarked, the score should be approximately $\gamma$ since a token has a probability $\gamma$ of being in the greenlist.
    If the text is watermarked, the score should be strictly higher than $\gamma$.
    
    \item For \citet{aaronson2023watermarking}:
    \begin{equation}
        W_{\textrm{score}} \left( x^{(t)} ;\, \vec{r}^{(t)} \right) = - \ln \left(1- \vec{r}^{(t)}_{x^{(t)}} \right),
    \end{equation}
    where $x^{(t)}$ and $\vec{r}^{(t)}$ are the $t^{\textrm{th}}$ token and its associated key vector.
    The total score presents the property that for non-watermarked text the expected score $\mathbb{E}(S_T) = T$, while otherwise $\displaystyle \mathbb{E}(S_T) \geq T +  \left( \frac{\pi^2}{6} -1 \right) H_T > T $, where $H_T = - \sum_{t=1}^T p_t\ln(p_t)$ is the entropy of the watermarked completion ($p_t = \vec{p}^{(t)}_{x^{(t)}}$).
    See proofs in App.~\ref{chap5/app:aaronson_score}.
\end{itemize}
Note that $\G^{(t)}$ and $\vec{r}^{(t)}$ only depend on the $k$ tokens that precede $x^{(t)}$ and the secret-key $\sk$.
A corollary is that the score computation does not require the model to be re-run, which is a significant advantage in terms of computational cost.

Moreover, the robustness of the watermark is better for small $k$, since changing a token in the watermark window will change the hash value and increment the score as if there was no watermark.
At the same time, the hash values are more likely to collide during generation when $k$ is low, which creates a bias in the distribution of generated tokens.
Therefore, the choice of $k$ is a trade-off between robustness and imperceptibility.
For instance, if $k=0$ the hash only takes one value, so the generation of some tokens is always favored, but the watermark is robust except if most of the tokens are modified.

\paragraph*{Statistical test for watermark detection.} 
The statistical hypothesis test distinguishes between the hypothesis $\H_0$: ``the text is natural'' and $\H_1$: ``the text has been generated with watermark''.
Originally, both approaches approximate the underlying distribution of the score $S_T$ by using a $Z$-test (\autoref{chapter:three-bricks} introduces statistical tests based on the true distribution of the score $S_T$).

The $Z$-test compares the observed score $S_T$ against its expected value under the hypothesis $\H_0$ (no watermark).
It is typically used for large sample sizes assuming a normal distribution under the null hypothesis thanks to the central limit theorem.  
The $Z$ statistic is computed as:
\begin{equation}
    Z = \frac{{S_T/T - \mu_0}}{{\sigma_0 / \sqrt{T}}},
\end{equation}
where $\mu_0$ and $\sigma_0$ are the expected mean and standard deviation of the score per token under $\H_0$.
A \pval\ is then calculated to determine the likelihood of observing a score as extreme as $S_T$ under $\H_0$:
\begin{equation}
    \text{p-value}(z) = \Prob(Z \geq z \mid \H_0) = 1 - \Phi(z),
\end{equation}
where $\Phi$ is the cumulative distribution function of the normal distribution. 
Texts are flagged as watermarked if the \pval\ is less than a fixed false positive rate (FPR).

%% file: chapter-1/main.tex
\chapter[Watermarking Images in Self-Supervised Latent Spaces]{Watermarking Images in Self-\\Supervised Latent Spaces}
\label{chapter:ssl-watermarking}

\newif\ifarxiv
\newif\ifnotarxiv

\arxivtrue
\notarxivfalse

This chapter is based on the paper \fullcite{fernandez2022sslwatermarking}.

In the context of content distribution on platforms, watermarking may improve content moderation by embedding a unique identifier -- a specific marker that can be used to verify the source -- into each piece of content as soon as it is uploaded. 
This allows for active tracing of its origin, ensuring greater control over the material shared.
In this chapter, we revisit watermarking techniques based on pre-trained deep networks, in the light of self-supervised approaches.
We present a way to embed both marks and binary messages into their latent spaces, leveraging data augmentation at marking time. 
The resulting method is very versatile: it can operate at any resolution, with variable payload and tradeoff between quality and robustness, which makes it suitable for a wide range of applications.
It creates watermarks robust to a broad range of transformations (rotations, crops, JPEG, contrast, etc). 
It significantly outperforms the previous zero-bit methods, and its performance on multi-bit watermarking is on par with state-of-the-art encoder/decoder architectures trained end-to-end for watermarking.
The code is available at \href{https://github.com/facebookresearch/ssl\_watermarking}{github.com/facebookresearch/ssl\_watermarking}

\newpage
\input{chapter-1/sections/intro.tex}

\input{chapter-1/sections/related.tex}

\input{chapter-1/sections/method.tex}
\input{chapter-1/sections/experiments.tex}

\input{chapter-1/sections/conclusion.tex}

%% file: chapter-1/sections/intro.tex
\section{Introduction}\label{chap1/sec:introduction}

\begin{figure*}[t]
    \centering
    \includegraphics[width=\textwidth, clip, trim=0 1cm 0 0]{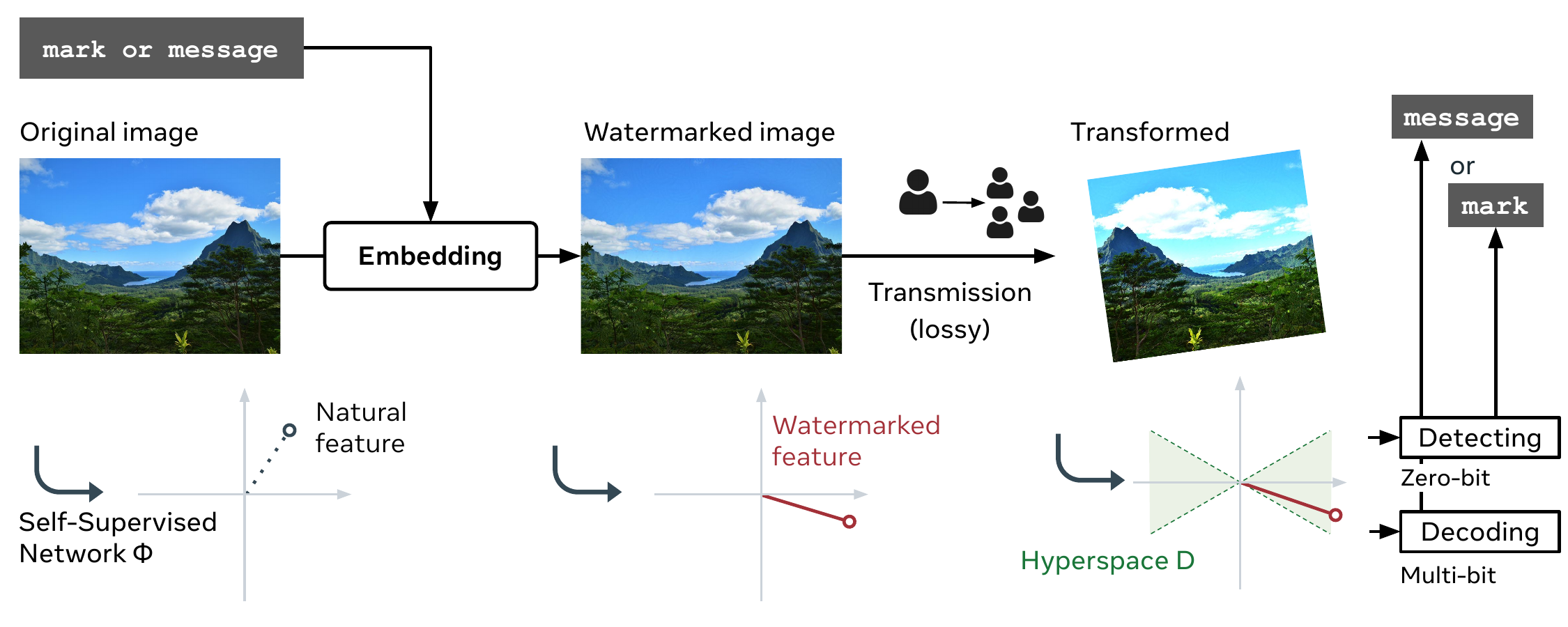}
    \caption{
    A self-supervised network trained with DINO \citep{caron2021dino} builds a latent space in which the watermark is embedded.
    Its effect is to shift the image's feature into a well-specified region of the latent space, such that transformations applied during transmission do not move much the feature. 
    The detection (zero-bit watermarking setup) or extraction (multi-bit watermarking setup) is performed in the same latent space.}
    \label{chap1/fig:splash}
\end{figure*}

The classic approach to image watermarking, coined \emph{TEmIt} (Transform, Embed, Inverse transform) by T. Kalker, embeds the watermark signal in the feature space of a transform (\eg, DFT, DCT, Wavelet). 
It provides coefficients that are supposedly reliable for watermarking by design because they are perceptually significant as conceptualized in~\cite[Sec.~8.1.3]{cox2007digital}. 
Deep learning based approaches for image watermarking improve the robustness to a broad range of alterations thanks to neural networks offering a reliable latent space where to embed the information.
Most of them explicitly train a watermarking network to be invariant to a set of image perturbations -- in this case, networks are usually encoder/decoder architectures trained end-to-end for watermarking~\citep{zhu2018hidden,wen2019romark, ahmadi2020redmark} -- while few others directly mark into the semantic space resulting from a supervised training over a given set of classes like ImageNet~\citep{vukotic2020classification}.

Our key insight is to leverage the properties of \emph{self-supervised} networks to watermark images.
Ideally, according to~\cite{cox2007digital}, a perceptually significant coefficient does not change unless the visual content of the image is different.
Similarly, some self-supervised methods aim to create representations invariant to augmentations, without explicit knowledge of the image semantics~\citep{caron2021dino,grill2020byol}.
These pre-trained networks offer us the desired embedding space ``for free'', saving us the heavy training of end-to-end architectures like HiDDeN~\citep{zhu2018hidden}.

In order to robustly embed in the latent spaces, gradient descent is performed over the pixels of the images. 
To further ensure both robustness and imperceptibility of the watermarks, we include data augmentation and image pre-processing at marking time.

Our contributions are the following:
\begin{itemize}
    \setlength\itemsep{0.1em}
    \item We provide a watermarking algorithm that can encode both marks and binary messages in the latent spaces of any pre-trained network;
    \item We leverage data augmentation at marking time;
    \item We experimentally show that networks trained with self-supervision provide excellent embedding spaces.
\end{itemize}

%% file: chapter-1/sections/related.tex
\section{Related work} \label{chap1/sec:relatedwork}

\paragraph*{Zero-bit watermarking with pre-trained neural networks.}
\cite{vukotic2018deep} mark images with a neural network pre-trained on supervised classification, instead of relying on encoder/decoder architectures like the one presented in Sec.~\ref{chap0/sec:deep learning-watermarking}.
The network plays the role of the transform in the TEmIt approach. 
Since it has no explicit inverse, a gradient descent to the image pixels ``pushes'' the image feature vector into a hypercone. 
The follow-up work~\citep{vukotic2020classification} increases the inherent robustness of the network by applying increasingly harder data augmentation at pre-training. 
It offers a guarantee on the false positive rate without requiring to train a network explicitly for watermarking, but no multi-bit version was proposed.

\paragraph*{Self-Supervised Learning}\label{chap1/par:ssl}
(SSL) does not use any labels and leverages the underlying structure of the data for supervision.
Its goal is to produce visual representations that contain high-level semantic information and are invariant to transformations of the input image.
Early SSL methods apply a transformation to the input image and train the network to predict the transformation.
For instance, \citet{doersch2015unsupervised} predict the relative position of two patches cropped from an image, 
\cite{gidaris2018unsupervised} predict a random rotation applied to an image.
Newer methods directly create features invariant to the image transformations~\citep{chen2020simclr,he2020moco}. 
They use contrastive learning to match positive pairs of images, that are transformations of each other; and keep apart negative pairs, that are different images of the batch. 
The challenges for these methods are (a) coverage, since it is impossible to cover the distribution of all the possible pairs and (b) mode collapse, when the features of all images become the same.
Non-contrastive methods avoid mode collapse in other ways\footnote{\cite{garrido2022duality} have later shown that these classes of methods are in fact very related.}.
Barlow Twins~\citep{zbontar2021barlow} uses a statistical prior of feature decorrelation to learn statistically independent features while preventing collapse.
BYOL~\citep{grill2020byol} uses an online network to predict the features of the teacher network, and does not use negative pairs.
In this work, we use DINO~\citep{caron2021dino}, which trains a student network to predict the features of the teacher network, in a similar way to BYOL.

%% file: chapter-1/sections/method.tex
\section{Watermarking with SSL networks}\label{chap1/section:0bit}

Our method adopts the framework of Vukoti\'c \etal\citep{vukotic2020classification}. 
We improve it by showing that networks build better watermarking features when trained with self-supervision, and by introducing data-augmentation at marking time. 
We also extend it to multi-bit watermarking.

\subsection{Using self-supervised networks as feature extractors}\label{chap1/subsec:ssl_as_feature_extractor}

\paragraph*{Motivation.}
We denote the image space by $\Ispace$ and the feature space by $\Fspace=\real^d$.
The feature extractor $\phi:\Ispace\to\Fspace$ must satisfy two properties: 
(1) geometric and valuemetric transformations on image $I$ should have minimal impact on feature $\phi(I)$,
(2) it should be possible to alter $\phi(I)$ by applying invisible perturbations to the image in order to embed the watermark signal.

We choose $\phi$ as a neural network pre-trained by self-supervised learning (SSL).
Our assumption is that SSL produces excellent marking spaces because 
(1) it explicitly trains features to be invariant to data augmentation; and 
(2) it does not suffer from the \textit{semantic collapse} of supervised classification, that gets rid off any information that is not necessary to assign classes~\citep{doersch2020crosstransformers}. 
From the recent SSL methods of the literature (contrastive learning~\citep{chen2020simclr,he2020moco}, statistical prior~\citep{zbontar2021barlow}, teacher/ student architecture~\citep{grill2020byol}), we select DINO~\citep{caron2021dino} for its training speed and content-based image retrieval performance.

\paragraph*{DINO pre-training.}
DINO~\citep{caron2021dino} pertains to the teacher/ student approach.
The teacher and the student share the same architecture.
Self distillation with no label trains the student network to match the outputs of the teacher
on different views of the same image. 
The student is updated by gradient descent while the teacher's parameters are updated as an exponential moving average of the student's parameters: $\theta_{t} \leftarrow \lambda \theta_{t} + (1-\lambda) \theta_{s}$, with $\lambda\lessapprox 1$.
The invariance of the features is ensured by random augmentations during the training: valuemetric (color jittering, Gaussian blur, solarization) and geometric (resized crops) transformations.
Furthermore, DINO encourages local-to-global correspondence by feeding only global views to the teacher while the student sees smaller crops.

\begin{figure}[bh]
    \centering
    \includegraphics[width=0.95\textwidth]{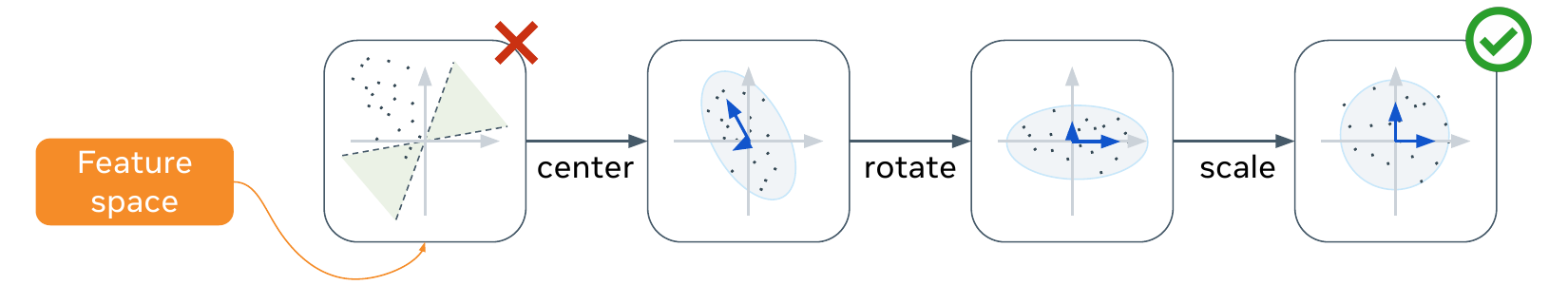}
    \caption{
        PCA-whitening. 
        The feature vectors are centered and decorrelated by a learned linear transformation.
        This is used to increase the performance of the watermarking method, and to ensure statistical guarantees in the zero-bit setup.
    }
    \label{chap1/fig:whitening}
\end{figure}

\paragraph*{Normalization layer.}\label{chap1/par:whitening} 
The watermark embedding must drive the feature to an arbitrary space region (defined by a secret key and the message to hide).
Therefore, it is essential that the features are not concentrated onto a manifold far away from this region. 
Moreover, for natural images, the features should be uniformly distributed on the hypersphere to provide for accurate false positive rates.
Although Wang \etal\citep{wang2020uniformity} show that contrastive learning optimizes the uniformity of the features on the hypersphere, it does not occur in practice with DINO.
To alleviate the issue, the output features are transformed by PCA-whitening (\aka, PCA-sphering). 
This learned linear transformation~\citep{jegou2012negative} outputs centered vectors with unit covariance of dimension $d=2048$.

\subsection{Embedding with back-propagation and augmentation}\label{chap1/subsec:embedding_algorithm}

The marking takes an original image $\Io$ and outputs a visually similar image $\Iw$.
In the image space $\Ispace$, the distortion is measured by $\Li:\Ispace\times\Ispace\to\real_+$. An example is the MSE:
$\Li(\Iw,\Io) = \|\Iw-\Io\|^2 / h/w$, but it could be replaced by perceptual image losses such as LPIPS~\citep{zhang2018unreasonable}.

In the feature space $\Fspace$, we define a region $\mathcal{D}$ which depends on a secret key (zero-bit and multi-bit setups) and the message to be hidden (only in multi-bit setup). 
Its definition is deferred to Sect.~\ref{chap1/sec:modulation} together with the loss $\Lw:\Fspace\to\real$ that captures how far away a feature $x\in\Fspace$ lies from $\mathcal{D}$. %
We also define a set $\mathcal T$ of augmentations, which include rotation, crops, blur, etc., each with a range of parameters.
$\Tr(I,t)\in\Ispace$ denotes the application of transformation $t\in\mathcal T$ to image $I$.

If the feature extractor is perfectly invariant, $\phi(\Tr(I,t)) \approx \phi(I) $ so $\phi(\Tr(I,t))$ lies inside $\mathcal{D}$ if $\phi(I)$ does. 
To ensure this, the watermarking uses data augmentation.
The losses $\Lw$ and $\Li$ are combined in:
\begin{equation}
\label{chap1/eq:global_loss}
    \L(I, \Io, t) := \lambda \Lw(\phi(\Tr(I,t))) + \Li(I,\Io).
\end{equation}

The term $\Lw$ aims to push the feature of \emph{any transformation} of $\Iw$ into $\mathcal D$, while the
term $\Li$ favors low distortion.
The training approach is typical for the adversarial attacks literature \citep{goodfellow2014adversarial, szegedy2013intriguing}: 
\begin{equation}
\label{chap1/eq:Watermarking}
    \Iw := \arg \min_{I\in \mathcal{C}(\Io)}
    \mathbb{E}_{t\sim \mathcal T} [\L(I, \Io, t)]
\end{equation}
where $\mathcal{C}(\Io)\subset\Ispace$ is the set of admissible images w.r.t. the original one. 
It is defined by two steps of normalization applied to the pixel-wise difference $\delta = I - \Io$:
(1) we apply a SSIM~\citep{wang2004ssim} heatmap attenuation, which scales $\delta$ pixel-wise to hide the information in perceptually less visible areas of the image;
(2) we set a minimum target PSNR and rescale $\delta$ if this target is exceeded.

The minimization is performed by stochastic gradient descent since the quality constraints, $\Tr$ and $\phi$ are differentiable w.r.t. the pixel values.
Stochasticity comes from the fact that
expectation $\mathbb{E}_{t\sim \mathcal T}$ is approximated by sampling according to a distribution over $\mathcal T$.
The final image is the rounded version of the update after $K$ iterations.

\subsection{Detection and decoding}\label{chap1/sec:modulation}

We consider two scenarios: zero-bit (detection only) and multi-bit watermarking (decoding the hidden message).  
Contrary to HiDDeN~\citep{zhu2018hidden} and followers, our decoding is mathematically sound.

\paragraph*{Zero-bit.}\label{chap1/hd:zero}
From a secret key $a\in\Fspace$ s.t. $\|a\|=1$, the detection region is the dual hypercone:
\begin{equation}
     \mathcal{D} := \left\{  x \in \R^d :  \abs{ x\T  a} > \norm{ x} \cos (\theta) \right\}.
\end{equation}
It is well grounded because the False Positive Rate (FPR) is given by:
\begin{align}
   \mathrm{FPR} :=&\; \mathbb{P}\left(\phi(I)\in \mathcal D \mid \text{``key } a \text{ uniformly distributed''}\right)\nonumber\\
            =&\; 1-I_{\cos^2(\theta)} \left(\frac{1}{2}, \frac{d-1}{2} \right)
   \label{chap1/eq:FPR}
\end{align}
where $I_\tau (\alpha, \beta)$ is the regularized Beta incomplete function (proof in App.~\ref{app/ssl-watermarking}).
The best embedding is obtained by increasing the following function under the distortion constraint:
\begin{equation}
   - \Lw(x) = (x^\top a)^2 -  \|x\|^2 \cos^2\theta.
\end{equation}
This quantity is negative when $x\notin\mathcal{D}$ and positive otherwise.
Cox~\etal originally called it the robustness estimate~\cite[Sec. 5.1.3]{cox2007digital}. 
This hypercone detector has been proved to be optimal under the asymptotical Gaussian setup~\citep{furon2019dualhypercone, merhav2008optimal}.

\paragraph*{Multi-bit}
We now assume that the message to be hidden is  $m = (m_1, ..., m_k) \in \{-1,1\}^k$. 
The decoder retrieves $\hat{m}=D(I)$.
Here, the secret key is a randomly sampled orthogonal family of carriers $a_1,...., a_k \in \R^d$. 
We modulate $m$ into the signs of the projection of the feature $\phi(I)$ against each of the carriers, so the decoder is: 
\begin{equation}
D(I) = \left[\mathrm{sign}\left(\phi(I)^\top a_1\right), ..., \mathrm{sign}\left(\phi(I)^\top a_k\right)\right].
\end{equation}

At marking time,  the functional is now defined as the hinge loss with margin $\mu\geq 0$ on the projections: 

\begin{equation}\label{chap1/eq:loss_multibit}
    \Lw(x) = \frac{1}{k} \sum_{i=1}^k \max \left( 0, \mu - (x^\top a_i).m_i \right).
\end{equation}

\subsection{Overview of the watermarking algorithm}

\begin{algorithm}
\caption{Watermarking algorithm}\label{chap1/alg:0bit}
\begin{algorithmic}
\item \textbf{Inputs}: $I_0 \in \mathcal{I}$, targeted PSNR in dB
    \item \quad if zero-bit: $ \textrm{FPR} \in [0,1]$, $a \in \R^d$
    \item \quad if multi-bit: $m \in \{-1,1\}^k$, $(a_i)_{i=1...k} \in \R^{k\times d}$
\item \textbf{Embedding}: 
    \item \quad if zero-bit: compute $\theta (\textrm{FPR})$
    \item \quad $I_w \gets I_0$
    \item \quad For $i=1, ..., n_{iter}$:
    \item \quad \quad $I_w \xleftarrow{\textrm{constraints}} I_w$ \Comment{impose constraints}
    \item \quad \quad $I_w \xleftarrow{t \sim \mathcal T} \textrm{Tr}(I_w, t)$ \Comment{sample \& apply augmentation}
    \item \quad \quad $x \gets \phi (I_w)$ \Comment{get feature of the image}
    \item \quad \quad $\mathcal{L} \gets \lambda \mathcal L_w(x) + \mathcal L_i(I_w, I_0) $ \Comment{compute loss}
    \item \quad \quad $I_w \gets I_w + \eta \times \mathrm{Adam}(\mathcal{L})$ \Comment{update the image}
    \item \quad Return $I_w$
\item \textbf{Detecting}: 
    \item \quad $x \gets \phi(I_m)$
    \item \quad if zero-bit:
    \item \quad \quad Detect if $x\in \mathcal D \Longleftrightarrow (x^Ta)^2-\norm{x}^2\cos^2\theta >0$
    \item \quad if multi-bit:
    \item \quad \quad Return $\left[\mathrm{sign}\left(x^\top a_1\right), ..., \mathrm{sign}\left(x^\top a_k\right)\right]$
\end{algorithmic}
\end{algorithm}

%% file: chapter-1/sections/experiments.tex
\newcommand{\rot}[1]{\rotatebox{60}{#1}}%

\section{Experiments and results}
\label{chap1/sec:experiments}

\subsection{Experimental setup and implementation details}\label{chap1/sec:exp_setup}

\paragraph*{Data.}
We evaluate our method on: 
1000 images of YFCC100M dataset~\citep{thomee2016yfcc100m} which is selected for the variety of its content, CLIC~\citep{2018clic} composed of 118 professional high-resolution images when comparing to~\citep{vukotic2020classification}, and 1000 images of MS-COCO~\citep{lin2014coco} composed of smaller images for comparison with~\citep{zhu2018hidden,luo2020distortion}.

\paragraph*{Backbone pre-training.}\label{chap1/par:backbone_pretraining} 
We use the ResNet-50 \citep{he2016deep} architecture as backbone model to extract features from its last convolutional layer ($d=2048$).
It is trained on ILSVRC2012~\citep{deng2009imagenet} without labels, using 200 epochs of DINO self-supervised learning with the default parameters~\citep{caron2021dino} and with rotation augmentation.
Models trained for classification come from the torchvision library~\citep{marcel2010torchvision}. 
The PCA-whitening is learned on 100k distinct images from YFCC (resp. COCO) when evaluating on YFCC and CLIC (resp. COCO).

\paragraph*{Embedding.}
We first set a desired FPR (which defines the hypercone angle $\theta$) and a target PSNR.
Watermarking~\eqref{chap1/eq:Watermarking} then uses the Adam optimizer~\citep{kingma2015adam} with learning rate $0.01$ over $100$ gradient descent iterations.
The weight in~\eqref{chap1/eq:global_loss} is set to $\lambda=1$ (zero-bit) or $\lambda=5\cdot 10^4$ (multi-bit). 
The margin of (\ref{chap1/eq:loss_multibit}) is set to $\mu=5$.

At each iteration, the preprocessing step performs the SSIM attenuation and clips the PSNR to the target. 
SSIM heatmaps are computed with $C_1=0.01^2$, $C_2=0.03^2$ and over $17\times 17$ tiles of the image's channels, then summed-up and clamped to be non negative, which generates a single heatmap per image. 
Then, a transformation $t$ is chosen randomly in $\mathcal T$ (identity, rotation, blur, crop or resize). 
The rotation angle $\alpha$ is sampled from a Von Mises distribution with $\mu=0$, $\kappa=1$ and divided by $2$. 
This generates angles in $\pi/2\times[-1,1]$ with a higher probability for small rotations, that are more frequent in practice. 
The crop and resize scales are chosen uniformly in $[0.2, 1.0]$. 
The crop aspect ratio is also randomly chosen between $3/4$ and $4/3$.
The blurring kernel size $b$ is randomly drawn from the odd numbers between $1$ and $15$ and $\sigma$ is set to $0.15 b+0.35$.
Finally, the image is flipped horizontally with probability $0.5$.

\paragraph*{Transformations.} 
\autoref{chap1/tab:transformations} presents the transformations used at pre-training, marking or evaluation stages. 
$p$ represents the cropping ratio in terms of area, $Q$ is the quality factor of the compression and $B$, $C$, $H$ are defined in~\citep{marcel2010torchvision}. 
``Meme format'' and ``Phone screenshot'' come from Augly~\citep{bitton2021augly}. 

\begin{table}[t]
    \centering
    \caption{Transformations used at pre-training, marking or evaluation stages.}\label{chap1/tab:transformations}
    \begin{tabular}{ l c|cc|cc }
        \toprule
                         &                  & \multicolumn{2}{c|}{Type} & \multicolumn{2}{c}{Used for}\\
        Transformations  & Parameter        & Geom. & Value   & Train & Mark\\ 
        \midrule
        Rotation         & angle $\alpha$    & \cmark & \xmark      & \cmark & \cmark \\
        Crop             & ratio $p$         & \cmark & \xmark      & \cmark & \cmark \\
        Resize           & scale $p$         & \cmark & \xmark      & \cmark & \cmark \\
        Gaussian Blur    & width $\sigma$    & \xmark & \cmark      & \cmark & \cmark \\
        Brightness       & $B$               & \xmark & \cmark      & \cmark & \xmark \\
        Contrast         & $C$               & \xmark & \cmark      & \cmark & \xmark \\
        Hue              & $H$               & \xmark & \cmark      & \cmark & \xmark \\
        JPEG             & quality $Q$       & \xmark & \cmark      & \xmark & \xmark \\
        Meme format      & -                 & \cmark & \cmark      & \xmark & \xmark \\
        Phone screenshot & -                 & \cmark & \cmark      & \xmark & \xmark \\
        \bottomrule
    \end{tabular}
\end{table}

\subsection{Qualitative results}\label{chap1/par:0bit_qualitative_results}

\begin{figure}[b!]
    \centering
    \includegraphics[width=0.8\linewidth]{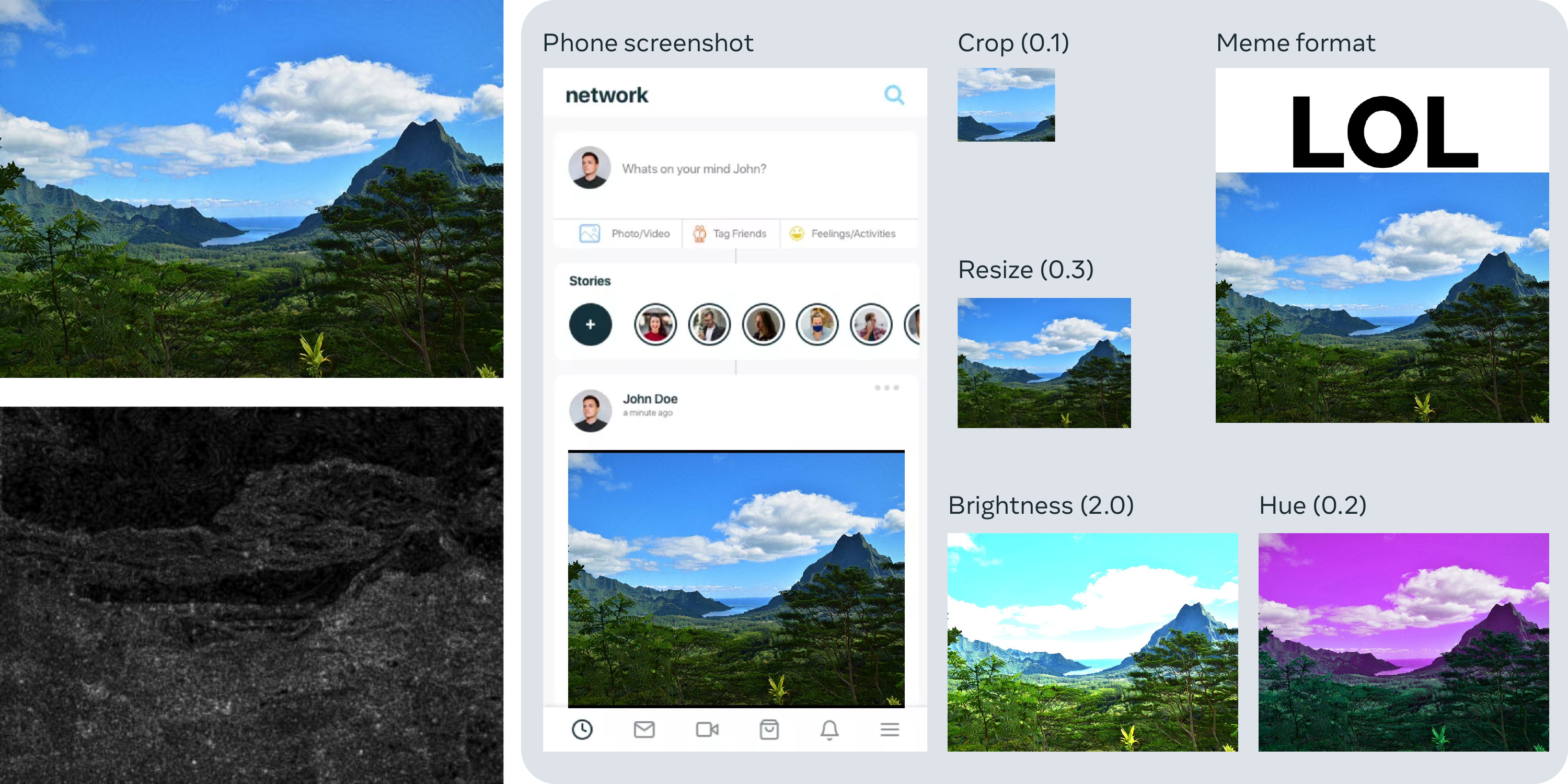}
    \caption{Example of an image ($800\times 600$) watermarked at PSNR 40~dB and FPR=$10^{-6}$, and some detected alterations. The black and white picture on the bottom left corner shows the scaled amplitude of the watermark signal.
    \label{chap1/fig:watermarked_imgs}}
\end{figure}

\begin{figure}[H]
    \centering
    \begin{subfigure}[b]{0.9\textwidth}
        \includegraphics[width=\textwidth]{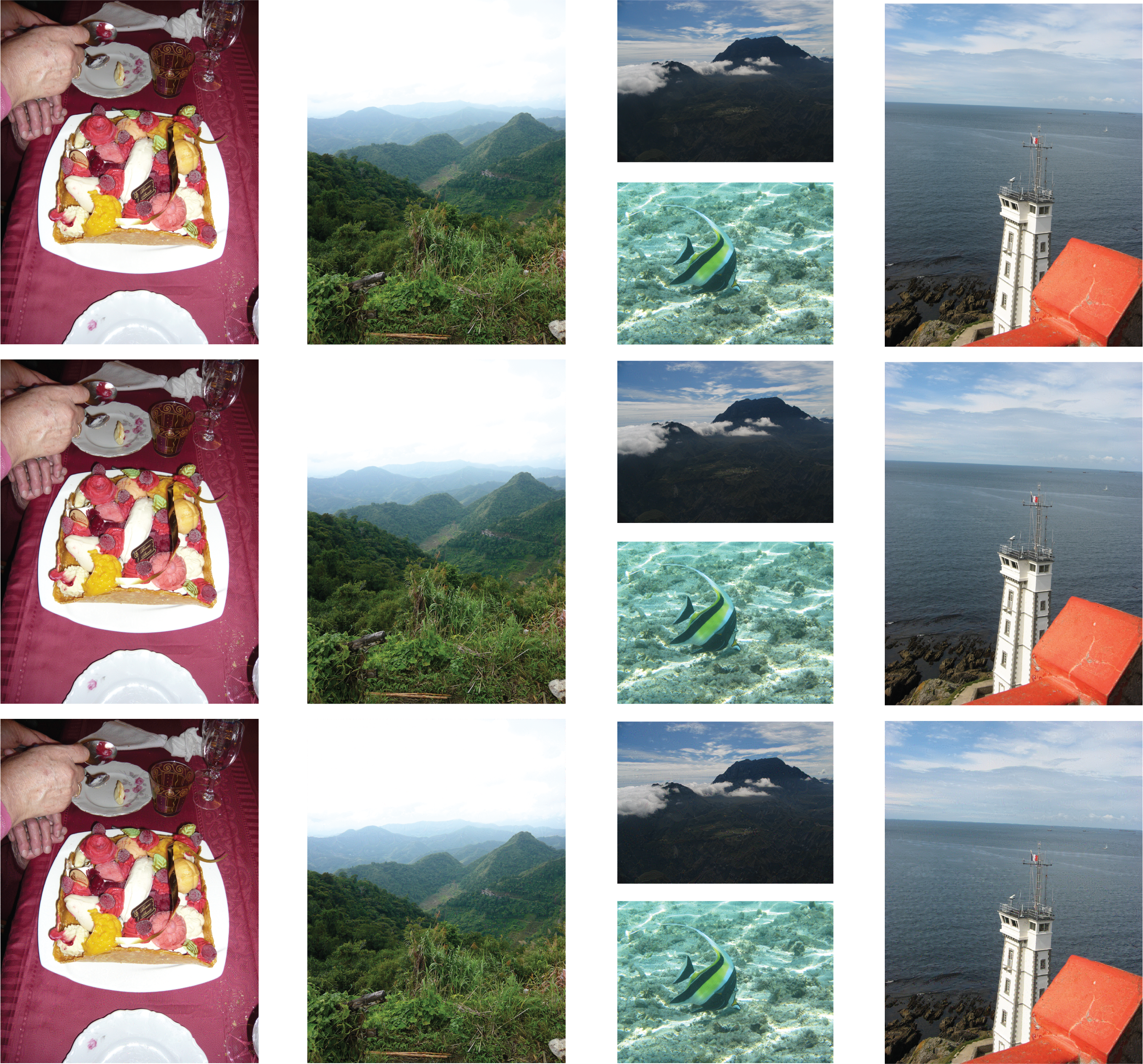}
        \caption{The watermark is added with our zero-bit watermarking method with an FPR of $10^{-6}$ and with different values for the PSNR: $52$dB (top row), $40$dB (middle row), $32$dB (bottom row)}
        \label{chap1/fig:holidays_0bit}
    \end{subfigure} \\[1em]
    \begin{subfigure}[b]{0.9\textwidth}
        \includegraphics[width=\textwidth]{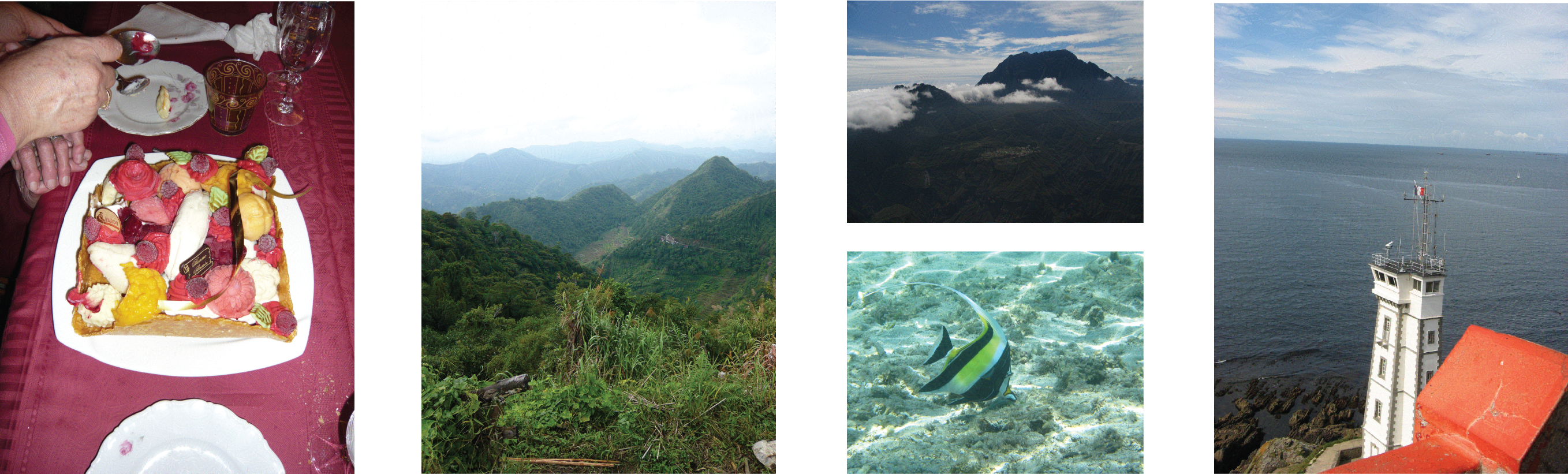}
        \caption{The watermark is added with our multi-bit watermarking method with a payload of $30$ bits with the PSNR at $32$dB.}
        \label{chap1/fig:holidays_multibit}
    \end{subfigure}
       \caption{Watermarked images from the Holidays dataset (resolution $\approx 2048 \times 1536$).}
       \label{chap1/fig:holidays}
\end{figure}

\begin{figure}[b!]
    \includegraphics[width=\textwidth]{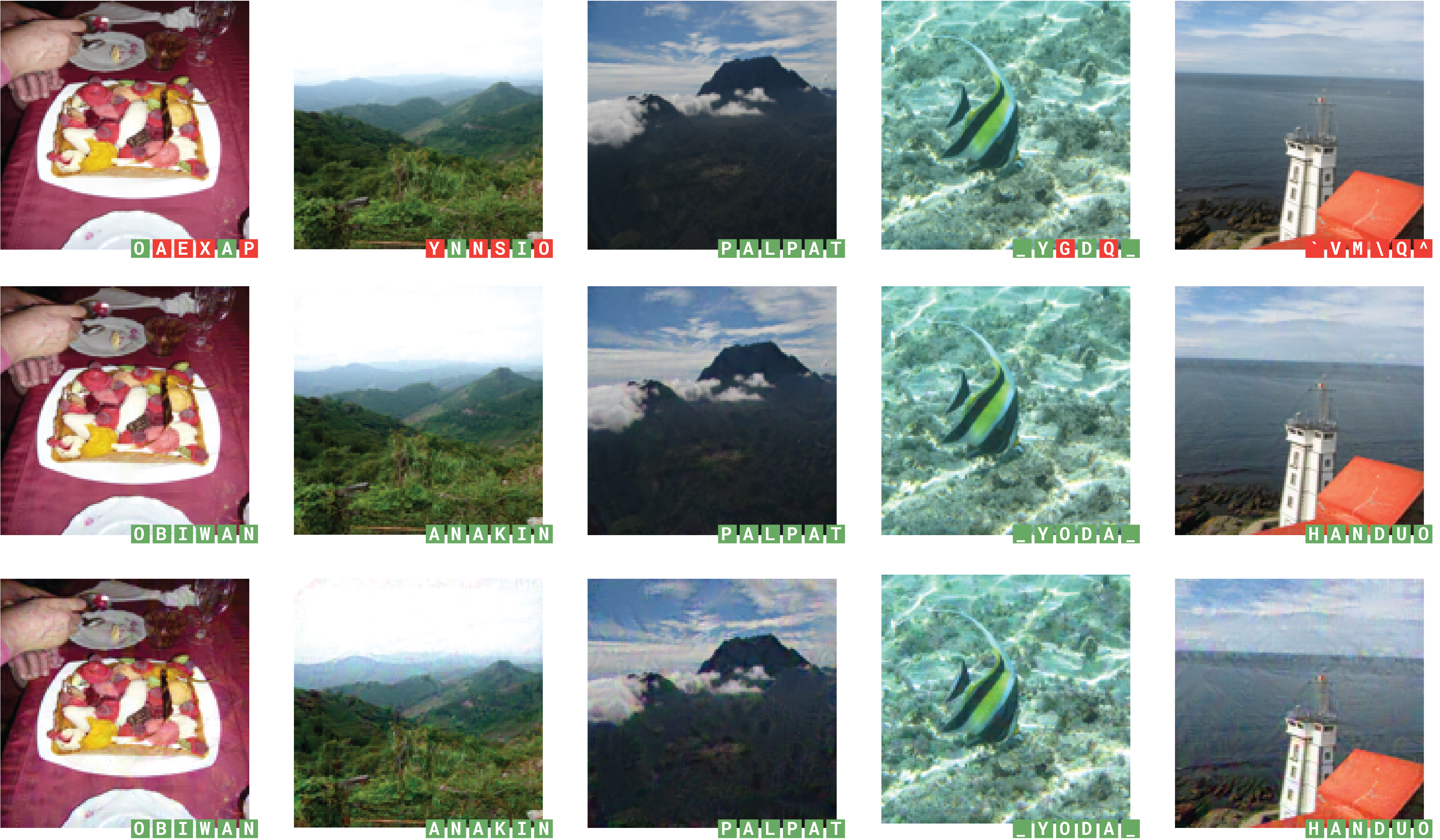}
    \caption{Watermarked images from the Inria Holidays dataset resized to $128 \times 128$. The watermark is added with our multi-bit watermarking method with a payload of $30$ bits and with different values for the target PSNR: $52$dB (top row), $40$dB (middle row), $32$dB (bottom row). 
    We use a 5-bits character encoding to encode and decode one message per image, and show the decoded messages for each image (without any transformation applied to the image). The higher the PSNR, the higher the decoding errors, and the less robust the decoding is to transformations.}
    \label{chap1/fig:holidays_resized_multibit}
\end{figure}

\autoref{chap1/fig:watermarked_imgs} presents an image watermarked at PSNR=40~dB and some detected alterations, as well as its pixel-wise difference w.r.t. the original image. 
The watermark is almost invisible even to the trained eye because it is added in the textured regions due to the perceptual SSIM normalization applied during watermarking.
We provide more examples of watermarked images in Fig.~\ref{chap1/fig:holidays_resized_multibit} and Fig.~\ref{chap1/fig:holidays}.

\subsection{Zero-bit watermarking}

\paragraph*{Comparison with the state of the art.} 
\autoref{chap1/tab:0bit} compares our method with~\citep{vukotic2020classification} on CLIC. 
In their setup, the FPR=$10^{-3}$ and PSNR must be $\geq 42$dB.
Overall, our method gives better results on CLIC than on YFCC because images have higher resolutions (hence more pixels can be used to convey the mark). We observe a strong improvement over \citep{vukotic2020classification}, especially for large rotations, crops and Gaussian blur where our method yields almost perfect detection over the 118 images. 

\begin{table}[t!]
    \centering
    \caption{TPR over various attacks, at FPR$=10^{-3}$ and PSNR$\geq 42$~dB.
    1\textsuperscript{st} setup: performance with SSL vs supervised ResNet-50 networks on YFCC. 
    2\textsuperscript{nd} setup: evaluation over CLIC.
    ($\star$) best results in~\citep{vukotic2020classification} (VGG-19 trained on hard attacks and RMAC aggregation), ($\star \star$) our implementation of \citep{vukotic2020classification} (with default pre-trained VGG-19).
    $\dagger$ denotes augmentations used at pre-training. 
    }
    \label{chap1/tab:0bit}
    \resizebox{1.0\linewidth}{!}{
    \begin{tabular}{ l | >{\columncolor{gray!10}}cc|>{\columncolor{gray!10}}ccc}
        \toprule
        \multicolumn{1}{c}{}&  \multicolumn{2}{c}{Setup 1: YFCC}    &  \multicolumn{3}{c}{Setup 2: CLIC} \\
        \cmidrule(lr){2-3} \cmidrule(lr){4-6}
        Transformations     & SSL              & Sup.               & Ours                        & \citep{vukotic2020classification} ($\star$) & \citep{vukotic2020classification} ($\star\star$) \\ \hline
        Identity            & 1.00$^\dagger$   & 1.00$^\dagger$     &   1.00$^\dagger$            & 1.0$^\dagger$            & 1.00$^\dagger$  \\
        Rotation (25)       & 0.97$^\dagger$   & 0.54$^\dagger$     &   \textbf{1.00}$^\dagger$   & $\approx 0.3$$^\dagger$ & 0.27$^\dagger$  \\
        Crop (0.5)          & 0.95$^\dagger$   & 0.79$^\dagger$     &   1.00$^\dagger$            & $\approx 0.1$$^\dagger$ & 1.00$^\dagger$  \\
        Crop (0.1)          & 0.39$^\dagger$   & 0.06$^\dagger$     &   \textbf{0.98}$^\dagger$   & $\approx 0.0$$^\dagger$ & 0.02$^\dagger$  \\
        Resize (0.7)        & 0.99$^\dagger$   & 0.85$^\dagger$     &   1.00$^\dagger$            & -                        & 1.00$^\dagger$  \\
        Blur (2.0)          & 0.99$^\dagger$   & 0.04               &   \textbf{1.00}$^\dagger$   & -                        & 0.25  \\
        JPEG (50)           & 0.81             & 0.20               &   0.97                      & $\approx 1.0$            & 0.96  \\
        Brightness (2.0)    & 0.94$^\dagger$   & 0.71               &   0.96$^\dagger$            & -                        & 0.99  \\
        Contrast (2.0)      & 0.96$^\dagger$   & 0.65               &   1.00$^\dagger$            & -                        & 1.00  \\
        Hue (0.25)          & 1.00$^\dagger$   & 0.46               &   1.00$^\dagger$            & -                        & 1.00  \\
        Meme                & 0.99             & 0.94               &   1.00                      & -                        & 0.98  \\
        Screenshot          & 0.76             & 0.18               &   \textbf{0.97}             & -                        & 0.86  \\
        \bottomrule
    \end{tabular}
    }
\end{table}

\begin{figure}[b!]
    \centering
    \begin{subfigure}[b]{0.95\linewidth}
        \includegraphics[width=\linewidth, trim={0 1.8em 0 0em},clip]{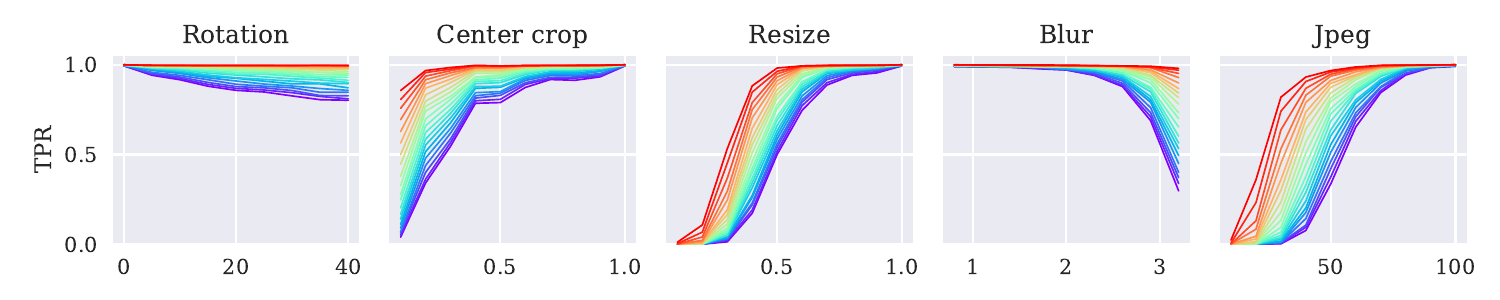}
    \end{subfigure}
    \begin{subfigure}[b]{0.95\linewidth}
        \includegraphics[width=\linewidth, trim={0 0em 0 2.em},clip]{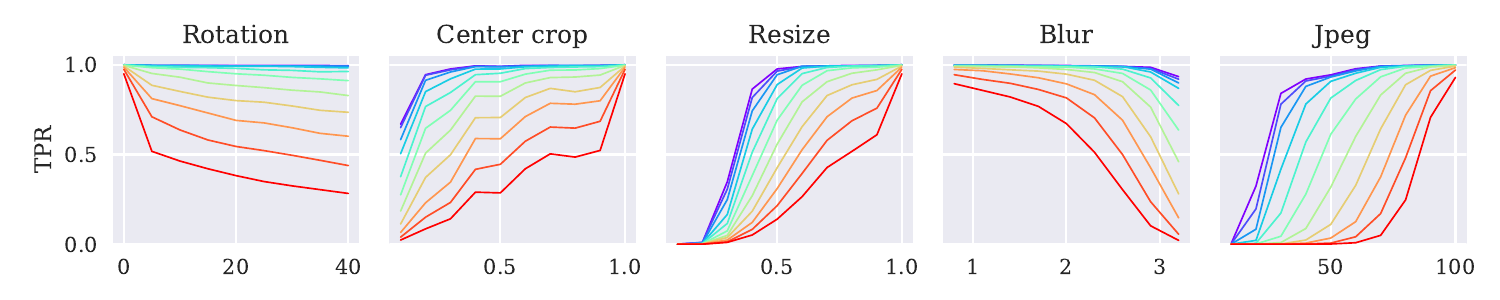}
    \end{subfigure}
    \caption{Robustness of the detection in the zero-bit setup against image transformations. \emph{Top:} PSNR set at 40~dB and FPR decreasing from \textcolor{BrickRed}{$10^{-2}$} (\textcolor{BrickRed}{red}) to \textcolor{Violet}{$10^{-12}$} (\textcolor{Violet}{blue}). \emph{Bottom:} FPR set at $10^{-6}$ and PSNR ranging from \textcolor{BrickRed}{52\,dB} to \textcolor{Violet}{32\,dB}.
    }
    \label{chap1/fig:0bit_trade_offs}
\end{figure}

\paragraph*{Trade-offs.}
The hypercone angle $\theta$ in~\eqref{chap1/eq:FPR} is given by the target FPR. %
A higher FPR implies a wider angle, making the method robust against more severe attacks, at the cost of detecting more false positives. 
The FPR is set to $10^{-6}$ in further experiments.
Large-scale applications usually operate at low FPR to avoid human verification.
As a sanity check we run detection on 100k natural images from YFCC, none of which are found to be marked. 
Similarly, there is only one false positive out of the 1,281,167 images of ImageNet.
Another trade-off lies in the imperceptibility, since allowing greater distortions (lower PSNR) improves the robustness.
It is illustrated in Fig.~\ref{chap1/fig:0bit_trade_offs}.

\begin{figure}[b!]
    \centering
    \includegraphics[width=0.9\linewidth,clip,trim={0 0em 15pt 0em}]{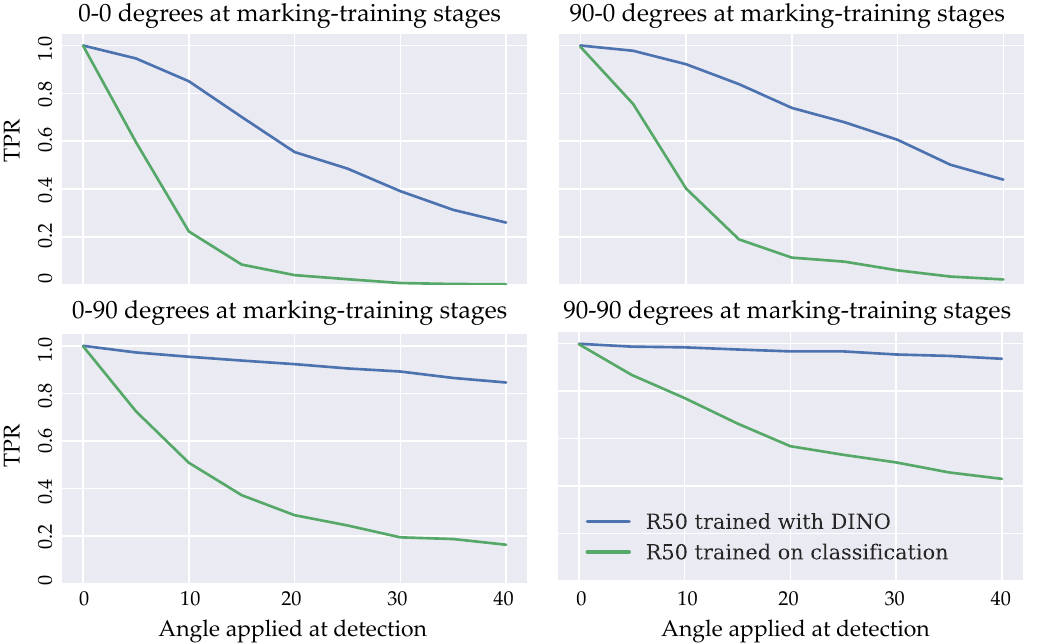}
    \caption{Robustness against rotation. Each row (column) represents different $\pm$amplitude of the rotation at training (resp. marking~\eqref{chap1/fig:0bit_rotations}). 
    }
    \label{chap1/fig:0bit_rotations}
\end{figure}

\paragraph*{Ablation studies.}\label{chap1/par:0bit_quantitative_results}
We showcase the influence of self supervision at pre-training and of augmentation at marking time.
The performance measure is the True Positive Rate (TPR), at a target PSNR=40\,dB, and FPR=$10^{-6}$.
Fig.~\ref{chap1/fig:0bit_rotations} evaluates the robustness for the specific case of the rotation. 
Rotation augmentation is needed both at pre-training and marking stages to achieve high robustness against it.
Comparison on a wider range of transformations is given in Tab. \ref{chap1/tab:0bit}.

\subsection{Multi-bit data hiding}\label{chap1/sec:multibit_exp}

\paragraph*{Quantitative results.}\label{chap1/par:multibit_quantitative_results}
We evaluate the method on YFCC, with  a target PSNR of 40~dB and a payload $k$ of $30$ random bits as in~\citep{zhu2018hidden,luo2020distortion}.
\autoref{chap1/tab:multibit_30} presents the Bit and Word Error Rate (BER and WER) over various attacks. The decoding achieves low rates over a wide range of geometric (rotation, crops, resize, etc.) and valuemetric (brightness, hue, contrast, etc.) attacks.
Rotation and Gaussian blur are particularly harmless since they are seen both at pre-training and at marking time.
Some images are harder to mark, which can be observed statistically on the empirical WER reported in Tab.~\ref{chap1/tab:multibit_30}. 
If all images were as difficult to mark, then BER would be equal for all images, and $\mathrm{WER} =1- (1-\mathrm{BER})^{k}$. 
Yet, the reported WER are significantly lower: \eg, for Brightness, $\mathrm{WER} = 0.607 < 1-(1-0.087)^{30} = 0.935$. 
Empirically, we see that images with little texture are harder to watermark than others, due to the SSIM normalization.
In practice, ECC can be used to achieve lower WERs.

\begin{table}[t]
    \centering
    \caption{\makebox{BER and WER (\%) for $30$-bits encoding at PSNR~40dB.}}
    \label{chap1/tab:multibit_30}
    \resizebox{1.0\linewidth}{!}{
    \begingroup
        \setlength{\tabcolsep}{3pt}
            \begin{tabular}{ c| *{12}{c}}
            \multicolumn{1}{c}{\rot{Transform.}} & \rot{Identity} & \rot{Rot. (25)} & \rot{Crop (0.5)} & \rot{Crop (0.1)} & \rot{Res. (0.7)} & \rot{Blur (2.0)} & \rot{JPEG (50)} & \rot{Bright. (2.0)} & \rot{Contr. (2.0)} & \rot{Hue (0.25)} & \rot{Meme} & \rot{Screenshot} \\ \midrule
            BER  & 0.1   & 3.3    & 4.8    & 28.9   & 2.1    & 0.5  & 20.8   & 8.7   & 8.4   & 2.5   & 6.4    & 23.9  \\
            WER  & 0.7   & 43.4   & 58.3   & 100    & 29.1   & 4.6  & 98.9   & 60.7  & 62.6  & 31.6  & 75.9   & 100   \\
            \bottomrule
    \end{tabular}
    \endgroup
    }
\end{table}

\soutx{\begin{figure}[b!]
    \centering
    \includegraphics[width=\textwidth]{chapter-1/figs/multibit_numbits.pdf}
    \caption{Illustration of the Capacity/Robustness trade-off. The length of the message increases from $10$ (purple) to $100$ (red): short messages are more robust to transformations than long ones.
    The target PSNR is set to $40$.}
    \label{chap1/fig:multibit_numbits}
\end{figure}}

\soutx{\begin{figure}[b!]
    \centering
    \includegraphics[width=\textwidth]{chapter-1/figs/multibit_psnrs.pdf}
    \caption{Illustration of the Quality/Robustness trade-off. The PSNR of the watermarked images increases from $32$ (purple) to $52$ (red): better robustness is achieved for lower quality images. The length of the message is set to $30$.}
    \label{chap1/fig:multibit_psnrs}
\end{figure}}

\paragraph*{Qualitative results.}
We notice that the watermark is perceptually less visible for multi-bit than for zero-bit watermarking at a fixed PSNR. 
Our explanation is that the energy put into the image feature is more spread-out across carriers in the multi-bit setup than on the zero-bit one where the feature is pushed as much as possible towards a single carrier.

\begin{table}[t!]
    \centering
    \caption{Comparison of BER. The first row uses original resolutions of COCO, while the others use a resized version (to $128 \times 128$). Results for \citep{zhu2018hidden, luo2020distortion} come from \citep{luo2020distortion}. $\dagger$ denotes transformations used in the watermarking process.
    }
    \vspace{-0.3cm}
    \label{chap1/tab:multibit_coco}
    \resizebox{1.0\linewidth}{!}{
    \begin{tabular}{ c|llllll}
        \multicolumn{1}{c}{Transformation} & \rotatebox{45}{Identity} & \rotatebox{45}{JPEG (50)} & \rotatebox{45}{Blur (1.0)} & \rotatebox{45}{Crop (0.1)}  & \rotatebox{45}{Resize (0.7)} & \rotatebox{45}{Hue (0.2)} \\ \midrule
        \rowcolor{Orchid!20} Ours                      & 0.00$^\dagger$ & 0.04           & 0.00$^\dagger$          & 0.18$^\dagger$            & 0.00$^\dagger$ & 0.03          \\ \midrule
        \rowcolor{gray!10} Ours, $128 \times 128$                          & 0.00$^\dagger$ & 0.16  & 0.01$^\dagger$ & 0.45$^\dagger$            & 0.18$^\dagger$ & 0.06 \\
        HiDDeN~\citep{zhu2018hidden}                     & 0.00$^\dagger$ & 0.23$^\dagger$ & 0.01$^\dagger$ & 0.00$^\dagger$   & 0.15           & 0.29          \\
        Dist. Agnostic~\citep{luo2020distortion}         & 0.00$^\dagger$ & 0.18$^\dagger$ & 0.07$^\dagger$          & 0.02$^\dagger$            & 0.12  & 0.06 \\
        \bottomrule
    \end{tabular}
    }
\end{table}

\paragraph*{Comparison with the state of the art.}
\autoref{chap1/tab:multibit_coco} compares against two deep data hiding methods~\citep{zhu2018hidden,luo2020distortion} using their setting: a payload of $30$ bits, a target PSNR of 33dB, over $1000$ images from COCO resized to $128 \times 128$. 
Overall, our method gives comparable results to the state of the art, except for the center crop transform where it fails to achieve high bit accuracies. 
Note also that the resize operation is not used as noise layer in neither of~\citep{zhu2018hidden,luo2020distortion}, which means that our method should have the advantage.
In contrast, while these methods are trained to be robust to JPEG compression with a differentiable approximation, our method achieves similar performance without specific training.

Furthermore, our method easily scales to higher resolution images, where it achieves lower BER for a fixed payload. 
We assume that \citep{zhu2018hidden,luo2020distortion} also scales but at the cost of a specific training, or at least a fine-tuning, for a given resolution. 
This training is more computationally expensive since the message is repeated at each pixel~\citep{zhu2018hidden}. 
It also needs a smaller batch-size to operate larger images, and new hyperparameters values.

%% file: chapter-1/sections/conclusion.tex
\section{Conclusion}

This chapter proposes a way to robustly and invisibly embed information into digital images, by watermarking onto latent spaces of off-the-shelf self-supervised networks. 
By incorporating data augmentation and constraints into the marking process, our zero-bit watermarking method greatly improves performance over the baseline \citep{vukotic2020classification}.
It is robust against a wide range of transformations while keeping high fidelity with regards to the original images, and ensuring a very low false positive rate for the detection.
When we extend the method to multi-bit watermarking, we obtain promising results, comparable to the state-of-the-art in deep data hiding, and even better with regards to some transformations of the image (\eg, JPEG compression or blur).

Most interestingly, networks trained with self-supervision naturally generate excellent watermarking spaces, without being explicitly trained to do so. 
They allow for a wide range of trade-offs between robustness and quality, and can be used at any resolution.
This makes them well-suited for real-life applications like content moderation on social networks, where the watermarking must work on images of varying resolution and under different operating points.
However, watermarking images with our method is expensive since it is not a single pass forward. 
On the opposite, more recent data hiding schemes like the one used in Chapters \ref{chapter:stable-signature} and \ref{chapter:audioseal}, based on end-to-end architectures for the most part, only need one forward pass to watermark images (at the cost of an expensive one-time-training process).
It is also worth noting that using foundational open-weights models opens the door to new adversarial attacks, as shown by~\citep{kinakh2024evaluation}.

%% file: chapter-2/main.tex
\chapter{Active Image Indexing}
\label{chapter:active-indexing}

This chapter is based on the paper \fullcite{fernandez2022active}.

Image copy detection and retrieval from large databases are particularly relevant for content moderation. 
They allow to automatically detect and remove an important fraction of harmful content.
This enables human moderators to focus their efforts on new instances of problematic material rather than repeatedly reviewing variations of already-flagged one.
This chapter improves the robustness of image copy detection with \emph{active indexing}, which draws inspiration from watermarking and the previous chapter to alter the image in a way that improves its retrieval.
Retrieval systems often leverage two components. 
First, a neural network maps an image to a vector representation, that is relatively robust to various transformations of the image. 
Second, an efficient but approximate similarity search algorithm trades scalability (size and speed) against quality of the search, thereby introducing a source of error. 
\emph{Active indexing} optimizes the interplay of these two components. 
We reduce the quantization loss of a given image representation by making imperceptible changes to the image before its release. 
The loss is back-propagated through the deep neural network back to the image, under perceptual constraints. These modifications make the image more retrievable. 
Our experiments show that the retrieval and copy detection of activated images is significantly improved. For instance, activation improves by $+40\%$ the Recall1@1 on various image transformations, and for several popular indexing structures based on product quantization and locality sensitivity hashing.
Code is available at \url{github.com/facebookresearch/active_indexing}.

\newpage
\input{chapter-2/sections/1-intro.tex}
\input{chapter-2/sections/2-related.tex}
\input{chapter-2/sections/3-method.tex}

\input{chapter-2/sections/5-analyses.tex}

\input{chapter-2/sections/4-results.tex}

\input{chapter-2/sections/6-conclusion.tex}

%% file: chapter-2/sections/1-intro.tex
\section{Introduction}\label{chap2/sec:intro}

\begin{figure}[t]
    \centering
    \includegraphics[width=1.0\textwidth]{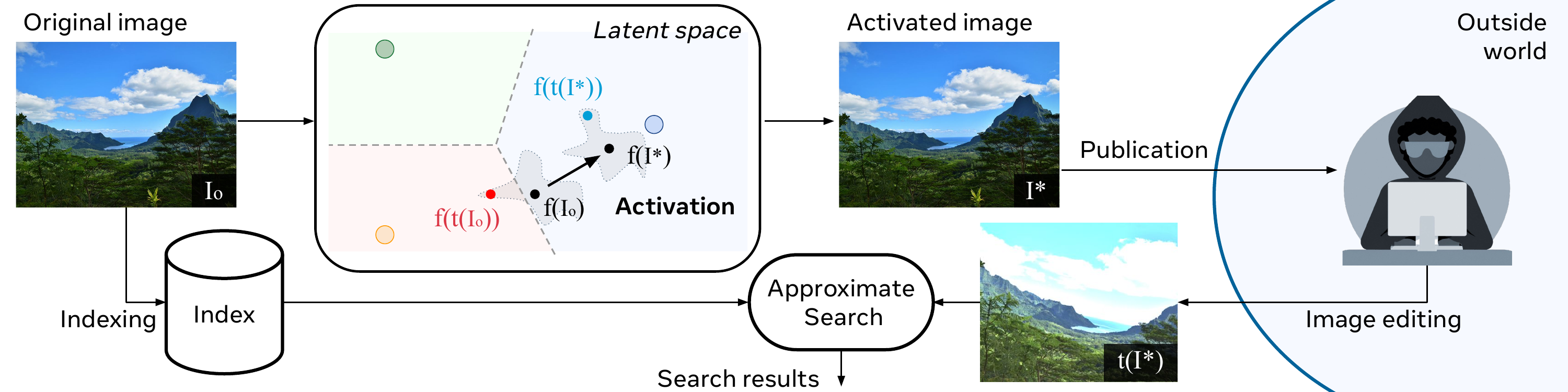}
    \caption{ 
    Overview of the method and 
        latent space representation. 
        We start from an original image $I_o$ that can be edited $t(\cdot)$ in various ways: its feature extraction $f(t(I_o))$ spawns the shaded region in the embedding space.
        The edited versions should be recoverable by nearest neighbor search on quantized representations.
        In the regular (non-active) case, $f(I_o)$ is quantized by the index as \includegraphics[width=0.5em]{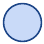}. 
        When the image is edited, $t(I_o)$ switches cells and the closest neighbor returned by the index is the wrong one \includegraphics[width=0.5em]{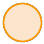}.
        In active indexing: $I_o$ is modified in an imperceptible way to generate $I^\acti$ such that $f(I^\acti)$ is further away from the boundary. 
        When edited copies $f(t(I^\acti))$ are queried, retrieval errors are significantly reduced.
    \label{chap2/fig:fig1}}
\end{figure}

The traceability of images on a media sharing platform is a challenge: 
they are widely used, easily edited and disseminated both inside and outside the platform.
In this chapter, we tackle the corresponding task of image \gls*{copy detection} (ICD), \ie, finding whether an image already exists in the database; and if so, give back its identifier.
ICD methods power reverse search engines, photography service providers checking copyrights, or media platforms moderating and tracking down malicious content (\eg, PhotoDNA~\citep{photodna} or NeuralHash~\citep{apple2021csamdetection}).
Image identification systems have to be robust to identify images that are edited (cropping, colorimetric change, JPEG compression \ldots) after their release~\citep{douze2021disc, wang2022benchmark}. 

The common approach for content-based image retrieval reduces images to high dimensional vectors, referred to as \emph{representations}. 
Early representations used for retrieval were hand-crafted features such as color histograms~\citep{swain1991color}, GIST~\citep{oliva2001modeling}, or Fisher Vectors~\citep{perronnin2010large}.
As of now, a large body of work on self-supervised learning focuses on producing discriminative representations with deep neural networks, which has inspired recent ICD systems. 
In fact, \emph{all} submissions to the NeurIPS2021 Image Similarity challenge~\citep{papakipos2022results} exploit neural networks.
They are trained to provide invariance to potential image transformations, akin to data augmentation in self-supervised learning.  

Scalability is another key requirement of image similarity search: searching must be fast on large-scale databases, which exhaustive vector comparisons cannot do.
In practice, ICD engines like \Gls*{FAISS} leverage approximate neighbor search algorithms, that trade search accuracy against scalability.
They speed up the search by \emph{not} computing the exact distance between all representations in the dataset~\citep{johnson2019faiss, guo2020scann}.
First they lower the number of scored items by partitioning the representation space, and evaluate the distances of only a few subsets. 
Second, they reduce the computational cost of similarity evaluation with quantization or binarization.
These mechanisms make indexing methods subject to the curse of dimensionality.
In particular, in high-dimensional spaces, vector representations lie close to boundaries of the partition~\citep{bohm2001searching}.
Since edited versions of an original image have noisy vector representations, they sometimes fall into different subsets or are not quantized the same way by the index. 
All in all, it makes approximate similarity search very sensitive to perturbations of the edited image representations, which causes images to evade detection.

In this chapter, we introduce a method that improves similarity search on large databases, provided that the platform or photo provider can modify the images before their release, as in watermarking (see Fig.~\ref{chap2/fig:fig1}). 
We put the popular saying ``attack is the best form of defense'' into practice by applying image perturbations and drawing inspiration from adversarial attacks. 
Indeed, representations produced with neural networks are subject to \emph{adversarial examples}~\citep{szegedy2013intriguing}:
small perturbations of the input image can lead to very different vector representations, making it possible to 
create adversarial queries that fool image retrieval systems~\citep{liu2019whos, tolias2019targeted,dolhansky2020adversarial}.
In contrast, we modify an image to make it \emph{more} indexing friendly.  
With minimal changes in the image domain, the image representation is pushed towards the center of the indexing partition, rising the odds that edited versions will remain in the same subset. 
This property is obtained by minimizing an indexation loss by gradient descent back to the image pixels, like for adversarial examples.
For indexing structures based on product quantization~\citep{jegou2010pq}, this strategy amounts to pushing the representation closer to its quantized codeword, in which case the indexation loss is simply measured by the reconstruction error.
Since the image quality is an important constraint here, the perturbation is shaped by perceptual filters to remain invisible to the human eye. 

Our contributions are:
\begin{itemize}[leftmargin=1cm,itemsep=0cm,topsep=-0.1cm]
    \item a new approach to improve ICD and retrieval, when images can be changed before release;
    \item an adversarial image optimization scheme that adds minimal perceptual perturbations to images in order to reduce reconstruction errors, and improve vector representation for indexing;
    \item experimental evidence that the method significantly improves index performance.
\end{itemize}

%% file: chapter-2/sections/2-related.tex
\section{Related work}\label{chap2/sec:related}

\paragraph*{Image watermarking} is an alternative technology for ICD. 
Our method bridges indexing and watermarking, where the image is modified before publication.
Regarding retrieval performance, active indexing is more robust than watermarking.
Indeed, the embedded signal reinforces the structure naturally present in the original image, whereas watermarking has to hide a large secret keyed signal independent of the original feature.
\autoref{chap2/sec:watermarking} provides a more thorough discussion and experimental results comparing indexing and watermarking, while \autoref{chap0/sec:watermarking} provides broader related works on image watermarking.

\paragraph*{Active \gls*{fingerprinting}} is more related to our work.
As far as we know, this concept was invented by \citet{voloshynovskiy2012active}.
They consider that the image $I\in \R^N$ is mapped to $x \in \R^N$ by an invertible transform $W$ such that $WW^\top$.
The binary fingerprint is obtained by taking the sign of the projections of $x$ against a set of vectors $b_1,., b_L \in \R^N$ (à la LSH).
Then, they change $x$ to strengthen the amplitude of these projections so that their signs become more robust to noise.
They recover $I^\acti$ with $W^\top$.
This scheme is applied to image patches in~\citep{7533094} where the performance is measured as a bit error rate after JPEG compression. 
This chapter adapts this idea from fingerprinting to indexing, with modern deep learning representations and state-of-the-art indexing techniques. 
The range of transformations is also much broader and includes geometric transforms.

%% file: chapter-2/sections/3-method.tex
\newcommand{\nprobe}{{k'}}
\def\qivf{q_\mathrm{c}}
\newcommand{\qcompressed}{q_\mathrm{f}}

\section[Preliminaries: representation learning and indexing]{Preliminaries: representation learning \\and indexing}

For the sake of simplicity, the exposure focuses on image representations from SSCD networks~\citep{pizzi2022sscd} and the indexing technique IVF-PQ~\citep{jegou2010pq}, since both are typically used for ICD.
Extensions to other methods can be found in Sec.~\ref{chap2/sec:generalization}.

\subsection{Deep descriptor learning}

Metric embedding learning aims to learn a mapping $f: \R^{c\times h\times w} \to \R^d$, such that measuring the similarity between images $I$ and $I'$ amounts to computing the distance $\norm{f(I) - f(I')}$. 
In recent works, $f$ is typically a neural network trained with self-supervision on raw data to learn metrically meaningful representations.
Methods include contrastive learning~\citep{chen2020simclr}, self-distillation~\citep{grill2020bootstrap, caron2021dino}, or masking random patches of images~\citep{he2022masked, assran2022masked}.
In particular, SSCD~\citep{pizzi2022sscd} is a training method specialized for ICD.
It employs the contrastive self-supervised method SimCLR~\citep{chen2020simclr} and entropy regularization~\citep{sablayrolles2018catalyser} to improve the distribution of the representations.
We provide more insights on these methods in Sec.~\ref{chap1/par:ssl}.

\subsection{\Gls*{indexing}}
Given a dataset $\mathcal X = \{x_i\}_{i=1}^{n}\subset \R^d$ of $d$-dimensional vector representations extracted from $n$ images and a query vector $x_q$, we consider the indexing task that addresses the problem: 
\begin{align}
    x^* := \mathop{\mathrm{argmin}}_{x \in \mathcal X} 
    \;  \norm{x - x_q}.  
\end{align}
This exact nearest neighbor search is not tractable over large-scale databases.
Approximate search algorithms lower the amount of scored items thanks to space partitioning and/or accelerate the computations of distances thanks to quantization and pre-computation. 

\paragraph{Space partitioning and cell-probe algorithms.}
As a first approximation, 
nearest neighbors are sought only within a fraction of $\mathcal{X}$:
at indexing time, $\mathcal{X}$ is partitioned into $\mathcal X = \bigcup_{i=1}^{b} \mathcal{X}_i$.
At search time, an algorithm $Q: \R^d \to \{1,..,b\}^\nprobe$ determines a subset of $\nprobe$ buckets in which to search, such that $\nprobe=|Q(x_q)| \ll b$, yielding the approximation: 
\begin{align}
    \mathop{\mathrm{argmin}}_{x \in \mathcal X} \; \norm{x-x_q}
    \approx 
    \mathop{\mathrm{argmin}}_{x \in \bigcup_{i\in Q(x_q)} \mathcal{X}_i} \; \norm{x-x_q}.
\end{align}
A well known partition is the KD-tree~\citep{bentley1975kdtree} that divides the space along predetermined directions.
Subsequently, locality sensitive hashing (LSH)~\citep{indyk1998lsh, gionis1999lsh} and derivative~\citep{datar2004lsh,pauleve2010locality} employ various hash functions for bucket assignment, which implicitly partitions the space.

We focus on the popular clustering and Inverted Files methods~\citep{sivic2003video}, herein denoted by IVF. 
They employ a codebook $\mathcal{C} = \{c_i\}_{i=1}^{k}\subset\R^d$ of $k$ centroids (also called ``visual words'' in a local descriptor context), for instance learned with k-means over a training set of representations. 
Then, $Q$ associates $x$ to its nearest centroid $\qivf(x)$ such that the induced partition is the set of the $k$ Voronoï cells.
When indexing $x$, the IVF stores $x$ in the bucket associated with $c_i=\qivf(x)$.
When querying $x_q$, IVF searches only the $\nprobe$ buckets associated to centroids $c_i$ nearest to $x_q$.

\paragraph{Efficient metric computation and product quantization.} 
Another approximation comes from compressed-domain distance estimation. 
Vector Quantization (VQ) maps a representation $x \in \mathbb{R}^d$ to a codeword $\qcompressed(x) \in \mathcal{C} = \{C_i\}_{i=1}^{K}$.
The function $\qcompressed$ is often referred to a \emph{quantizer} and $C_i$ as a \emph{reproduction value}.
The vector $x$ is then stored as an integer in $\{1, .., K\}$ corresponding to $\qcompressed(x)$.
The distance between $x$ and query $x_q$ is approximated by $\norm{\qcompressed(x) - x_q}$, which is an ``asymmetric'' distance computation (ADC) because the query is not compressed. 
This leads to: 
\begin{align}
    \mathop{\mathrm{argmin}}_{x \in \mathcal X} \;  \norm{x-x_q}  
    \approx 
    \mathop{\mathrm{argmin}}_{x \in \mathcal X} \;  \norm{\qcompressed(x)- x_q} . 
\end{align}
Binary quantizers (\aka, sketches, \citep{charikar2002similarity} lead to efficient computations but inaccurate distance estimates~\citep{weiss2008spectral}.
Product Quantization (PQ)~\citep{jegou2010pq} or derivatives \citep{ge2013optimized} offer better estimates.
In PQ, a vector $x\in \R^d$ is split into $m$ subvectors in $\R^{d/m}$: $x=(x^1, \ldots, x^m)$.
The product quantizer then quantizes the subvectors: $\qcompressed: x \mapsto (q^1(x^1), \ldots, q^m(x^m))$. 
If each subquantizer $q^j$ has $K_s$ reproduction values, the resulting quantizer $\qcompressed$ has a high  $K=(K_s)^m$. 
The squared distance estimate is decomposed as:
\begin{align}
    \norm{\qcompressed(x)-x_q}^2 = \sum_{j=1}^m \norm{q^j(x^j)-x_q^j}^2.
\end{align}
This is efficient since $x$ is stored by the index as $\qcompressed (x)$ which has $m\log_2 K_s$ bits, and since summands can be precomputed without requiring decompression at search time.

\section{Active indexing}\label{section:method}

Active indexing takes as input an image $I_o$, adds the image representation to the index and outputs an activated image $I^\acti$ with better traceability properties for the index.
It makes the feature representation produced by the neural network more compliant with the indexing structure. %
The activated image is the one that is disseminated on the platform, therefore the alteration must not degrade the perceived quality of the image.

Images are activated by an optimization on their pixels. 
The general optimization problem reads:
\vspace*{-1em}
\begin{align}
    I^\acti := \mathop{\mathrm{argmin}}_{I \in \mathcal{C}(I_o)} 
    \; \mathcal{L}\left(I;I_o\right),
    \label{eq:active_image}
\end{align}
where $\mathcal{L}$ is an indexation loss dependent on the indexing structure, $\mathcal{C}(I_o)$ is the set of images perceptually close to $I_o$.
We provide and overview of the active image indexing optimization in Alg.~\ref{alg:1} and in Fig.~\ref{chap2/fig:fig1}.

\begin{wrapfigure}{R}{0.45\textwidth}
\vspace{-0.7cm}
\resizebox{1.0\linewidth}{!}{
\begin{minipage}{0.5\textwidth}
    \begin{algorithm}[H]
    \caption{Active indexing for IVF-PQ}
    \label{alg:1}
        \begin{algorithmic}
            \State \textbf{Input}: $I_o$: original image; $f$: feature extractor;
            \State Add $x_o = f(I_o)$ to Index, get $q(x_o)$;
            \State Initialize $\delta_0 = 0_{(c\times h\times w)}$;
            \For{$t = 0, ..., N-1$}
            \State $I_t \gets I_o + \alpha \,.\, H_{\mathrm{JND}}(I_o) \odot \mathrm{tanh}(\delta_t)$ 
            \State $x_{t}\gets f(I_{t})$
            \State $\mathcal{L} \gets \Lf(x_{t}, q(x_o)) + \lambda \Li(\delta_t)$
            \State $\delta_{t+1} \gets \delta_t + \eta \times \mathrm{Adam}(\mathcal{L})$
            \EndFor
            \State \textbf{Output}: $I^\acti=I_N$ activated image 
        \end{algorithmic}
    \end{algorithm}
\end{minipage}
}
\end{wrapfigure}

\subsection{Image optimization dedicated to IVF-PQ (``activation'')}

The indexing structure IVF-PQ involves a coarse quantizer $\qivf$ built with k-means clustering for space partitioning, and a fine product quantizer $\qcompressed$ on the residual vectors, such that a vector $x \in \R^d$ is approximated by $q(x) = \qivf(x) + \qcompressed\left( x-\qivf(x) \right)$.

We solve the optimization problem~\eqref{eq:active_image} by iterative gradient descent, back-propagating through the neural network back to the image.
The method is classically used in adversarial example generation~\citep{szegedy2013intriguing, carlini2017c&w} and watermarking~\citep{vukotic2020classification, fernandez2022sslwatermarking}.

Given an original image $I_o$, the loss is an aggregation of the following objectives: 
\begin{align} \label{eq:objective}
    & \Lf(x,q(x_o)) = \norm{x - q(x_o)}^2
    \textrm{\qquad  with } x_o = f(I_o) ,\, x = f(I) \\
    & \Li(I,I_o) = \norm{I - I_o}^2.
\end{align}
$\Li$ is a regularization on the image distortion.
$\Lf$ is the indexation loss that operates on the representation space.
$\Lf$ is the Euclidean distance between $x$ and the target $q(x_o)$ and its goal is to push the image feature towards $q(x_o)$.
With IVF-PQ as index, the representation of the activated image gets closer to the quantized version of the original representation, but also closer to the coarse centroid.
Finally, the losses are combined as 
$\mathcal{L}(I;I_o) = \Lf(x,q(x_o)) + \lambda \Li(I,I_o)$.

\subsection{Perceptual attenuation}

\paragraph*{Overview.}
It is common to optimize a perturbation $\delta$ added to the image, rather than the image itself. 
The adversarial example literature often considers perceptual constraints in the form of an $\ell_p$-norm bound applied on $\delta$ (\citet{madry2017towards} use $\norm{\delta}_\infty < \varepsilon = 8/255$).
Although a smaller $\varepsilon$ makes the perturbation less visible, this constraint is not optimal for the human visual system (HVS), \eg, perturbations are more noticeable on flat than on textured areas of the image.

We employ a handcrafted perceptual attenuation model
based on a Just Noticeable Difference (JND) map~\citep{wu2017enhanced}, that adjusts the perturbation intensity according to luminance and contrast masking.
Given an image $I$, the JND map $H_{\mathrm{JND}}(I)\in \R^{c\times h \times w}$ models the minimum difference perceivable by the HVS at each pixel and additionally rescales the perturbation channel-wise since the human eye is more sensible to red and green than blue color shift.

The relation that links the image $I$ sent to $f$, $\delta$ being optimized and the original $I_o$, reads:
\vspace*{-1em}
\begin{align}
    I = I_o + \alpha \,.\, H_{\mathrm{JND}}(I_o) \odot \mathrm{tanh}(\delta),
    \label{eq:scaling}
\end{align}
with $\alpha$ a global scaling parameter that controls the strength of the perturbation and $\odot$ the pointwise multiplication. 
Since the hyperbolic tangent imposes a bound on the distortion $\delta$, $\alpha$ is directly linked to a lower bound on the PSNR.
Coupled with the regularization $\Li$~\eqref{eq:objective}, it enforces that the activated image is perceptually similar, \ie, $I^\acti\in \mathcal{C}(I_o)$ as required in~\eqref{eq:active_image}.
Note that in the above equation, $H_{\mathrm{JND}}$ does not need to be differentiable since it is computed on the original image $I_o$ and not $I$, therefore it is independent of $\delta$.

\paragraph*{Just Noticeable Difference map}

\begin{figure}[b!]
    \centering
    \hspace{1em}
    \includegraphics[width= 0.3\textwidth]{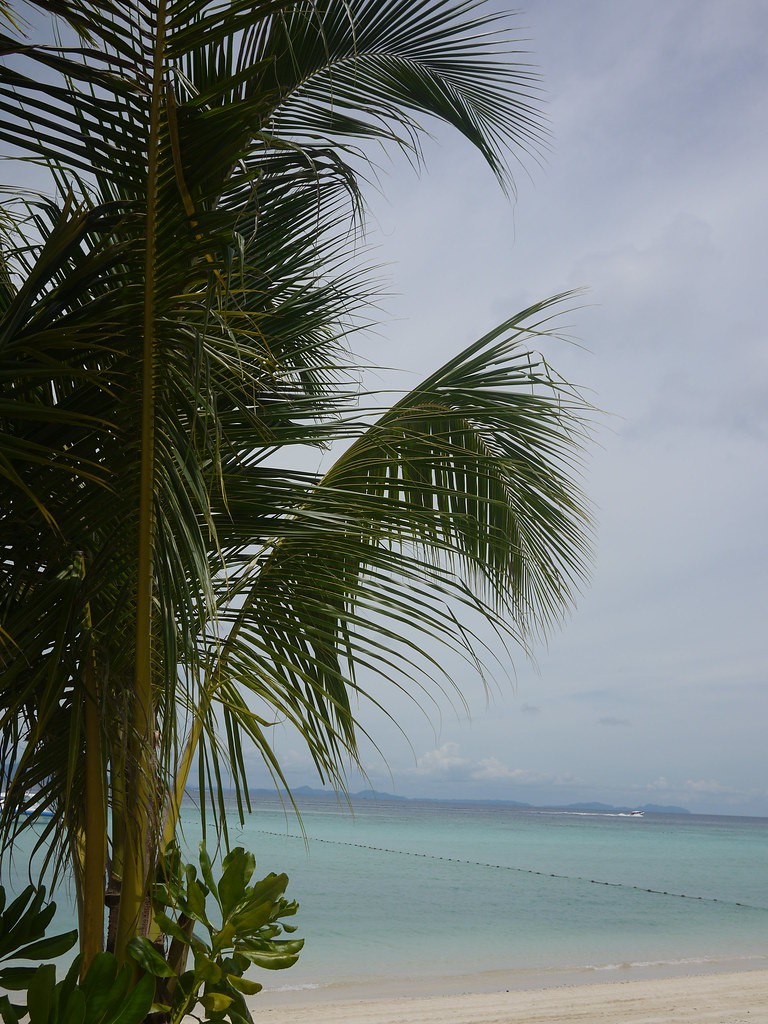}
    \hspace{0.1\textwidth}
    \captionsetup{font=small}
    \includegraphics[width= 0.3\textwidth]{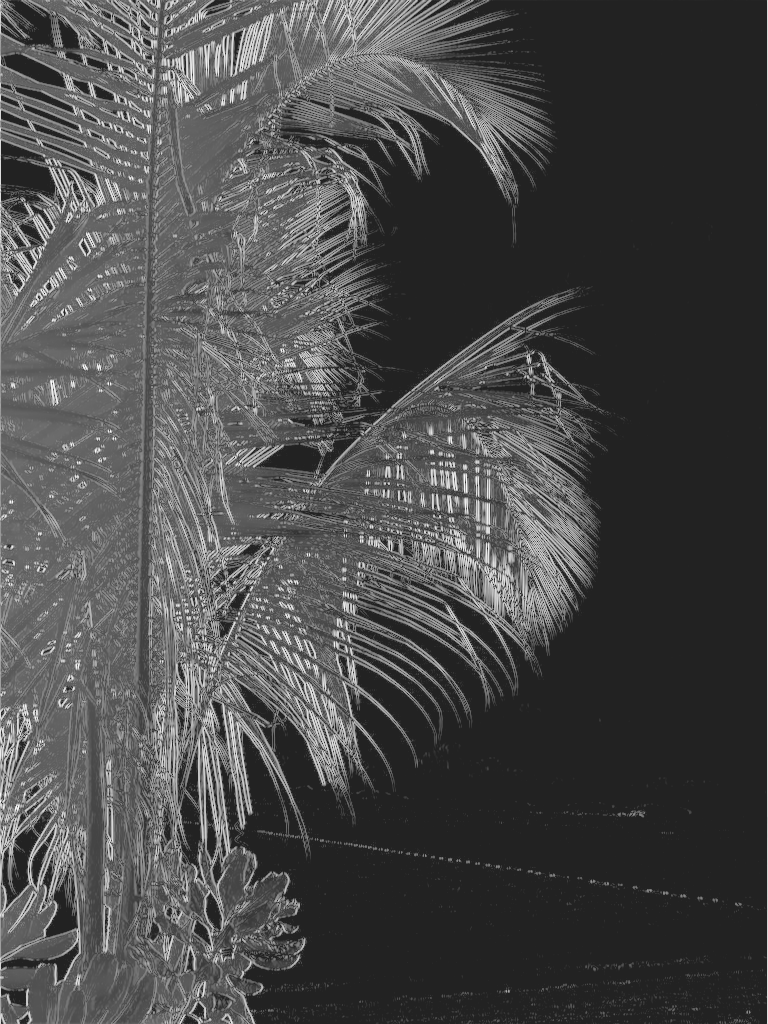}
    \caption[Caption]{A reference image $I$ from DISC21 (\href{http://www.flickr.com/photos/61368956@N00/5060849004/}{R002815.jpg}), and the associated perceptual heatmap $H_{\mathrm{JND}}(I)$.}
    \label{chap2/fig:heatmap}
\end{figure}

The maximum change that the human visual system (HVS) cannot perceive is sometimes referred to as the just noticeable difference (JND)~\citep{krueger1989reconciling}. 
It is used in many applications, such as image/video watemarking, compression, quality assessment (JND is also used in audio). 

JND models in pixel domain directly calculate the JND at each pixel location (\ie, how much pixel difference is perceivable by the HVS). The JND map that we use is based on the work of \citep{chou1995perceptually}.
We use this model for its simplicity, its efficiency and its good qualitative results.
More complex HVS models could also be used if even higher imperceptibility is needed (\citep{watson1993dct, yang2005just, zhang2008just, jiang2022jnd} to cite a few).
The JND map takes into account two characteristics of the HVS, namely the luminance adaptation (LA) and the contrast masking (CM) phenomena. We follow the same notations as \citep{wu2017enhanced}.

The CM map $\mathcal{M}_C$ is a function of the image gradient magnitude $\mathcal{C}_l$ (the Sobel filter of the image):
\begin{equation}
    \mathcal{M}_C(x) = 0.115 \times 
    \frac{\alpha \cdot \mathcal{C}_l(x)^{2.4}}
    { \mathcal{C}_l(x)^{2} + \beta^2}
    \textrm{\quad , with \,}
    \mathcal{C}_l = \sqrt{ \nabla_x I(x)^2 + \nabla_y I(x)^2},
\end{equation}
where $x$ is the pixel location, $I(x)$ the image intensity, $\alpha = 16$, and $\beta = 26$. 
It is an increasing function of $\mathcal{C}_l$, meaning that the stronger the gradient is at $x$, the more the image is masking a local perturbation, and the higher the noticeable pixel difference is.

LA takes into account the fact that the HVS presents different sensitivity to background luminance (\eg, it is less sensible in dark backgrounds).
It is modeled as:
\begin{align}
    \mathcal{L}_A (x) =
    \begin{cases}
        \displaystyle 17 \times \left( 1-\sqrt{\frac{B(x)}{127}} \right) & \textrm{\quad if\,} B(x)<127 \\
        \displaystyle \frac{3 \times \left( B(x) - 127 \right)}{128} +3 & \textrm{\quad if\,}B(x)\geq 127, 
    \end{cases}
\end{align}
where $B(x)$ is the background luminance, which is calculated as the mean luminance value of a local patch centered on $x$.

Finally, both effects are combined with a nonlinear additivity model:
\begin{equation}
    H_{\mathrm{JND}} = \mathcal{L}_A + \mathcal{M}_C - C \cdot \min \{ \mathcal{L}_A, \mathcal{M}_C \},
\end{equation}
where $C$ is set to $0.3$ and determines the overlapping effect. 
For color images, the final RGB heatmap is $H_{\mathrm{JND}} = [\alpha_R H, \alpha_G H, \alpha_B H]$, where $(\alpha_{R}, \alpha_{G}, \alpha_{B})$ are inversely proportional to the mixing coefficients for the luminance: $(\alpha_{R}, \alpha_{G}, \alpha_{B}) = 0.072 / (0.299, 0.587, 0.114)$.

\subsection{Impact on the indexing performance}

\autoref{chap2/fig:fig1} illustrates that the representation of the activated image gets closer to the reproduction value $q(f(I_o))$, and farther away from the Voronoï boundary.
This is expected to make image similarity search more robust because 
(1) it decreases the probability that $x=f(t(I_o))$ ``falls'' outside the bucket; and
(2) it lowers the distance between $x$ and $q(x)$,  improving the PQ distance estimate.

Besides, by design, the representation stored by the index is invariant to the activation.
Formally stated, consider two images $I$, $J$, and one activated version $J^\acti$ together with their representations $x,y,y^\acti$. 
When querying $x=f(I)$, the distance estimate is $\norm{q(y^\acti)- x} = \norm{q(y)- x}$, so the index is oblivious to the change $J\rightarrow J^\acti$.
This means that the structure can index passive and activated images at the same time.
Retrieval of activated images is more accurate but the performance on passive images does not change. 
This compatibility property makes it possible to select only a subset of images to activate, but also to activate already-indexed images at any time.

%% file: chapter-2/sections/5-analyses.tex
\section{Analyses}\label{chap2/sec:analyses}

We provide insights on the method for IVF-PQ, considering the effects of quantization and space partitioning.
For an image $I$ whose representation is $x=f(I)\in\R^d$, 
$\hat{x}$ denotes the representation of a transformed version: $\hat{x} = f(t(I))\in\R^d$, and $x^\acti$ the representation of the activated image $I^\acti$.
For details on the images and the implementation used in the experimental validations, see Sec. \ref{chap2/sec:experimental}.

\begin{figure}[b!]
   \begin{minipage}{0.45\textwidth}
        \centering
        \vspace{3pt}
        \includegraphics[width=1.05\linewidth, trim={0 0.8em 0 0em},clip]{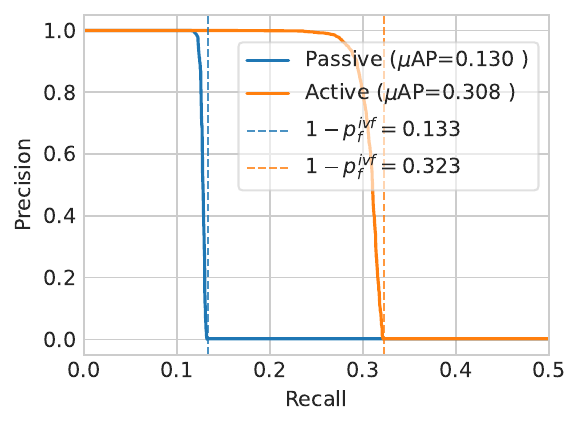}
        \caption{Precision-Recall curve for ICD with 50k queries and 1M reference images (more details for the experimental setup in Sec. \ref{chap2/sec:experimental}). $p_\mathrm{f}^{\mathrm{ivf}}$ is the probability of failure of the IVF (Sec. \ref{chap2/sec:space_partitioning}).
        }
        \label{chap2/fig:prc}
   \end{minipage}\hfill
      \begin{minipage}{0.51\textwidth}
        \centering
        \includegraphics[width=0.8\textwidth, trim={0 0.15cm 0 0.23cm}, clip]{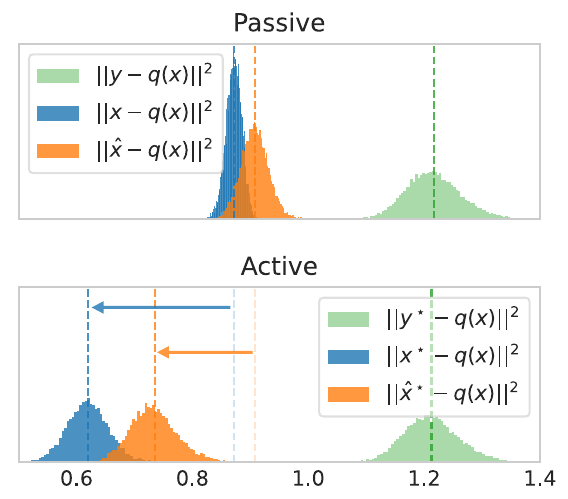}
        \caption{
            Distance estimates histograms (sec. \ref{chap2/sec:quantization}).
            With active indexing, $\|x- q(x)\|^2$ is reduced ({\color{blue}$\leftarrow$}), inducing a shift ({\color{orange}$\leftarrow$}) in the distribution of $\|\hat{x}- q(x)\|^2$, where $t(I)$ a hue-shifted version of $I$.
            $y$ is a random query.
        }
        \label{chap2/fig:dists}
   \end{minipage}
\end{figure}

\subsection{Product quantization: impact on distance estimate}\label{chap2/sec:quantization}

We start by analyzing the distance estimate considered by the index:
\begin{equation}
    \|\hat{x}-q(x)\|^2 = \|x-q(x)\|^2 + \|\hat{x}-x\|^2
    + 2 (x-q(x))^\top (\hat{x}-x).
    \label{eq:distance}
\end{equation}
The activation aims to reduce the first term, \ie, the quantization error $\|x- q(x)\|^2$, which in turn reduces $\|\hat{x}- q(x)\|^2$. 
Figure~\ref{chap2/fig:dists} shows in blue the empirical distributions of $\|x- q(x)\|^2$ (passive) and $\|x^\acti- q(x)\|^2$ (activated). 
As expected the latter has a lower mean, but also a stronger variance.
The variation of the following factors may explain this: \emph{i)} the strength of the perturbation (due to the HVS modeled by $H_{\mathrm{JND}}$ in~\eqref{eq:scaling}), \emph{ii)} the sensitivity of the feature extractor $\|\nabla_x f(x)\|$ (some features are easier to push than others), \emph{iii)} the shapes and sizes of the Voronoï cells of PQ.

The second term of \eqref{eq:distance} models the impact of the image transformation in the feature space. 
Comparing the orange and blue distributions in Fig.~\ref{chap2/fig:dists}, we see that it has a positive mean, but the shift is bigger for activated images.
We can assume that the third term has null expectation for two reasons: \emph{i)} the noise $\hat{x}-x$ is independent of $q(x)$ and centered around 0, \emph{ii)} in the high definition regime, quantification noise $x-q(x)$ is independent of $x$ and centered on 0. 
Thus, this term only increases the variance. 
Since $x^\acti-q(x)$ has smaller norm, this increase is smaller for activated images.

All in all, $\|\hat{x}^\acti-q(x)\|^2$ has a lower mean but a stronger variance than its passive counterpart $\|\hat{x}-q(x)\|^2$.
Nevertheless, the decrease of the mean is so large that it compensates the larger variance. 
The orange distribution in active indexing is further away from the green distribution for negative pairs, \ie, the distance between an indexed vector $q(x)$ and an independent query $y$.

\subsection[Space partitioning: impact on the IVF probability of failure]{Space partitioning: impact on the IVF \\probability of failure}\label{chap2/sec:space_partitioning}

We denote by $p_\mathrm{f} := \mathbb{P}(\qivf(x) \neq \qivf (\hat{x}) )$ the probability that $\hat{x}$ is assigned to a wrong bucket by IVF assignment $\qivf$.
In the single-probe search ($k'=1$), the recall (probability that a pair is detected when it is a true match, for a given threshold $\tau$ on the distance) is upper-bounded by $1 - p_\mathrm{f}$:
\begin{align}
   R_\tau 
   = \mathbb{P} \left(\{ \qivf(\hat{x}) = \qivf(x) \} \cap \{ \| \hat{x}-q(x)\| < \tau \} \right) 
   \leq \mathbb{P} \left(\{ \qivf(\hat{x}) = \qivf(x) \}  \right) 
   =1 - p_\mathrm{f}.
\end{align}
In other terms, even with a high threshold $\tau \rightarrow \infty$ (and low precision), the detection misses representations that ought to be matched, with probability $p_\mathrm{f}$. It explains the sharp drop at recall $R=0.13$ in Fig.~\ref{chap2/fig:prc}.
This is why it is crucial to decrease $p_\mathrm{f}$.
The effect of active indexing is to reduce $\|\hat{x}-\qivf(x)\|$
therefore reducing $p_\mathrm{f}$ and increasing the upper-bound for $R$:
the drop shifts towards $R=0.32$. 

This explanation suggests that pushing $x$ towards $\qivf(x)$ decreases even more efficiently $p_\mathrm{f}$. 
This makes the IVF more robust to transformation but this may jeopardize the PQ search because features of activated images are packed altogether.
In a way, our strategy, which pushes $x$ towards $q(x)$, dispatches the improvement over the IVF and the PQ search.

%% file: chapter-2/sections/4-results.tex
\renewcommand{\rot}[1]{\rotatebox{45}{#1}\hspace*{-0.6cm}}

\section{Experimental results}

\subsection{Experimental setup}\label{chap2/sec:experimental}

\paragraph*{Dataset.}
We use DISC21~\citep{douze2021disc} a dataset dedicated to ICD.
The dataset can be freely downloaded on \href{https://ai.facebook.com/datasets/disc21-dataset/}{its webpage}.
It includes 1M reference images and 50k query images, 10k of which are true copies from reference images. 
A disjoint 1M-image set with same distribution as the reference images is given for training. 
Images resolutions range from 200$\times$200 to 1024$\times$1024 pixels (most of the images are around 1024$\times$768 pixels).

The queries used in our experiments are \emph{not} the queries in DISC21, since we need to control the image transformations in our experiments, and most transformations of DISC21 were done manually so they are not reproducible. 
Our queries are recreated transformations of images \emph{after active indexing}. 
These transformations range from simple attacks like rotation to more realistic social network transformations which created the original DISC21 queries\footnote{
    We create a first test set ``Ref10k'' by selecting the 10k images from the reference set that were originally used to generate the queries (the ``dev queries'' from the downloadable version).
    We also re-create a query set ``Query50k''. 
    To be as close as possible, we use the same images that were used for generating queries in DISC.
    Edited images are generated using the AugLy library~\citep{papakipos2022augly}, following the guidelines given in the ``Automatic Transformations" section of the DISC paper.
    Therefore, the main difference between the query set used in our experiments and the original one is that ours do not have manual augmentations.
}.

\paragraph*{Metrics.}
For retrieval, our main metric is Recall $1$@$1$ (\rone for simplicity), which corresponds to the proportion of positive queries where the top-1 retrieved results is the reference image.
For copy detection, we use the same metric as the NeurIPS Image Similarity Challenge~\citep{douze2021disc}.
We retrieve the $k=10$ most similar database features for every query; and we declare a pair is a match if the distance is lower than a threshold $\tau$.
To evaluate detection efficiency, we use the 10k matching queries above-mentioned together with 40k negative queries (\ie, not included in the database). 
We use precision and recall, as well as the area under the precision-recall curve, which is equivalent to the micro average precision (\emph{$\mu$AP}).
While \rone only measures ranking quality of the index, $\mu$AP takes into account the confidence of a match.
As for image quality metric, we use the Peak Signal-to-Noise Ratio (PSNR) which is defined as $10\log_{10} \left( 255^2 / \mathrm{MSE}(I, I')^2 \right)$, as well as SSIM~\citep{wang2004ssim} and the norm $\norm{I-I'}_\infty$.

\paragraph*{Transformations seen at test time}\label{chap2/sec:transformations}

\begin{table}[t!]
    \centering
    \captionsetup{font=small}
    \caption{Illustration of all transformations evaluated in Tab.~\ref{chap2/tab:act_vs_pas_retrieval}.}
    \label{chap2/fig:all_transformations}
    \begin{tabular}{*{5}{l}}
         Identity & Contrast 0.5 & Contrast 2.0 & Brightness 0.5 & Brightness 2.0 \\
         \begin{minipage}{.16\linewidth}\includegraphics[width=\linewidth]{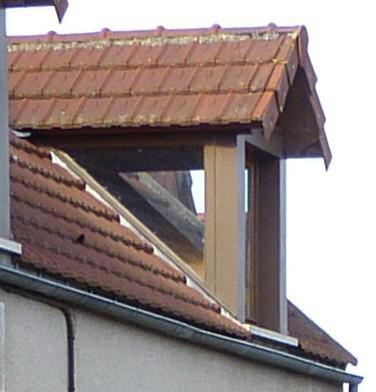}\end{minipage} &  
         \begin{minipage}{.16\linewidth}\includegraphics[width=\linewidth]{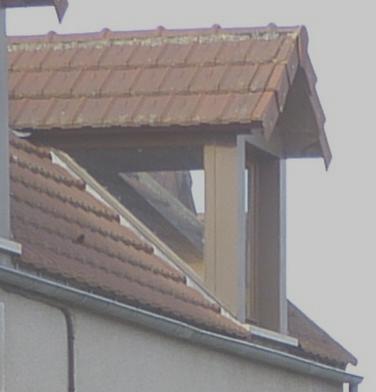}\end{minipage} &  
         \begin{minipage}{.16\linewidth}\includegraphics[width=\linewidth]{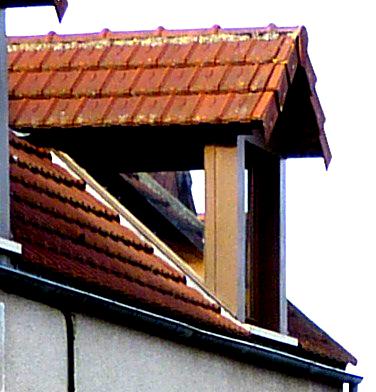}\end{minipage} &  
         \begin{minipage}{.16\linewidth}\includegraphics[width=\linewidth]{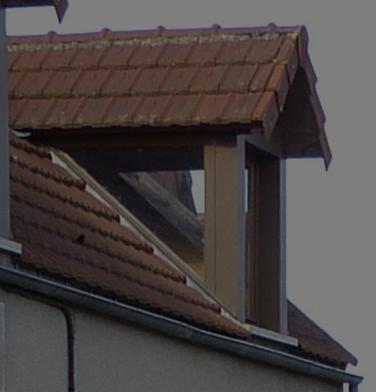}\end{minipage} &  
         \begin{minipage}{.16\linewidth}\includegraphics[width=\linewidth]{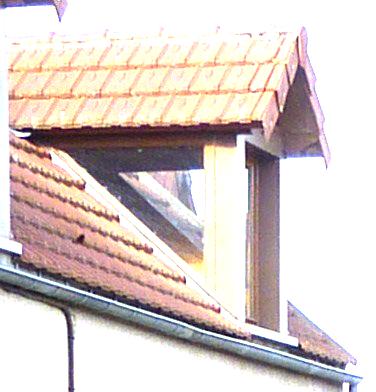}\end{minipage} 
         \\ \\
          Hue 0.2 & Blur 2.0 & JPEG 50 & Rotation 25 & Rotation 90 \\
         \begin{minipage}{.16\linewidth}\includegraphics[width=\linewidth]{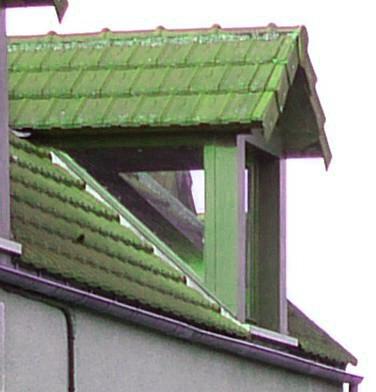}\end{minipage} &  
         \begin{minipage}{.16\linewidth}\includegraphics[width=\linewidth]{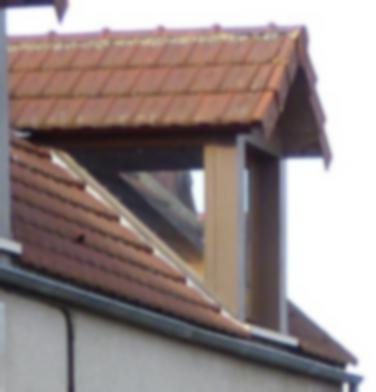}\end{minipage} &  
         \begin{minipage}{.16\linewidth}\includegraphics[width=\linewidth]{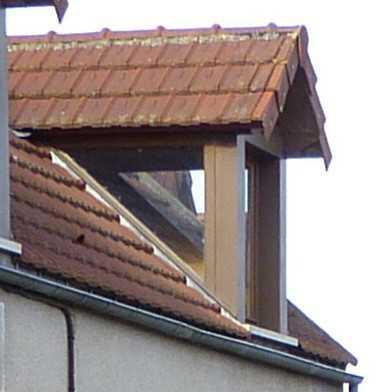}\end{minipage} &  
         \begin{minipage}{.16\linewidth}\includegraphics[width=\linewidth]{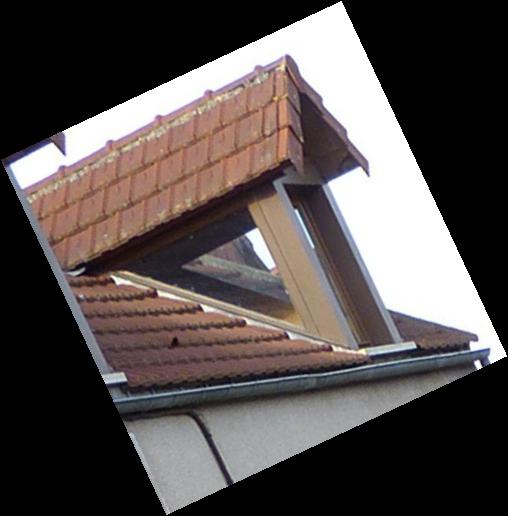}\end{minipage} &  
         \begin{minipage}{.16\linewidth}\includegraphics[width=\linewidth]{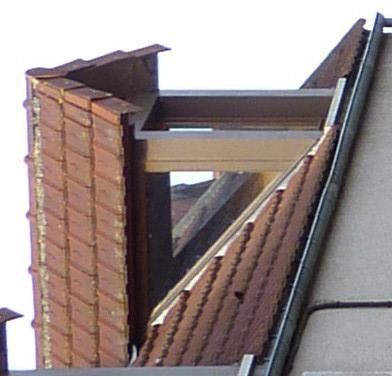}\end{minipage} 
         \\ \\
          Crop 0.5 & Resize 0.5 & Meme & Random \\
         \begin{minipage}{.16\linewidth}\includegraphics[width=\linewidth]{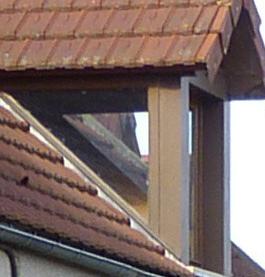}\end{minipage} &  
         \begin{minipage}{.16\linewidth}\includegraphics[width=\linewidth]{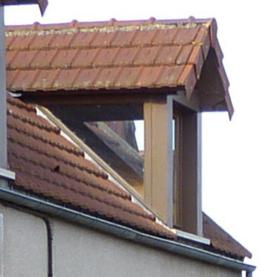}\end{minipage} &  
         \begin{minipage}{.12\linewidth}\includegraphics[width=\linewidth]{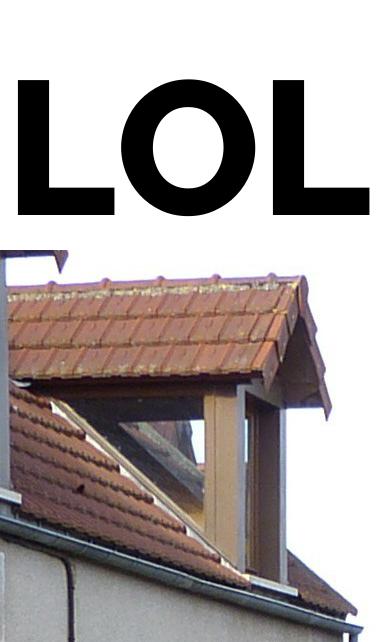}\end{minipage} &  
         \begin{minipage}{.16\linewidth}\includegraphics[width=1.7\linewidth]{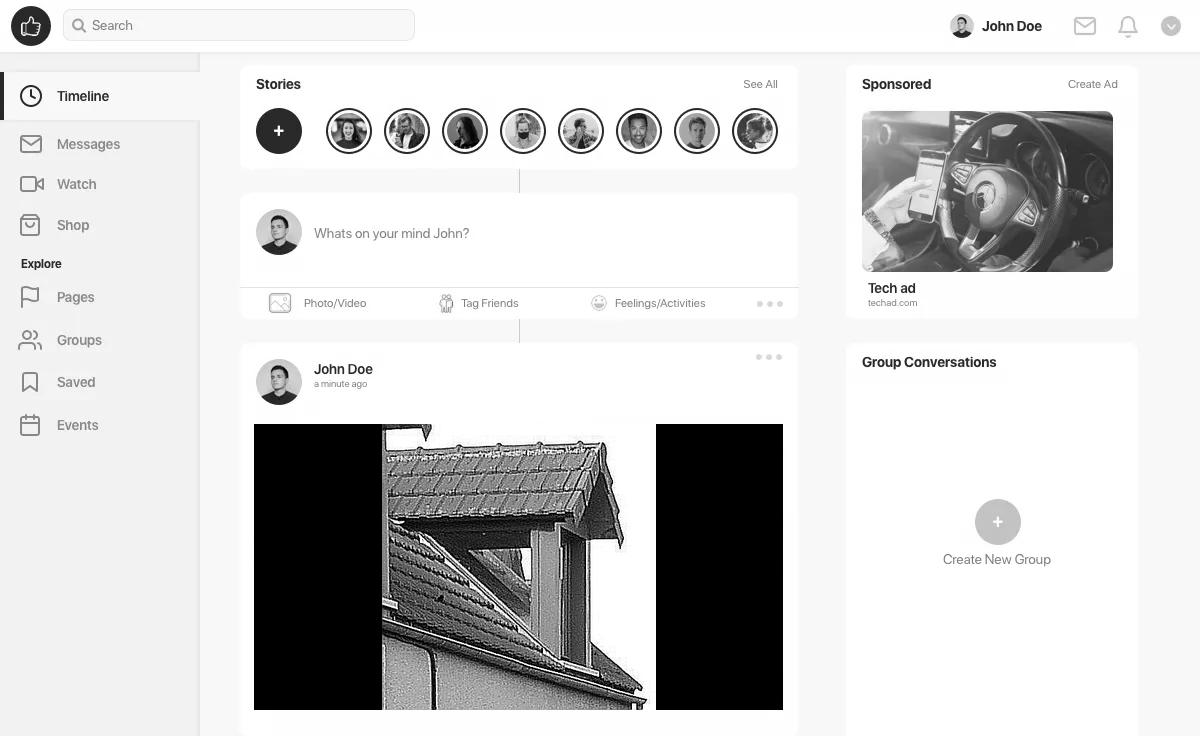}\end{minipage} &  
         \\
         
    \end{tabular}
\end{table}

They cover both spatial transformations (crops, rotation, etc.), pixel-value transformations (contrast, hue, jpeg, etc.) and ``everyday life'' transformations with the AugLy augmentations. 
All transformations are illustrated in Fig.~\ref{chap2/fig:all_transformations}.
The parameters for all transformations are the ones of the torchvision library~\citep{marcel2010torchvision}, except for the crop and resize that represent area ratios. For the Gaussian blur transformation we use alternatively $\sigma$, the scaling factor in the exponential, or the kernel size $k_b$ (in torchvision $k_b = (\sigma-0.35)/0.15$). 
The ``Random'' transformation is the one used to develop the 50k query set. 
A series of simple 1-4 AugLy transformations are picked at random, with skewed probability for a higher number.
Among the possible transformations, there are pixel-level, geometric ones, as well as embedding the image as a screenshot of a social network GUI.

\paragraph*{Implementation details.}\label{chap2/sec:details}
The evaluation procedure is: (1) we train an index on the 1M training images, (2) index the 1M reference images, (3) activate (or not) 10k images from this reference set.
(4) At search time, we use the index to get closest neighbors (and their distances) of transformed versions from a query set made of the 10k images.

Unless stated otherwise, we use a IVF4096,PQ8x8 index (IVF quantizer with 4096 centroids, and PQ with 8 subquantizers of $2^8$ centroids), and use only one probe on IVF search for shortlist selection ($k'=1$). 
Compared to a realistic setting, we voluntarily use an indexing method that severely degrades learned representations to showcase and analyze the effect of the active indexing.
For feature extraction, we use an SSCD model with a ResNet50 trunk~\citep{he2016resnet}. 
It takes image resized to 288$\times$288 and generates normalized representations in $\R^{512}$.
Optimization~\eqref{eq:active_image} is done with the Adam optimizer~\citep{kingma2014adam}, the learning rate  is set to $1$, the number of iterations to $N=10$ and the regularization to $\lambda=1$.
In~\eqref{eq:scaling}, the distortion scaling  is set to $\alpha=3$ (leading to an average PNSR around $43$~dB).
In this setup, activating 128 images takes around 6s ($\approx$ 40ms/image) with a 32GB GPU. 
It can be sped-up at the cost of some accuracy (see Sec.~\ref{chap2/sec:speedup}).

Models used for feature extraction in Sec.~\ref{chap2/sec:generalization} (\href{https://github.com/facebookresearch/sscd-copy-detection/}{SSCD}, \href{https://github.com/facebookresearch/dino}{DINO}, \href{https://github.com/lyakaap/ISC21-Descriptor-Track-1st}{ISC-dt1}) can be downloaded in their respective repositories.
Our implementation builds upon the open-source Pytorch~\citep{paszke2019pytorch} and FAISS~\citep{johnson2019faiss} libraries.

\subsection{Active vs. passive}\label{chap2/sec:act_vs_passive}
This section compares retrieval performance of active and passive indexing.
We evaluate $R$@1 when different transformations are applied to the 10k reference images before search.
The ``Passive'' lines of Tab.~\ref{chap2/tab:act_vs_pas_retrieval} show how the IVF-PQ degrades the recall. 
This is expected, but the IVF-PQ also accelerates search 500$\times$ and the index is 256$\times$ more compact, which is necessary for large-scale applications. 
Edited images are retrieved more often when they were activated for the index:
increase of up to $+60$ \rone for strong brightness and contrast changes, close to results of the brute-force search.
We also notice that the performance of the active IVF-PQ$^{k'=1}$ is approximately the same as the one of the passive IVF-PQ$^{k'=16}$, meaning that the search can be made more efficient at equal performance.
For the IVF-PQ$^\dagger$ that does less approximation in the search (but is slower and takes more memory), retrieval on activated images is also improved, though to a lesser extent. 

\begin{table}[t]
    \centering
    \caption{
        Comparison of the index performance between activated and passive images. 
        The search is done on a 1M image set and $R$@1 is averaged over 10k query images submitted to different transformations before search.  
        \emph{Random}: randomly apply 1 to 4 transformations.
        \emph{Avg.}: average on the transformations presented in the table (details in Sec.~\ref{chap2/sec:transformations}).
        \emph{No index}: exhaustive brute-force nearest neighbor search. \emph{IVF-PQ}: \textsc{IVF4096,PQ8x8} index with $k'$=1 (16 for \emph{IVF-PQ}$^{16}$).
        \emph{IVF-PQ}$^\dagger$: \textsc{IVF512,PQ32x8} with $k'=32$.
    }
    \label{chap2/tab:act_vs_pas_retrieval}
    \resizebox{1.0\linewidth}{!}{
    \begingroup
        \setlength{\tabcolsep}{4pt}
        \def\arraystretch{1.1}
        \begin{tabular}{ l |l| l| c| *{15}{p{0.04\textwidth}}}
            \multicolumn{1}{c}{} & \multicolumn{1}{c}{\rot{Search (ms)}} & \multicolumn{1}{c}{\rot{Bytes/vector}} & \multicolumn{1}{c}{\rot{Activated}} & \rot{Identity} & \rot{Contr. 0.5} & \rot{Contr. 2.0} & \rot{Bright. 0.5} & \rot{Bright. 2.0} & \rot{Hue 0.2} & \rot{Blur 2.0}  & \rot{JPEG 50}  & \rot{Rot. 25} & \rot{Rot. 90} & \rot{Crop 0.5} & \rot{Resi. 0.5} & \rot{Meme} & \rot{Random} & \rot{Avg.} \\ \midrule 
            No index & 252 & 2048 & \xmarkg & 1.00 & 1.00 & 1.00 & 1.00 & 1.00 & 1.00 & 1.00 & 1.00 & 1.00 & 1.00 & 1.00 & 1.00 & 1.00 & 0.90 & 0.99 \\ \midrule
            & & & \xmarkg  & 1.00 & 0.73 & 0.39 & 0.73 & 0.28 & 0.62 & 0.48 & 0.72 & 0.07 & 0.14 & 0.14 & 0.72 & 0.14 & 0.13 & 0.45 \\
            \rowcolor{gray!15} \cellcolor{white!0} \multirow{-2}{*}{IVF-PQ} & \multirow{-2}{*}{0.38} \cellcolor{white!0} & \multirow{-2}{*}{8} \cellcolor{white!0}
            & \checkmark & 1.00 & 1.00 & 0.96 & 1.00 & 0.92 & 1.00 & 0.96 & 0.99 & 0.10 & 0.50 & 0.29 & 1.00 & 0.43 & 0.32 & 0.75 \\ 
            \midrule
            & & & \xmarkg & 1.00 & 1.00 & 0.90 & 1.00 & 0.78 & 0.99 & 0.95 & 0.99 & 0.35 & 0.57 & 0.57 & 1.00 & 0.56 & 0.39 & 0.79 \\
            \rowcolor{gray!15} \cellcolor{white!0} \multirow{-2}{*}{IVF-PQ$^{16}$} & \multirow{-2}{*}{0.42} \cellcolor{white!0} & \multirow{-2}{*}{8} \cellcolor{white!0} & 
            \checkmark & 1.00 & 1.00 & 1.00 & 1.00 & 0.98 & 1.00 & 1.00 & 1.00 & 0.43 & 0.88 & 0.75 & 1.00 & 0.84 & 0.50 & 0.88 \\
            \midrule
            & & & \xmarkg & 1.00 & 1.00 & 0.99 & 1.00 & 0.95 & 1.00 & 0.99 & 1.00 & 0.72 & 0.87 & 0.88 & 1.00 & 0.87 & 0.61 & 0.92 \\
            \rowcolor{gray!15} \cellcolor{white!0} \multirow{-2}{*}{IVF-PQ$^\dagger$} & \multirow{-2}{*}{1.9} \cellcolor{white!0} & \multirow{-2}{*}{32} \cellcolor{white!0} &
            \checkmark & 1.00 & 1.00 & 0.99 & 1.00 & 0.98 & 1.00 & 1.00 & 1.00 & 0.75 & 0.92 & 0.91 & 1.00 & 0.92 & 0.63 & 0.94 \\
            \bottomrule
        \end{tabular}
    \endgroup
    }
\end{table}

As for copy detection, \autoref{chap2/fig:prc} gives the precision-recall curves obtained for a sliding value of $\tau$, and corresponding $\mu$AP. 
Again, we observe a significant increase ($\times 2$) in $\mu$AP with active indexing.
Note that the detection performance is much weaker than the brute-force search even in the active case because of the strong approximation made by space partitioning (more details in Sec.~\ref{chap2/sec:space_partitioning}).

\paragraph*{Detailed metrics on different image transformations}

\begin{figure}[b!]
    \centering
    \includegraphics[width=0.99\linewidth,trim={0 0.4cm 0 0.2cm}, clip]{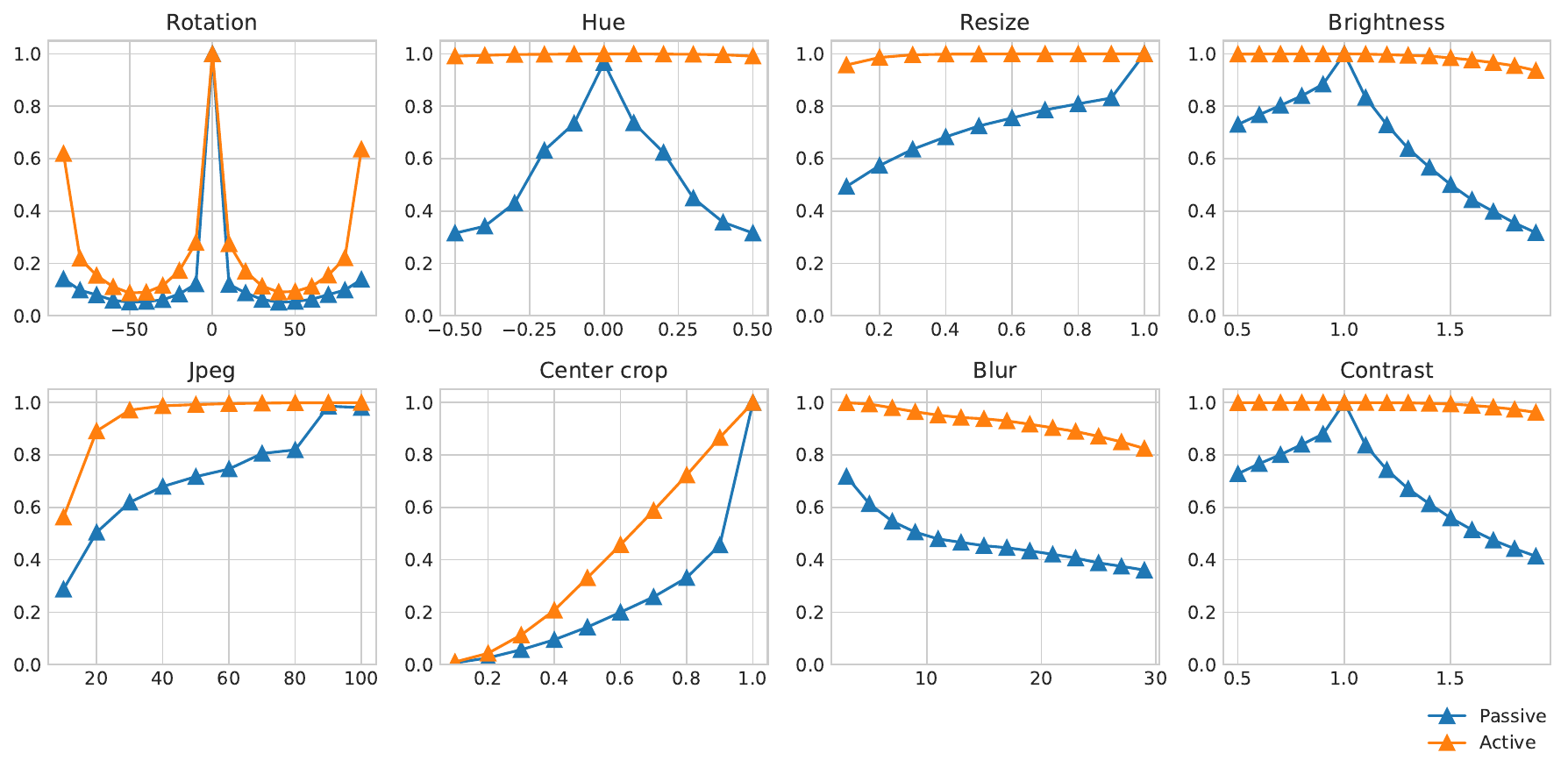}
    \captionsetup{font=small}
    \caption{
    Average $R$@1 over 10k images indexed with IVF-PQ.
   }
    \label{chap2/fig:tranformations}
\end{figure}

On Fig.~\ref{chap2/fig:tranformations}, we evaluate the average \rone over the 10k images from the reference dataset.
The experimental setup is the same as for Tab.~\ref{chap2/tab:act_vs_pas_retrieval} with the IVF-PQ, but a higher number of transformation parameters are evaluated.
As expected, the higher the strength of the transformation, the lower the retrieval performance is.
The decrease in performance is significantly reduced with activated images.

\subsection{Generalization}\label{chap2/sec:generalization}

\paragraph*{Generalization to other neural feature extractors.}

We first reproduce the experiment of Sec.~\ref{chap2/sec:details} with different extractors, that cover distinct training methods and architectures.
Among them, we evaluate a ResNext101~\citep{xie2017aggregated} trained with SSCD~\citep{pizzi2022sscd}, a larger network than the ResNet50 used in our main experiments ; 
the winner of the descriptor track of the NeurIPS ISC, \textsc{Lyakaap}-dt1~\citep{yokoo2021isc}, that uses an EfficientNetv2 architecture~\citep{tan2021efficientnetv2} ; networks from DINO~\citep{caron2021dino}, either based on ResNet50 or ViT~\citep{dosovitskiy2020vit}, like the ViT-S model~\citep{touvron2021training}.

\autoref{chap2/tab:extractors} presents the \rone obtained on 10k activated images when applying different transformations before search.
The \rone is better for activated images for all transformations and all neural networks. The average improvement on all transformations ranges from $+12\%$ for DINO ViT-s to $+30\%$ for SSCD ResNet50.

\begin{table}[t]
    \centering
    \caption{
        \rone for different transformations before search. 
        We use our method to activate images for indexing with IVF-PQ, with different neural networks used as feature extractors.
    }
    \label{chap2/tab:extractors}
    \resizebox{1.0\linewidth}{!}{
    \begingroup
        \setlength{\tabcolsep}{4pt}
        \def\arraystretch{1.1}
        \begin{tabular}{ l| l| c| *{15}{p{0.04\textwidth}}}
            \multicolumn{1}{c}{\rot{Name}} & \multicolumn{1}{c}{\rot{Arch.}} & \multicolumn{1}{l}{\rot{Activated}} & \rot{Identity} & \rot{Contr. 0.5} & \rot{Contr. 2.0} & \rot{Bright. 0.5} & \rot{Bright. 2.0} & \rot{Hue 0.2} & \rot{Blur 2.0}  & \rot{JPEG 50}  & \rot{Rot. 25} & \rot{Rot. 90} & \rot{Crop 0.5} & \rot{Resi. 0.5} & \rot{Meme} & \rot{Random} & \rot{Avg.} \\ \midrule 
             &  & \xmarkg & 1.00 & 0.73 & 0.39 & 0.73 & 0.28 & 0.62 & 0.48 & 0.72 & 0.07 & 0.14 & 0.14 & 0.72 & 0.14 & 0.13 & 0.45  \\ 
            \rowcolor{gray!15} \cellcolor{white!0} & \multirow{-2}{*}{ResNet50} \cellcolor{white!0}& \checkmark & 1.00 & 1.00 & 0.96 & 1.00 & 0.92 & 1.00 & 0.96 & 0.99 & 0.10 & 0.50 & 0.29 & 1.00 & 0.43 & 0.32 & 0.75 \\ \cmidrule{2-18}
             &  & \xmarkg & 1.00 & 0.88 & 0.68 & 0.88 & 0.57 & 0.84 & 0.46 & 0.79 & 0.46 & 0.63 & 0.53 & 0.80 & 0.48 & 0.28 & 0.66 \\ 
             \rowcolor{gray!15} \cellcolor{white!0} \multirow{-4}{*}{SSCD} & \multirow{-2}{*}{ResNext101} \cellcolor{white!0} & \checkmark & 1.00 & 1.00 & 0.96 & 1.00 & 0.90 & 0.99 & 0.77 & 0.97 & 0.53 & 0.85 & 0.64 & 1.00 & 0.74 & 0.37 & 0.84 \\ \midrule
             &  & \xmarkg & 1.00 & 0.66 & 0.65 & 0.65 & 0.52 & 0.71 & 0.52 & 0.82 & 0.07 & 0.20 & 0.51 & 0.84 & 0.62 & 0.18 & 0.57 \\ 
            \rowcolor{gray!15} \cellcolor{white!0} & \multirow{-2}{*}{ResNet50} \cellcolor{white!0} & \checkmark & 1.00 & 0.99 & 0.88 & 0.99 & 0.75 & 0.93 & 0.72 & 0.94 & 0.08 & 0.25 & 0.57 & 0.99 & 0.82 & 0.23 & 0.72 \\ \cmidrule{2-18}
             &  & \xmarkg & 1.00 & 0.89 & 0.71 & 0.86 & 0.64 & 0.75 & 0.74 & 0.90 & 0.14 & 0.18 & 0.57 & 0.88 & 0.61 & 0.25 & 0.65 \\ 
             \rowcolor{gray!15} \cellcolor{white!0} \multirow{-4}{*}{DINO} & \multirow{-2}{*}{ViT-s} \cellcolor{white!0} & \checkmark & 1.00 & 0.99 & 0.94 & 0.99 & 0.92 & 0.98 & 0.89 & 0.99 & 0.15 & 0.28 & 0.63 & 0.99 & 0.77 & 0.32 & 0.77 \\ \midrule
              &  & \xmarkg & 1.00 & 0.25 & 0.08 & 0.16 & 0.01 & 0.51 & 0.54 & 0.84 & 0.18 & 0.16 & 0.23 & 0.79 & 0.16 & 0.18 & 0.36 \\ 
              \rowcolor{gray!15} \cellcolor{white!0} \multirow{-2}{*}{ISC-dt1} & \multirow{-2}{*}{EffNetv2} \cellcolor{white!0} & \checkmark & 1.00 & 0.57 & 0.16 & 0.33 & 0.01 & 0.88 & 0.79 & 0.97 & 0.20 & 0.24 & 0.29 & 0.97 & 0.26 & 0.26 & 0.49 \\ 
            \bottomrule
        \end{tabular}
    \endgroup
    }
\end{table}

\paragraph*{Generalization to other indexes.}
The method easily generalizes to other types of indexing structures, the only difference being in the indexation loss $\Lf$~\eqref{eq:objective}. 
We present some of them below:
\begin{itemize}[leftmargin=0.5cm,itemsep=0cm,topsep=-0.1cm]
    \item \textbf{PQ and OPQ}.\quad 
        In PQ~\citep{jegou2010pq}, a vector $x \in \R^d$ is approximated by $\qcompressed(x)$. 
        $\Lf$ reads $\norm{x-\qcompressed(x_o)}$. 
        In OPQ~\citep{ge2013optimized}, vectors are rotated by matrix $R$ before codeword assignment, such that $RR^\top = I$. 
        $\Lf$ becomes $\norm{x-R^\top\qcompressed(Rx_o)}$. 
    \item \textbf{IVF.} \quad 
        Here, we only do space partitioning. 
        Employing $\Lf = \norm{x- \qivf (x_o)}$ (``pushing towards the cluster centroid'') decreases the odds of $x$ falling in the wrong cell (see Sec.~\ref{chap2/sec:space_partitioning}).
        In this case, an issue can be that similar representations are all pushed together to a same centroid, which makes them less discriminate. 
        Empirically, we found that this does not happen because perceptual constraint in the image domain prevents features from getting too close.
    \item \textbf{LSH.} \quad 
        Locality Sensitive Hashing maps $x\in \R^d$ to a binary hash $b(x)\in \R^L$.
        It is commonly done with projections against a set of vectors, which give for $j \in [1,..,L]$, $b_j(x) = \mathrm{sign} (w_j^\top x)$.
        The objective $\Lf = -1/L \sum_{j} \mathrm{sign}(b(x_o))\cdot w_j^\top x$,  allows to push $x$ along the LSH directions and to improve the robustness of the hash.
\end{itemize}
\autoref{chap2/tab:indexes} presents the \rone and \muap obtained on the 50k query set. 
Again, results are always better in the active scenario.
We remark that active indexing has more impact on space partitioning techniques: the improvement for IVF is higher than with PQ and the LSH binary sketches. As to be expected, the impact is smaller when the indexing method is more accurate. 

\begin{table}[t]
    \caption{
        \rone averaged on transformations presented in Tab.~\ref{chap2/tab:act_vs_pas_retrieval} and \muap for different indexing structures
    }\label{chap2/tab:indexes}
    \centering
    \resizebox{0.7\linewidth}{!}{
    \begingroup
        \setlength{\tabcolsep}{3pt}
        \begin{tabular}{ c|c |cc|cc}
                \toprule
                 \multirow{2}{*}{Index} & \multirow{2}{*}{Search time} & \multicolumn{2}{c|}{\rone avg.} & \multicolumn{2}{c}{\muap} \\ 
                & & Passive & Activated & Passive & Activated     \\ \midrule
                IVF 1024 & 0.32 ms & 0.47 & \textbf{0.83} & 0.16 & \textbf{0.43} \\
                OPQ 8x8   & 5.71 ms & 0.92 & \textbf{0.94} & 0.48 & \textbf{0.55} \\
                PCA64, LSH & 0.99 ms & 0.72 & \textbf{0.83} & 0.25 & \textbf{0.39} \\
                \bottomrule
        \end{tabular}
    \endgroup
    }
\end{table}

\subsection{Qualitative results}

\begin{figure}[H]
    \centering
    \includegraphics[width=0.9\textwidth, trim={0 0.4cm 0 0.4cm}, clip]{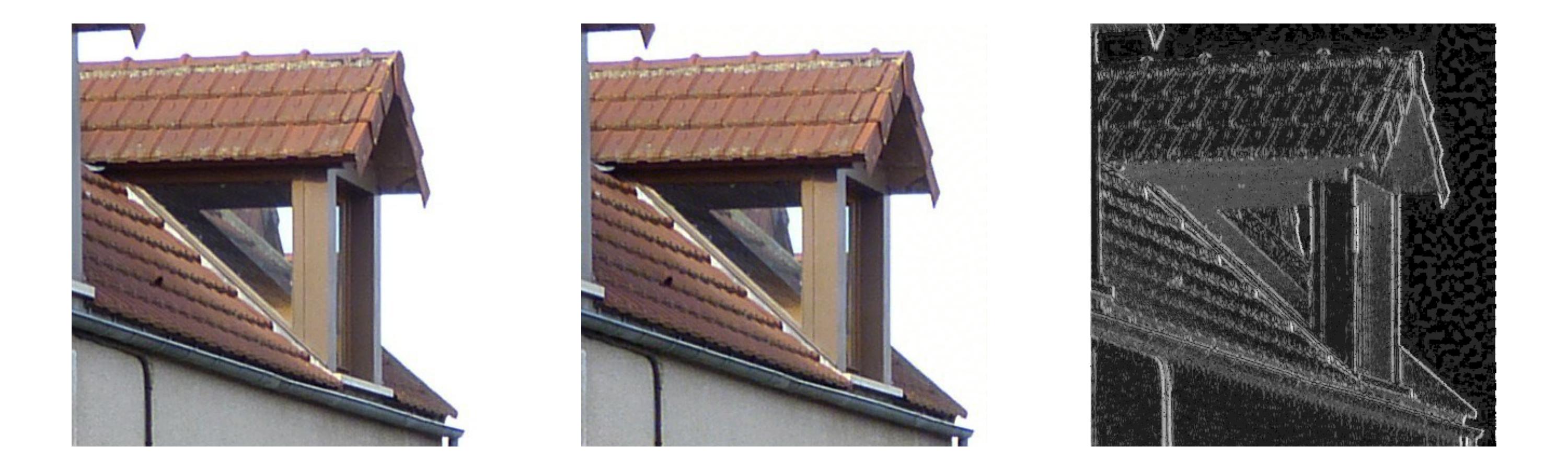}

    \includegraphics[width=0.9\textwidth, trim={0 0.4cm 0 0.4cm}, clip]{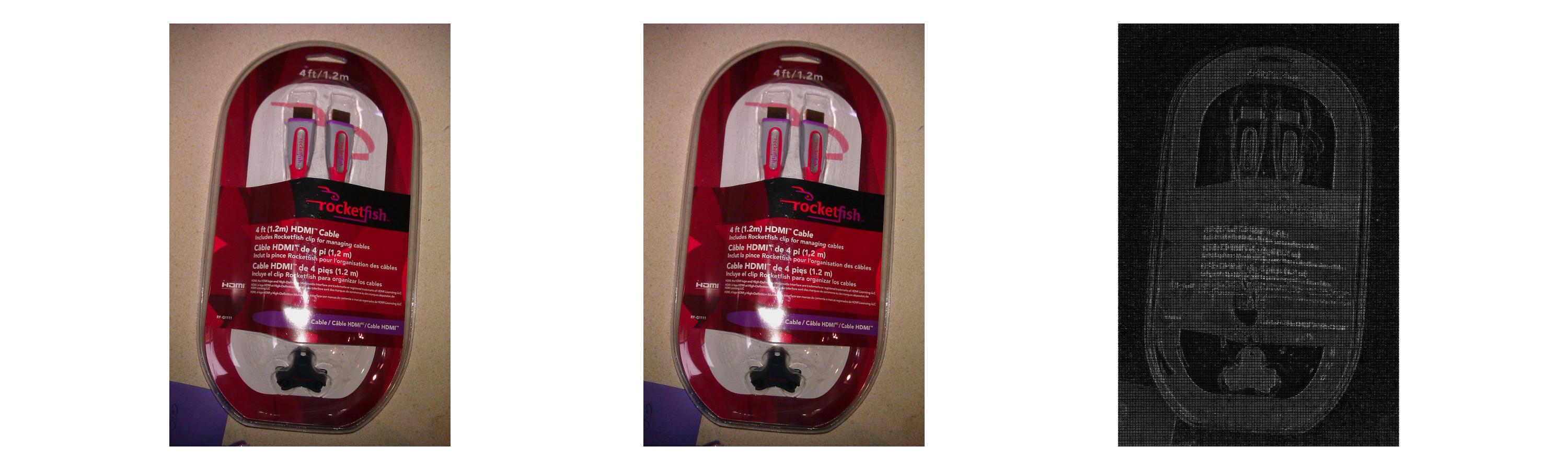}
    \includegraphics[width=0.9\textwidth, trim={0 0.4cm 0 0.4cm}, clip]{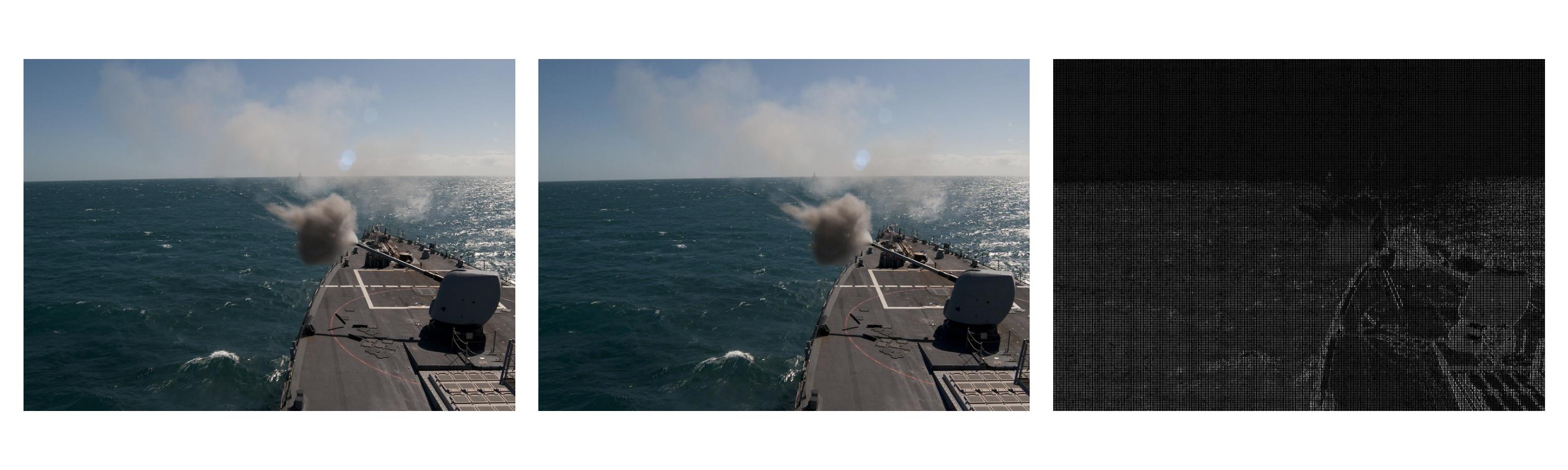}
    \includegraphics[width=0.9\textwidth, trim={0 0.4cm 0 0.4cm}, clip]{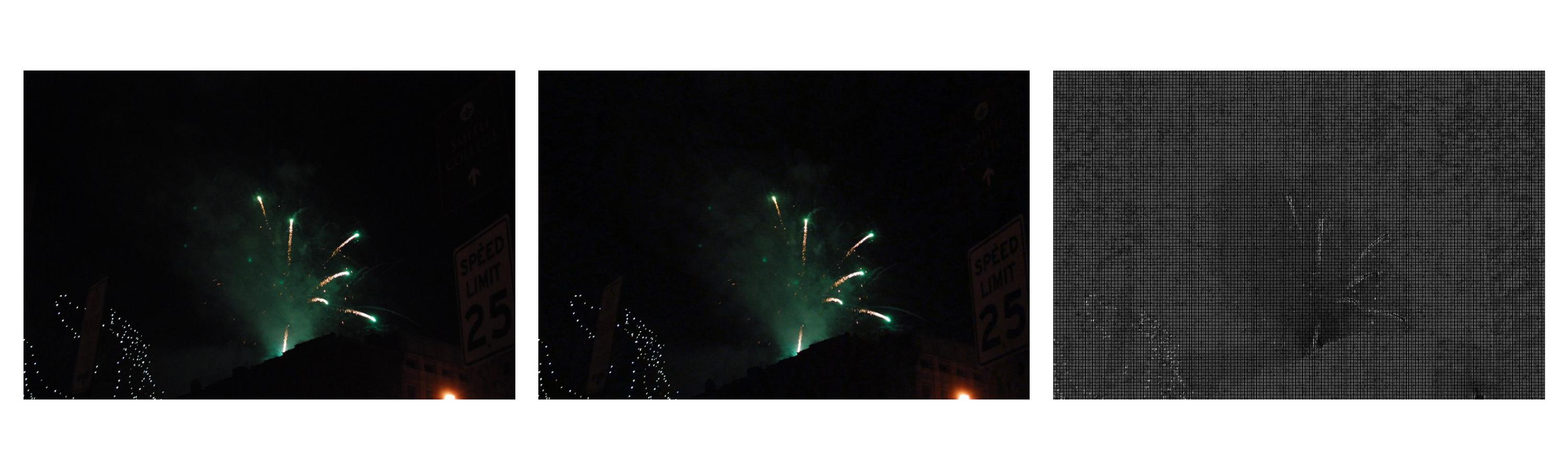}
    \includegraphics[width=0.9\textwidth, trim={0 0.4cm 0 0.4cm}, clip]{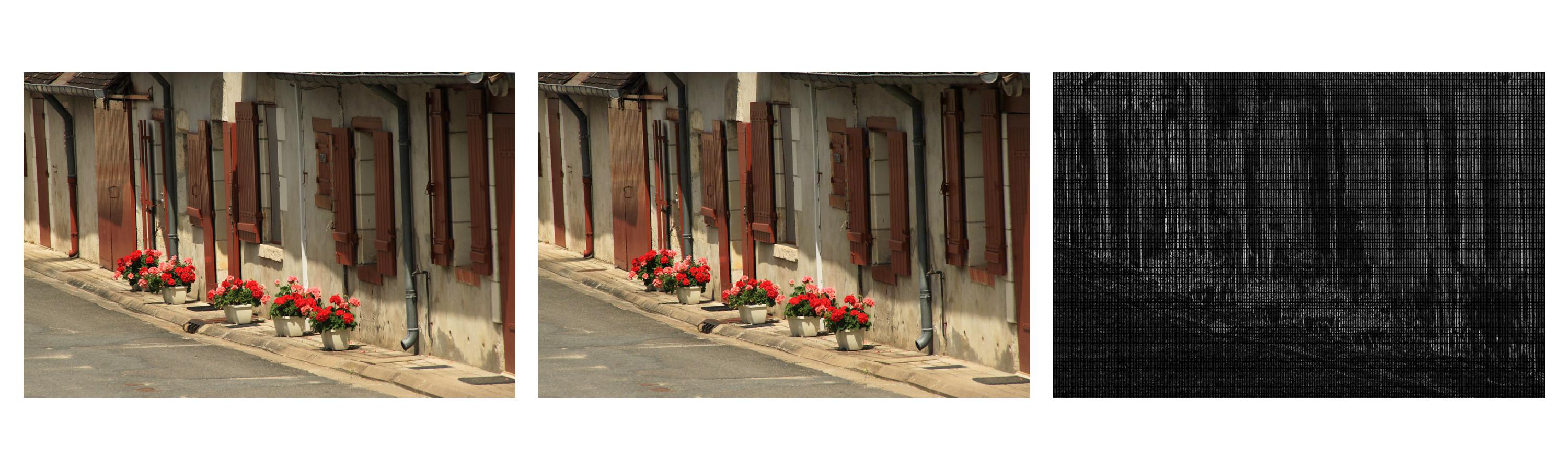}
    \captionsetup{font=small}
    \caption{More activated images. (Left) original images, (Middle) activated images, (Right) pixel-wise difference. Images are 
    \href{http://www.flickr.com/photos/87738888@N03/8039927199/}{R000005.jpg}, 
    \href{http://www.flickr.com/photos/79653482@N00/7766743380/}{R000045.jpg}, 
    \href{http://www.flickr.com/photos/12420018@N03/5452964454/}{R000076.jpg}, 
    \href{http://www.flickr.com/photos/56594044@N06/5933951717/}{R000172.jpg} and 
    \href{http://www.flickr.com/photos/31369133@N04/5570867046/}{R000396.jpg}.
    }
    \label{chap2/fig:more_qualitative2}
\end{figure}

Example of activated images for the IVF-PQ are given in Fig.~\ref{chap2/fig:more_qualitative2} while the image metrics are as follows: PSNR$=43.8\pm 2.2$~dB, SSIM$=0.98 \pm 0.01$, and $\norm{I-I'}_\infty=14.5 \pm 1.2$.
These results are computed on 10k images, the $\pm$ indicates the standard deviation.
The perturbation is very hard to notice (if not invisible), even in flat areas of the images because the perceptual model focuses on textures.
We also see that the perturbation forms a regular pattern.
This is due to the image (bilinear) resize that happens before feature extraction.

\section{Ablations}

\subsection{Comparison with $\ell_\infty$ constraint embedding}\label{chap2/sec:linf}
\begin{figure}[b!]
    \centering
    \begin{subfigure}[b]{0.49\textwidth}
        \centering
        \includegraphics[width= 0.48\textwidth]{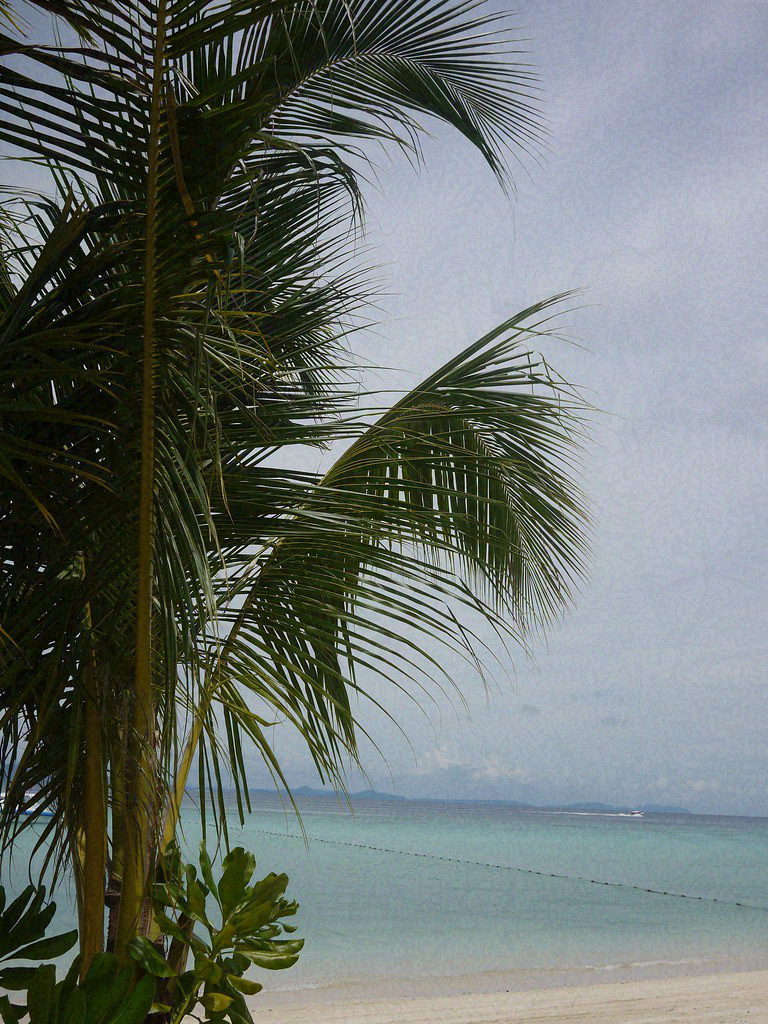}
        \includegraphics[width= 0.48\textwidth]{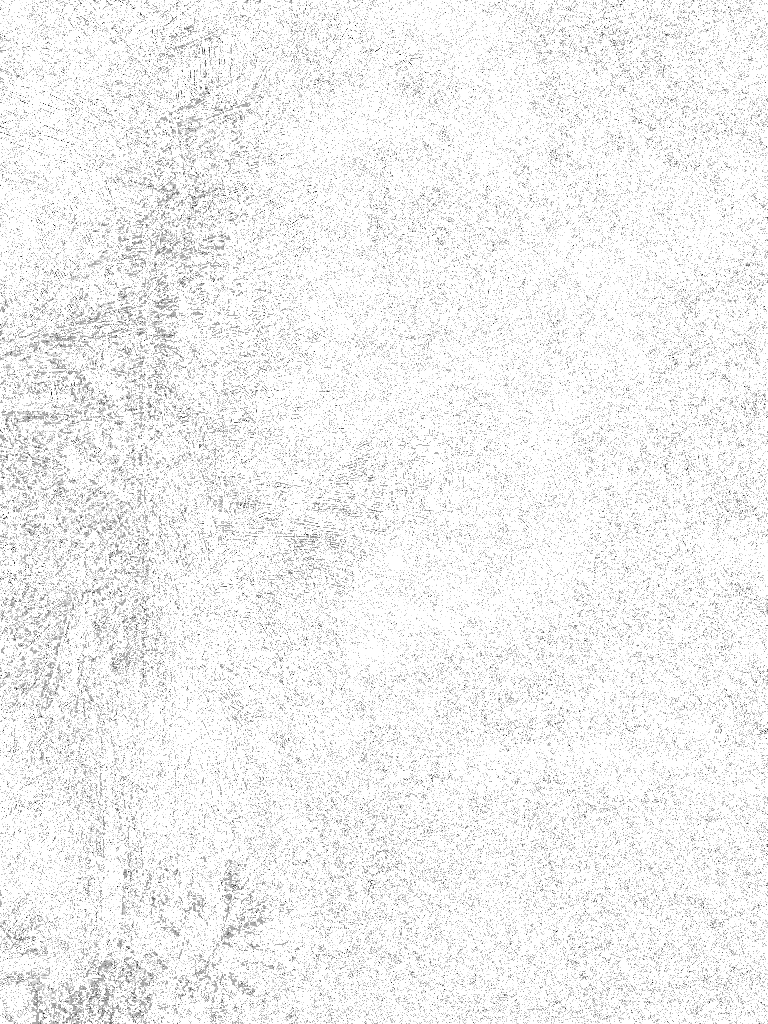}
        \captionsetup{font=small}
        \caption{$\ell_\infty=4$, $\mathrm{PSNR}=36.4$~dB, $\mathrm{SSIM}=0.91$}
    \end{subfigure}
    \hfill
    \begin{subfigure}[b]{0.49\textwidth}
        \centering
        \includegraphics[width= 0.48\textwidth]{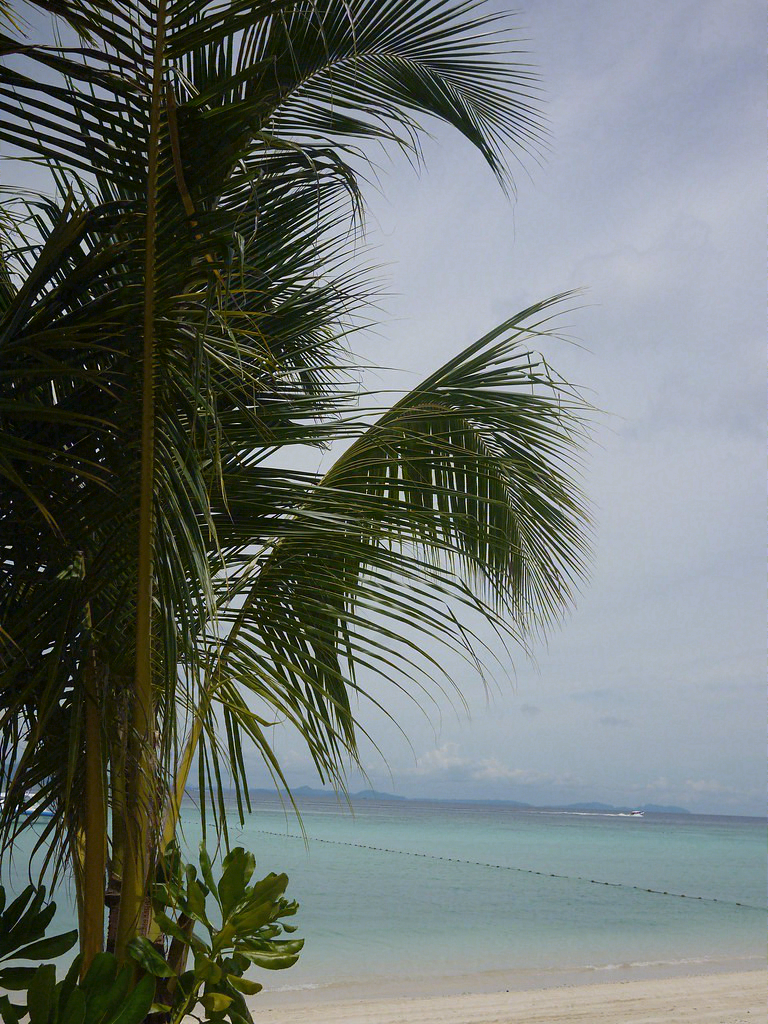}
        \includegraphics[width= 0.48\textwidth]{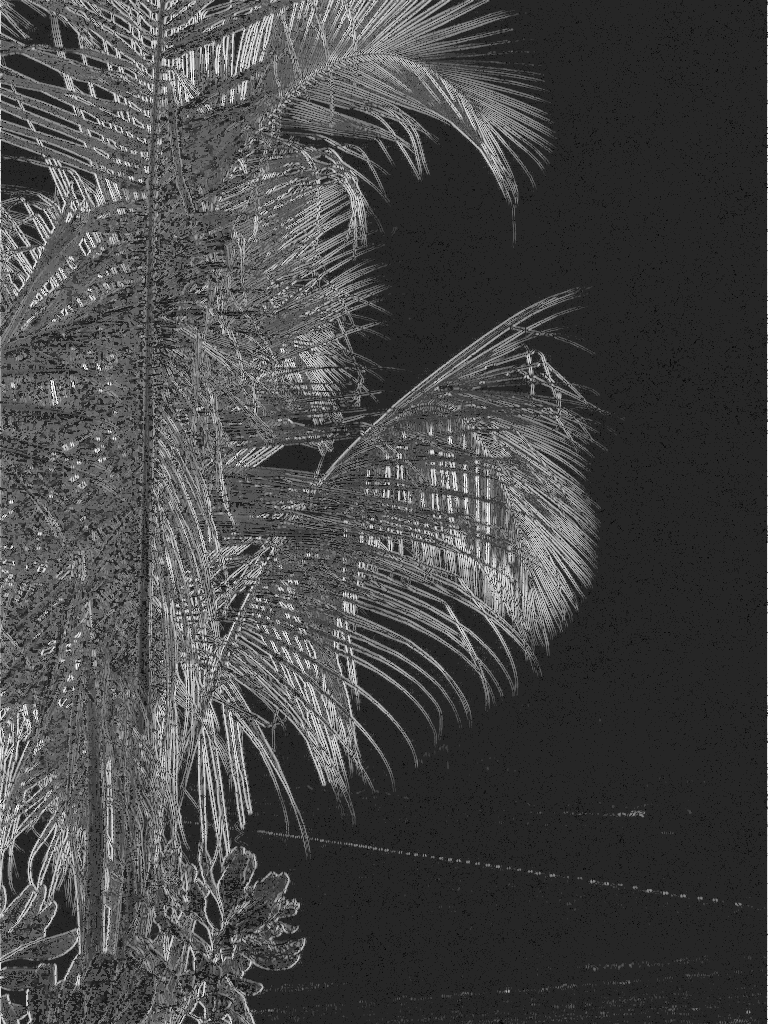}
        \captionsetup{font=small}
        \caption{$\ell_\infty=23$, $\mathrm{PSNR}=34.4$~dB, $\mathrm{SSIM}=0.94$}
    \end{subfigure}
    \captionsetup{font=small}
    \caption{
        Activated images, either with (a) the $\ell_\infty \leq 4$ constraint or with (b) our perceptual model (best viewed on screen).
        We give the corresponding measures between the original and the protected image, as well as the pixel-wise difference.
        The perturbation on the right is much less perceptible thanks to the perceptual model, even though its $\ell_\infty$ distance with the original image is much higher.
    }
    \label{chap2/fig:linf_vs_perc}
\end{figure}

\autoref{chap2/fig:linf_vs_perc} shows the same image activated using either the $\ell_\infty$ constraint (commonly used in the adversarial attack literature) or our perceptual constraint based on the JND model explained above.
Even with very small $\varepsilon$ ($4$ over 255 in the example bellow), the perturbation is visible  especially in the flat regions of the images, such as the sea or sky.

\citep{laidlaw2021perceptual} also show that the $\ell_\infty$ is not a good perceptual constraint.
They use the LPIPS loss~\citep{zhang2018unreasonable} as a surrogate for the HVS to develop more imperceptible adversarial attacks.
Although a similar approach could be used here, we found that at this small level of image distortion the LPIPS did not capture CM and LA as well as the handcrafted perceptual models present in the compression and watermarking literature.

\subsection{Image quality tradeoff}

\begin{figure}[b!]
        \centering
        \includegraphics[width=0.55\textwidth]{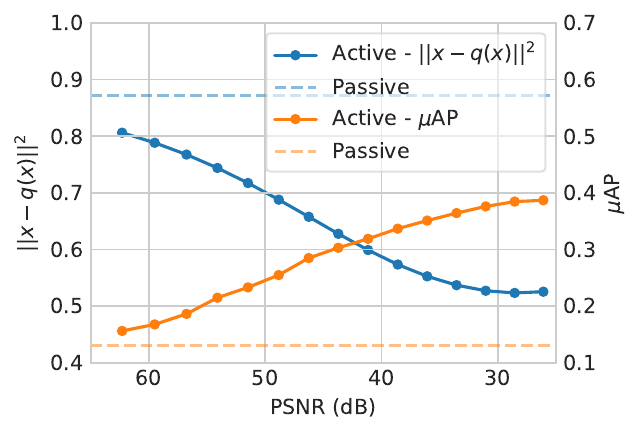}
        \caption{PSNR trade-off. As the PSNR decreases, the {\color{orange}\muap (orange)} gets better, because the {\color{blue} distance (blue)} between activated representations $x$ and $q(x)$ decreases.}
        \label{chap2/fig:psnr}
\end{figure}

For a fixed index and neural extractor, the performance of active indexing mainly depends on the scaling $\alpha$ that controls the activated image quality.
In Fig. \ref{chap2/fig:psnr}, we repeat the previous experiment for different values of $\alpha$ and plot the $\mu$AP against the average PSNR. 
As expected, lower PSNR implies better $\mu$AP. For instance, at PSNR 30~dB, the $\mu$AP is augmented threefold compared to the passive case.
Indeed, for strong perturbations the objective function of \eqref{eq:objective} can be further lowered, reducing even more the gap between representations and their quantized counterparts.

\begin{figure}[b!]
    \centering
    \includegraphics[width=0.9\textwidth]{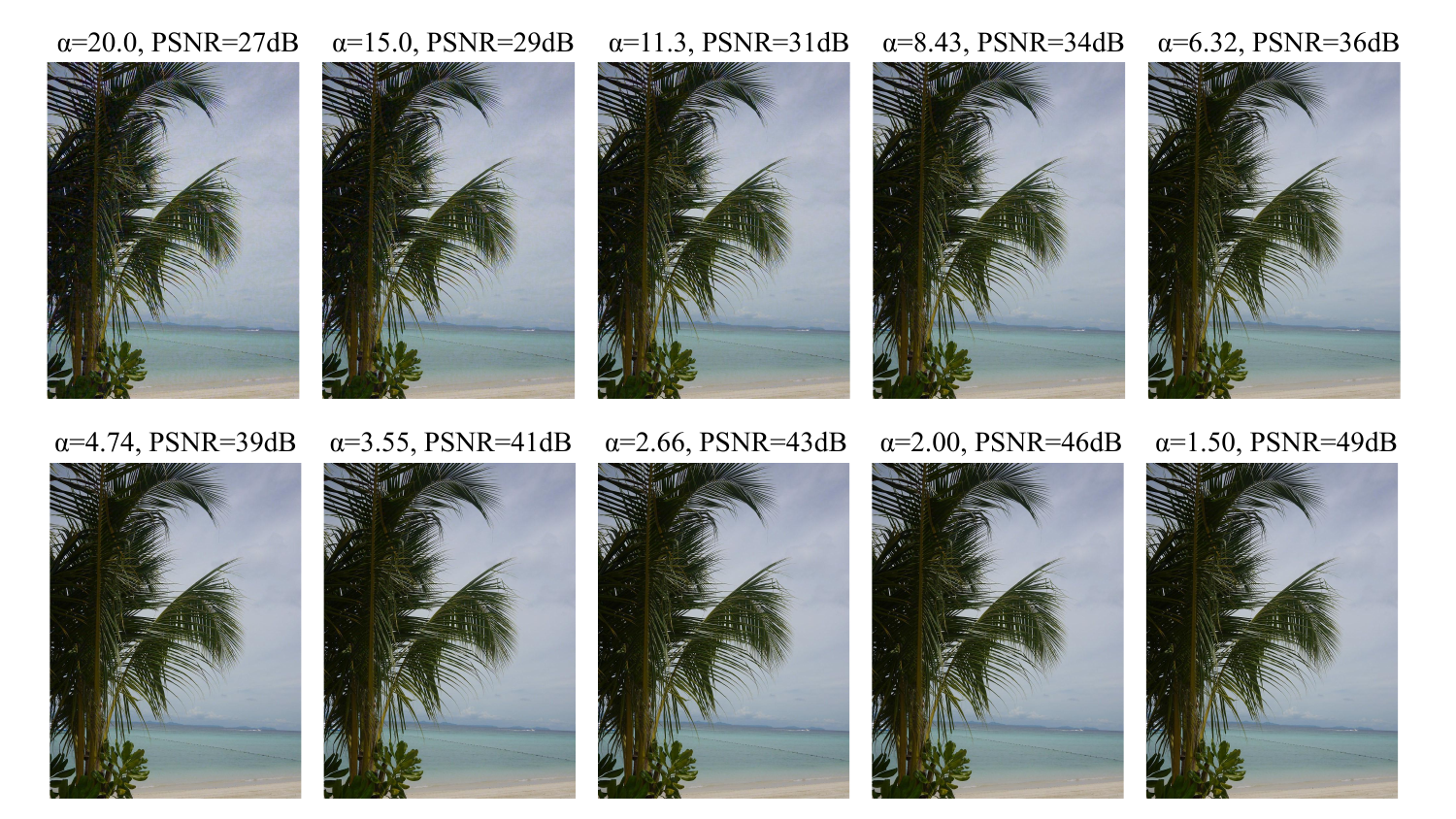}
    \captionsetup{font=small}
    \caption{Example of one activated image at different levels of $\alpha$.}
    \label{chap2/fig:more_qualitative}
\end{figure}

\autoref{chap2/fig:more_qualitative} gives example of an image activated at several values of perturbation strength $\alpha$ of Eq.~\eqref{eq:scaling} 
(for instance, for $\alpha=20$ the image has PSNR $27$dB and for $\alpha=1$ the image has PSNR $49$dB).
The higher the $\alpha$, the more visible the perturbation induced by the activation is.
Nevertheless, even with low PSNR values ($<35$dB), it is hard to notice if an image is activated or not.

\subsection{Speeding-up the optimization}\label{chap2/sec:speedup}

In our experiments, the optimization is done using 10 iterations of gradient descent, which takes approximately 40ms/image.
If the indexation time is important (often, this is not the case and only the search time is), it can be reduced at the cost of some accuracy.

We activated 10k reference images, with the same IVF-PQ indexed presented in Sec.~\ref{chap2/sec:act_vs_passive} with only one step of gradient descent with a higher learning rate.
Activation times are computed on average. 
The \rone results in Tab.~\ref{chap2/tab:speedup} indicate that the speed-up in the image optimization has a small cost in retrieval accuracy.
Specifically, it reduces the \rone for unedited images. 
The reason is that the learning rate is too high: it can cause the representation to be pushed too far and to leave the indexing cell. 
This is why a higher number number of steps and a lower learning rate are used in practice.
If activation time is a bottleneck, it can however be useful to use less optimization steps.

\begin{table}[t!]
    \centering
    \captionsetup{font=small}
    \caption{\rone for different transformations applied before search, with either 1 step at learning rate 10, or 10 steps at learning rate 1.}
    \label{chap2/tab:speedup}
    \resizebox{\linewidth}{!}{
    \begingroup
        \setlength{\tabcolsep}{4pt}
        \def\arraystretch{1.1}
            \begin{tabular}{ l|l| *{15}{c}}
            \multicolumn{1}{c}{}           & \multicolumn{1}{l}{\rot{Activation}} & \rot{Identity} & \rot{Contr. 0.5} & \rot{Contr. 2.0} & \rot{Bright. 0.5} & \rot{Bright. 2.0} & \rot{Hue 0.2} & \rot{Blur 2.0}  & \rot{JPEG 50}  & \rot{Rot. 25} & \rot{Rot. 90} & \rot{Crop 0.5} & \rot{Resi. 0.5} & \rot{Meme} & \rot{Random} & \rot{Avg.} \\ \midrule
            Passive & - & 1.00 & 0.73 & 0.39 & 0.73 & 0.28 & 0.62 & 0.48 & 0.72 & 0.07 & 0.14 & 0.14 & 0.72 & 0.14 & 0.13 & 0.45 \\
            \rowcolor{gray!15} lr=1 - 10 steps & 39.8 ms/img & 1.00 & 1.00 & 0.96 & 1.00 & 0.92 & 1.00 & 0.96 & 0.99 & 0.10 & 0.50 & 0.29 & 1.00 & 0.43 & 0.32 & 0.75 \\
            lr=10 - 1 step &  4.3 ms/img & 0.99 & 0.99 & 0.92 & 0.99 & 0.84 & 0.99 & 0.95 & 0.99 & 0.10 & 0.39 & 0.25 & 0.99 & 0.36 & 0.27 & 0.72  \\
            \bottomrule
    \end{tabular}
    \endgroup
    }

\end{table}

\subsection{Data augmentation at indexing time and EoT}

Expectation over Transformations~\citep{athalye2018eot} was originally designed to create adversarial attacks robust to a set of image transformations.
We follow a similar approach to improve robustness of the marked image against a set of augmentations $\mathcal{T}$. 
At each optimization step, we randomly sample $A$ augmentations $\{t_i\}_{i=1}^A$ in $\mathcal{T}$ and consider the average loss: 
$\Lf = \sum_{i=1}^{A} \mathcal{L}(I,t_i;I_o) /A $.
In our experiments, $\mathcal{T}$ encompasses rotations, Gaussian blurs, color jitters and a differentiable approximation of the JPEG compression~\citep{shin2017jpeg}. 
$A$ is set to $8$ and we always take the un-augmented image in the chosen set of augmentations. 

We activated 10k reference images, with the same IVF-PQ as Sec.~\ref{chap2/sec:act_vs_passive} with or without using EoT.
Table \ref{chap2/tab:eot} shows the average \rone performance over the images submitted to different transformations before search. 
EoT brings a small improvement, specifically on transformations where base performance is low (\eg, rotation or crops here).
However, it comes at a higher computational cost since each gradient descent iteration needs $A$ passes through the network, and since fewer images can be jointly activated due to GPU memory limitations (we need to store and back-propagate through $A$ transformations for every image).
If the time needed to index or activate an image is not a bottleneck, using EoT can therefore be useful. 
Otherwise, it is not worth the computational cost.

\begin{table}[t!]
    \centering
    \captionsetup{font=small}
    \caption{\rone for different transformations applied before search, with or without EoT when activating the images.}
    \label{chap2/tab:eot}
    \resizebox{\linewidth}{!}{
    \begingroup
        \setlength{\tabcolsep}{4pt}
        \def\arraystretch{1.1}
            \begin{tabular}{ l |l| *{15}{c}}
            \multicolumn{1}{c}{}          & \multicolumn{1}{l}{\rot{Activation}}  & \rot{Identity} & \rot{Contr. 0.5} & \rot{Contr. 2.0} & \rot{Bright. 0.5} & \rot{Bright. 2.0} & \rot{Hue 0.2} & \rot{Blur 2.0}  & \rot{JPEG 50}  & \rot{Rot. 25} & \rot{Rot. 90} & \rot{Crop 0.5} & \rot{Resi. 0.5} & \rot{Meme} & \rot{Random} & \rot{Avg.} \\ \midrule
                Without EOT & 40 ms & 1.00 & 1.00 & 0.96 & 1.00 & 0.92 & 1.00 & 0.96 & 0.99 & 0.10 & 0.50 & 0.29 & 1.00 & 0.43 & 0.32 & 0.75 \\
               \rowcolor{gray!15} 
                With EOT & 870 ms & 1.00 & 1.00 & 0.95 & 1.00 & 0.92 & 1.00 & 0.95 & 0.99 & 0.14 & 0.64 & 0.33 & 1.00 & 0.45 & 0.33 & 0.76 \\
            \bottomrule
    \end{tabular}
    \endgroup
    }
\end{table}

\section{Active indexing vs. watermarking}\label{chap2/sec:watermarking}

\paragraph*{Discussion.}
Watermarking and active indexing both modify images for tracing and authentication, however there are significant differences between them.
Watermarking embeds arbitrary information into the image. The information can be a message, a copyright, a user ID, etc. 
In contrast, active indexing modifies it to improve the efficiency of the search engine.
Watermarking also focuses on the control over the False Positive Rate of copyright detection, \ie, a bound on the probability that a random image has the same message as the watermarked one (up to a certain distance).

Although watermarking considers different settings than indexing methods, it could also be leveraged to facilitate the re-identification of near-duplicate images. In this supplemental section, we consider it to address a use-case similar to the one we address in this chapter with our active indexing approach. 
In this scenario, the watermark encoder embeds binary identifiers into database images.
The decoded identifier is then directly mapped to the image (as the index of a list of images).

\paragraph*{Experimental setup.}
In the rest of the section, we compare active indexing against recent watermarking techniques based on deep learning. 
\begin{itemize}[leftmargin=0.5cm,itemsep=0cm,topsep=-0.1cm]
    \item For indexing, we use the same setting as in Sec.~\ref{chap2/sec:experimental} (IVF-PQ index with 1M reference images).
    When searching for an image, we look up the closest neighbor with the help of the index.
    \item For watermarking, we encode $20$-bit messages into images, which allows to represent $2^{20}\approx 10^6$ images (the number of reference images).
    When searching for an image, we use the watermark decoder to get back an identifier and the corresponding image in the database.
\end{itemize}
Like before, we use \rone as evaluation metric.
For indexing, it corresponds to the accuracy of the top-1 search result. 
For watermarking, the \rone also corresponds to the word accuracy of the decoding, that is the proportion of images where the message is perfectly decoded. 
Indeed, with $20$-bit encoding almost all messages have an associated image in the reference set, so an error on a single bit causes a mis-identification (there is no error correction\footnote{In order to provide error correction capabilities, one needs longer messages. This makes it more difficult to insert bits: in our experiments, with 64 bits we observe a drastic increase of the watermarking bit error rate.}).

We use two state-of-the-art watermarking methods based on deep learning:
SSL Watermarking presented in Chap.~\ref{chapter:ssl-watermarking}, which also uses an adversarial-like optimization to embed messages, and \Gls*{HiDDeN}~\citep{zhu2018hidden}, which encodes and decodes messages thanks to Conv-BN-ReLU networks.
The only difference with the original methods is that their perturbation $\delta$ is modulated by the handcrafted perceptual attenuation model. 
This approximately gives the same image quality, thereby allowing for a direct comparison between active indexing and watermarking.
To better compare with Chap.~\ref{chapter:ssl-watermarking} we also make the image optimization as similar as possible to Active indexing (10 steps with no augmentation).

\paragraph*{Results.}
\autoref{chap2/tab:watermarking} compares the \rone when different transformations are applied before search or decoding.
Our active indexing method is overall the best by a large margin. 
For some transformations, watermarking methods are not as effective as passive indexing, yet for some others, like crops for \Gls*{HiDDeN}, the watermarks are more robust.

\begin{table}[t]
    \centering
    \caption{\rone for different transformations applied before search, when using either watermarking or active indexing. Results are averaged on 1k images. Best result is in \textbf{bold} and second best in \textit{italic}. }
    \label{chap2/tab:watermarking}
    \resizebox{1.0\linewidth}{!}{
        \setlength{\tabcolsep}{4pt}
        \def\arraystretch{1.1}
            \begin{tabular}{ p{4.5cm}| *{14}{c}c}
            \multicolumn{1}{c}{}          & \rot{Identity} & \rot{Contr. 0.5} & \rot{Contr. 2.0} & \rot{Bright. 0.5} & \rot{Bright. 2.0} & \rot{Hue 0.2} & \rot{Blur 2.0}  & \rot{JPEG 50}  & \rot{Rot. 25} & \rot{Rot. 90} & \rot{Crop 0.5} & \rot{Resi. 0.5} & \rot{Meme} & \rot{Random} & \rot{Avg.} \\ \midrule
    Passive indexing       & \bf 1.00 & 0.73 & 0.39 & 0.73 & 0.28 & 0.62 & 0.48 & \it 0.72 & \it 0.07 & 0.14 & 0.14 & \it  0.72 & 0.14 &  0.13 & 0.45  \\ 
        Active indexing (ours) & \bf 1.00 & \bf 1.00 & \bf 0.96 & \bf 1.00 & \bf 0.92 & \bf 1.00 & \bf 0.96 & \bf 0.99 & \bf 0.10 & \bf 0.50 & \it 0.29 & \bf 1.00 & 0.43 & \bf 0.32 & \bf 0.75 \\ \midrule
    \parbox{4.5cm}{SSL Watermarking \\ \citep{fernandez2022sslwatermarking}} & \bf 1.00 & \it 0.98 & \it 0.53 & \it 0.98 & \it 0.63 & \it 0.85 & 0.13 & 0.00 & 0.00 & \it 0.15 & 0.11 & 0.00 & \it 0.46 & 0.07 & 0.42 \\ \midrule
             \parbox{4.5cm}{\Gls*{HiDDeN}\footnote{} \\ \citep{zhu2018hidden}} & \it 0.94 & 0.87 & 0.36 & 0.85 & 0.55 & 0.00 & \it 0.81 & 0.00 & 0.00 & 0.00 & \bf 0.92 & 0.44 & \bf 0.77 & \it 0.16 & \it 0.48 \\ 
            \bottomrule
    \end{tabular}
    }
\end{table}

\footnotetext{Our implementation. 
As reported in other papers from the literature, results of the original paper are hard to reproduce.
Therefore to make it work better, our model is trained on higher resolution images (224$\times$224), with a payload of $20$-bits, instead of 30 bits embedded into 128$\times$128. 
Afterwards, the same network is used on images of arbitrary resolutions, to predict the image distortion which is later rescaled as in Eq.~\eqref{eq:scaling}.
In this setting the watermark can not always be inserted (6\% failure).}

%% file: chapter-2/sections/6-conclusion.tex
\section{Conclusion}

We introduce a way to improve ICD in large-scale settings, when images can be changed before release.
This limits the application but is not unrealistic, for instance this would be the case for generative AI providers, that control what is output by the APIs, or for big social platforms that could provide modified images after they have been uploaded for the first time.
It leverages an optimization scheme, similar to adversarial examples, that modifies images so that (1) their representations are better suited for indexing, (2) the perturbation is invisible to the human eye.
We provide grounded analyses on the method and show that it significantly improves retrieval performance of activated images, on a number of neural extractors and indexing structures.

The method has several limitations. 
(1) Activating images takes time (in the order of 10~ms/image) but one advantage is that the database may contain both active and passive images: active indexing does not spoil the performance of passive indexing and vice-versa. This is good for legacy compliance and also opens the door to flexible digital asset management strategies (actively indexing images of particular importance).
(2) The method is not agnostic to the indexing structure and extractor that are used by the similarity search.
(3) an adversary could break the indexing system in several ways.
In a black-box setting (no knowledge of the indexing structure and neural network extractor), adversarial purification~\citep{shi2021online} could get rid of the perturbation that activated the image.
In a semi-white-box setting (knowledge of the feature extractor), targeted mismatch attacks against passive indexing like~\citep{tolias2019targeted} may also work. 
Adversarial training~\citep{madry2017towards} could be a defense.
For instance, it is interesting to know if adversarial training prevents active indexing, or if the perceptual perturbation that is used in our method is still able to push features in the latent space of a robust and defended neural network.

%% file: chapter-3/main.tex
\chapter[Rooting Watermarks in Latent Diffusion Models]{Rooting Watermarks in Latent Diffusion Models}\label{chapter:stable-signature}

This chapter is based on the paper \fullcite{fernandez2023stable}.

Generative image modeling enables a wide range of applications but raises ethical concerns about responsible deployment.
In particular, it is challenging to trace the origin of generated content, which is important for content moderation, intellectual property, and accountability, in a context where deepfakes and manipulated images are increasingly common.
This chapter introduces Stable Signature, an active content tracing method combining image watermarking and Latent Diffusion Models. 
The goal is for all generated images to conceal an invisible watermark allowing for future detection and/or identification.
The method quickly fine-tunes the latent decoder of the image generator, conditioned on a binary signature.
A pre-trained watermark extractor recovers the hidden signature from any generated image and a statistical test then determines whether it comes from the generative model. 
We evaluate the invisibility of the watermarks on a variety of generation tasks, and show that they are robust to strong image modifications.
For instance, it detects the origin of an image generated from a text prompt, then cropped to keep 10\% of the content, with 90+\% accuracy at a false positive rate below 10$^{-6}$.
Code is available at \url{github.com/facebookresearch/stable_signature}.

\newpage
\input{chapter-3/sections/1-intro.tex}

\input{chapter-3/sections/7-related.tex}

\input{chapter-3/sections/2-statement.tex}

\input{chapter-3/sections/3-method.tex}

\input{chapter-3/sections/4-applications.tex}

\input{chapter-3/sections/5-results.tex}

\input{chapter-3/sections/6-ablations.tex}

\input{chapter-3/sections/8-conclu.tex}

%% file: chapter-3/sections/1-intro.tex
\section{Introduction}

Recent progress in generative modeling and natural language processing enable easy creation and manipulation of photo-realistic images.
They have given birth to many image edition tools like ControlNet~\citep{zhang2023adding}, Instruct-Pix2Pix~\citep{brooks2022instructpix2pix}, and others~\citep{couairon2022diffedit, gal2022image, ruiz2022dreambooth}, that are becoming mainstream creative tools for artists, designers, and the general public.
While this is a great step forward for image generative AI, it also undermines confidence in the authenticity or veracity of photo-realistic images. 
Indeed, methods for photo-realistic image edition existed before, but generative AI significantly lowers the barriers to convincing synthetic image generation and edition (\eg, a generated picture recently won an art competition~\citep{gault2022vice}).
This raises new risks like deep fakes, impersonation or copyright usurpation~\citep{brundage2018malicious, denton2021ethical}.
A tool to determine that images are AI-generated would make it easier to ensure their compliance with ethical standards and to remove them from certain platforms. 
    
A baseline solution to identify generated images is forensics, \ie, methods to detect generated/manipulated images passively (the image is not modified for identification).
An active baseline is to apply existing watermarking methods after the image generation, like the ones presented in Sec.~\ref{chap0/sec:deep learning-watermarking} and in Chap.~\ref{chapter:ssl-watermarking},~\ref{chapter:active-indexing} and \ref{chapter:audioseal}.
This has several drawbacks. 
If the model leaks or is open-sourced, the post-generation watermarking can be removed trivially.
The open source Stable Diffusion~\citep{2022stablediffusion} is a case in point, since removing the watermark amounts to commenting out a single line in the source code.

\begin{figure}[t]
    \centering 
    \includegraphics[width=\linewidth, trim={0cm 0cm 1.5cm 0cm}, clip]{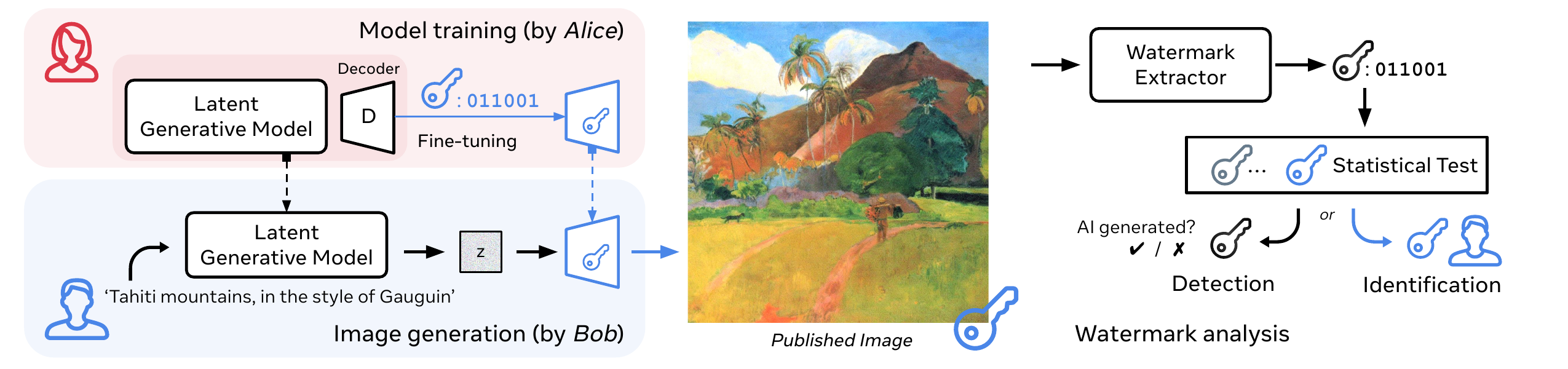}
    \caption{
        Overview of Stable Signature. 
        The latent decoder of the latent diffusion model is fine-tuned to preemptively embed a signature into all generated images.
    }
    \label{chap3/fig:fig1}
\end{figure}

Our method, called Stable Signature, merges watermarking into the generation process itself, without any architectural change.
It adjusts the pre-trained generative model such that all the images it produces conceal a given watermark.
There are several advantages to this approach~\citep{lin2022cycleganwm, yu2022responsible}.
It does not require additional processing of the generated image, which makes the watermarking computationally lighter, straightforward, and secure.
Model providers could deploy their models to different user groups with a unique watermark, and monitor that they are used in a responsible manner.
They could give art platforms, news outlets and other sharing platforms %
the ability to detect when an image has been generated by their AI.

We focus on Latent Diffusion Models (LDM)~\citep{rombach2022high} that can perform a wide range of generative tasks.
We show that simply fine-tuning a small part of the generative model -- the decoder that generates images from the latent vectors -- is enough to natively embed a watermark into generated images.
Stable Signature does not require an architectural change and does not modify the diffusion process. 
Hence it is to our knowledge compatible with all LDM-based generative methods~\citep{brooks2022instructpix2pix, couairon2022diffedit, peebles2022dit, ruiz2022dreambooth, zhang2023adding}.
The fine-tuning stage is performed by back-propagating a combination of a perceptual image loss and a message decoding loss from a watermark extractor back to the LDM decoder.
We pre-train the extractor with a simplified version of the deep watermarking method HiDDeN~\citep{zhu2018hidden}, as presented in Chap.~\ref{chapter:technical-background}, Sec.~\ref{chap0/sec:deep learning-watermarking}.

We create an evaluation benchmark close to real world situations where images may be edited.
The tasks are: detection of AI generated images, tracing models from their generations.
For instance, we detect $90\%$ of images generated with the generative model, even if they are cropped to $10\%$ of their original size, while flagging only one false positive every $10^6$ images. 
To ensure that the model's utility is not weakened, we show that the FID~\citep{heusel2017gans} score of the generation is not affected and that the generated images are perceptually indistinguishable from the ones produced by the original model. 
This is done over several tasks involving LDM (text-to-image, inpainting, edition, etc.).

As a summary, 
(1) we efficiently merge watermarking into the generation process of LDMs, in a way that is compatible with most of the LDM-based generative methods;
(2) we demonstrate how it can be used to detect and trace generated images, through a real-world evaluation benchmark;
(3) we compare to post-hoc watermarking methods, showing that it is competitive while being more secure and efficient, and (4) evaluate robustness to intentional attacks.

%% file: chapter-3/sections/7-related.tex
\section{Related work}

We here provide a brief overview of the literature on watermarking, generative models, and detection of AI-generated/manipulated images that is relevant to this chapter. 
Please refer to Chap.~\ref{chapter:related-work} for a more comprehensive review.

\paragraph{Latent generative models.} 
They are a class of generative models that first generate a latent representation of the image to create, and then decode it to the image space.
Most of state-of-the-art models are based on this principle, since it allows for the generation to happen in a lower-dimensional, sometimes quantized, space, instead of the pixel space:
\begin{itemize}
    \item Diffusion-based models -- of which Stable Diffusion~\citep{rombach2022high} is the flagship -- generate images by diffusing a noise vector in the latent space of a perceptual autoencoder trained as VQGAN~\citep{esser2021taming}.
    These models can also perform conditional image generation like inpainting or text-guided image editing by fine-tuning the diffusion model with additional conditioning, \eg, masked input image, segmentation map, etc.~\citep{lugmayr2022repaint, saharia2022palette}.
    Because of their iterative denoising algorithm, diffusion models can also be adapted for image editing in a zero-shot fashion by guiding the generative process~\citep{couairon2022diffedit, hertz2022prompt, kawar2022imagic, mokady2022null, valevski2022unitune, wu2022unifying}.
    \item Sequence-based models treat the generation process as a sequence-to-sequence problem~\citep{ramesh2021zero, ding2021cogview, esser2021taming, gafni2022make, singer2022makeavideo, yu2022scaling, team2024chameleon}. 
    They are based on image tokenizers (\eg, VQGAN~\citep{esser2021taming}, ViT-VQGAN~\citep{yu2021vector}), which are models that convert images into discrete tokens, and vice versa.
    A transformer model is used to generate the tokens given some conditioning (from text, image, etc.), which are decoded back into an image.
    This methodology aligns with the advancements in language models and leverages the scaling capabilities known from large language models.
\end{itemize}

We present these methods together because they operate in the latent space, requiring a latent decoder to produce an RGB image. 
It presents two interesting properties.
First, it makes the watermarking method presented in the chapter directly applicable to these models.
Second, the watermarking method is robust to changes in the generative model, as long as the latent decoder is kept unchanged. 
This is particularly helpful since, most of the time, only the latent generative model is fine-tuned to give new capabilities~\citep{zhang2023adding, brooks2022instructpix2pix}.
 
\paragraph{Detection of AI-generated/manipulated images.}
Purely relying on forensics and passive detection is limiting, \eg, the best performing method to our knowledge~\citep{corvi2022detection} is able to detect $50\%$ of generated images for a false positive rate around $1$/$100$:
if a user-generated content platform were to receive $1$ billion images every day, it would wrongly flag $10$ million images to detect only half of the generated images.
Besides, passive techniques cannot trace images from different versions of the same model, in contrast with watermarking.

\paragraph{Watermarking image generative models.}
The first works aim to watermark the training set on which the generative model is learned~\citep{yu2021artificial}.
It is highly inefficient since every new message to embed requires a new training pipeline.
Merging the watermarking and the generative process is a recent idea~\citep{fei2022supervised, lin2022cycleganwm, nie2023attributing, qiao2023novel, wu2020watermarking, yu2022responsible, zhang2020model}, that is closer to the model watermarking literature (see Chap.~\ref{chapter:invariants} for more details).
They suffer from two strong limitations.
First, these methods only apply to GANs, while LDM are progressively replacing them for most applications. 
Second, watermarking is incorporated in the training process of the GAN from the start. 
This strategy is unsustainable because the generative model training is more and more costly\footnote{Stable Diffusion training costs $\sim$\$600k of cloud compute (\href{https://en.wikipedia.org/wiki/Stable_Diffusion}{Wikipedia}).}.
Our work shows that a quick fine-tuning of the latent decoder part of the generative model is enough to achieve a good watermarking performance, provided that the watermark extractor is well chosen.

%% file: chapter-3/sections/2-statement.tex
\section{Problem statement and background}

\autoref{chap3/fig:fig1} shows a model provider \emph{Alice} who deploys a latent diffusion model to users \emph{Bobs}.
Stable Signature embeds a binary signature into the generated images. 
This section derives how Alice can use this signature for two scenarios:
\begin{itemize}
    \item 
    \emph{Detection: ``Is it generated by my model?''.} 
    Alice detects if an image was generated by her model.
    Generated images should be flagged as reliably as possible, while controlling the probability of flagging a natural image. 
    \item 
    \emph{Identification: ``Who generated this image?''.} 
    Alice monitors who created each image, while avoiding to mistakenly identifying a Bob. %
\end{itemize}

\subsection{Image watermarking for detection}
Alice embeds a $k$-bit binary signature into the generated images.
The watermark extractor then decodes messages from the images it receives and detects when the message is close to Alice's signature.
An example application is to block AI-generated images on a content sharing platform.

\paragraph{Statistical test.}\label{chap3/subsec:statistical-test}
Let $m\in \{ 0,1 \}^{k}$ be Alice's signature. 
We extract the message $m'$ from an image $x$ and compare it to $m$.
As done in previous works~\citep{lin2022cycleganwm, yu2021artificial},
the detection test relies on the number of matching bits $M(m,m')$: if
\begin{equation} 
M\left(m,m'\right) \geq \tau \,\,\textrm{ where }\,\, \tau\in \{0,\ldots,k\},
\label{chap3/eq:detectiontest}
\end{equation}
then the image is flagged.
This provides a level of robustness to imperfections of the watermarking. 
This is a typical use case of the binomial test presented in Sec.~\ref{chap0/sec:test}.

Formally, we test the statistical hypothesis $\H_1$: ``$x$ was generated by Alice's model'' 
against the null hypothesis $\H_0$: ``$x$ was not generated by Alice's model''.
Under $\H_0$ (\ie, for vanilla images), we assume that bits $m'_1,\ldots, m'_k$ are (i.i.d.) Bernoulli random variables with parameter $0.5$.
Then $M(m, m')$ follows a binomial distribution with parameters ($k$, $0.5$).
We verify this assumption experimentally in Sec.~\ref{chap3/app:assumption}.
The False Positive Rate (FPR) is the probability that $M(m, m')$ takes a value bigger than the threshold $\tau$.
It is obtained from the c.d.f. of the binomial distribution, and a closed-form can be written with the regularized incomplete beta function $I_x(a;b)$ (more in~Sec.~\ref{chap0/sec:test}):
\begin{align}\label{chap3/eq:p-value}
    \text{FPR}(\tau) & = \mathbb{P}\left(M \geq \tau \mid \H_0\right) = I_{1/2}(\tau, k - \tau +1).
\end{align}

\subsection{Image watermarking for identification}
Alice now embeds a signature $m^{(i)}$ drawn randomly from $\{0,1\}^k$ into %
the model distributed to
Bob$^{(i)}$ (for $i=1\cdots N$, with $N$ the number of Bobs).
Alice can trace any misuse of her model: generated images violating her policy (gore content, deepfakes) are linked back to the specific Bob by comparing the extracted message to Bobs' signatures.

\paragraph{Statistical test.}\label{chap3/subsec:statistical-test-identification}
We compare the message $m'$ from the watermark extractor to  $m^{(1)}$, $\dots$, $m^{(N)}$. 
There are now $N$ detection hypotheses to test.
If the $N$ hypotheses are rejected, we conclude that the image was not generated by any of the models.
Otherwise, we attribute the image to $\textrm{argmax}_{i=1..N} M\left(m', m^{(i)}\right)$.
With regards to the detection task, false positives are more likely since there are $N$ tests. 
The global FPR at a given threshold $\tau$ is:
\begin{equation}\label{chap3/eq:globalFPR}
    \text{FPR}(\tau,N) = 1-(1-\text{FPR}(\tau))^N\approx N.\text{FPR}(\tau).
\end{equation}

Equation~\eqref{chap3/eq:globalFPR} (resp.~\eqref{chap3/eq:p-value}), is used reversely: we find threshold $\tau$ to achieve a required FPR for identification (resp.~detection). 
Note that these formulae hold only under the assumption of i.i.d. Bernoulli bits extracted from vanilla images. 
This condition is enforced in the next section.

%% file: chapter-3/sections/3-method.tex
\begin{figure}[b!]
    \centering
    \includegraphics[width=1.0\textwidth, trim={0cm 0.7cm 0cm 0cm}, clip]{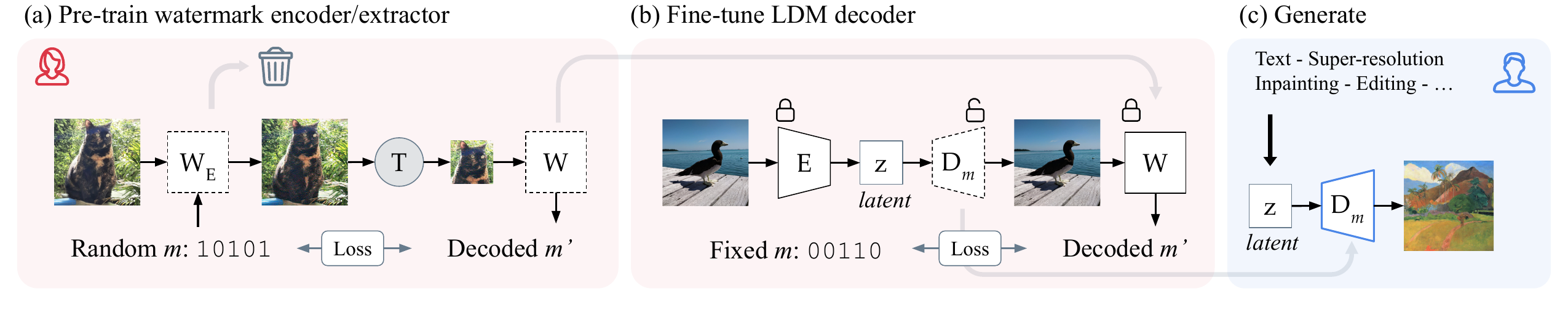}
    \caption{
        Steps of the method.
        (a) We pre-train a watermark encoder $\mathcal{W}_E$ and extractor $\mathcal{W}$, to extract binary messages.
        (b) We fine-tune the decoder $\mathcal{D}$ of the LDM's autoencoder with a fixed signature $m$ such that all the generated images (c) lead to $m$ through $\mathcal{W}$.
        }
    \label{chap3/fig:method}
\end{figure}

\section{Method}

Stable Signature modifies the generative network so that the generated images have a given signature through a fixed watermark extractor.
It is trained in two phases.
First, we create the watermark extractor network $\mathcal{W}$.
We then fine-tune the LDM decoder $\mathcal{D}$, such that all generated images yield a given signature through $\mathcal{W}$.

\subsection{Pre-training the watermark extractor}\label{chap3/subsec:pre-training}

We use \Gls*{HiDDeN}~\citep{zhu2018hidden}, a classical method in the deep watermarking literature.
It operates as presented in~\nameref{chap0/sec:deep learning-watermarking}.
We here briefly describe the method (and change the notations of the embedder and extractor to reduce possible confusions with the generative model).

We jointly optimize an encoder $\mathcal{W}_E$ and an extractor $\mathcal{W}$ to embed $\payload$-bit messages into images. 
$\mathcal{W}_E$ takes as inputs a cover image $\im_o$ and a $\payload$-bit message $m$.
$\mathcal{W}_E$ outputs a residual $\delta$ of the same size as $\im_o$, that is multiplied by a $\alpha$ to produce thewatermarked image $\im_w = \im_o + \alpha \delta$.
At each optimization step an image transformation $T$ is sampled from a set that includes common image processing operations. 
For the non-differentiable JPEG compression, we use the forward attack simulation layer as done by~\cite{zhang2021asl}. 
The soft message is extracted from the transformed image: $m' = \mathcal{W}(T(\im_w))$, and the \emph{message loss} is the Binary Cross Entropy (BCE) between $m$ and the sigmoid $\sigma (m')$.

The network architectures are kept simple to ease the LDM fine-tuning in the second phase.
They are the same as HiDDeN (see Sec.~\ref{chap3/app:archi-hidden}) with two changes. 
First, we discard $\mathcal{W}_E$ after training, since only $\mathcal{W}$ serves our purpose. 
Therefore, its perceptual quality is not as important, so the perceptual loss and the adversarial network are not needed. 
Instead, the distortion is constrained by a $\mathrm{tanh}$ function on output of $\mathcal{W}_E$ and by the scaling factor $\alpha$.
This improves the bit accuracy of the recovered message and makes it possible to increase its size $k$.
Second, we observed that $\mathcal{W}$'s output bits for vanilla images are correlated and highly biased, which violates the assumptions of Sec.~\ref{chap3/subsec:statistical-test}. 
Therefore, to get closer to i.i.d. dimensions we remove the bias and decorrelate the outputs of $\mathcal{W}$ by applying a PCA whitening transformation (more details in Sec.~\ref{chap3/app:hidden-centering}).

\subsection{Fine-tuning the generative model}\label{chap3/subsec:finetuning}

In \Gls*{LDM}, the diffusion happens in the latent space of an autoencoder.
The latent vector $z$ obtained at the end of the diffusion is input to decoder $\mathcal{D}$ to produce an image.
Here we fine-tune $\mathcal{D}$ such that the image contains a given message $m$ that can be extracted by $\mathcal{W}$.
Stable Signature is compatible with many generative tasks, since modifying only $\mathcal{D}$ does not affect the diffusion process.
We first fix the signature $m=(m_1,\ldots, m_\payload) \in \{0,1\}^k$. 
The fine-tuning of $\mathcal{D}$ into $\mathcal{D}_m$ is inspired by the original training of the autoencoder in LDM~\citep{rombach2022high}.

Training image $\im \in \mathbb{R}^{H\times W\times 3}$ is fed to the LDM encoder $\mathcal{E}$ 
that outputs activation map $z = \mathcal{E}(\im) \in \mathbb{R}^{h\times w\times c}$, downsampled by a power-of-two factor $f = H/h = W/w $.
The decoder reconstructs an image $\im' = \mathcal{D}_m(z)$ and the extractor recovers $m' = \mathcal{W} (\im')$.
The \emph{message loss} is the BCE between $m'$ and the original $m$: $\mathcal{L}_m = \mathrm{BCE}(\sigma \left( m' \right), m)$.

In addition, the original decoder $\mathcal{D}$ reconstructs the image without watermark: $\im'_o = \mathcal{D}(z)$. 
The \emph{image perceptual loss} $\mathcal{L}_\mathrm{i}$ between $\im'$ and $\im'_o$, controls the distortion.
We use the Watson-VGG perceptual loss introduced by~\cite{czolbe2020loss}, an improved version of LPIPS~\citep{zhang2018unreasonable}.
It is essential that the decoder learns luminance and contrast masking to add less visible watermarks. %
The weights of $\mathcal{D}_m$ are optimized to minimize:
\begin{equation}\label{chap3/eq:loss2}
    \mathcal{L} = \mathcal{L}_\mathrm{m} + \lambda_\mathrm{i}~ \mathcal{L}_\mathrm{i}.
\end{equation}
\noindent
This is done over $100$ iterations with the AdamW optimizer~\citep{loshchilov2017decoupled} and batch of size $4$, \ie, the fine-tuning sees \emph{fewer than 500 images} and takes \emph{one minute on a single GPU}.
The learning rate follows a cosine annealing schedule with $20$ iterations of linear warmup to $10^{-4}$ and decays to $10^{-6}$.
$\lambda_\mathrm{i}$ in Eq.~\eqref{chap3/eq:loss2} is set to $0.2$ by default.

\subsection{Implementation details and parameters}\label{chap3/app:method-details}

\paragraph*{Architectures of the watermark encoder/extractor.}\label{chap3/app:archi-hidden}
We keep the same architecture as in HiDDeN~\citep{zhu2018hidden}, which is a simple convolutional encoder and extractor.
The encoder consist of $4$ Conv-BN-ReLU blocks, with $64$ output filters, $3\times 3$ kernels, stride $1$ and padding $1$.
The extractor has $7$ blocks, followed by a block with $k$ output filters ($k$ being the number of bits to hide), an average pooling layer, and a $k\times k$ linear layer.
For more details, we refer the reader to the original paper.

\paragraph*{Watermark extractor training.}
We train on the MS-COCO dataset~\citep{lin2014microsoft}, with $256 \times 256$ images.
The number of bits is $k=48$, and the scaling factor is $\alpha=0.3$.
The optimization is carried out for $300$ epochs on $8$ GPUs, with the Lamb optimizer~\citep{you2019lamb} (it takes around a day). 
The learning rate follows a cosine annealing schedule with $5$ epochs of linear warmup to $10^{-2}$, and decays to $10^{-6}$.
The batch size per GPU is $64$.

\paragraph*{Attack simulation layer.}\label{chap3/app:hidden-attack}
The attack layer produces edited versions of the watermarked image to improve robustness.
It takes as input the image output by the watermark encoder $x_w = \mathcal{W}_E(x_o)$ and outputs a new image $x'$ that is fed to the decoder $\mathcal{W}$.
This layer is made of cropping, resizing, or identity chosen at random, unless otherwise stated.
The parameter for the crop or resize is set to $0.3$ or $0.7$ with equal probability.
This is followed by a JPEG compression with probability $0.5$.
The parameter for the compression is set to $50$ or $80$ with equal probability.
This last layer is not differentiable, therefore we back-propagate only through the difference between the uncompressed and compressed images:
$x'= x_{\mathrm{aug}} + \mathrm{nograd}(x_{\mathrm{aug}, \mathrm{JPEG}} - x_{\mathrm{aug}})$~\citep{zhang2021asl}, as explained in Sec.~\ref{chap0/sec:deep learning-watermarking}.

\paragraph*{Whitening.}\label{chap3/app:hidden-centering}
At the end of the training, we whiten the output of the extractor to make the hard thresholded bits i.i.d. Bernoulli on vanilla images -- so that the assumption of the statistical test~\ref{chap3/subsec:statistical-test} holds better, see Sec.~\ref{chap3/app:assumption}.
We perform the PCA of the output of the watermark extractor on a set of $10$k images from COCO, and get the mean $\mu$ and eigendecomposition of the covariance matrix $\Sigma= U\Lambda U^T$. 
The whitening is applied as a linear layer with bias $-\Lambda^{-1/2}U^T\mu$ and weight $\Lambda^{-1/2}U^T$, appended to the extractor. 
See Fig.~\ref{chap1/fig:whitening} for a visual representation of the whitening process.

\section{Evaluated tasks and baselines}\label{chap3/app:implementation-details}

We hereby describe the generative tasks and watermarking methods used in the experiments of the following sections.
Feel free to skip and refer to it later if needed.

\subsection{Generative tasks}\label{chap3/app:generative-tasks}

Since our method only involves the LDM decoder, it is compatible with many generative tasks. 
We evaluate text-to-image generation and image edition on the validation set of MS-COCO~\citep{lin2014microsoft}, super-resolution and inpainting on the validation set of ImageNet~\citep{deng2009imagenet}.

\paragraph*{Text-to-image.}
In text-to-image generation, the diffusion process is guided by a text prompt. 
We follow the standard protocol in the literature~\citep{ramesh2022hierarchical, ramesh2021zero, rombach2022high, saharia2022photorealistic} and evaluate the generation on prompts from the validation set of MS-COCO~\citep{lin2014microsoft}.
To do so, we first retrieve all the captions from the validation set, keep only the first one for each image, and select the first $1000$ or $5000$ captions depending on the evaluation protocol.
We use guidance scale $3.0$ and $50$ diffusion steps.
If not specified, the generation is done for $5000$ images.
The FID is computed over the validation set of MS-COCO, resized to $512\times 512$.

\paragraph*{Image edition.}
DiffEdit~\citep{couairon2022diffedit} takes as input an image, a text describing the image and a novel description that the edited image should match. 
First, a mask is computed to identify which regions of the image should be edited. 
Then, mask-based generation is performed in the latent space, before converting the output back to RGB space with the image decoder. 
We use the default parameters used in the original paper, with an encoding ratio of 90\%, and compute a set of $5000$ images from the COCO dataset, edited with the same prompts as the paper~\citep{couairon2022diffedit}.
The FID is computed over the validation set of MS-COCO, resized to $512\times 512$.

\paragraph*{Inpainting.}
We follow the protocol of LaMa~\citep{suvorov2022resolution}, and generate $5000$ masks with the ``thick'' setting, at resolution $512\times 512$, each mask covering $1-50\%$ of the initial image (with an average of $27\%$).
For the diffusion-based inpainting, we use the inference-time algorithm presented in \citep{song2020score}, also used in Glide~\citep{nichol2021glide}, which corrects intermediate estimations of the final generated image with the ground truth pixel values outside the inpainting mask. 
For latent diffusion models, the same algorithm can be applied in latent space. 
In this case, we consider $2$ different variations: (1) inpainting is performed in the latent space and the final image is obtained by simply decoding the latent image; and (2) the same procedure is applied, but after decoding, ground truth pixel values from outside the inpainting mask are copy-pasted from the original image. 
The latter allows to keep the rest of the image perfectly identical to the original one, at the cost of introducing copy-paste artifacts, visible in the borders. 
Image quality is measured with an FID score, computed over the validation set of ImageNet~\citep{deng2009imagenet}, resized to $512\times 512$.

\paragraph*{Super-resolution.}
We follow the protocol suggested by Saharia~\etal~\citep{saharia2022image}.
We first resize $5000$ random images from the validation set of ImageNet to $128\times 128$ using bicubic interpolation, and upscale them to $512\times 512$.
The FID is computed over the validation set of ImageNet, cropped and resized to $512\times 512$.

\subsection{Watermarking methods}\label{chap3/app:watermarking}
For DCT-DWT, we use the implementation of \url{https://github.com/ShieldMnt/invisible-watermark} (the one used in Stable Diffusion).
For SSL Watermark~\citep{fernandez2022sslwatermarking} and FNNS~\citep{kishore2021fixed} the watermark is embedded by optimizing the image, such that the output of a pre-trained model is close to the given key -- like in adversarial examples~\citep{goodfellow2014adversarial}, see Chap.~\ref{chapter:ssl-watermarking} for all the details.
The difference between the two is that in SSL Watermark we use a model pre-trained with DINO~\citep{caron2021dino}, while FNNS uses a watermark or stenography model.
For FNNS we use the HiDDeN extractor used in all our experiments, and not SteganoGAN~\citep{zhang2019steganogan} as in the original paper, because we want to extract watermarks from images of different sizes.
We use the image optimization scheme of Chap.~\ref{chapter:active-indexing}, \ie, we optimize the distortion image for $10$ iterations, and modulate it with a perceptual just noticeable difference (JND) mask.
This avoids visible artifacts and gives a PSNR comparable with our method ($\approx 30$dB).
For HiDDeN, we use the watermark encoder and extractor from our pre-training phase, but the extractor is not whitened and we modulate the encoder output with the same JND mask.
Note that in all cases we watermark images one by one for simplicity. 
In practice the watermarking could be done by batch, which would be more efficient.

\subsection{Evaluation metrics}\label{chap3/app:metrics}

We evaluate the image distortion with the Peak Signal-to-Noise Ratio (PSNR), which is defined as $\mathrm{PSNR}(x,x') = -10\cdot \log_{10} (\mathrm{MSE}(x,x'))$, for $x,x'\in [0,1]^{c\times h\times w}$, as well as Structural Similarity score (SSIM)~\citep{wang2004image}.
They compare images generated with and without watermark. 
On the other hand, we evaluate the diversity and quality of the generated images with the Fr\'echet Inception Distance (FID)~\citep{heusel2017gans}.

The \emph{bit accuracy} -- the percentage of bits correctly decoded -- evaluates the robustness of the watermark after transformations.
For detection, we use 
the \emph{True Positive Rate} (\Gls*{TPR}), \ie, the probability of flagging a generated image, 
the \emph{False Positive Rate} (\Gls*{FPR}), \ie, the probability of flagging a vanilla image,
and \emph{Receiver Operating Characteristic} (\Gls*{ROC}) curve, which plots TPR against FPR.

\begin{figure}[b!]
    \centering
    \footnotesize
    \begin{tabular}{*{5}{l}}
        Crop 0.1 & JPEG 50 & Resize 0.7 & Brightness 2.0 & Contrast 2.0 \\ 
        \begin{minipage}{.16\linewidth}\centering \includegraphics[width=0.3\linewidth]{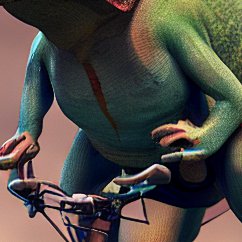}\end{minipage} &  
        \begin{minipage}{.16\linewidth}\includegraphics[width=\linewidth]{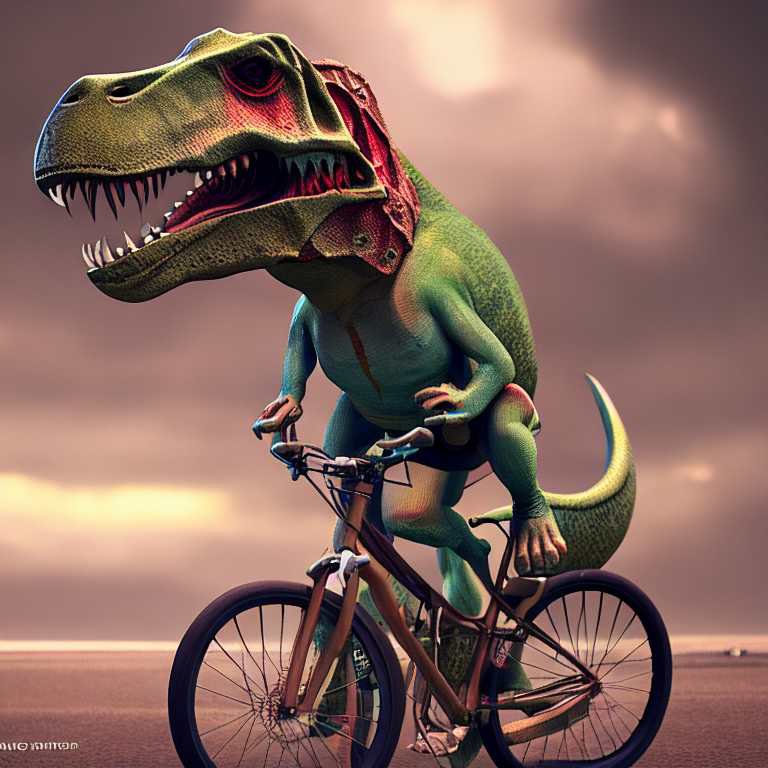}\end{minipage} &  
        \begin{minipage}{.16\linewidth}\centering\includegraphics[width=0.8\linewidth]{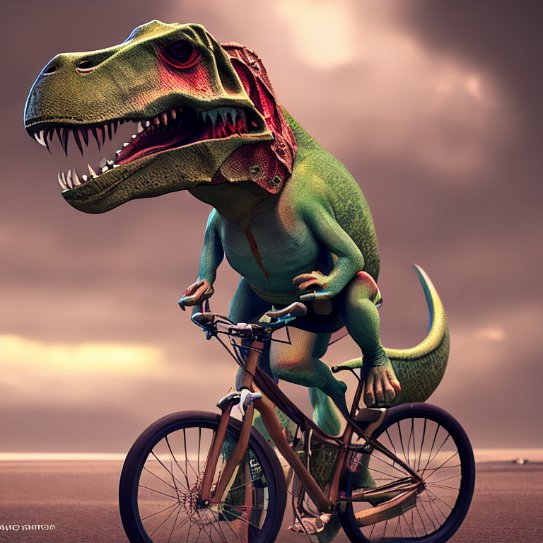}\end{minipage} &  
        \begin{minipage}{.16\linewidth}\includegraphics[width=\linewidth]{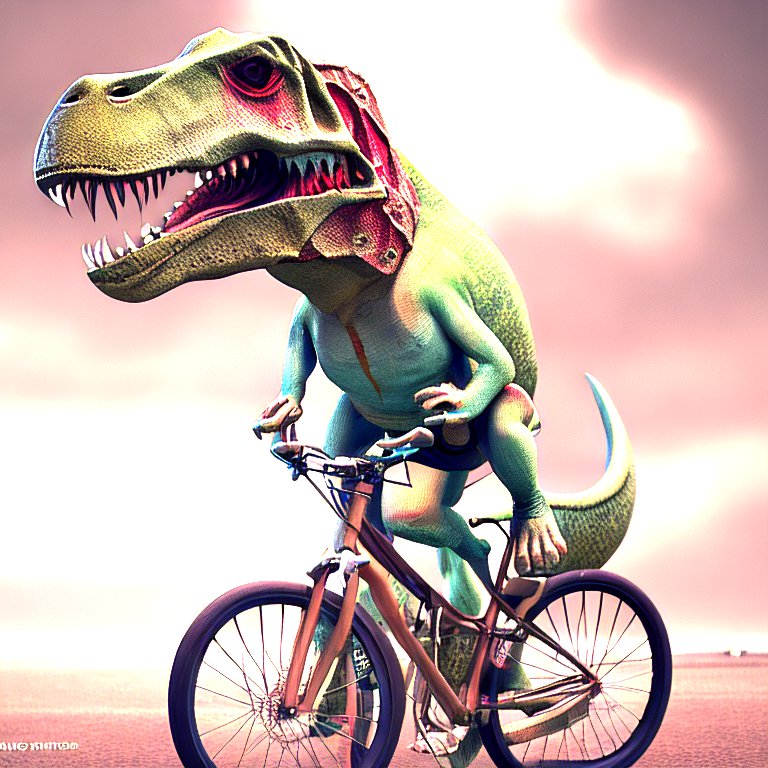}\end{minipage} &
        \begin{minipage}{.16\linewidth}\includegraphics[width=\linewidth]{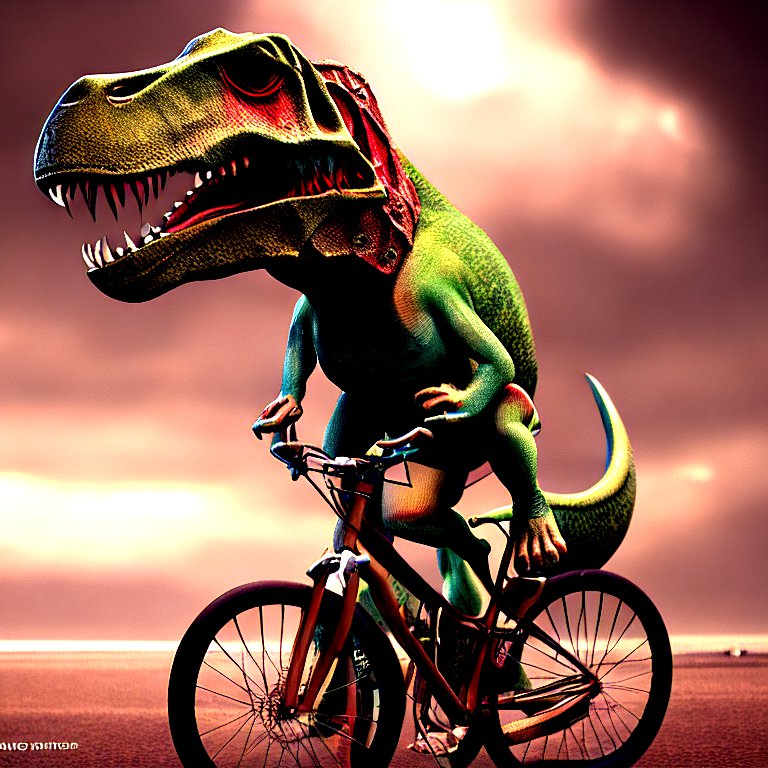}\end{minipage} 
        \\ \\
        Saturation 2.0 & Sharpness 2.0 & Rotation $90$ & Text overlay & Combined \\
        \begin{minipage}{.16\linewidth}\includegraphics[width=\linewidth]{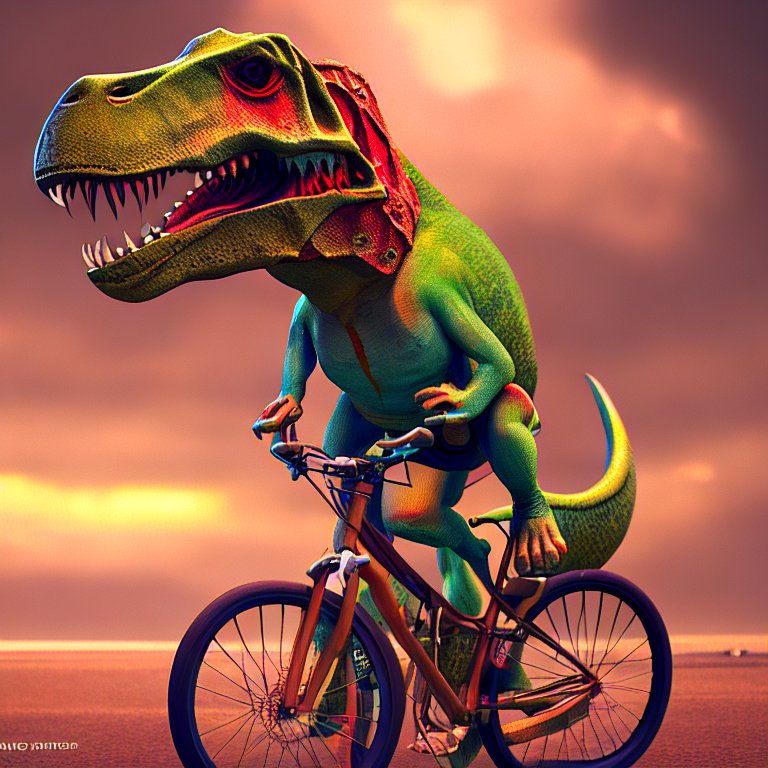}\end{minipage} &  
        \begin{minipage}{.16\linewidth}\includegraphics[width=\linewidth]{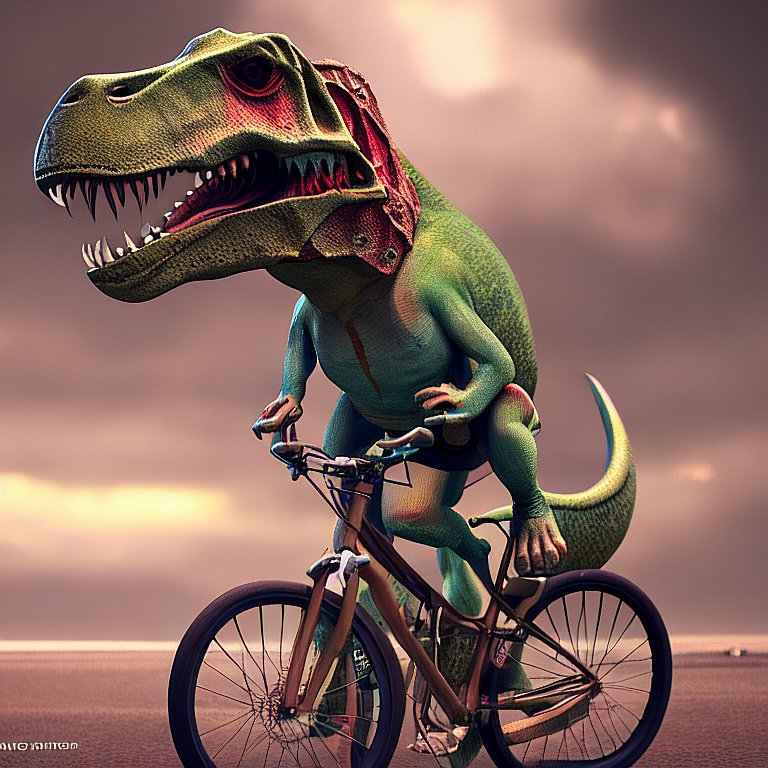}\end{minipage} &  
        \begin{minipage}{.16\linewidth}\includegraphics[width=\linewidth]{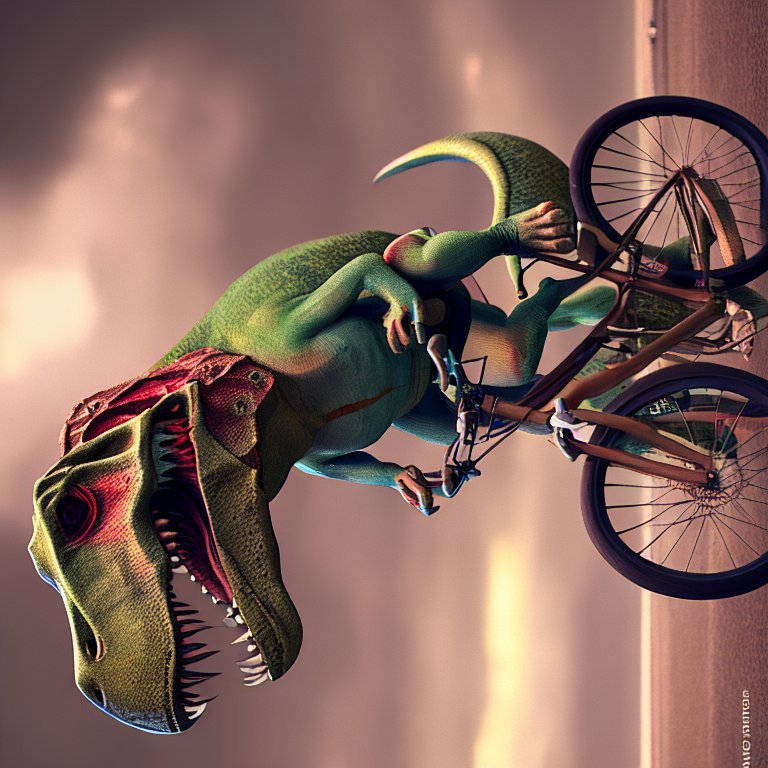}\end{minipage} &  
        \begin{minipage}{.16\linewidth}\includegraphics[width=\linewidth]{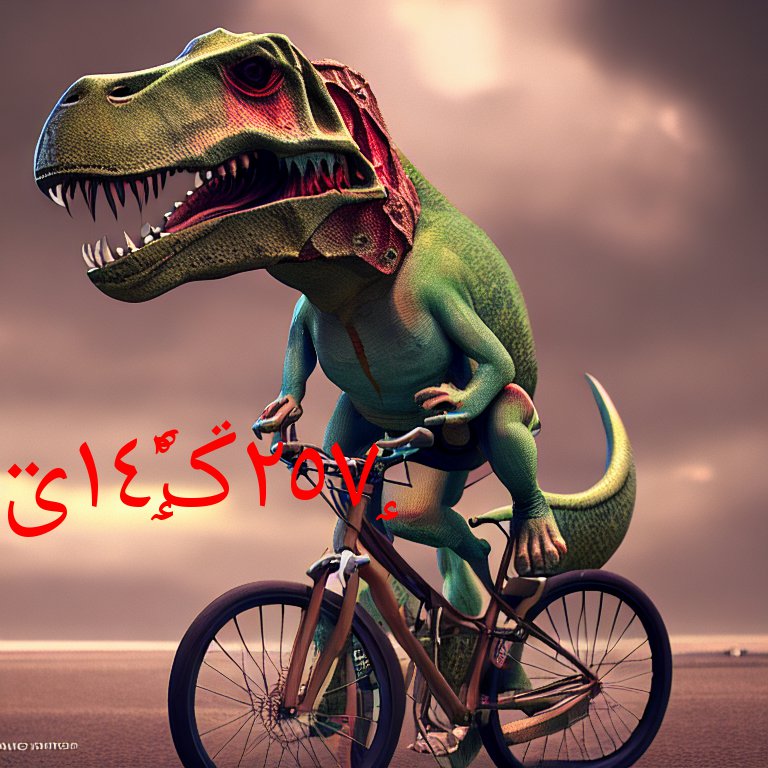}\end{minipage} &  
        \begin{minipage}{.16\linewidth}\includegraphics[width=\linewidth]{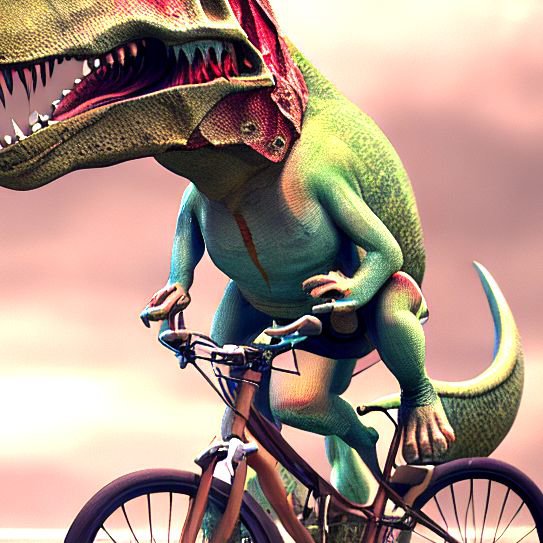}\end{minipage} 
        \\ \\
    \end{tabular}
\caption{Illustration of all transformations evaluated in sections~\ref{chap3/sec:application} and \ref{chap3/sec:experiments}.}
\label{chap3/fig:all-transformations}
\end{figure}

\subsection{Image transformations}\label{chap3/app:transformations}
We evaluate the robustness of the watermark to a set of transformations in sections~\ref{chap3/sec:application} and \ref{chap3/sec:experiments}.
They simulate image processing steps that are commonly used in image editing software.
We illustrate them in \autoref{chap3/fig:all-transformations}.
For crop and resize, the parameter is the ratio of the new area to the original area.
For rotation, the parameter is the angle in degrees.
For JPEG compression, the parameter is the quality factor (in general 90\% or higher is considered high quality, 80\%-90\% is medium, and 70\%-80\% is low).
For brightness, contrast, saturation, and sharpness, the parameter is the default factor used in the PIL and Torchvision~\citep{marcel2010torchvision} libraries.
The text overlay is made through the AugLy library~\citep{papakipos2022augly}, and adds a text at a random position in the image.
The combined transformation is a combination of a crop $0.5$, a brightness change $1.5$, and a JPEG $80$ compression.

%% file: chapter-3/sections/4-applications.tex
\section{Text-to-image watermarking performance }\label{chap3/sec:application}
This section discusses our method's detection and identification capability for images generated by a large latent diffusion model~\citep{rombach2022high}\footnote{
    We refrain from experimenting with pre-existing third-party generative models, such as Stable Diffusion or LDMs, and instead use a large diffusion model (2.2B parameters) trained on an internal dataset of 330M licensed image-text pairs.
    Although the diffusion-based generative model has been trained on an internal dataset of licensed images, we use the KL autoencoder from LDM~\citep{rombach2022high} with compression factor $f=8$.
    This is the one used by open-source alternatives.
}.
We apply generative models watermarked with $48$-bit signatures on prompts of the MS-COCO~\citep{lin2014microsoft} validation set.
We evaluate detection and identification on the outputs, as illustrated in~\autoref{chap3/fig:fig1}.
The detection rates are partly obtained from experiments and partly by extrapolating small-scale measurements.

For simplicity, we restrict the robustness evaluation to the following transformations:
strong cropping ($10\%$ of the image remaining), 
brightness shift (strength factor $2.0$), 
as well as a combination of crop $50\%$, brightness shift $1.5$ and JPEG $80$. 
This covers typical geometric and photometric edits (see Fig.~\ref{chap3/fig:all-transformations} for visual examples).

\begin{figure}[b!]
    \begin{minipage}{0.47\textwidth}
        \centering
        \includegraphics[width=1.0\textwidth, trim={0cm 0cm 0cm 0cm}, clip]{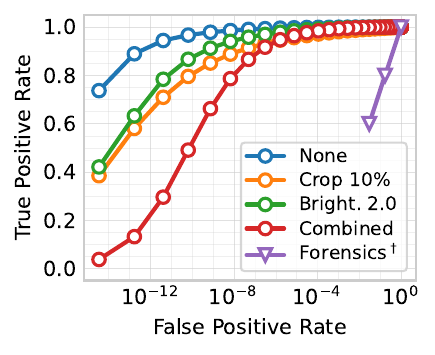}
        \caption{
            Detection results. \Gls*{ROC} curve of the detection under different transformations.
            Forensics$^\dagger$ indicates passive detection~\citep{corvi2022detection}.
        }\label{chap3/fig:tpr-fpr}
    \end{minipage}\hfill
    \begin{minipage}{0.47\textwidth}
        \centering
        \includegraphics[width=1.0\textwidth, trim={0cm 0cm 0cm 0cm}, clip]{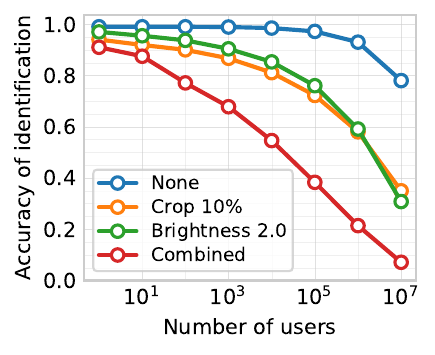}
        \caption{
            Identification results. 
            Proportion of well-identified Bobs.
            Detection with FPR=$10^{-6}$ is run beforehand, and we consider it an error if the image is not flagged.
        }\label{chap3/fig:identification}
    \end{minipage}
\end{figure}

\subsection{Detection results}
For detection, we fine-tune the decoder of the LDM with a random key $m$, generate $1000$ images and use the test of Eq.~\eqref{chap3/eq:detectiontest}.
We report the tradeoff between TPR and the FPR, while varying $\tau\in \{0, .. ,48\}$.
For instance, for $\tau=0$, we flag all images so $\textrm{FPR}=1$, and $\textrm{TPR}=1$.
The TPR is measured directly, while the FPR is inferred from Eq.~\eqref{chap3/eq:p-value}, because it would otherwise be too small to be measured on reasonably sized problems 
(this approximation is validated experimentally in Sec.~\ref{chap3/app:fpr-check}).
The experiment is run on $10$ random signatures and we report averaged results.

\autoref{chap3/fig:tpr-fpr} shows the tradeoff under image transformations.
For example, when the generated images are not modified, Stable Signature detects $99\%$ of them, while only $1$ vanilla image out of $10^9$ is flagged.
At the same $\textrm{FPR}=10^{-9}$, Stable Signature detects $84\%$ of generated images for a crop that keeps $10\%$ of the image, 
and $65\%$ for a transformation that combines a crop, a color shift, and a JPEG compression.
For comparison, we report results of a state-of-the-art passive method~\citep{corvi2022detection}, applied on resized and compressed images.
As expected, we observe that these baseline results have orders of magnitudes larger FPR than Stable Signature, which actively marks the content.

\subsection{Identification results}
Each Bob has its own copy of the generative model. 
Given an image, the goal is to find if any of the $N$ Bobs created it (detection) and if so, which one (identification).
There are $3$ types of error: 
\emph{false positive}: flag a vanilla image; 
\emph{false negative}: miss a generated image; 
\emph{false accusation}: flag a generated image but identify the wrong Bob.

For evaluation, we fine-tune $N'=1000$ models with random signatures.
Each model generates $100$ images.
For each of these $100$k watermarked images, we extract the Stable Signature message, compute the matching score with all $N$ signatures and select the Bob with the highest score. 
The image is predicted to be generated by that Bob if this score is above threshold $\tau$.
We determined $\tau$ such that $\textrm{FPR}=10^{-6}$, see Eq.~\eqref{chap3/eq:globalFPR}. 
For example, for $N=1$, $\tau=41$ and for $N=1000$, $\tau=44$.
Accuracy is extrapolated beyond the $N'$ Bobs by adding additional signatures and having $N > N'$ (\eg, Bobs that have not generated any images).

\autoref{chap3/fig:identification} reports the per-transformation identification accuracy.
For example, we identify a Bob among $N$=$10^{5}$ with $98\%$ accuracy when the image is not modified.
Note that for the combined edit, this becomes $40\%$.
This may still be dissuasive:
if a Bob generates $3$ images, he will be identified $80\%$ of the time.
We observe that at this scale, the false accusation rate is zero, \ie, we never identify the wrong Bob.
This is because $\tau$ is set high to avoid FPs, which also makes false accusations unlikely.
We observe that the identification accuracy decreases when $N$ increases, because the threshold $\tau$ required to avoid false positives is higher when $N$ increases, as pointed out by the approximation in Eq.~\eqref{chap3/eq:globalFPR}.
In a nutshell, by distributing more models, Alice trades some accuracy of detection against the ability to identify Bobs.

%% file: chapter-3/sections/5-results.tex
\section{Experimental results}\label{chap3/sec:experiments}

The previous section showed how to leverage watermarks for detection and identification of images generated from text prompts.
We now present more general results on robustness and image quality for different generative tasks.
We also compare Stable Signature to post-generation watermarking algorithms.

\subsection{Image generation quality}

\begin{figure}[b!]
    \centering
    \setlength{\tabcolsep}{0pt}
    \resizebox{0.85\linewidth}{!}{
    \begin{tabular}{c @{\hspace{0.1cm}} c @{\hspace{0.1cm}} c}
        \toprule
        Generated with original & Generated with watermark & Pixel-wise difference ($\times 10$) \\
        \midrule
        \includegraphics[width=0.33\linewidth]{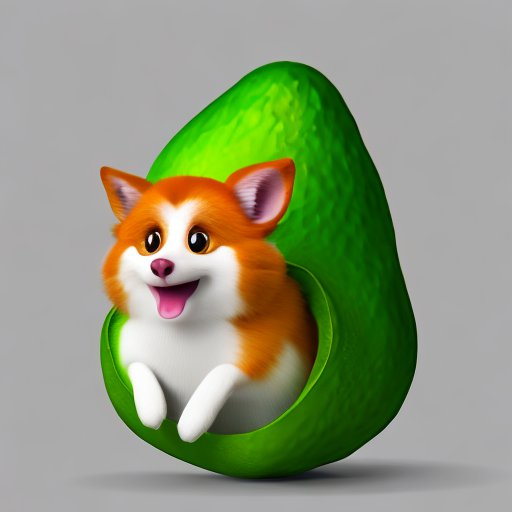} &
        \includegraphics[width=0.33\linewidth]{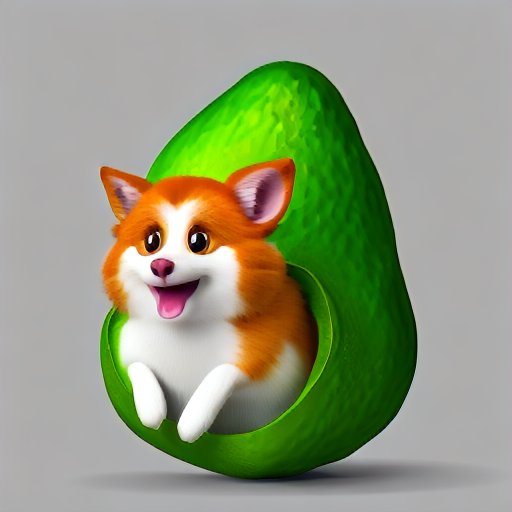} &
        \includegraphics[width=0.33\linewidth]{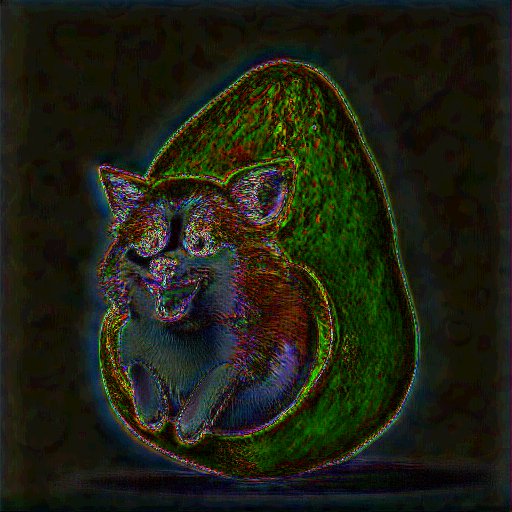} \\ 
        \rule{0pt}{0.2cm}
        \includegraphics[width=0.33\linewidth]{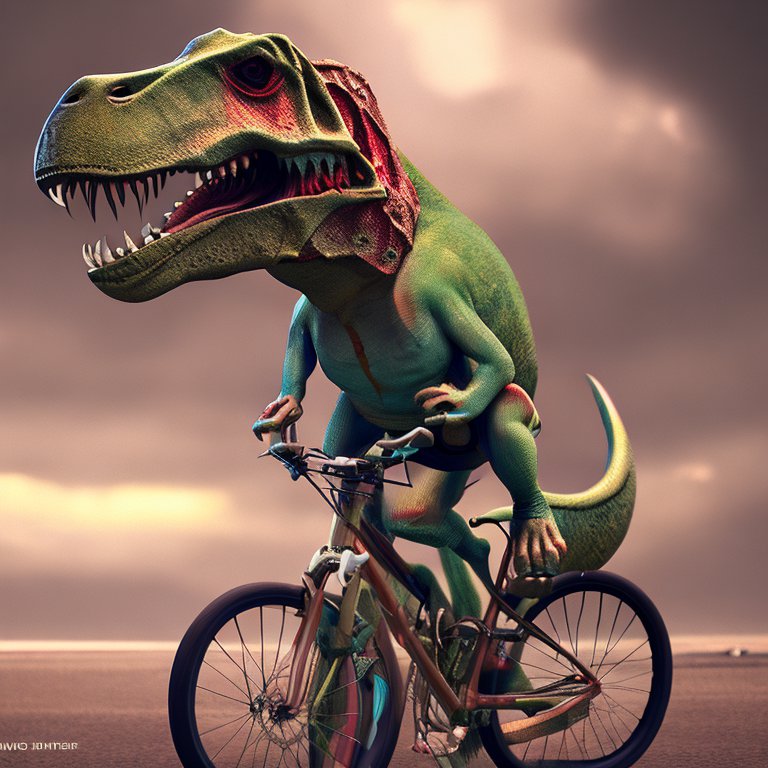} &
        \includegraphics[width=0.33\linewidth]{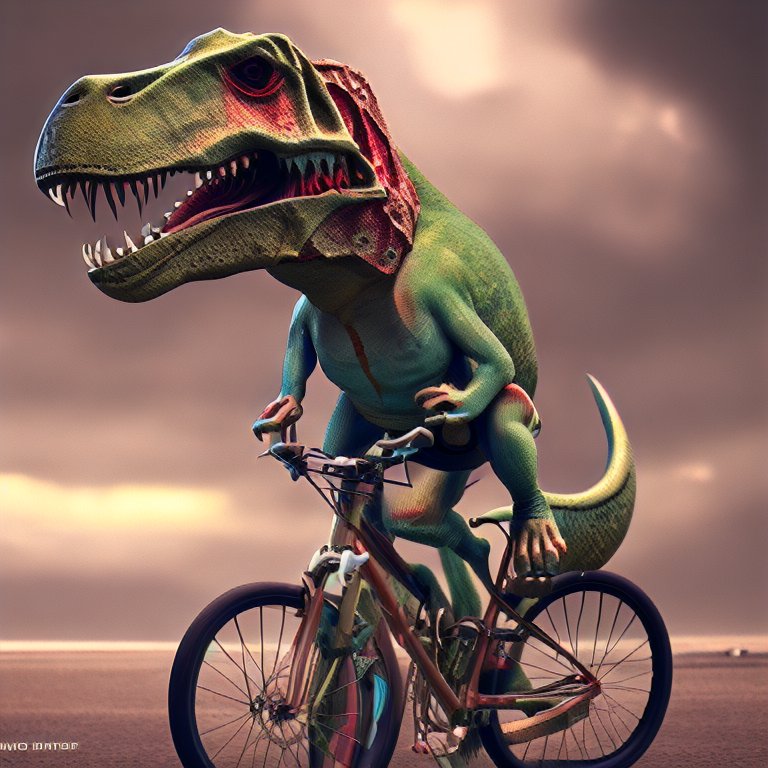} &
        \includegraphics[width=0.33\linewidth]{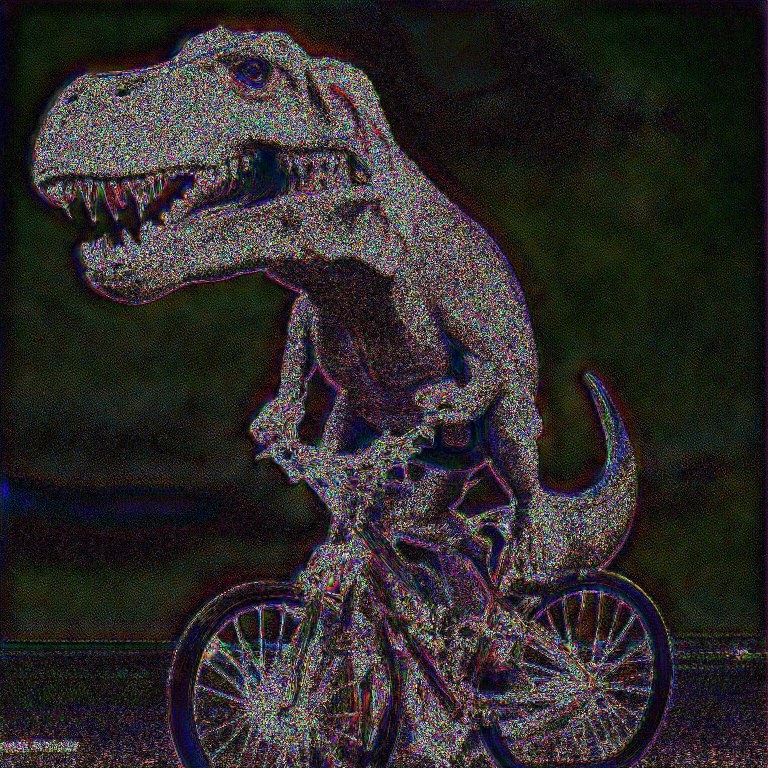} \\       
        \bottomrule\\
    \end{tabular}
    }
    \caption{
    Images generated from text prompts. 
    The PSNR is $35.4$\,dB in the first row and $28.6$\,dB in the second.
    Images generated with Stable Signature look natural because modified areas are located where the eye is not sensitive.
    }
    \label{chap3/fig:qualitative}
\end{figure}

\begin{figure}[H]
\centering
    \scriptsize
    \newcommand{\imwidth}{0.165\textwidth}
        \setlength{\tabcolsep}{0pt}
        \begin{tabular}{c@{\hskip 2pt}ccccc}
        \toprule
        Original & Stable Signature & Dct-Dwt & SSL Watermark & FNNS & HiDDeN \\
        \midrule
        \includegraphics[width=\imwidth]{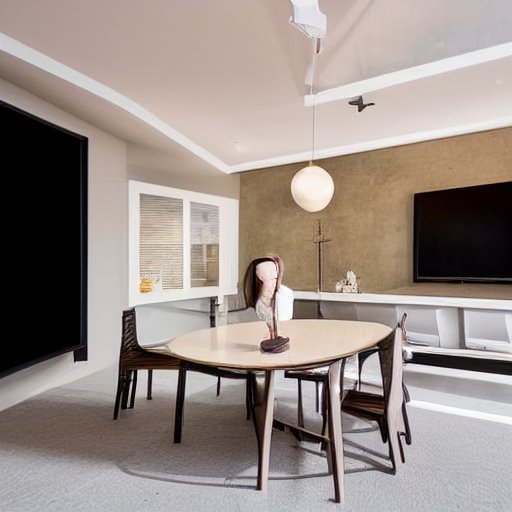} &
        \includegraphics[width=\imwidth]{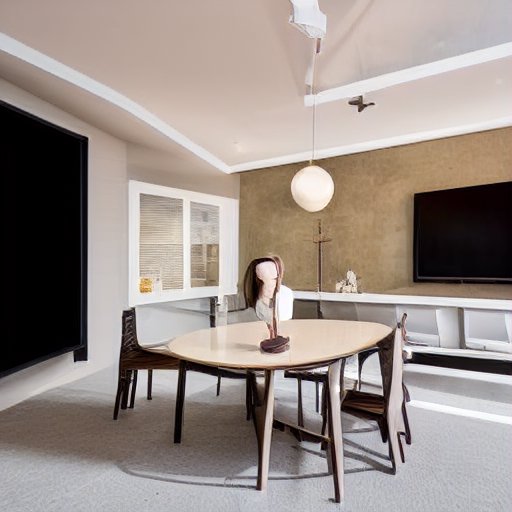} &
        \includegraphics[width=\imwidth]{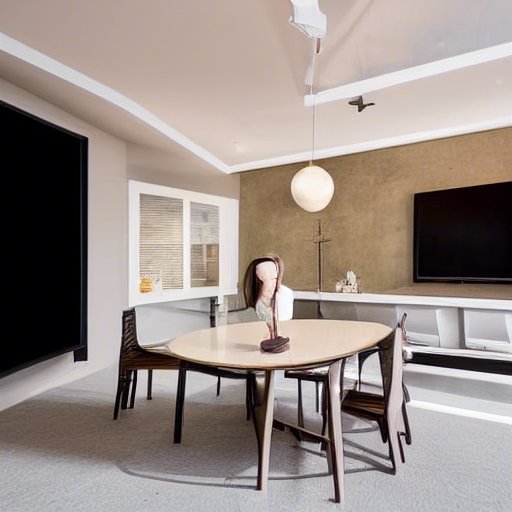} &
        \includegraphics[width=\imwidth]{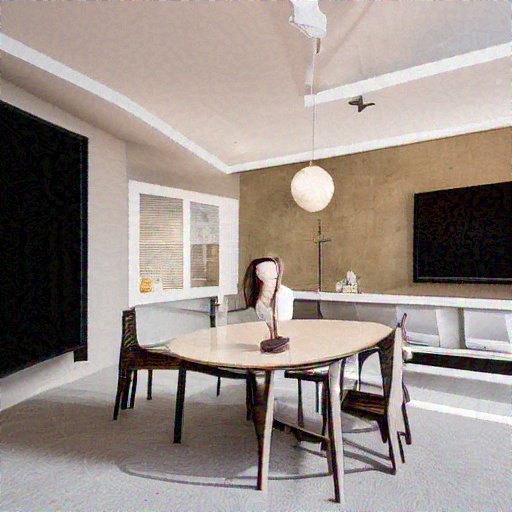} &
        \includegraphics[width=\imwidth]{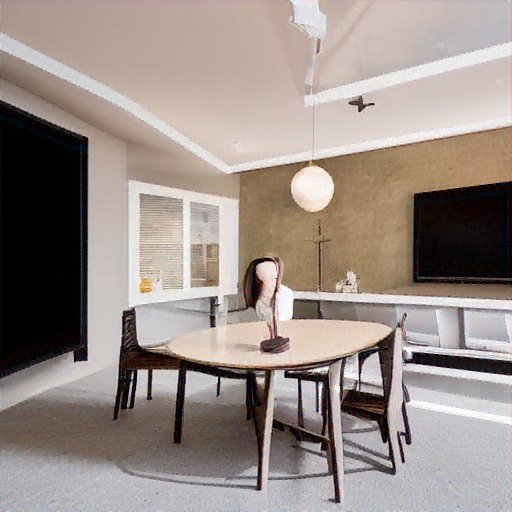} &
        \includegraphics[width=\imwidth]{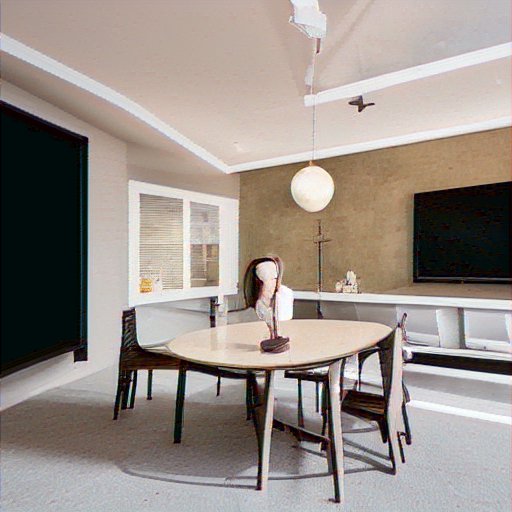} \\

            &
        \includegraphics[width=\imwidth]{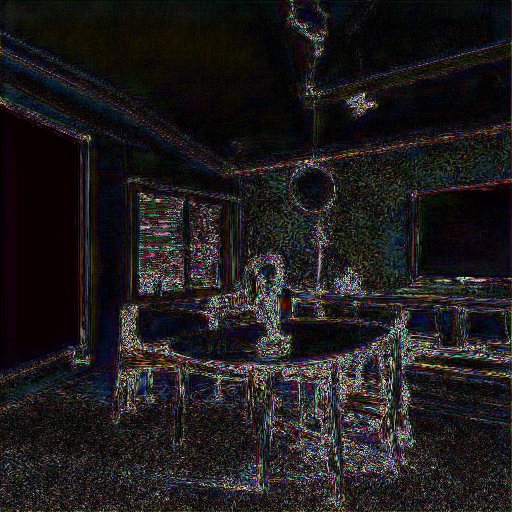} &
        \includegraphics[width=\imwidth]{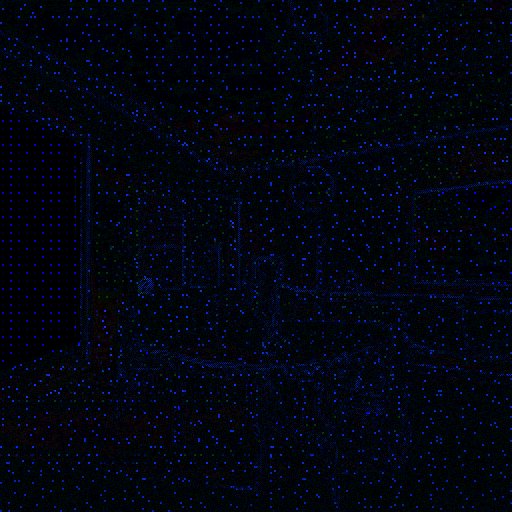} &
        \includegraphics[width=\imwidth]{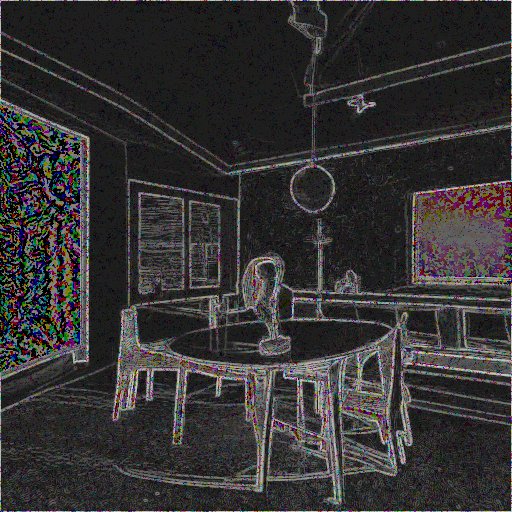} &
        \includegraphics[width=\imwidth]{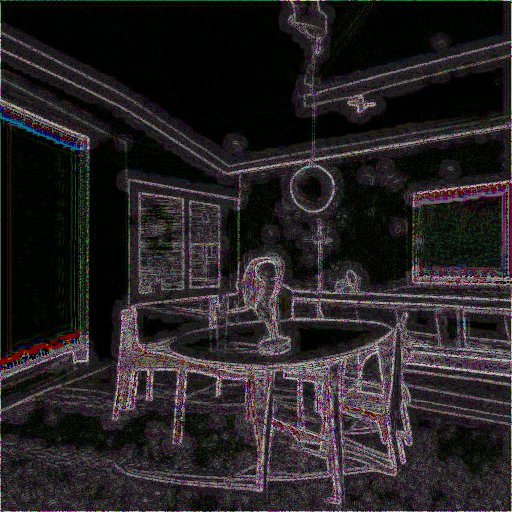} &
        \includegraphics[width=\imwidth]{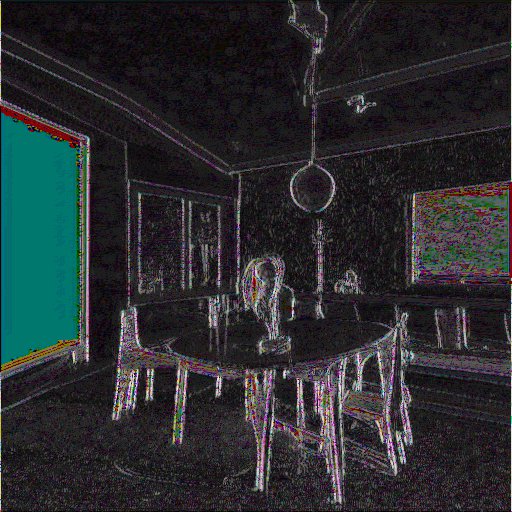} \\
        \rule{0pt}{6ex}%

        \includegraphics[width=\imwidth]{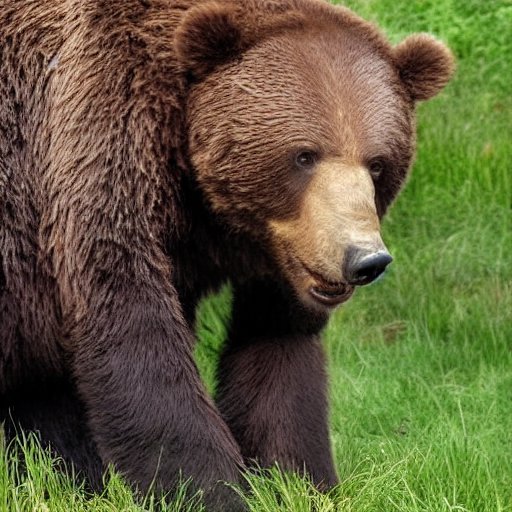} &
        \includegraphics[width=\imwidth]{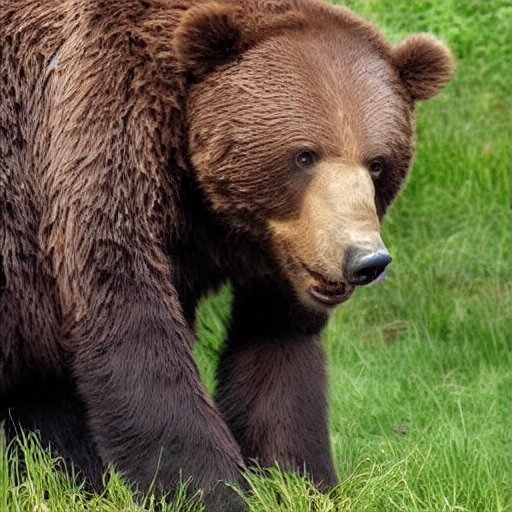} &
        \includegraphics[width=\imwidth]{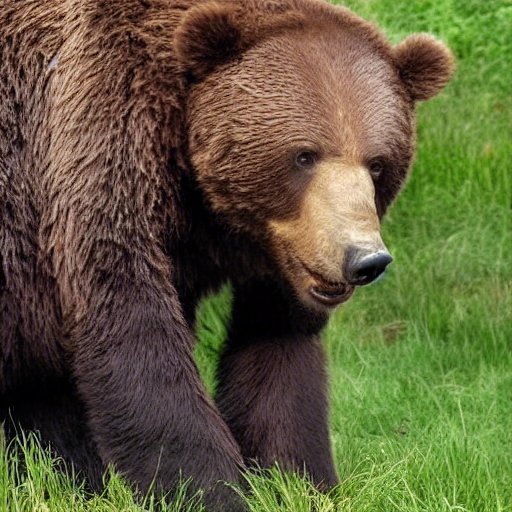} &
        \includegraphics[width=\imwidth]{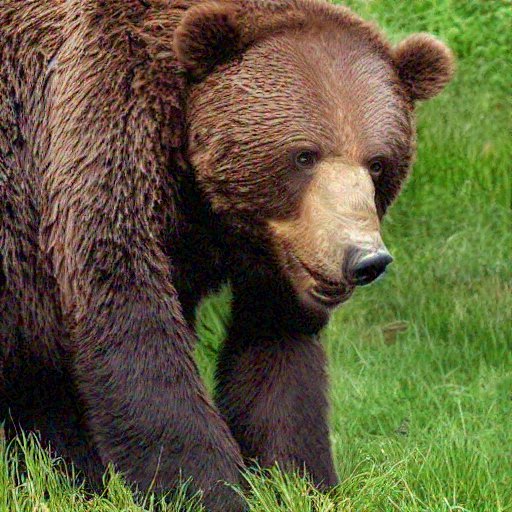} &
        \includegraphics[width=\imwidth]{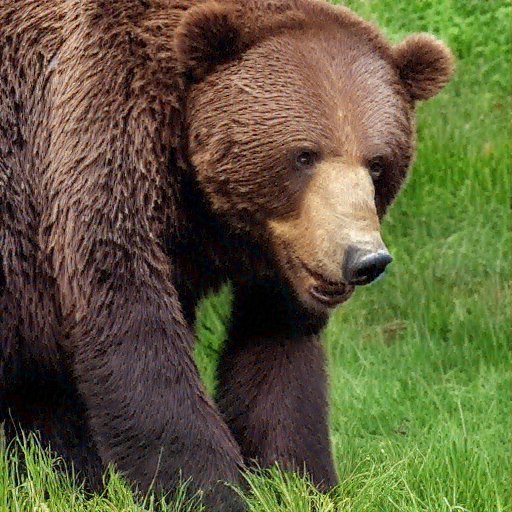} &
        \includegraphics[width=\imwidth]{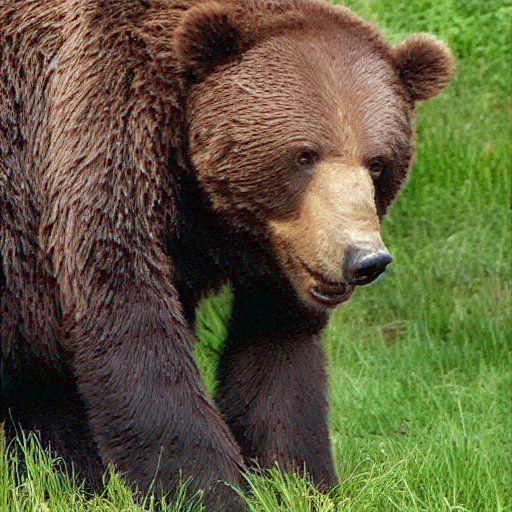} \\

            &
        \includegraphics[width=\imwidth]{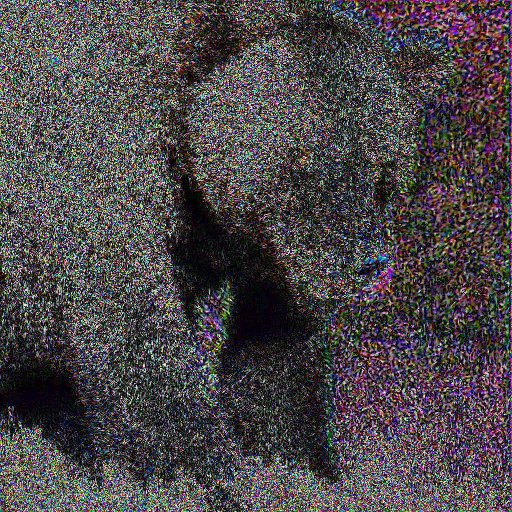} &
        \includegraphics[width=\imwidth]{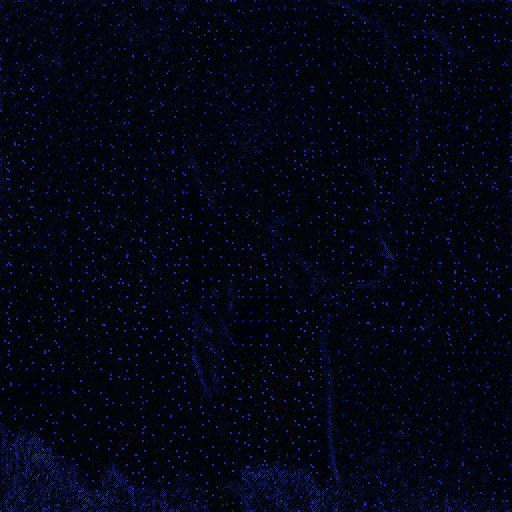} &
        \includegraphics[width=\imwidth]{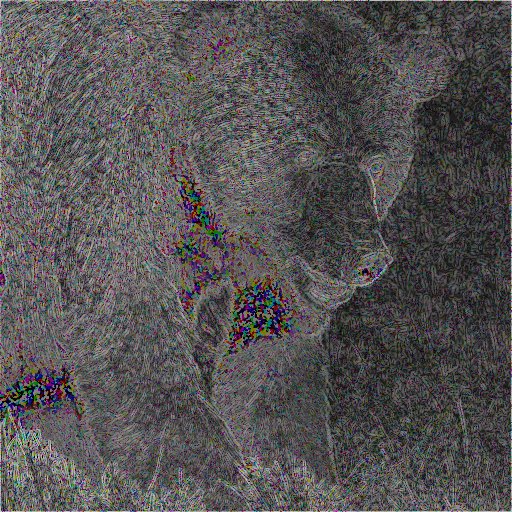} &
        \includegraphics[width=\imwidth]{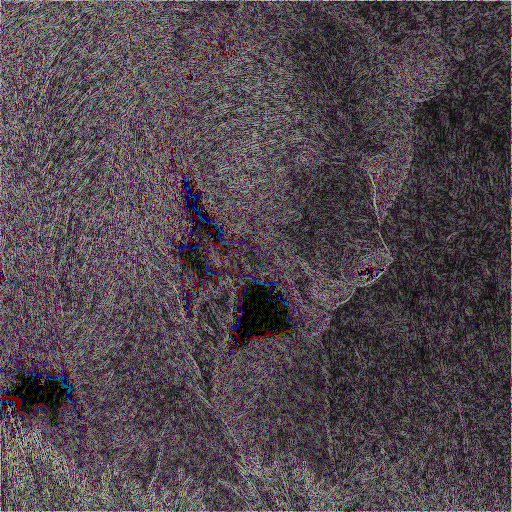} &
        \includegraphics[width=\imwidth]{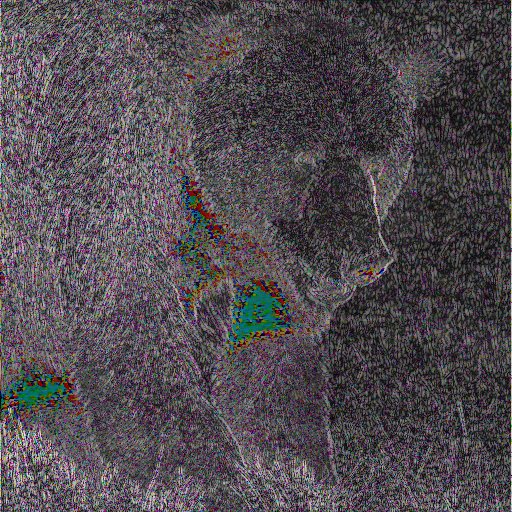} \\
    \\
    \end{tabular}
\caption{\label{chap3/fig:supp-watermark} Qualitative results for different watermarking methods on generated images at resolution $512$.}
\end{figure}

\begin{figure}[H]
\centering
    \scriptsize
    \newcommand{\imwidth}{0.165\textwidth}
        \setlength{\tabcolsep}{0pt}
        \begin{tabular}{ccc@{\hskip 2pt}ccc}
        \toprule
        Original & Watermarked & Difference & Original & Watermarked & Difference \\
        \midrule
        \includegraphics[width=\imwidth]{chapter-3/figs/supp/txt2img/00_nw.jpg} &
        \includegraphics[width=\imwidth]{chapter-3/figs/supp/txt2img/00_w.jpg} &
        \includegraphics[width=\imwidth]{chapter-3/figs/supp/txt2img/00_diff.jpg} &
        \includegraphics[width=\imwidth]{chapter-3/figs/supp/txt2img/01_nw.jpg} &
        \includegraphics[width=\imwidth]{chapter-3/figs/supp/txt2img/01_w.jpg} &
        \includegraphics[width=\imwidth]{chapter-3/figs/supp/txt2img/01_diff.jpg} \\
        \rule{0pt}{6ex}%

        \includegraphics[width=\imwidth]{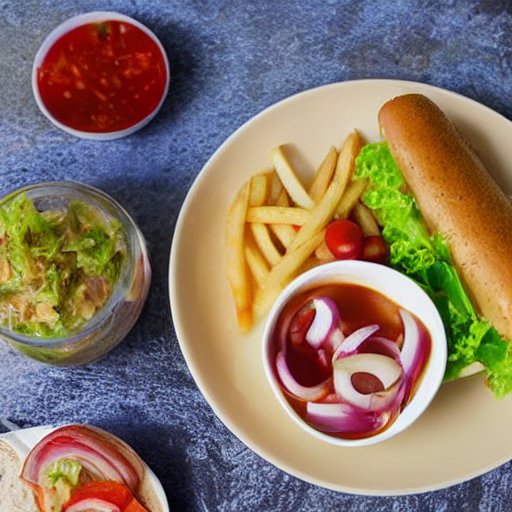} &
        \includegraphics[width=\imwidth]{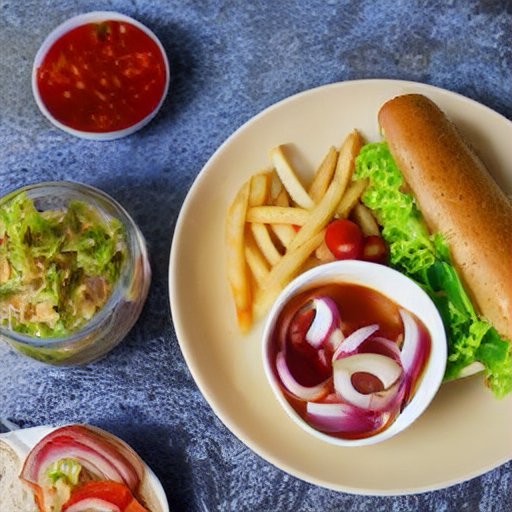} &
        \includegraphics[width=\imwidth]{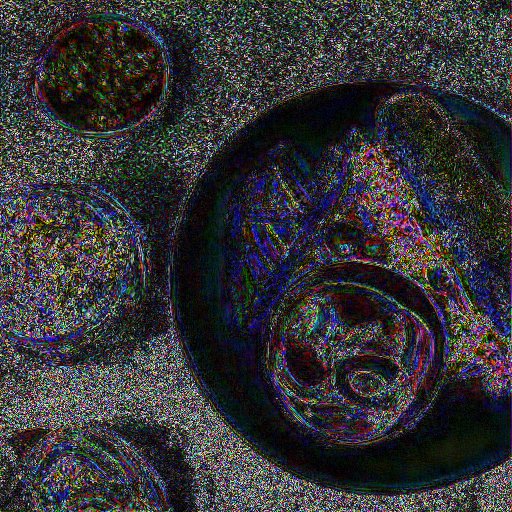} &
        \includegraphics[width=\imwidth]{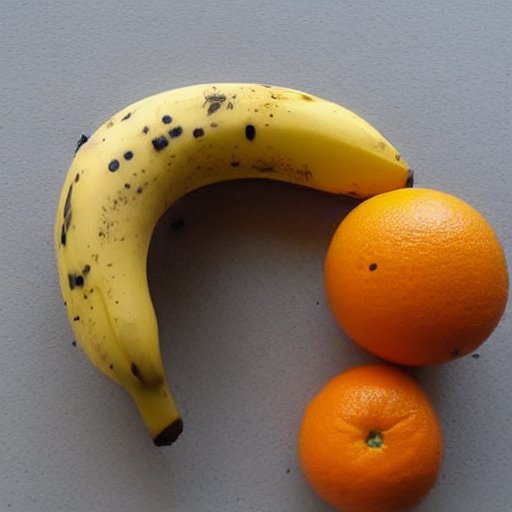} &
        \includegraphics[width=\imwidth]{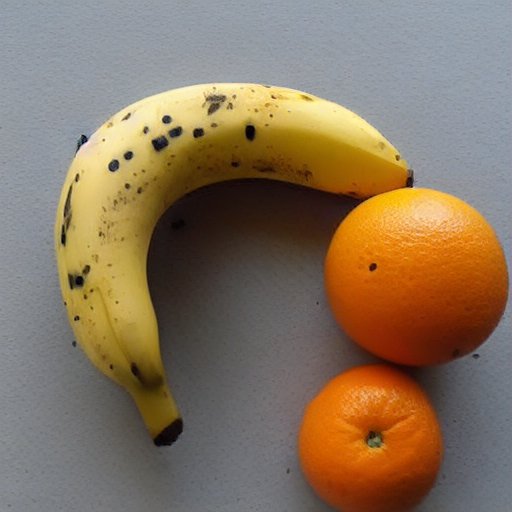} &
        \includegraphics[width=\imwidth]{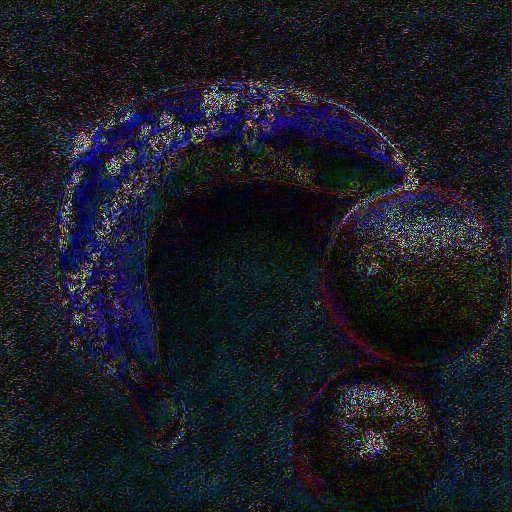} \\
        \rule{0pt}{6ex}%

        \includegraphics[width=\imwidth]{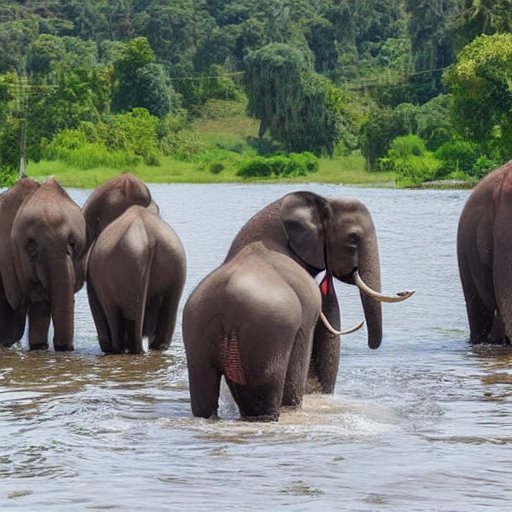} &
        \includegraphics[width=\imwidth]{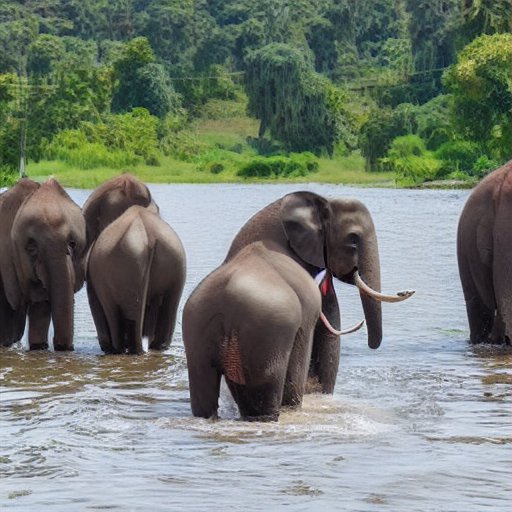} &
        \includegraphics[width=\imwidth]{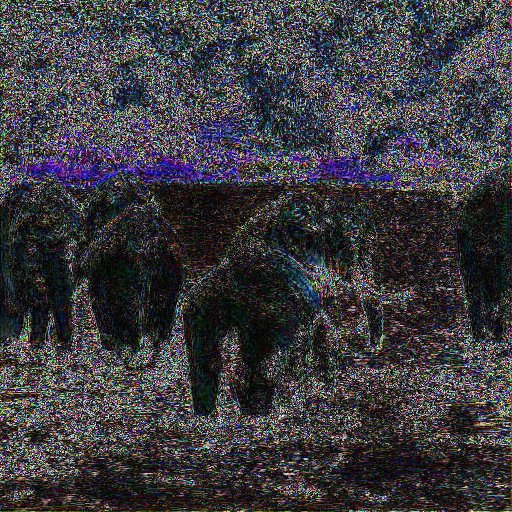} &
        \includegraphics[width=\imwidth]{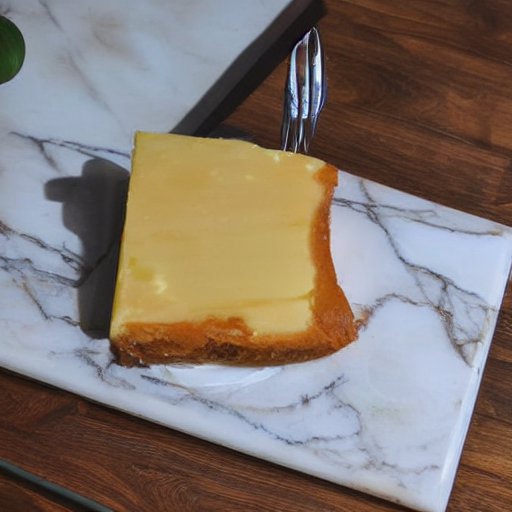} &
        \includegraphics[width=\imwidth]{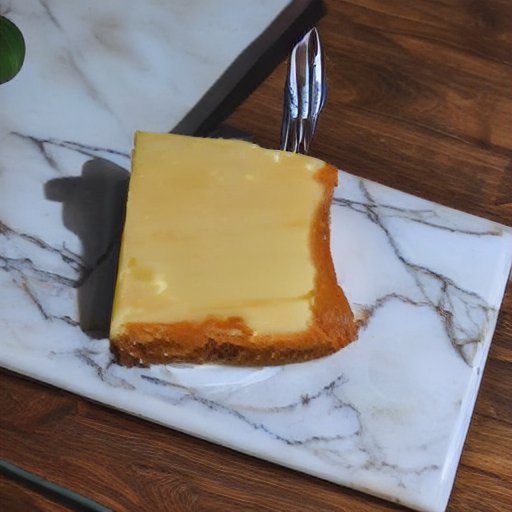} &
        \includegraphics[width=\imwidth]{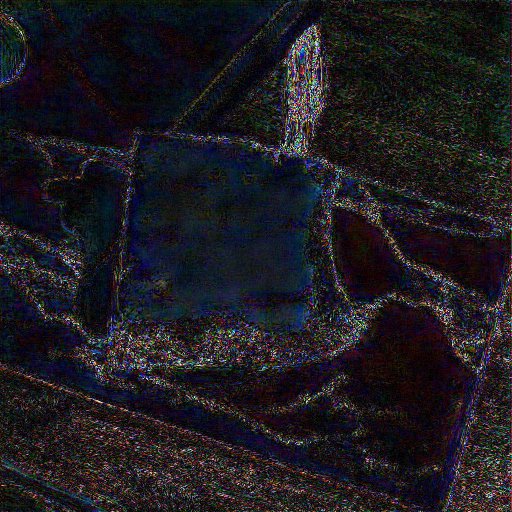} \\
        \rule{0pt}{6ex}%

        \includegraphics[width=\imwidth]{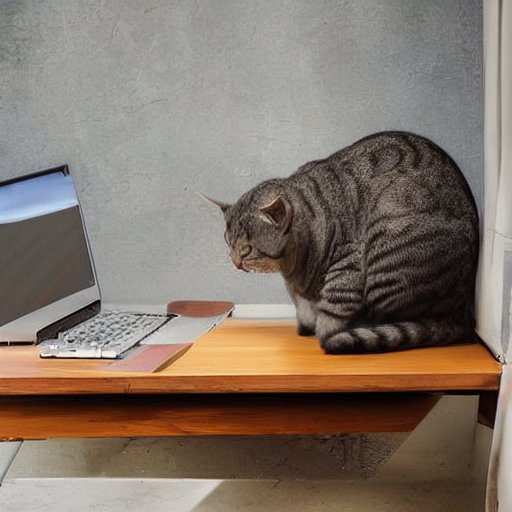} &
        \includegraphics[width=\imwidth]{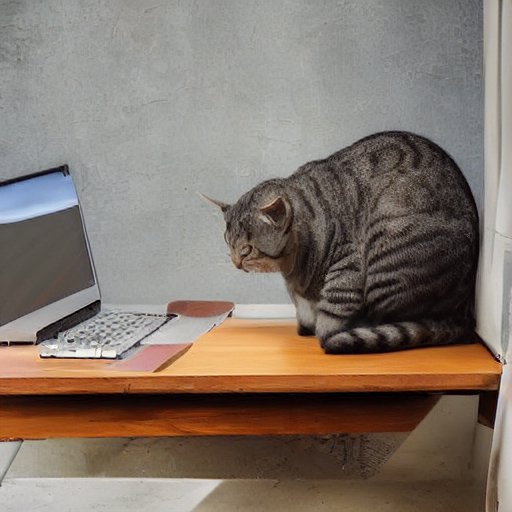} &
        \includegraphics[width=\imwidth]{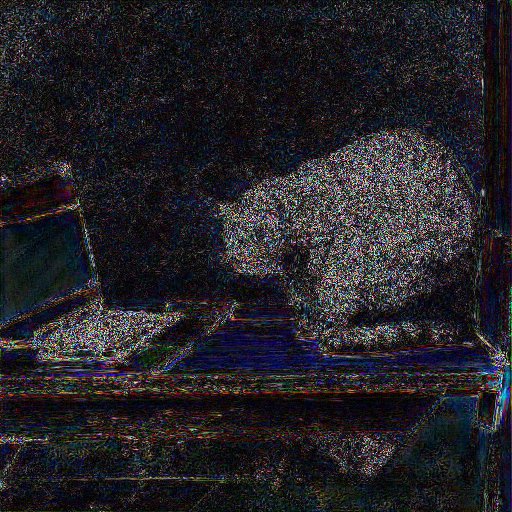} &
        \includegraphics[width=\imwidth]{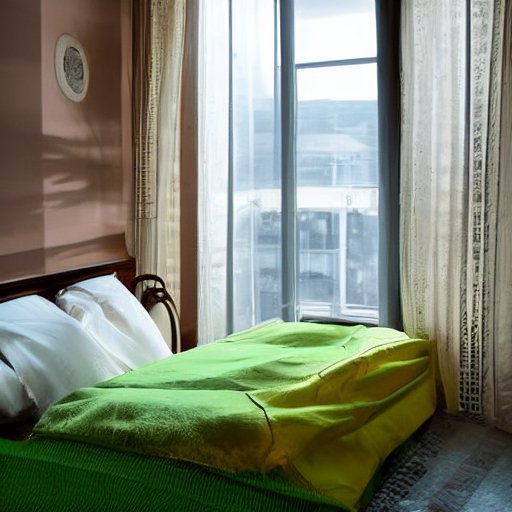} &
        \includegraphics[width=\imwidth]{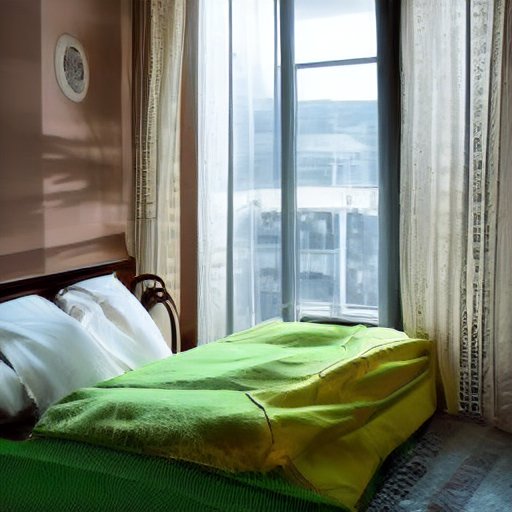} &
        \includegraphics[width=\imwidth]{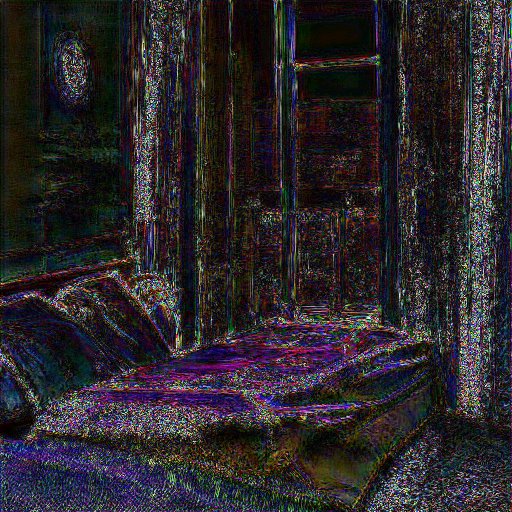} \\
        \rule{0pt}{6ex}%
        
        \includegraphics[width=\imwidth]{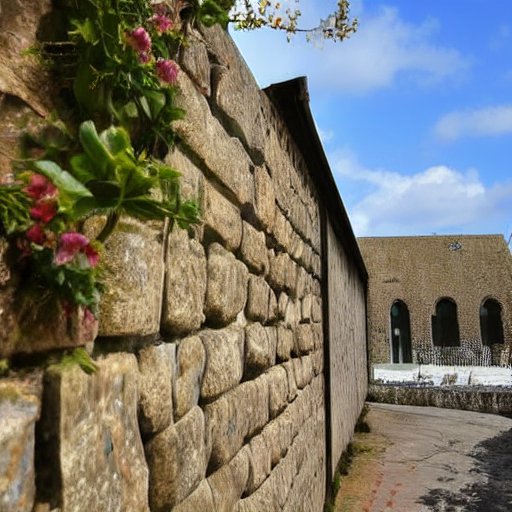} &
        \includegraphics[width=\imwidth]{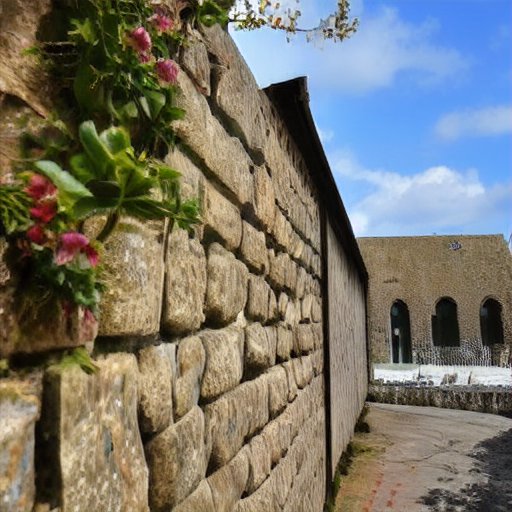} &
        \includegraphics[width=\imwidth]{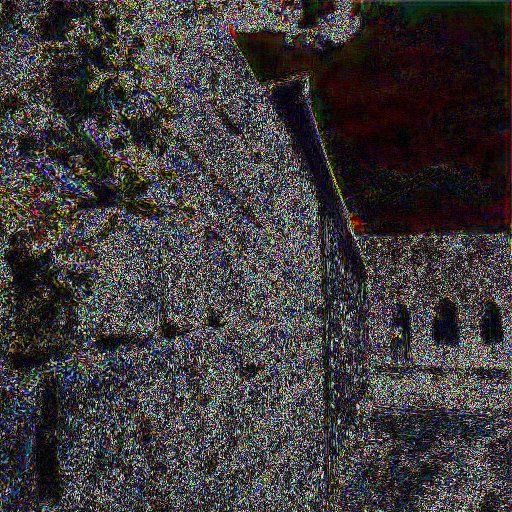} &
        \includegraphics[width=\imwidth]{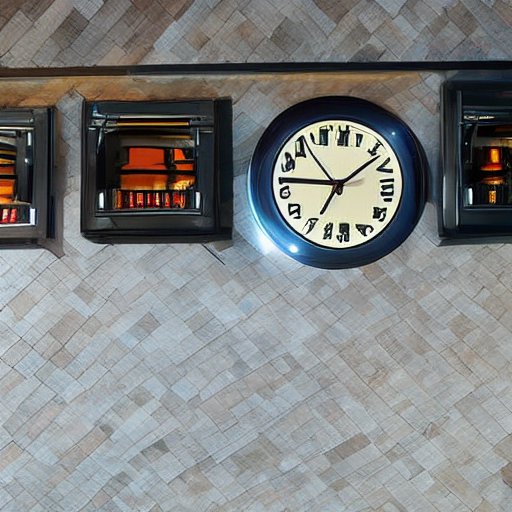} &
        \includegraphics[width=\imwidth]{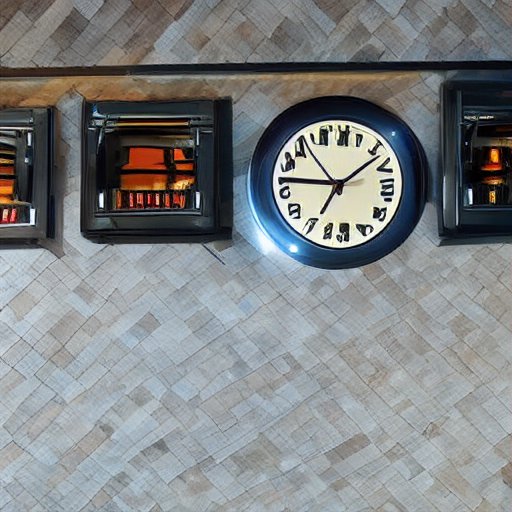} &
        \includegraphics[width=\imwidth]{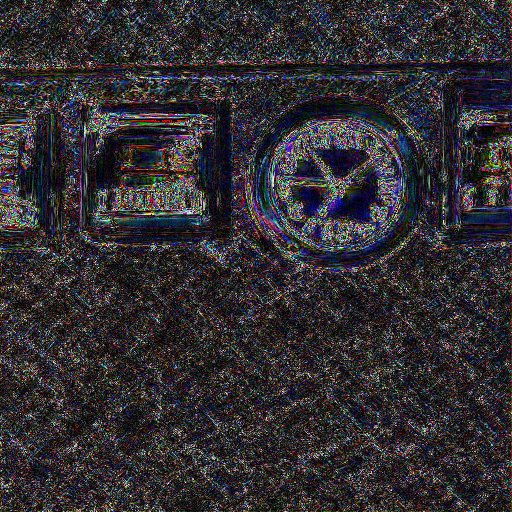} \\
        \rule{0pt}{6ex}%

        \includegraphics[width=\imwidth]{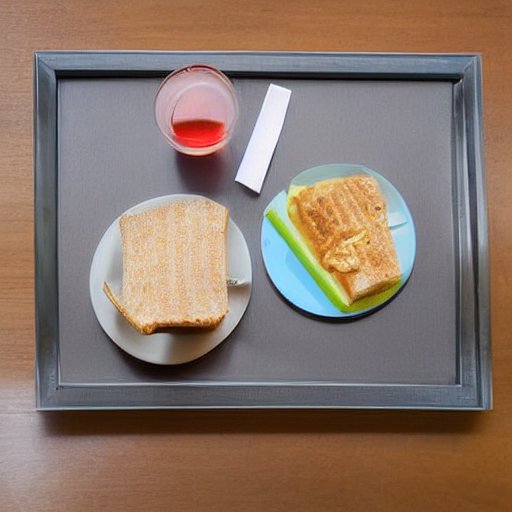} &
        \includegraphics[width=\imwidth]{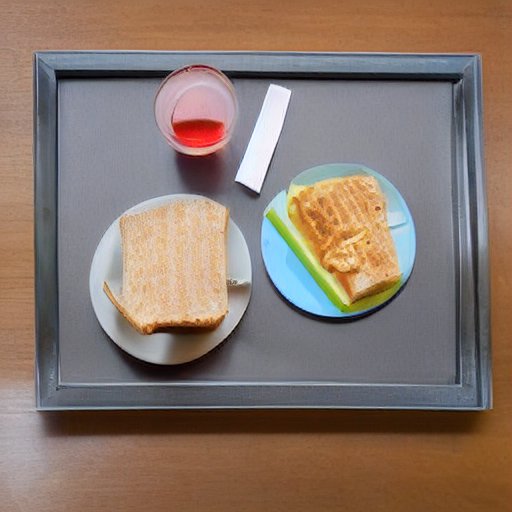} &
        \includegraphics[width=\imwidth]{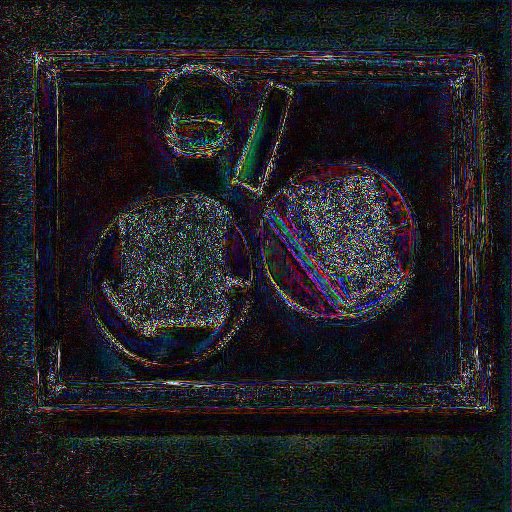} &
        \includegraphics[width=\imwidth]{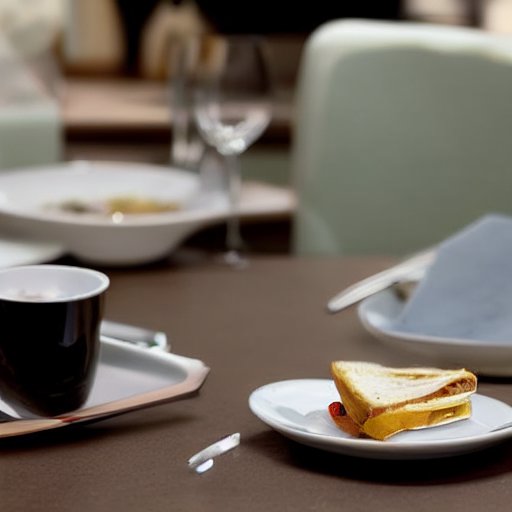} &
        \includegraphics[width=\imwidth]{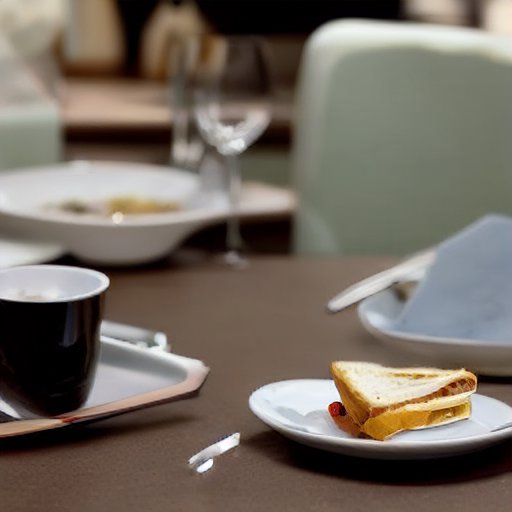} &
        \includegraphics[width=\imwidth]{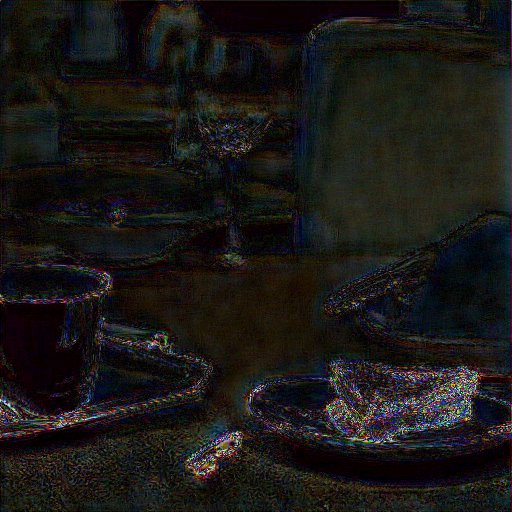} \\
    \bottomrule \\
    \end{tabular}
\caption{\label{chap3/fig:supp-txt2img} Qualitative results on prompts of the validation set of MS-COCO, at resolution $512$ and for a $48$-bits signature.
Images are generated from the same latents, with original or watermarked generative models.}
\end{figure}

\begin{figure}[H]
    \centering
    \scriptsize
    \newcommand{\imwidth}{0.124\textwidth}
    \setlength{\tabcolsep}{0pt}
    \begin{tabular}{cc@{\hskip 2pt}ccc@{\hskip 2pt}ccc}
        \toprule
        Image to inpaint & Mask & Original & Watermarked & Difference & Original & Watermarked & Difference \\
        \midrule
        \hspace{0pt}
        \includegraphics[width=\imwidth]{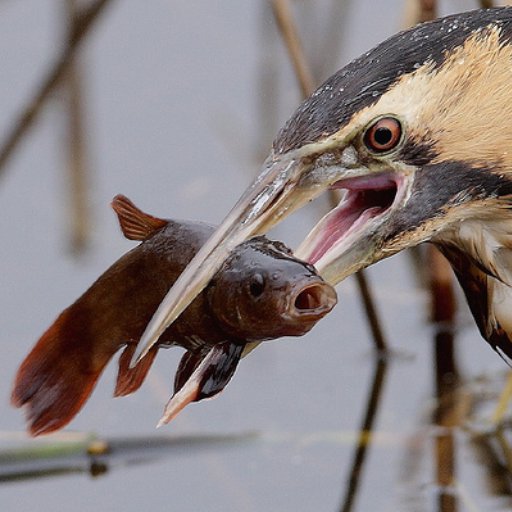} &
        \includegraphics[width=\imwidth]{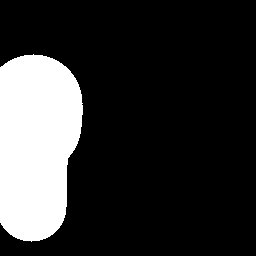} &
        \includegraphics[width=\imwidth]{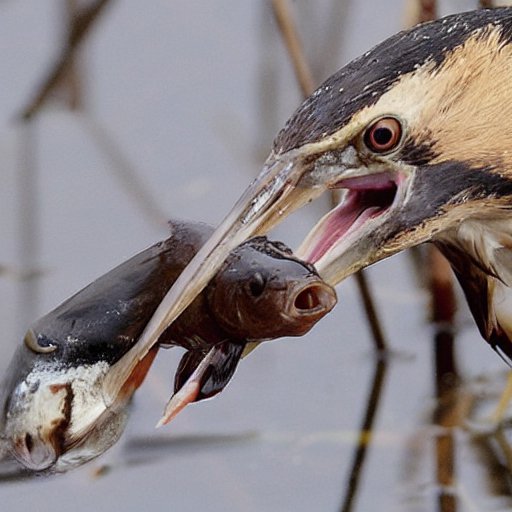} &
        \includegraphics[width=\imwidth]{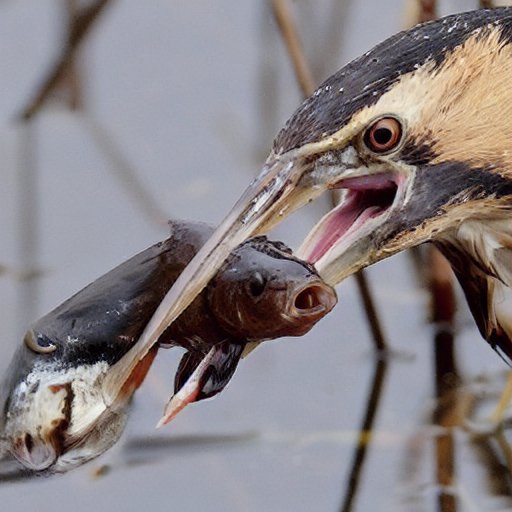} &
        \includegraphics[width=\imwidth]{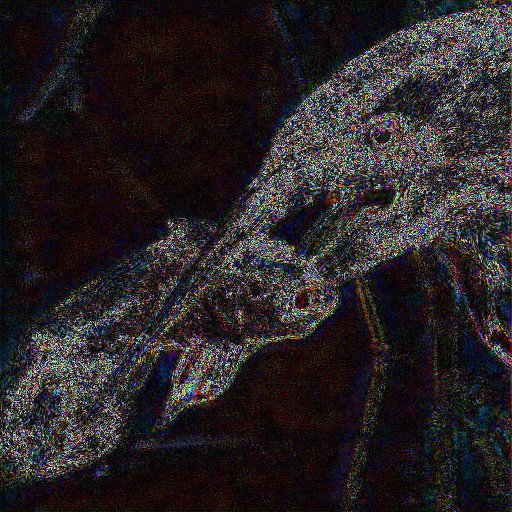} &
        \includegraphics[width=\imwidth]{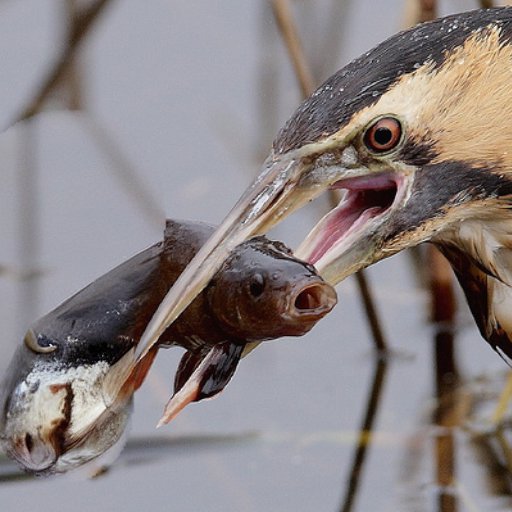} &
        \includegraphics[width=\imwidth]{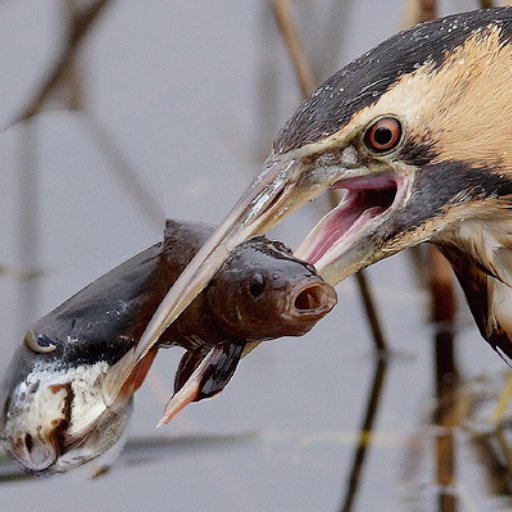} &
        \includegraphics[width=\imwidth]{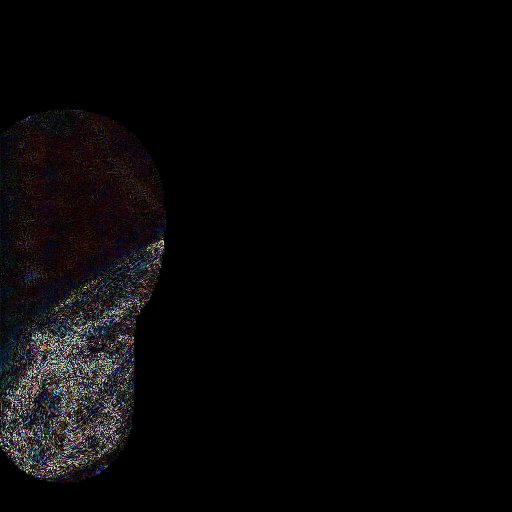} \\
        \rule{0pt}{6ex}

        \includegraphics[width=\imwidth]{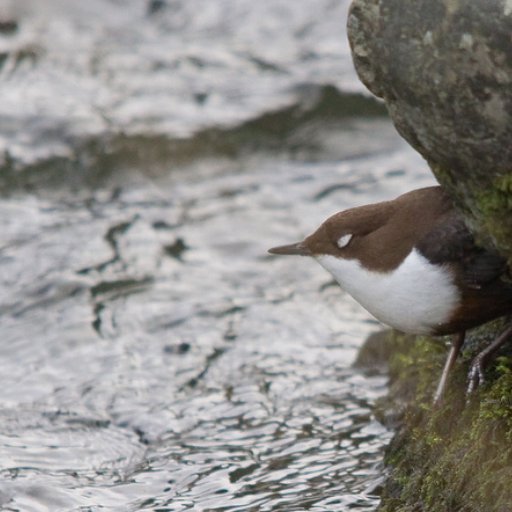} &
        \includegraphics[width=\imwidth]{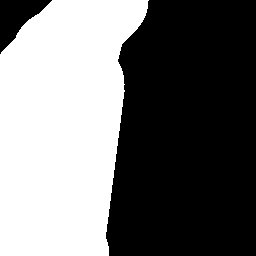} &
        \includegraphics[width=\imwidth]{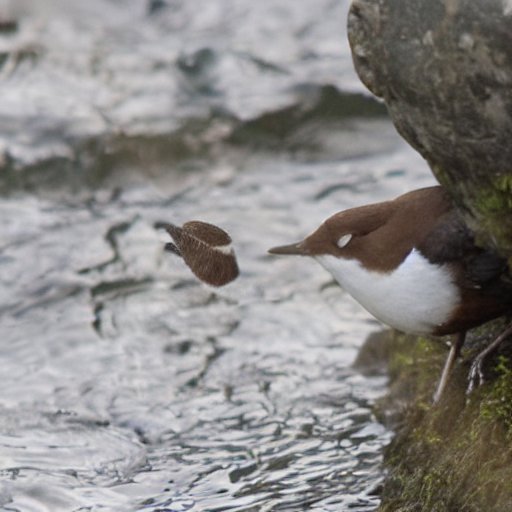} &
        \includegraphics[width=\imwidth]{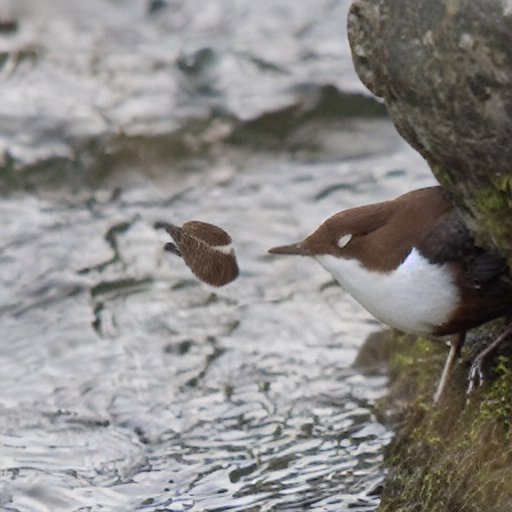} &
        \includegraphics[width=\imwidth]{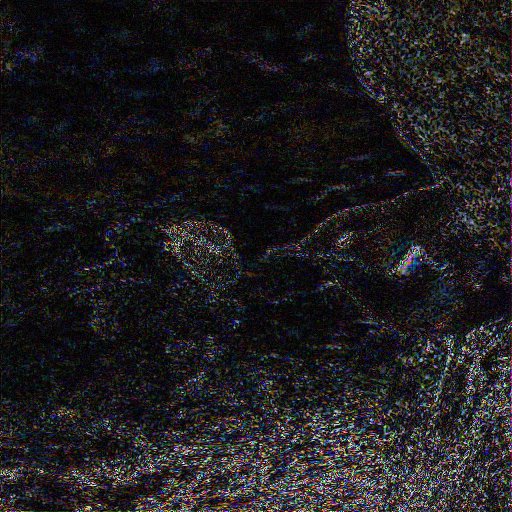} &
        \includegraphics[width=\imwidth]{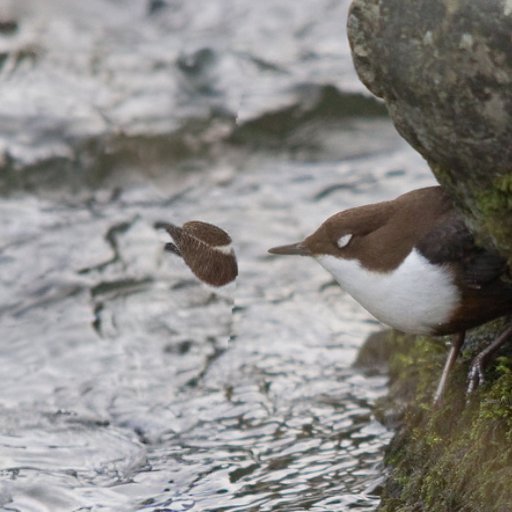} &
        \includegraphics[width=\imwidth]{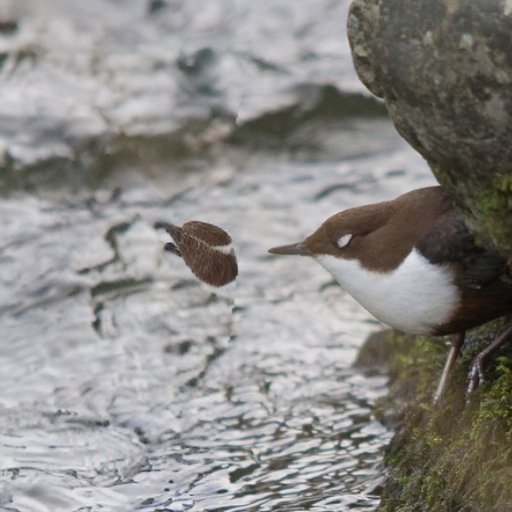} &
        \includegraphics[width=\imwidth]{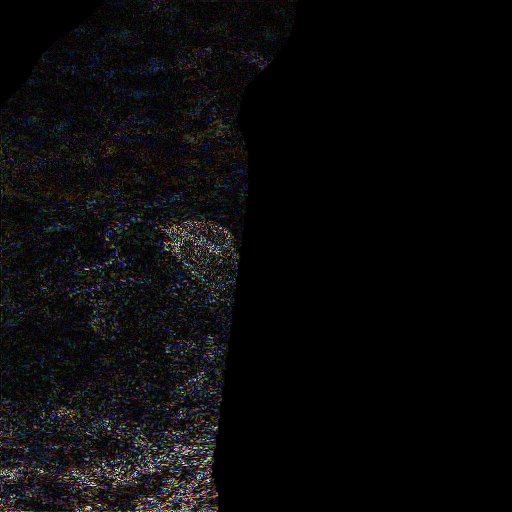} \\
        \rule{0pt}{6ex}

        \includegraphics[width=\imwidth]{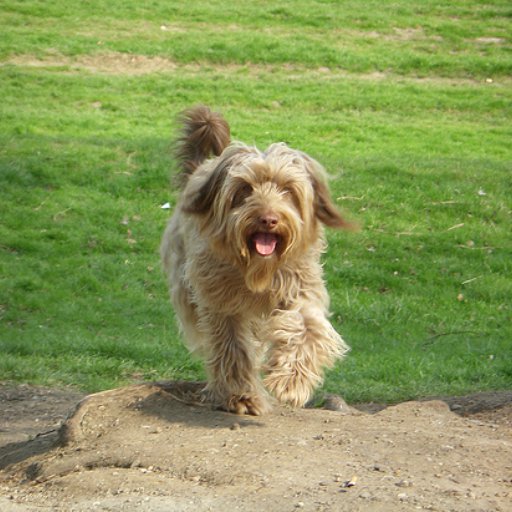} &
        \includegraphics[width=\imwidth]{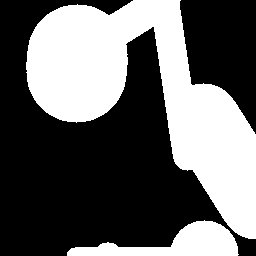} &
        \includegraphics[width=\imwidth]{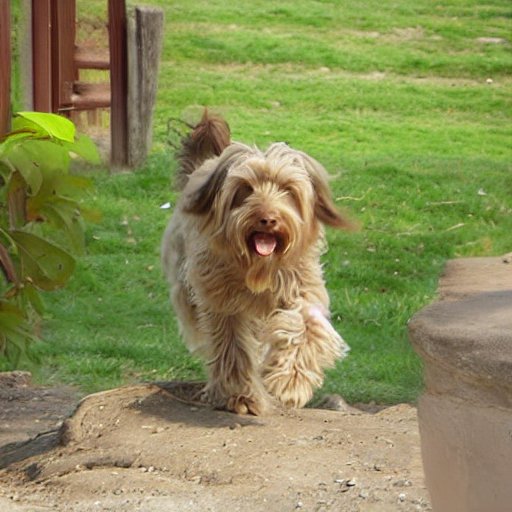} &
        \includegraphics[width=\imwidth]{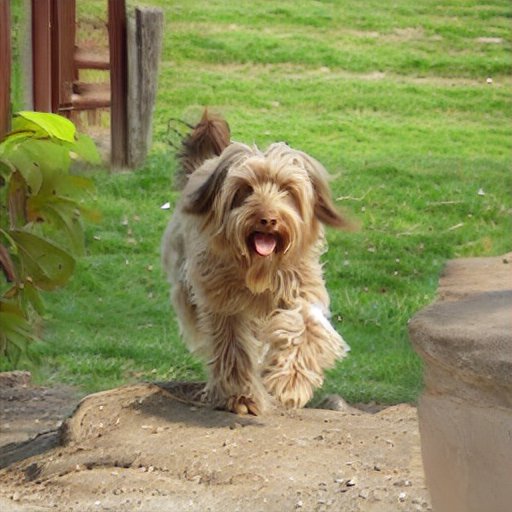} &
        \includegraphics[width=\imwidth]{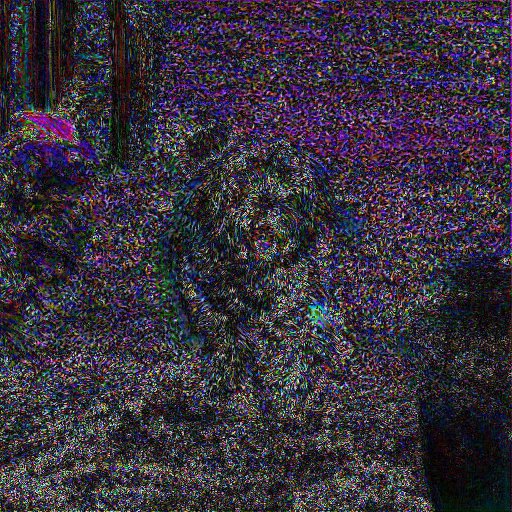} &
        \includegraphics[width=\imwidth]{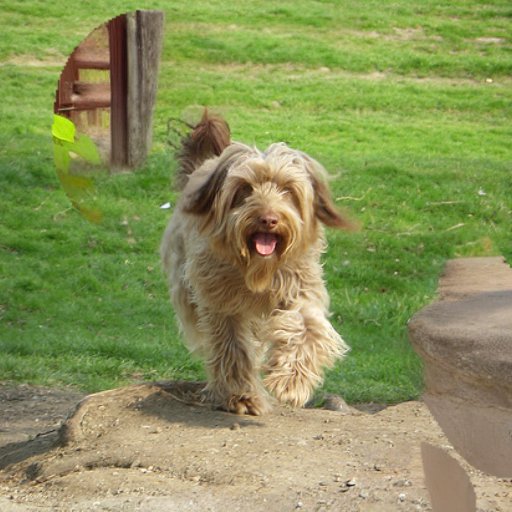} &
        \includegraphics[width=\imwidth]{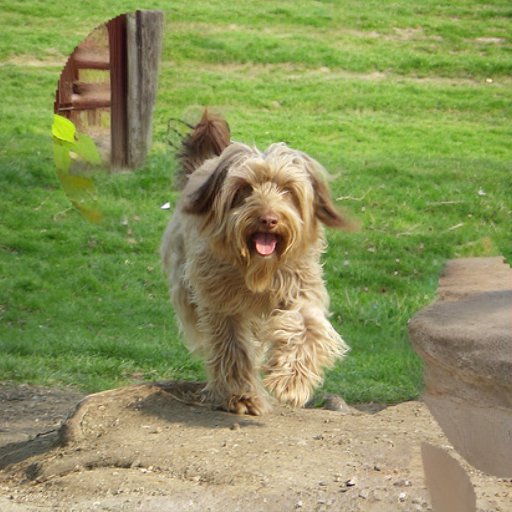} &
        \includegraphics[width=\imwidth]{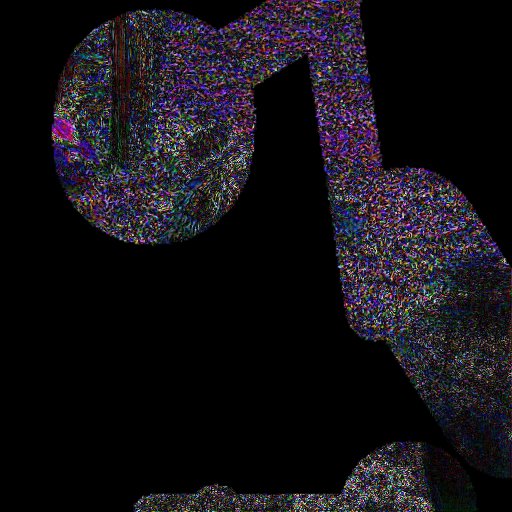} \\
        \rule{0pt}{6ex}

        \includegraphics[width=\imwidth]{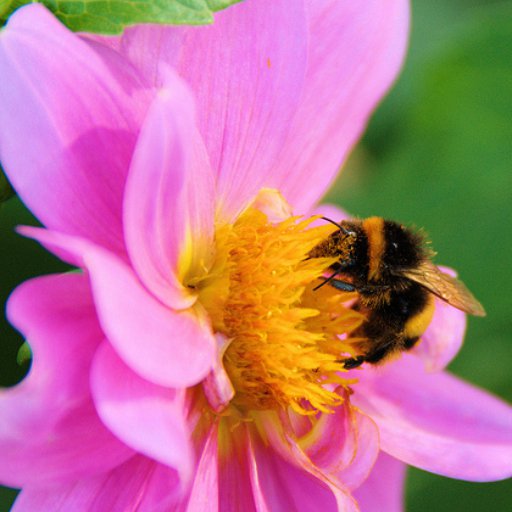} &
        \includegraphics[width=\imwidth]{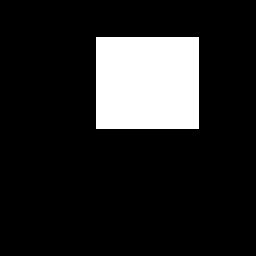} &
        \includegraphics[width=\imwidth]{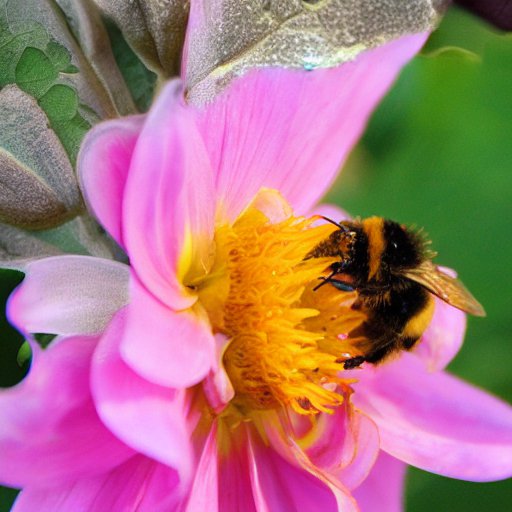} &
        \includegraphics[width=\imwidth]{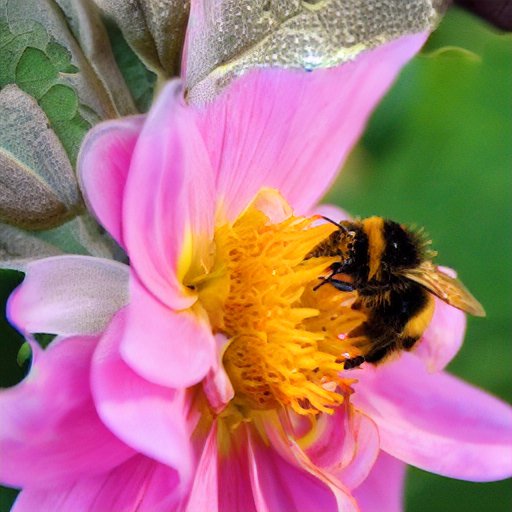} &
        \includegraphics[width=\imwidth]{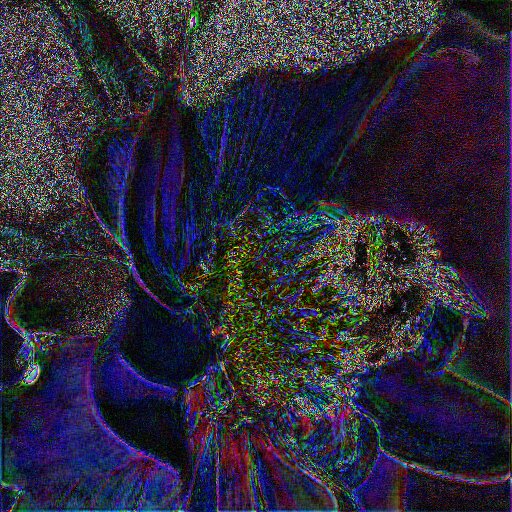} &
        \includegraphics[width=\imwidth]{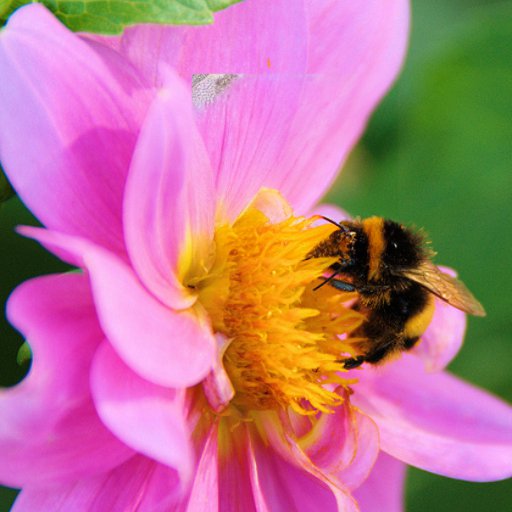} &
        \includegraphics[width=\imwidth]{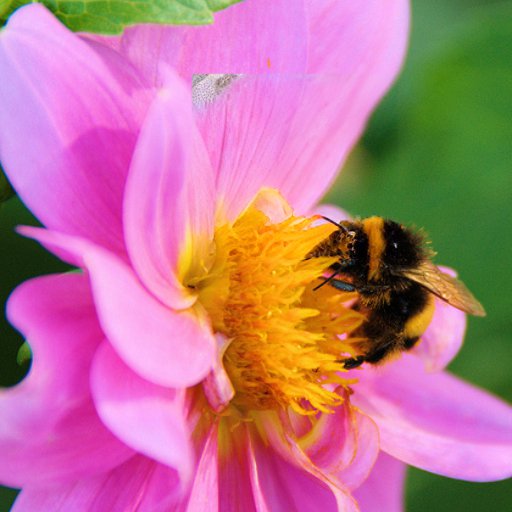} &
        \includegraphics[width=\imwidth]{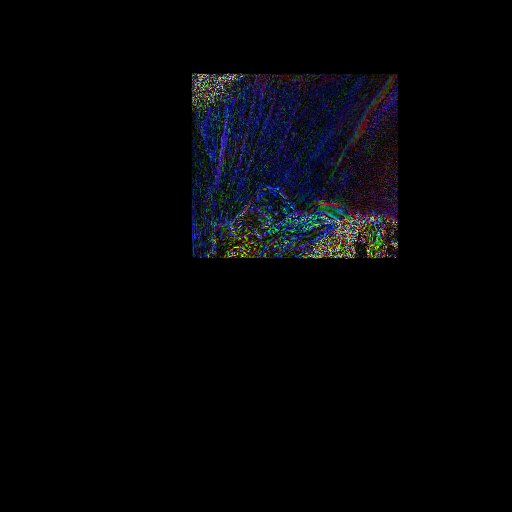} \\
        \rule{0pt}{6ex}

        \includegraphics[width=\imwidth]{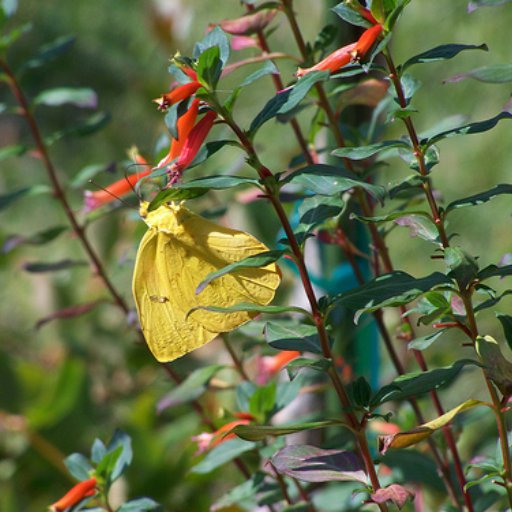} &
        \includegraphics[width=\imwidth]{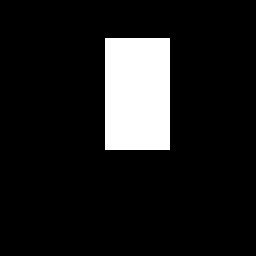} &
        \includegraphics[width=\imwidth]{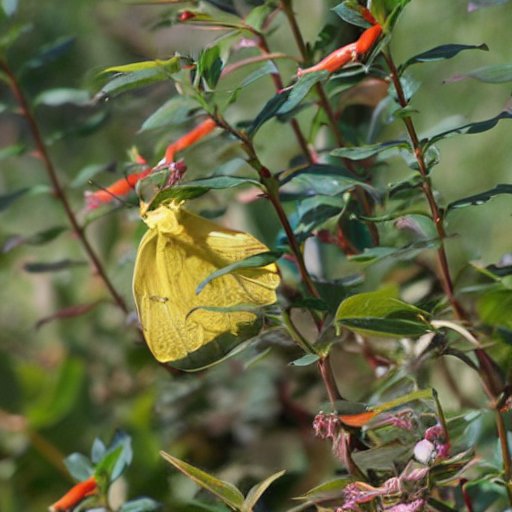} &
        \includegraphics[width=\imwidth]{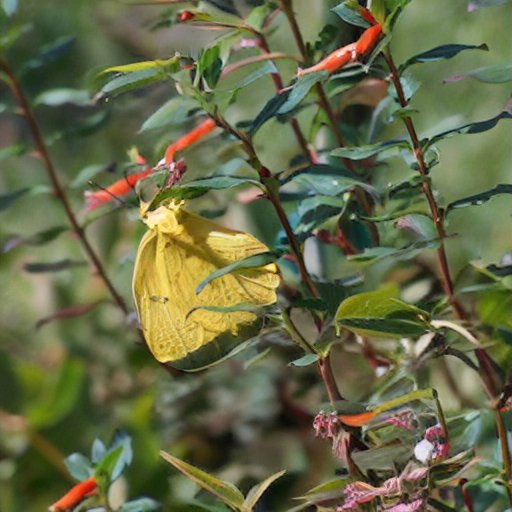} &
        \includegraphics[width=\imwidth]{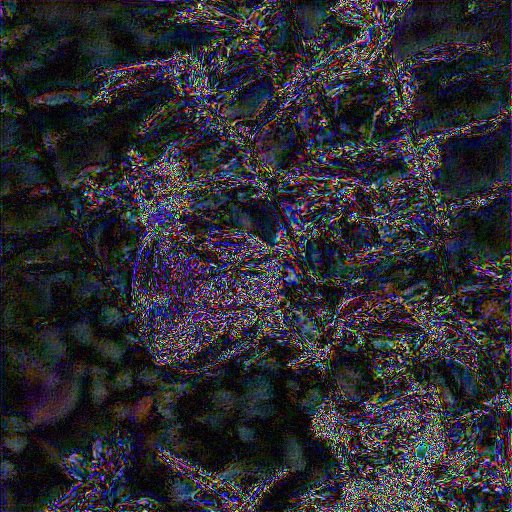} &
        \includegraphics[width=\imwidth]{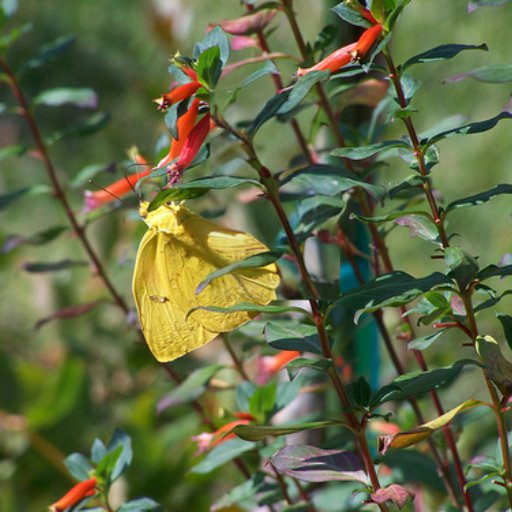} &
        \includegraphics[width=\imwidth]{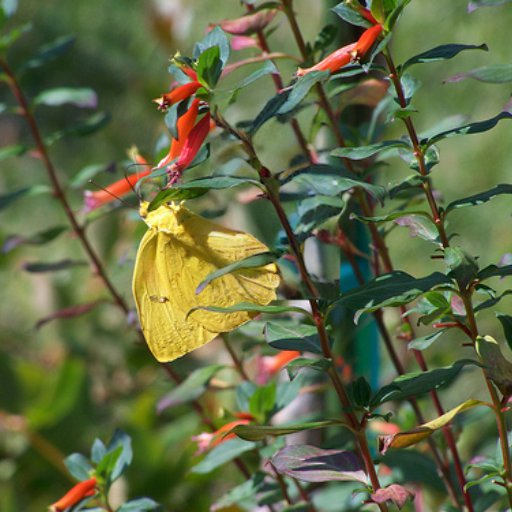} &
        \includegraphics[width=\imwidth]{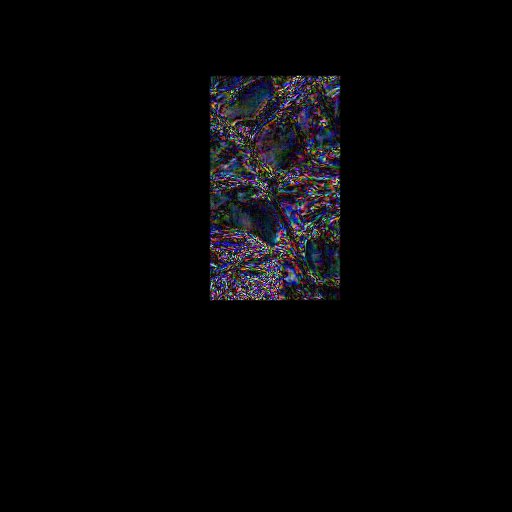} \\
        \rule{0pt}{6ex}

        \includegraphics[width=\imwidth]{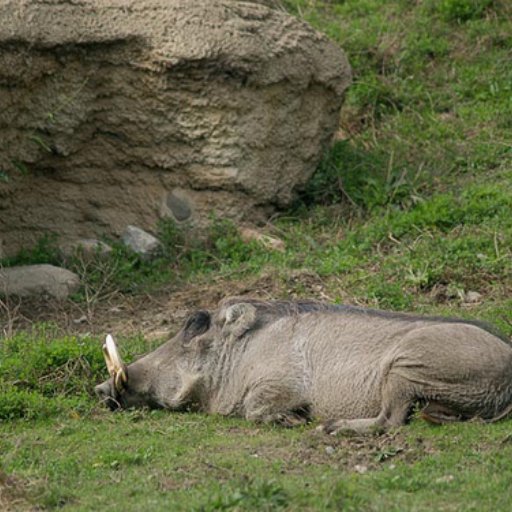} &
        \includegraphics[width=\imwidth]{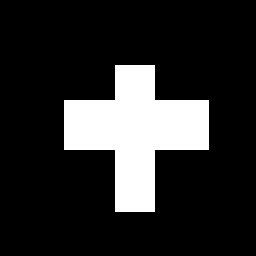} &
        \includegraphics[width=\imwidth]{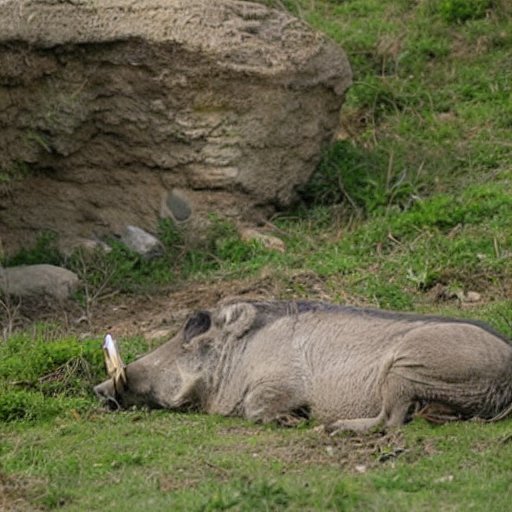} &
        \includegraphics[width=\imwidth]{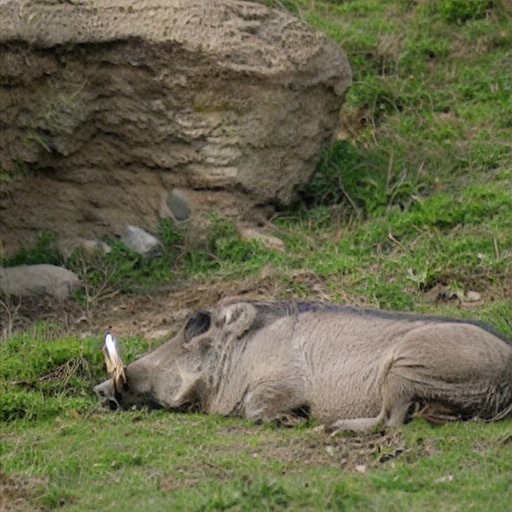} &
        \includegraphics[width=\imwidth]{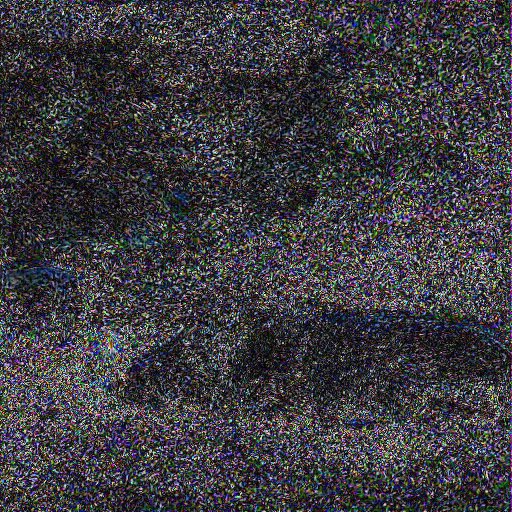} &
        \includegraphics[width=\imwidth]{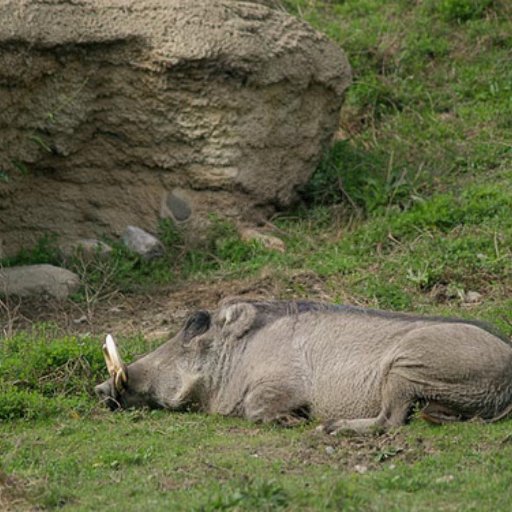} &
        \includegraphics[width=\imwidth]{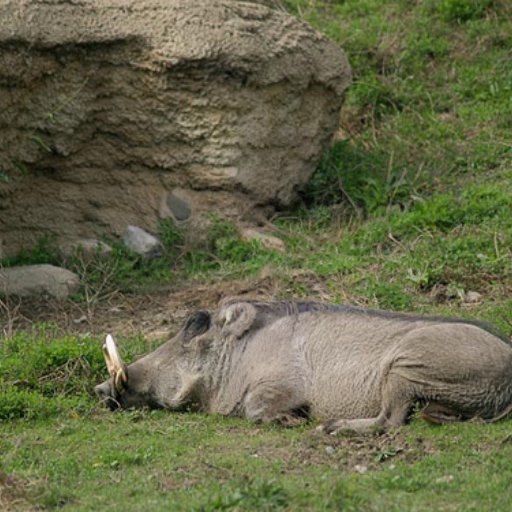} &
        \includegraphics[width=\imwidth]{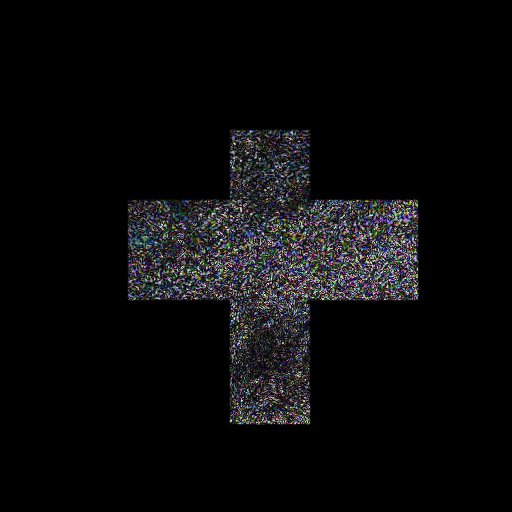} \\
        \bottomrule \\
    \end{tabular}
    \caption{
        \label{chap3/fig:supp-inpaint} Qualitative results for inpainting on ImageNet with original or watermarked generative models.
        We consider 2 scenarios: 
        (middle) the full image is modified to fill the masked area, 
        (right) only the masked area is filled.
        Since our model is not fine-tuned for inpainting, the last scenario introduces copy-paste artifacts.
        From a watermarking point of view, it is also the more interesting, since the watermark signal is only present in the masked area (and erased wherever the image to inpaint is copied).
        Even in this case, the watermark extractor achieves bit accuracy significantly higher than random.
    }
\end{figure}

\begin{figure}[H]
    \centering
    \scriptsize
    \newcommand{\imwidth}{0.124\textwidth}
    \setlength{\tabcolsep}{0pt}
    \begin{tabular}{cccc@{\hskip 2pt}cccc}
        \toprule
        Low resolution & Original & Watermarked & Difference & Low resolution & Original & Watermarked & Difference \\
        \midrule
        \hspace{0pt}
        \includegraphics[width=\imwidth]{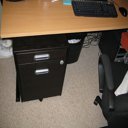} &
        \includegraphics[width=\imwidth]{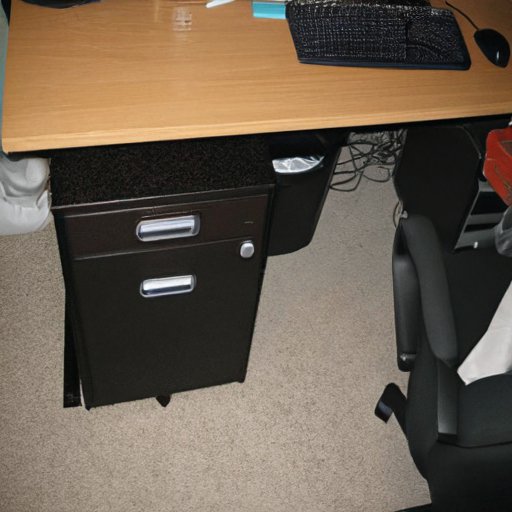} &
        \includegraphics[width=\imwidth]{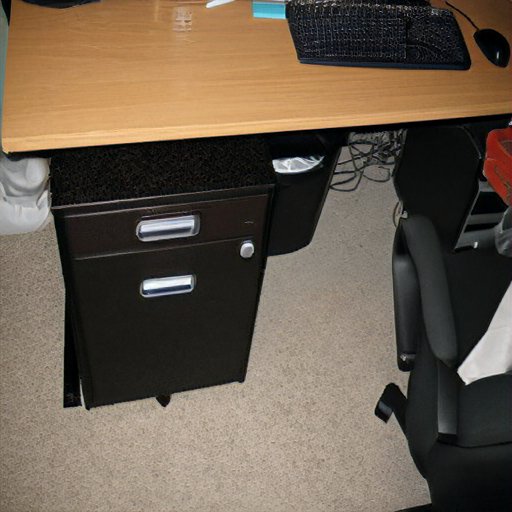} &
        \includegraphics[width=\imwidth]{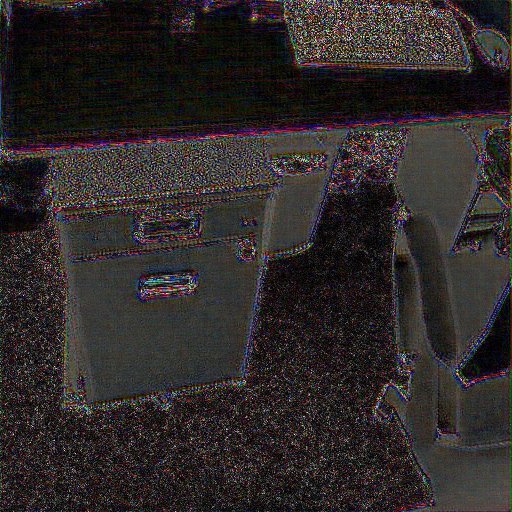} &
        \includegraphics[width=\imwidth]{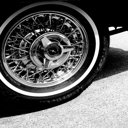} &
        \includegraphics[width=\imwidth]{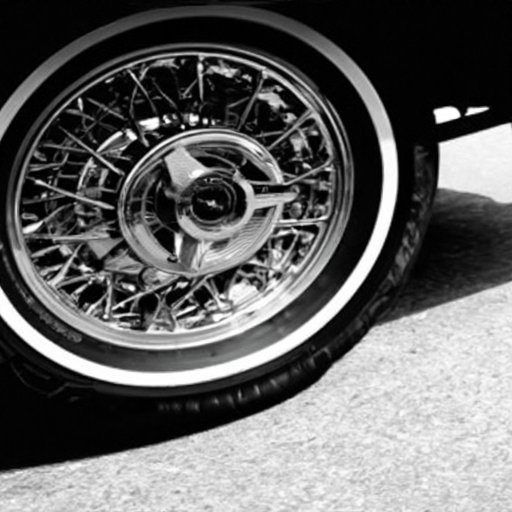} &
        \includegraphics[width=\imwidth]{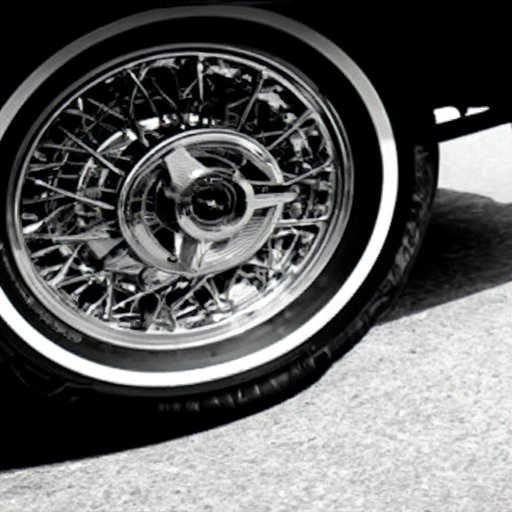} &
        \includegraphics[width=\imwidth]{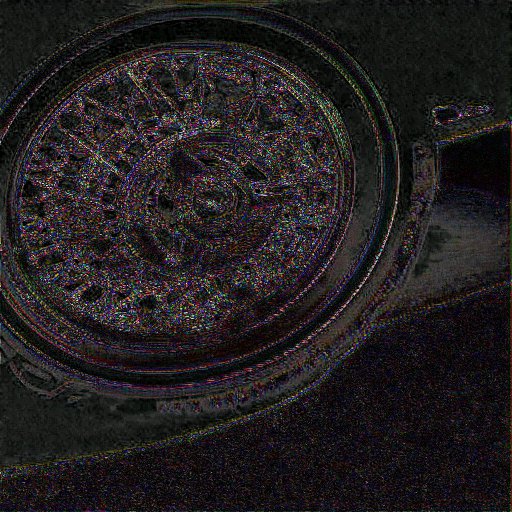} \\
        \rule{0pt}{6ex}

        \includegraphics[width=\imwidth]{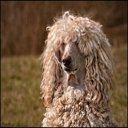} &
        \includegraphics[width=\imwidth]{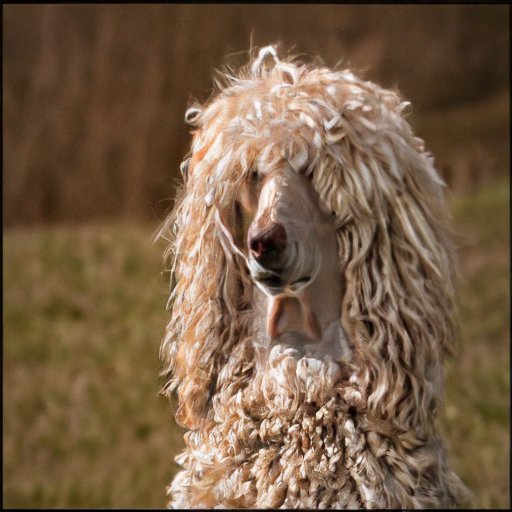} &
        \includegraphics[width=\imwidth]{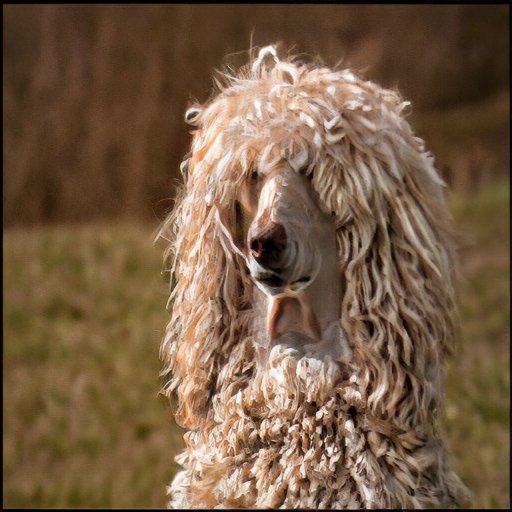} &
        \includegraphics[width=\imwidth]{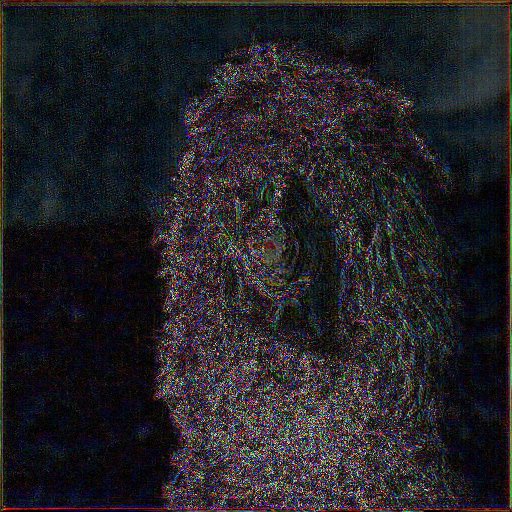} &
        \includegraphics[width=\imwidth]{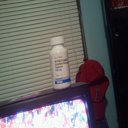} &
        \includegraphics[width=\imwidth]{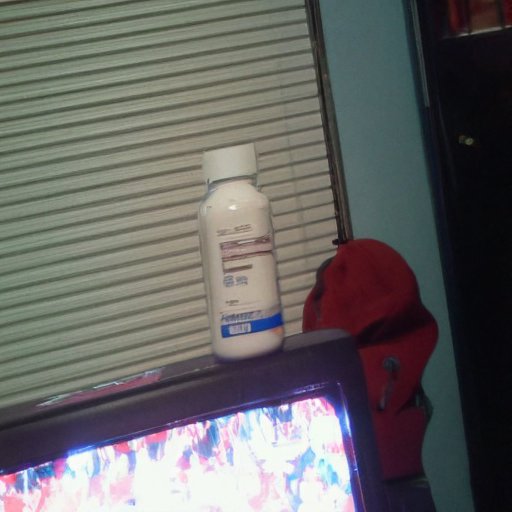} &
        \includegraphics[width=\imwidth]{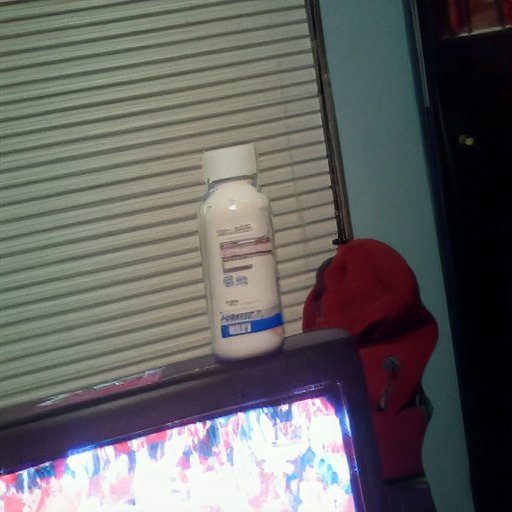} &
        \includegraphics[width=\imwidth]{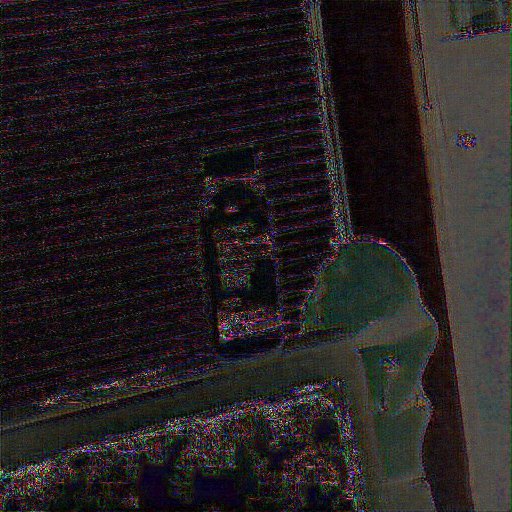} \\
        \rule{0pt}{6ex}

        \includegraphics[width=\imwidth]{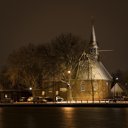} &
        \includegraphics[width=\imwidth]{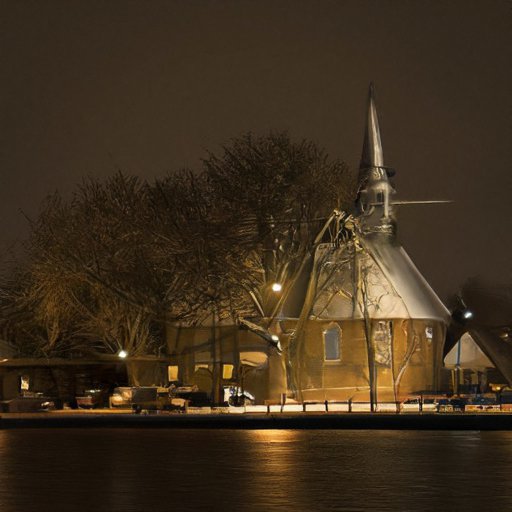} &
        \includegraphics[width=\imwidth]{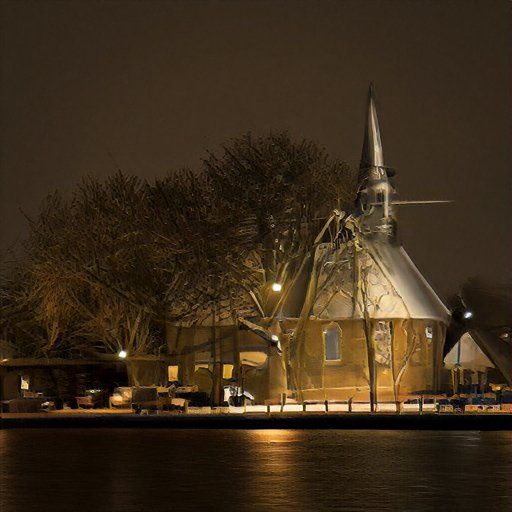} &
        \includegraphics[width=\imwidth]{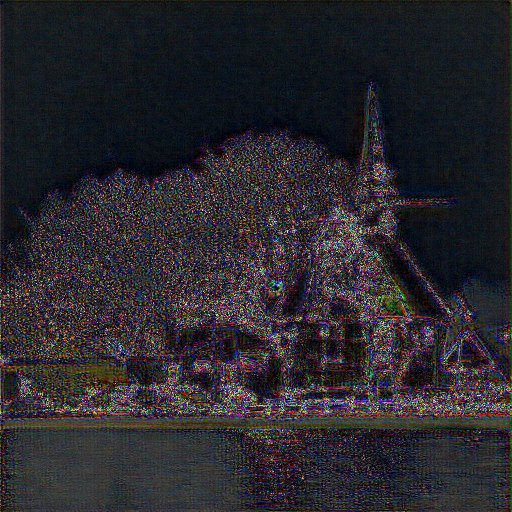} &
        \includegraphics[width=\imwidth]{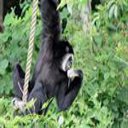} &
        \includegraphics[width=\imwidth]{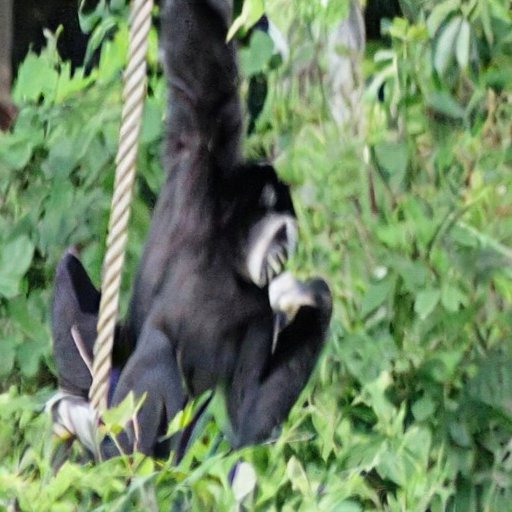} &
        \includegraphics[width=\imwidth]{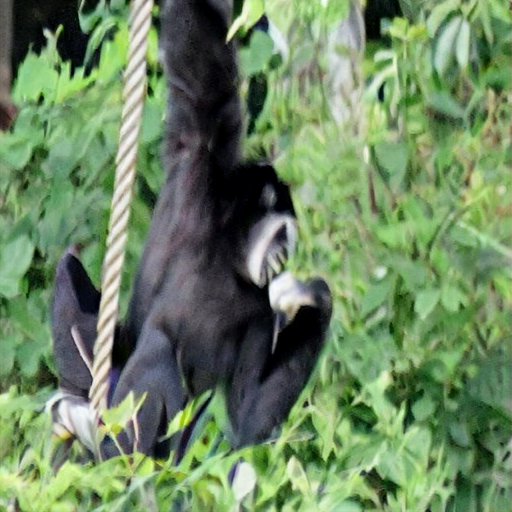} &
        \includegraphics[width=\imwidth]{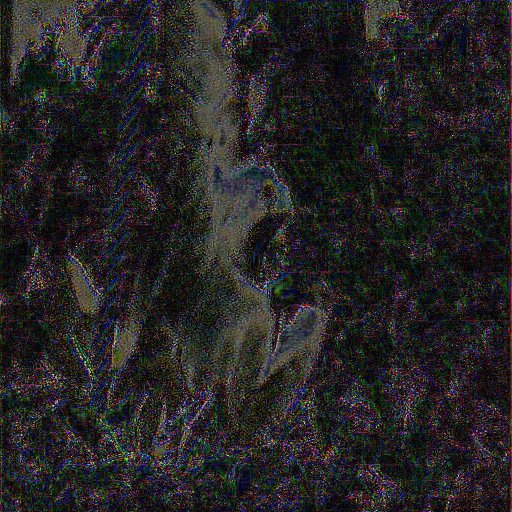} \\
        \rule{0pt}{6ex}

        \includegraphics[width=\imwidth]{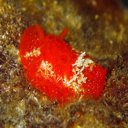} &
        \includegraphics[width=\imwidth]{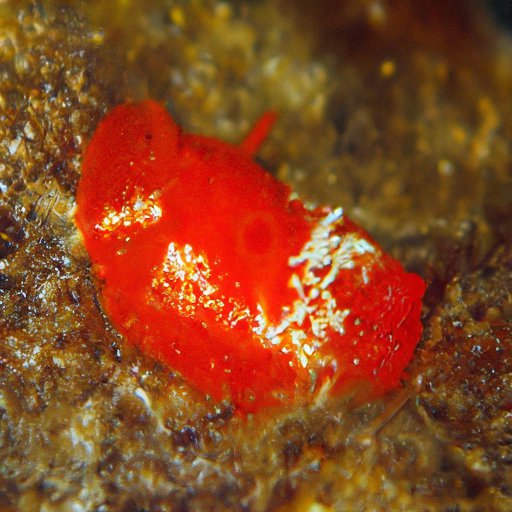} &
        \includegraphics[width=\imwidth]{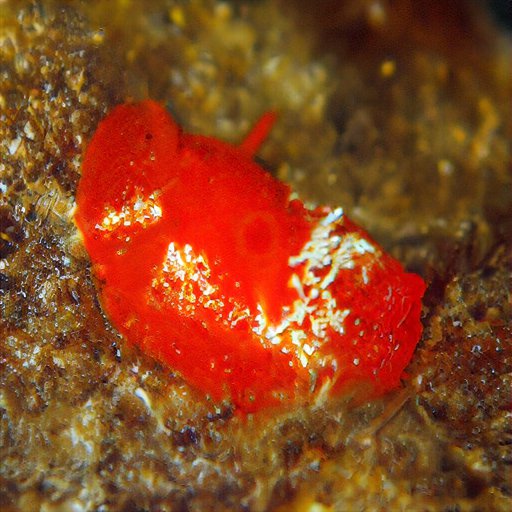} &
        \includegraphics[width=\imwidth]{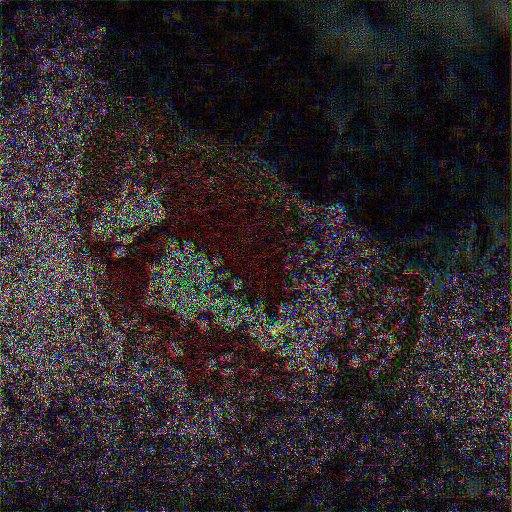} &
        \includegraphics[width=\imwidth]{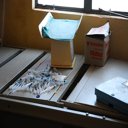} &
        \includegraphics[width=\imwidth]{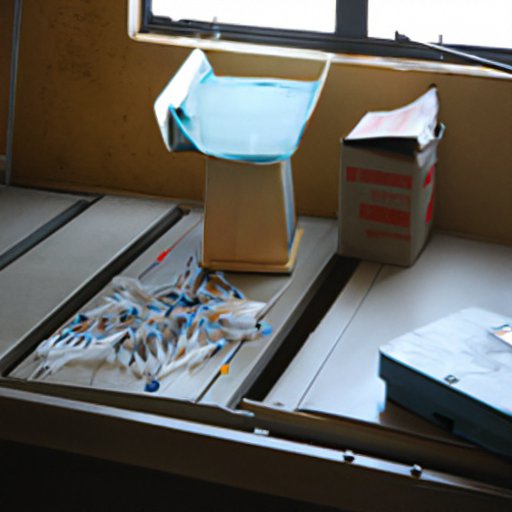} &
        \includegraphics[width=\imwidth]{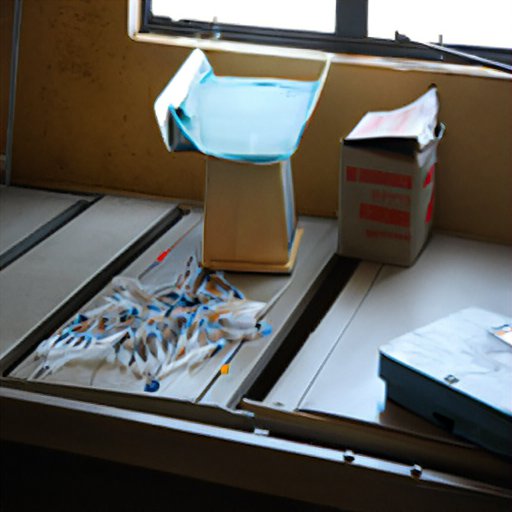} &
        \includegraphics[width=\imwidth]{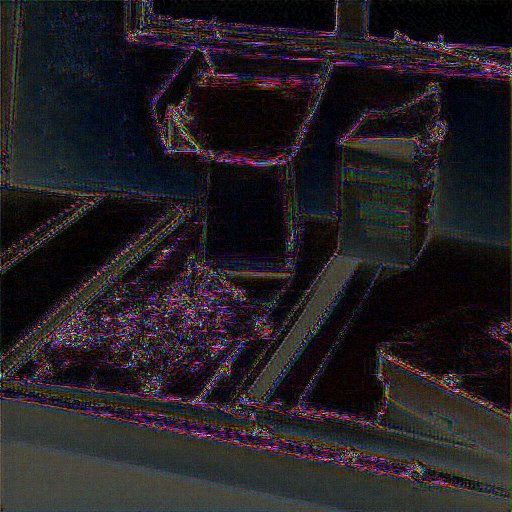} \\
        \bottomrule \\
    \end{tabular}
    \caption{
        \label{chap3/fig:supp-sr} Qualitative results for super-resolution on ImageNet, with original and watermarked generative models.
        Low resolution images are $128\times128$, and upscaled to $512\times512$ with an upscaling factor $f=4$.
    }
\end{figure}

\autoref{chap3/fig:qualitative} shows qualitative examples of how the image generation is altered by the latent decoder's fine-tuning. 
The difference is very hard to perceive even for a trained eye. 
This is surprising for such a low PSNR, especially since the watermark embedding is not constrained by any human visual system like in professional watermarking techniques. 
Most interestingly, the LDM decoder has indeed learned to add the watermark signal only over textured areas where the human eyes are not sensitive, while the uniform backgrounds are kept intact (see the pixel-wise difference).
Figures~\ref{chap3/fig:supp-watermark}, \ref{chap3/fig:supp-txt2img}, \ref{chap3/fig:supp-inpaint} and \ref{chap3/fig:supp-sr} provide additional qualitative results for different watermarking methods, text-to-image generation, inpainting and super-resolution tasks.

\autoref{chap3/tab:quality-watermarking} presents a quantitative evaluation of image generation on the different tasks.
We report the FID, and the average PSNR and SSIM that are computed between the images generated by the fine-tuned LDM and the original one.
The results show that no matter the task, the watermarking has very small impact on the FID of the generation.

The average PSNR is around $30$~dB and SSIM around $0.9$ between images generated by the original and a  watermarked model.
This is a bit low from a watermarking perspective, because we do not explicitly optimize for them.
However, in a real world scenario, one would only have the watermarked version of the image. 
Therefore it is not as important to be extremely close to the original image, we just need to generate artifacts-free images. Without access to the image generated by the original LDM, it is very hard to tell whether a watermark is present or not.

\begin{table}[t!]
    \centering
    \caption{
        Generation quality and comparison to post-hoc watermarking on 512$\times$512 images and $48$-bit signatures.
        PSNR and SSIM are computed between generations of the original and watermarked generators.
        For FID, we show in {\color{orange} (color)} the difference with regards to original.
        Post-hoc watermarking is evaluated on text-generated images.
        Overall, Stable Signature has minimal impact on generation quality. 
        It has comparable robustness to post-hoc methods while being rooted in the generation itself.
        \autoref{chap3/app:implementation-details} gives additional details on methods, baselines, metrics and transformations.
    }\label{chap3/tab:quality-watermarking} 
    \footnotesize
    \setlength{\tabcolsep}{4pt}
        \begin{tabular}{ c l @{\hspace{2pt}} l  *{2}{c} *{4}{p{25pt}}}
        \toprule
       & & \multirow{2}{*}{}          & \multirow{2}{*}{PSNR / SSIM $\uparrow$} & \multirow{2}{*}{FID $\downarrow$} &\multicolumn{4}{c}{Bit accuracy $\uparrow$ on:} \\ 

    &                                         &                           &           &                                            &  None & Crop & Brigh. & Comb.  \\ \midrule
\multirow{6}{*}{\rotatebox[origin=c]{90}{Tasks}}  
        & Text-to-Image                      & LDM             & $30.0$ / $0.89$   & $19.6$ \color{orange}{($-0.3$)} & $0.99$ & $0.95$ & $0.97$ & $0.92$ \\ \cmidrule{2-9}
       & Image Edition                       & DiffEdit                  & $31.2$ / $0.92$ & $15.0$ \color{orange}{($-0.3$)}       & $0.99$ & $0.95$ & $0.98$ & $0.94$ \\ \cmidrule{2-9}
       & Inpainting - Full          & \multirow{2}{*}{Glide}  & $31.1$ / $0.91$  & $16.8$ \color{orange}{($+0.6$)} & $0.99$ & $0.97$ & $0.98$ & $0.93$ \\ 
       & {\color{white}Inpa} - Mask only          &                                   & $37.8$ / $0.98$  & $9.0$~~ \color{orange}{($+0.1$)} & $0.89$ & $0.76$ & $0.84$ & $0.78$\\ \cmidrule{2-9}
       & Super-Resolution & LDM  & $34.0$ / $0.94$ & $11.6$ \color{orange}{($+0.0$)}      & $0.98$ & $0.93$ & $0.96$ & $0.92$ \\ 
    \midrule \rule{0pt}{8pt} \rule{0pt}{8pt} 
\multirow{7}{*}{\rotatebox[origin=c]{90}{WM Methods}} 
           & \emph{Post generation} \\
           & DCT-DWT                      & $0.14$ (s/img)      &  $39.5$ / $0.97$  & $19.5$ \color{orange}{($-0.4$)} & $0.86$ & $0.52$ & $0.51$ & $0.51$ \\ 
           & SSL Watermark  & $0.45$ (s/img)      &  $31.1$ / $0.86$  & $20.6$ \color{orange}{($+0.7$)} & $1.00$ & $0.73$ & $0.93$ & $0.66$ \\ 
           & FNNS          & $0.28$ (s/img)      &  $32.1$ / $0.90$  & $19.0$ \color{orange}{($-0.9$)} & $0.93$ & $0.93$ & $0.91$ & $0.93$ \\ 
           & HiDDeN                        & $0.11$ (s/img)      &  $32.0$ / $0.88$  & $19.7$ \color{orange}{($-0.2$)} & $0.99$ & $0.97$ & $0.99$ & $0.98$ \\ \cmidrule{2-9}
           & \emph{Merged in generation} \\
           & Stable Signature                            & $0.00$ (s/img)             &  $30.0$ / $0.89$ & $19.6$ \color{orange}{($-0.3$)}     & $0.99$ & $0.95$ & $0.97$ & $0.92$ \\ 
        \bottomrule
    \end{tabular}
\end{table}

\subsection{Watermark robustness}\label{chap3/subsec:robustness}

We evaluate the robustness of the watermark to different image transformations applied before extraction.
For each task, we generate $1$k images with $10$ models fine-tuned for different messages, and report the average bit accuracy in Tab.~\ref{chap3/tab:quality-watermarking} (as a reminder, the watermark is a $48$-bit binary key).
Additionally, \autoref{chap3/tab:robustness} reports results on more image transformations for images generated from COCO prompts, and \autoref{chap3/tab:supp-robustness} reports robustness results on more tasks and transformations.

\begin{table}[t!]
    \centering
    \caption{
        Watermark robustness to transformations.
        We report the bit accuracy, averaged over $10\times1$k images generated from COCO prompts with $10$ different keys.
    }\label{chap3/tab:robustness}
    \footnotesize
    \setlength{\tabcolsep}{4pt}
    \resizebox{0.96\linewidth}{!}{
        \begin{tabular}{ll|ll|ll|ll}
            \toprule
            \bf{Attack}             & \bf{Bit acc.}     & \bf{Attack}         & \bf{Bit acc.} & \bf{Attack} & \bf{Bit acc.} & \bf{Attack} & \bf{Bit acc.} \\
            \midrule
            None                    & $0.99$            & Crop $0.1$          & $0.95$        & JPEG $50$   & $0.88$        & Comb.       & $0.92$ \\
            Sharpness $2.0$         & $0.99$            & Bright. $2.0$       & $0.97$        & Cont. $2.0$ & $0.98$        & Sat. $2.0$  & $0.99$ \\
            Med. Filter $k$=7       & $0.94$            & Resize $0.7$        & $0.91$        & Text overlay & $0.99$       &             &        \\
            \bottomrule
        \end{tabular}
    }
\end{table}

\newcommand{\rota}[1]{\rotatebox{25}{\footnotesize\hspace{-0cm}#1}}
\begin{table}[t]
\centering
\caption{
    Watermark robustness on different tasks and image transformations applied before decoding.
    We report the bit accuracy, averaged over $10\times1$k images generated with $10$ different keys.
    The combined transformation is a combination Crop $50\%$, Brightness $1.5$ and JPEG $80$.
    More detail on the evaluation is available in Sec.~~\ref{chap3/app:generative-tasks}. \\
}\label{chap3/tab:supp-robustness}
\vspace*{-0.5cm}
\resizebox{1.0\linewidth}{!}{
    \footnotesize
    \begin{tabular}{ *{2}{l} *{11}{p{1.0cm}} }
    \toprule
    \multirow{2}{*}{Task}                       & \multirow{2}{*}{}       & \multicolumn{9}{c}{Image transformation} \\ \cmidrule{3-12}
                                                &                       &  \rota{None} & \rota{Crop $0.1$} & \rota{JPEG $50$} & \rota{Resi. $0.7$} & \rota{Bright. $2.0$} & \rota{Cont. $2.0$} & \rota{Sat. $2.0$} & \rota{Sharp. $2.0$} & \rota{Text over.} & \rota{Comb.}\\ \midrule
     Text-to-Image          & LDM                         & $0.99$ & $0.95$ & $0.88$ & $0.91$ & $0.97$ & $0.98$ & $0.99$ & $0.99$ & $0.99$ & $0.92$ \\ \midrule
     Image Edition          & DiffEdit & $0.99$ & $0.95$ & $0.90$ & $0.91$ & $0.98$ & $0.98$ & $0.99$ & $0.99$ & $0.99$ & $0.94$\\ \midrule
     Inpainting - Full      & \multirow{2}{*}{Glide}     & $0.99$ & $0.97$ & $0.88$ & $0.90$ & $0.98$ & $0.99$ & $0.99$ & $1.00$ & $0.99$ & $0.93$ \\ 
    {\color{white}Inpa} - Mask only   &                                         & $0.89$ & $0.76$ & $0.73$ & $0.77$ & $0.84$ & $0.86$ & $0.89$ & $0.91$ & $0.89$ & $0.78$ \\  \midrule
     Super-Resolution & LDM & $0.98$ & $0.93$ & $0.86$ & $0.85$ & $0.96$ & $0.96$ & $0.97$ & $0.98$ & $0.98$ & $0.92$\\ 
    \bottomrule
    \end{tabular}
}
\end{table}

We see that the watermark is indeed robust for several tasks and across transformations, and most often yields above $0.9$ bit accuracy.
\emph{With regards to the tasks}, the bit accuracy is not really altered, except for inpainting, when replacing only the masked region of the image (between $1-50$\% of the image, with an average of $27\%$ across masks).
This is expected, as the watermark is not present in the masked region.
\emph{With regards to the transformations}, the resize and JPEG $50$ transformations seem to be the most challenging ones, and sometimes get bellow $0.9$.
Note that the crop location is not important but the visual content of the crop is, \eg, there is no way to decode the watermark on crops of blue sky (this is the reason we only show center crop).
Besides, the bit accuracy is not perfect even without edition, mainly because there are images that are hard to watermark (\eg, the ones that are very uniform, like the background in Fig.~\ref{chap3/fig:qualitative}) and for which the accuracy is lower.

Note that the robustness comes even without any transformation during the LDM fine-tuning phase:
it is due to the watermark extractor.
If the watermark embedding pipeline is learned to be robust against an augmentation, then the LDM learns how to produce watermarks that are robust against it during fine-tuning.

\subsection{Comparison to post-hoc watermarking}\label{chap3/subsec:watermarking}

An alternative way to watermark generated images is to process them after the generation (post-hoc). 
This may be simpler, but less secure and efficient than Stable Signature.
We compare our method to a frequency based method, DCT-DWT~\cite{invisible-watermark},
iterative approaches (SSL Watermark~\citep{fernandez2022sslwatermarking} and FNNS~\citep{kishore2021fixed}), and an encoder/decoder one like HiDDeN~\citep{zhu2018hidden}.
We choose DCT-DWT since it is employed by the Stable Diffusion repository~\citep{2022stablediffusion}, and the other methods because of their performance and their ability to handle arbitrary image sizes and number of bits.
We use our implementations (see details in Sec.~\ref{chap3/app:watermarking}).

\autoref{chap3/tab:quality-watermarking} compares the generation quality and the robustness over $5$k generated images.
Overall, we achieve comparable results in terms of robustness. 
HiDDeN's performance is a bit higher but its output bits are not i.i.d. meaning that it cannot be used with the same guarantees as the other methods.
We also observe that our post-hoc watermarking tend to present artifacts (see Fig.~\ref{chap3/fig:supp-watermark}).
One explanation is that Stable Signature is merged into the high-quality generation process with the LDM autoencoder model, which is able to modify images in a more subtle way.

%% file: chapter-3/sections/6-ablations.tex
\section{Attacks on Stable Signature's watermarks}\label{chap3/sec:attacks}

We examine the watermark's resistance to intentional tampering, 
as opposed to distortions that happen without bad intentions like crops or compression (discussed in Sec.~\ref{chap3/sec:application}). 
We consider two threat models: one  is typical for many image watermarking methods~\citep{cox2007digital} and operates at the image level, and another targets the generative model level. 
For image-level attacks, we evaluate on $5$k images generated from COCO prompts.

\subsection{Image-level attacks}\label{chap3/subsec:image-level-attacks}

\begin{figure}[b!]
    \centering
    \includegraphics[width=0.6\linewidth, trim={0 0 0 0}, clip]{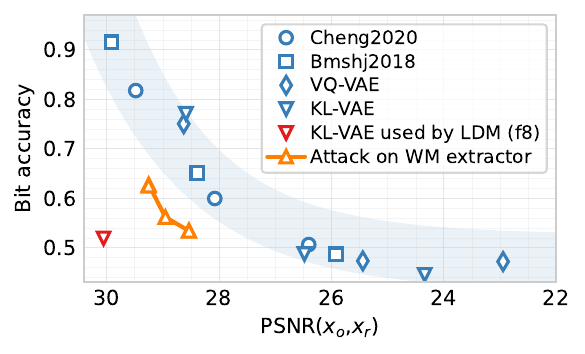}
    \caption{Removal attacks.
    $x_o$ is the image produced by the original generator, 
    $x_r$ is the version produced by the watermarked generator and then attacked.
    Bit accuracy is on the watermark extracted from $x_r$.
    Neural autoencoders~\citep{balle2018variational, cheng2020learned, esser2021taming} follow the \colorbox[HTML]{ebf2f8}{same trend}, except for the one used by LDM (`KL-f8' for our LDM).
    When access to the watermark extractor is granted, adversarial attacks also remove the watermark at lower PSNR budget.
    }
    \label{chap3/fig:purification}
\end{figure}

\paragraph{Watermark removal.}
Bob alters the image to remove the watermark with deep learning techniques, like methods used for adversarial purification~\citep{shi2021online, yoon2021adversarial} or neural autoencoders~\citep{abdelnabi2021adversarial, liu2020defending}.
This kind of attacks had not yet been explored at the time of writing this chapter.

The perceptual autoencoders aim to create compressed latent representations of images.
We select $2$ state-of-the-art autoencoders from the CompressAI library zoo~\citep{begaint2020compressai}: the factorized prior model~\citep{balle2018variational} and the anchor model variant~\citep{cheng2020learned}.
We also select the autoencoders from~\citet{esser2021taming} and~\citet{rombach2022high}.
For all models, we use different compression factors to observe the trade-off between quality degradation and removal robustness.
For \texttt{bmshj2018}: $1$, $4$ and $8$, for \texttt{cheng2020}: $1$, $3$ and $6$, for \texttt{esser2021}: VQ-$4$, $8$ and $16$, for \texttt{rombach2022} KL-$4$, $8$, $16$ and $32$ (KL-$8$ being the one used by SD v1.4).
We generate $1$k images from text prompts with our LDM watermarked with a $48$-bits key.
We then try to remove the watermark using the autoencoders, and compute the bit accuracy on the extracted watermark.
The PSNR is computed between the original image and the reconstructed one, which explains why the PSNR does not exceed $30$dB (since the watermarked image already has a PNSR of $30$dB).
If we compared between the watermarked image and the image reconstructed by the autoencoder instead, the curves would show the same trend but the PSNR would be $2$-$3$ points higher.

We evaluate the robustness of the watermark against neural autoencoders~\citep{balle2018variational, cheng2020learned, esser2021taming, rombach2022high} at different compression rates, and show the outcome in Fig.~\ref{chap3/fig:purification}.
To reduce the bit accuracy closer to random (50\%), the image distortion needs to be strong (PSNR$<$26).
However, assuming the attack is \emph{informed on the generative model}, \ie, the autoencoder is the same as the one used to generate the images, the attack becomes much more  effective.
It erases the watermark while achieving high quality (PSNR$>$29).
This is because the image is modified precisely in the bandwidth where the watermark is embedded.
Note that this assumption is strong, because Alice does not need to distribute the original generator.

\paragraph{Watermark removal \& embedding (white-box).}
To go further, we assume that the attack is \emph{informed on the watermark extractor} -- \eg, because it has leaked.
Bob can use an adversarial attack to remove the watermark by optimizing the image under a PSNR constraint.
The adversarial attack is performed by optimizing the image in the same manner as in chapters~\ref{chapter:ssl-watermarking} and~\ref{chapter:active-indexing}.
The objective is a MSE loss between the output of the extractor and a random binary message fixed beforehand. 
The attack is performed for $10$ iterations with the Adam optimizer~\citep{kingma2014adam} with learning rate $0.1$.
It makes it possible to erase the watermark with a lower distortion budget, as seen in Fig.~\ref{chap3/fig:purification}.
Instead of removing the watermark, an attacker could embed a signature into vanilla images (unauthorized embedding~\citep{cox2007digital}) to impersonate another Bob of whom they have a generated image. 
It highlights the importance of keeping the watermark extractor private.

\subsection{Network-level attacks}\label{chap3/subsec:network-level-attacks}

\paragraph{Quantization and pruning.}
We first study the impact of quantizing or pruning the watermarked generative model -- namely the LDM decoder -- on the watermark extraction on generated images.
These methods were developed for memory or efficiency requirements, but they may also be used to attack the watermark\footnote{Note that quantization and pruning could happen unintentionally, and be seen as a specific case of model-level robustness.}.
Quantization is performed naively, by rounding the weights to the closest quantized value in the min-max range of every weight matrix.
Pruning is done using PyTorch~\citep{paszke2019pytorch} pruning API, with the L1 norm as criterion.
Here is the bit accuracy on the generated images after network attacks, observed over 10$\times$1k images generated from text prompts: Quantization (8-bits) 0.99, (4-bits) 0.99; Pruning L1 (30\%) 0.99, (60\%) 0.95.
We observe that the network generation quality degrades faster than watermark robustness. 
To reduce bit accuracy lower than 98\%, quantization degrades the PSNR $<$25dB, and pruning $<$20dB.

\begin{figure}[b!]
    \centering
    \includegraphics[width=0.6\linewidth, trim={0 0 0 0}, clip]{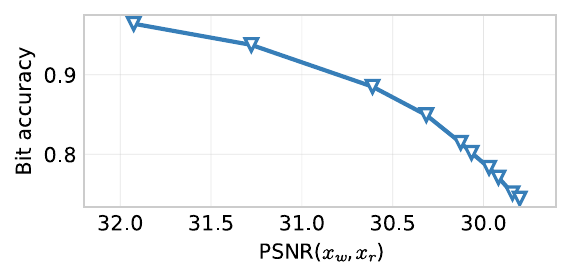}
    \caption{
        Robustness to model purification, \ie, fine-tuning the model to remove watermarks. 
         $x_w$ is the watermarked image, $x_{r}$ is generated with the purified model at different steps of the process.
    }\label{chap3/fig:purification-model}
\end{figure}

\paragraph{Model purification.} 
Bob gets Alice's generative model and uses a fine-tuning process akin to Sec.~\ref{chap3/subsec:finetuning} to eliminate the watermark embedding -- that we coin \emph{model purification}. 
This involves removing the message loss $\mathcal L_\mathrm{m}$, and shifting the focus to the perceptual loss $\mathcal L_\mathrm{i}$ between the original image and the one reconstructed by the LDM autoencoder.

We use the same fine-tuning procedure as in Sec.~\ref{chap3/subsec:finetuning}.
This is done for different numbers of steps, namely $100$, $200$, and every multiple of $200$ up to $1600$.
The bit accuracy and the reported PSNR are computed on $1$k images of the validation set of COCO, for the auto-encoding task.

\autoref{chap3/fig:purification-model} shows the results of this attack for the MSE loss.
The PSNR between the watermarked and purified images is plotted at various stages of fine-tuning.
Empirically, it is difficult to significantly reduce the bit accuracy without compromising the image quality: artifacts start to appear during the purification\footnote{
    \cite{hu2024stable} show, a posteriori, that through a more sophisticated fine-tuning process, the watermark can be removed without compromising much the image quality.
}.

\paragraph{Model collusion.}
Users may collude by aggregating their models.
For instance, Bob$^{(i)}$ and Bob$^{(j)}$ can average the weights of their models (like Model soups~\citep{wortsman2022model}) creating a new model to deceive identification.
We found that the bit at position $\ell$ output by the extractor will be $0$ (resp. $1$) when the $\ell$-th bits of Bob$^{(i)}$ and Bob$^{(j)}$ are both $0$ (resp. $1$), and that the extracted bit is random when their bits disagree.

The goal is to observe the decoded watermarks on the generation when $2$ models are averaged together.
We fine-tune the LDM decoder for $10$ different $48$-bits keys (representing $10$ Bobs).
We then randomly sample a pair of Bobs and average the $2$ models, with which we generate $100$ images.
We then extract the watermark from the generated images and compare them to the $2$ original keys.
We repeat this experiment $10$ times, meaning that we observe $10\times 100 \times 48=48000$ decoded bits.

We show the distributions of the soft bits (before thresholding) output by the watermark extractor on images generated by the average model. 
The $\ell$-th output is labeled by bits of Bob$^{(i)}$ and Bob$^{(j)}$ (\texttt{00} means both have \texttt{0} at position $\ell$):
\begin{center}
    \includegraphics[width=0.5\linewidth, trim={0 0.7cm 0 0cm}, clip]{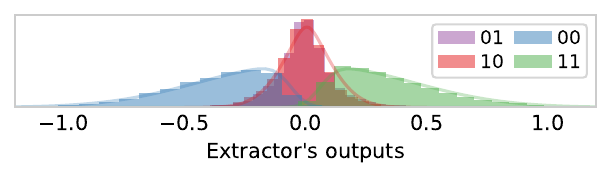}   
\end{center}
The rightmost skewed normal is fitted with the Scipy library and the corresponding parameters are $a:6.96, e:0.06, w:0.38$. 
This is done over all bits where Bobs both have a $1$.
The same observation holds when there is no collusion, with approximately the same parameters.
When the bit is not the same between Bobs, we denote by $m_1^{(i)}$ the random variable representing the output of the extractor in the case where the generative model only comes from Bob$^{(i)}$,
and by $m_2$ the random variable representing the output of the extractor in the case where the generative model comes from the average of the two Bobs.
Then in our model $m_2 = 0.5 \cdot ( m_1^{(i)} + m_1^{(j)})$, and the pdf of $m_2$ is the convolution of the pdf of $m_1^{(i)}$ and the pdf of $m_1^{(j)}$, rescaled in the x axis because of the factor $0.5$.

This \emph{marking assumption} plays a crucial role in traitor tracing literature~\citep{furon:hal-00757152,meerwald:hal-00740964, tardos2008optimal}.
Surprisingly, it holds even though our watermarking process is not explicitly designed for it.
The study has room for improvement, such as creating user identifiers with more powerful traitor tracing codes~\citep{tardos2008optimal} and using more powerful traitor accusation algorithms~\citep{furon:hal-00757152,meerwald:hal-00740964}.
Importantly, we found the precedent remarks also hold if the colluders operate at the image level.

\section{Ablation studies}\label{chap3/sec:ablations}

\subsection{Quality-robustness tradeoff}\label{chap3/subsec:quality-tradeoff}

We can choose to maximize the image quality or the robustness of the watermark thanks to the weight $\lambda_\mathrm{i}$ of the perceptual loss in Eq.~\eqref{chap3/eq:loss2}.
We report the average PSNR of $1$k generated images, as well as the bit accuracy obtained on the extracted message for the `Combined' editing applied before detection (qualitative results are in Sec.~\ref{chap3/sec:supp-percep-loss}).
A higher $\lambda_\mathrm{i}$ leads to an image closer to the original one, but to lower bit accuracies on the extracted message, see \autoref{chap3/tab:tradeoff}.

\begin{table}[t!]
    \centering
    \caption{Quality-robustness trade-off during fine-tuning.}\label{chap3/tab:tradeoff}
    \footnotesize
    \begin{tabular}{l *{6}{c@{\hspace*{8pt}}}}
        \toprule
        $\lambda_i$ for fine-tuning     & $0.8$ & $0.4$ & $0.2$ & $0.1$ & $0.05$ \\ \midrule
        \rule{0pt}{2ex}
        PSNR $\uparrow$ & $31.4$ & $30.6$ & $29.7$ & $28.5$ & $26.8$\\ 
        \rule{0pt}{2ex}
        Bit acc. $\uparrow$ on `comb.' & $0.85$ & $0.88$ & $0.90$ & $0.92$ & $0.94$ \\ 
        \bottomrule 
    \end{tabular}
\end{table}

\subsection{Perceptual loss}\label{chap3/sec:supp-percep-loss}

\begin{figure}[b!]
    \centering
    \scriptsize
    \newcommand{\imwidth}{0.165\textwidth}
    \setlength{\tabcolsep}{0pt}
    \begin{tabular}{cc@{\hskip 2pt}cc@{\hskip 2pt}cc}
        \toprule
        $\lambda_i = 0.025$ &  & $\lambda_i = 0.05$ &  & $\lambda_i = 0.1$ &  \\
        \midrule
        \includegraphics[width=\imwidth]{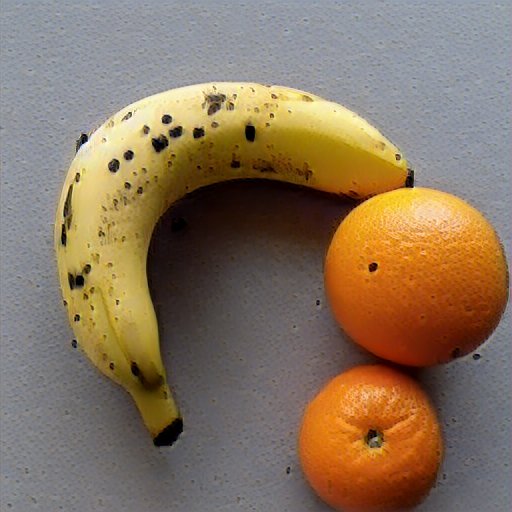} &
        \includegraphics[width=\imwidth]{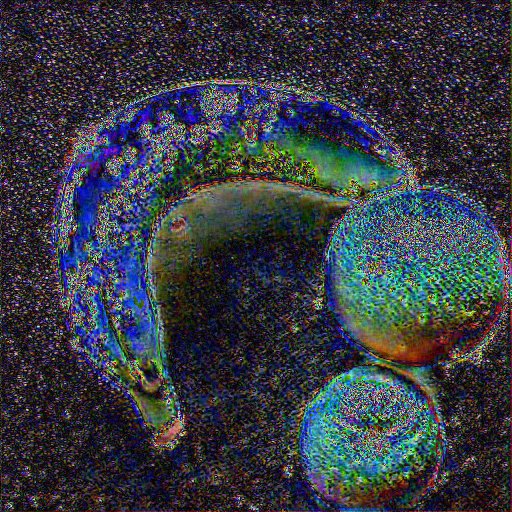} &
        \includegraphics[width=\imwidth]{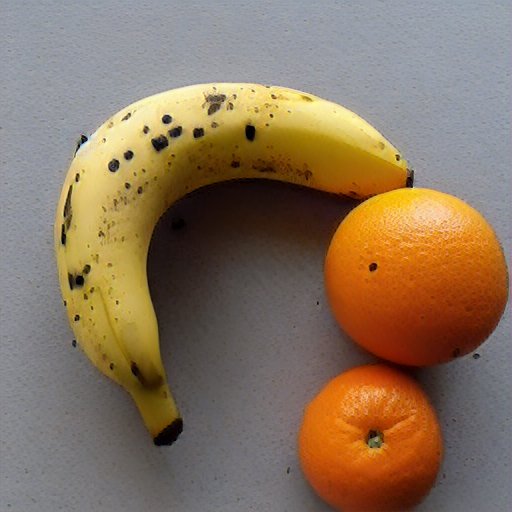} &
        \includegraphics[width=\imwidth]{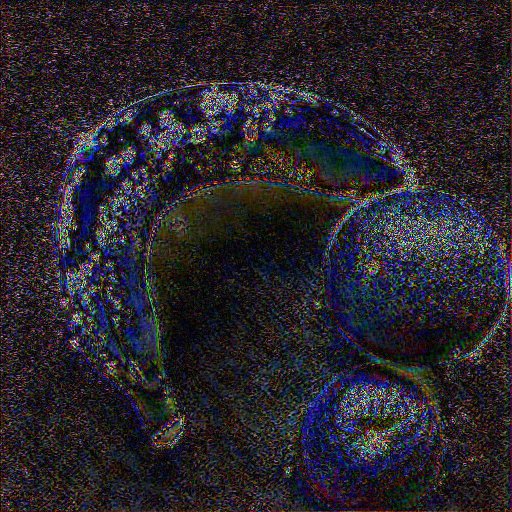} &
        \includegraphics[width=\imwidth]{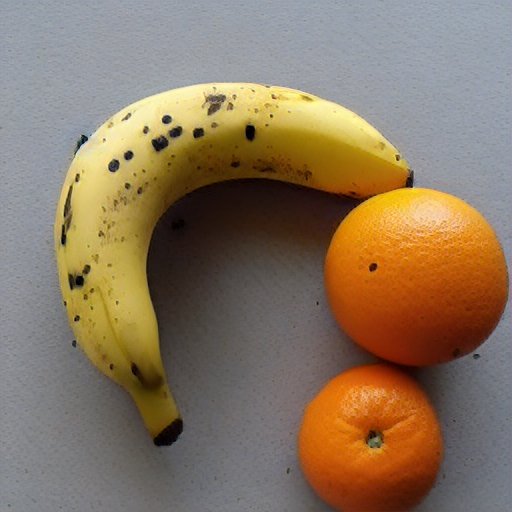} &
        \includegraphics[width=\imwidth]{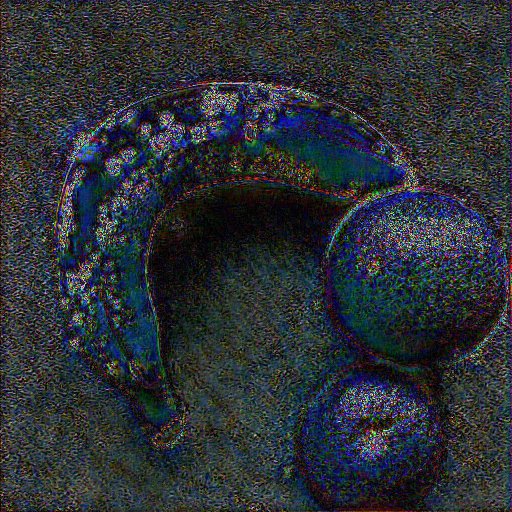} \\
    \end{tabular}
    \begin{tabular}{c@{\hskip 2pt}ccccc}
        \toprule
        Original & Watson-VGG & Watson-DFT & LPIPS & MSE & LPIPS + $0.1\cdot$MSE \\
        \midrule
        \includegraphics[width=\imwidth]{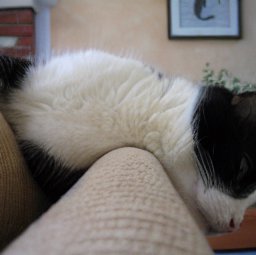} &
        \includegraphics[width=\imwidth]{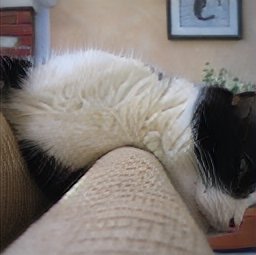} &
        \includegraphics[width=\imwidth]{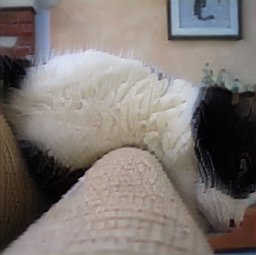} &
        \includegraphics[width=\imwidth]{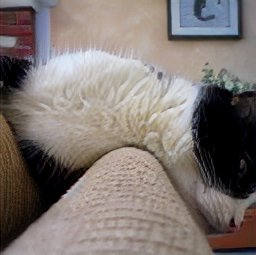} &
        \includegraphics[width=\imwidth]{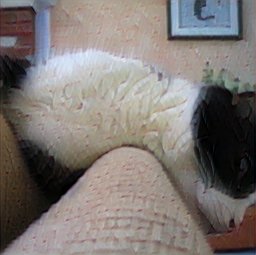} &
        \includegraphics[width=\imwidth]{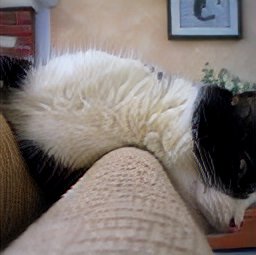} \\
        \bottomrule \\
    \end{tabular}
    \caption{
        Qualitative influence of the perceptual loss during LDM fine-tuning. 
        (Top): we show images generated with the LDM autoencoder fine-tuned with different $\lambda_i$, and the pixel-wise difference ($\times 10$) with regards to the image obtained with the original model.
        PSNR are $24$dB, $26$dB, $28$dB from left to right. 
        (Bottom): we change the perceptual loss and fix $\lambda_i$ to have approximately the same bit accuracy of $0.95$ on the ``combined'' augmentation. 
    }
    \label{chap3/fig:supp-lossi}
\end{figure}

The perceptual loss of Eq.~\eqref{chap3/eq:loss2} affects the image quality.
\autoref{chap3/fig:supp-lossi} shows how the parameter $\lambda_i$ affects the image quality.
For high values, the image quality is very good. %
For low values, artifacts mainly appear in textured area of the image. 
It is interesting to note that this begins to be problematic only for low PSNR values (around 25 dB).

\autoref{chap3/fig:supp-lossi} shows an example of a watermarked image for different perceptual losses: Watson-VGG~\citep{czolbe2020loss}, Watson-DFT~\citep{czolbe2020loss}, LPIPS~\citep{zhang2018unreasonable}, MSE, and LPIPS+MSE.
We set the weight $\lambda_i$ of the perceptual loss so that the watermark performance is approximately the same for all types of loss, and such that the degradation of the image quality is strong enough to be seen.
Overall, we observe that the Watson-VGG loss gave the most eye-pleasing results, closely followed by the LPIPS.

\subsection{Are the decoded bits i.i.d. Bernoulli random variables?}\label{chap3/app:assumption}

\begin{figure}[b]
    \begin{minipage}{0.67\textwidth}
        \centering
        \scriptsize
        \newcommand{\imheight}{0.33\textwidth}
        \setlength{\tabcolsep}{0pt}
        \begin{tabular}{lll}
            Before whitening: & After whitening: & Bernoulli simulation: \\
            \includegraphics[height=\imheight, trim=2cm 0.5cm 3.7cm 1cm, clip]{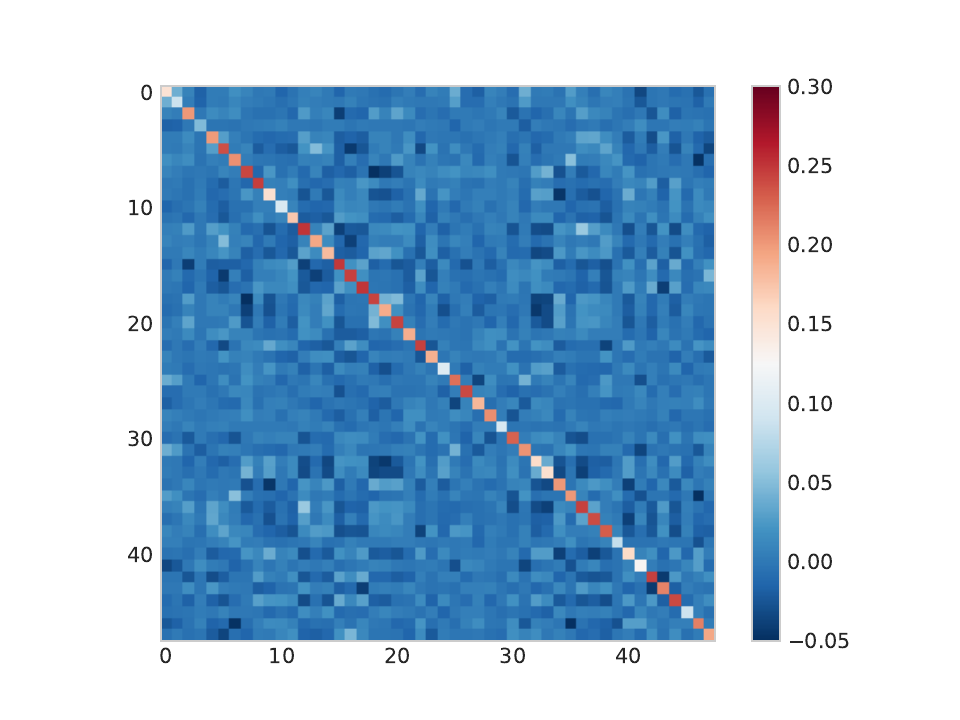} &
            \includegraphics[height=\imheight, trim=2cm 0.5cm 3.7cm 1cm, clip]{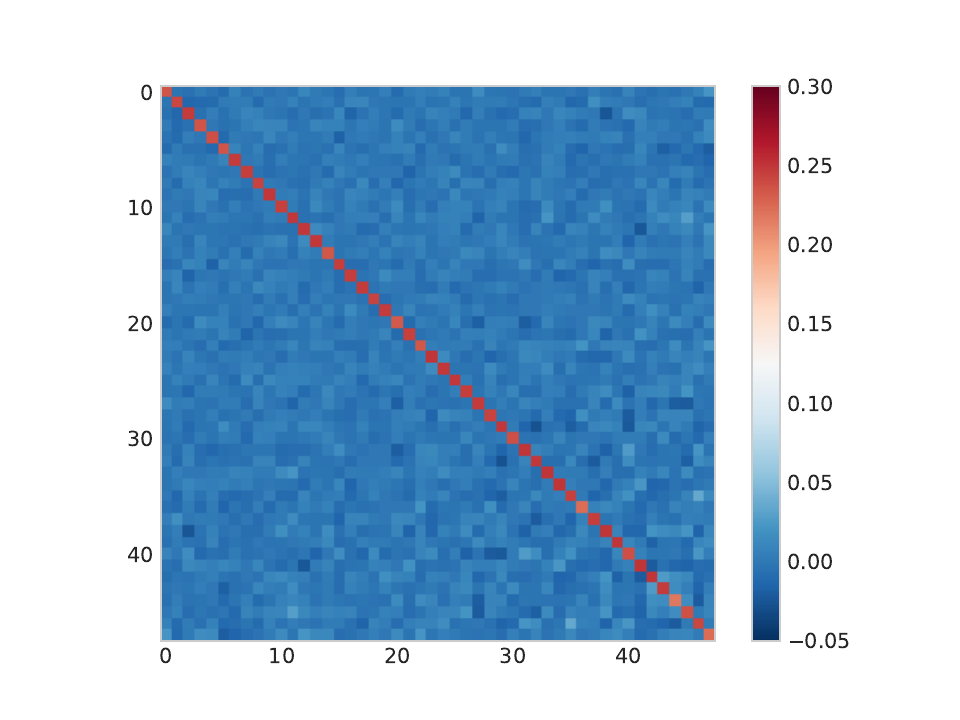} &
            \includegraphics[height=\imheight, trim=2cm 0.5cm 1.5cm 1cm, clip]{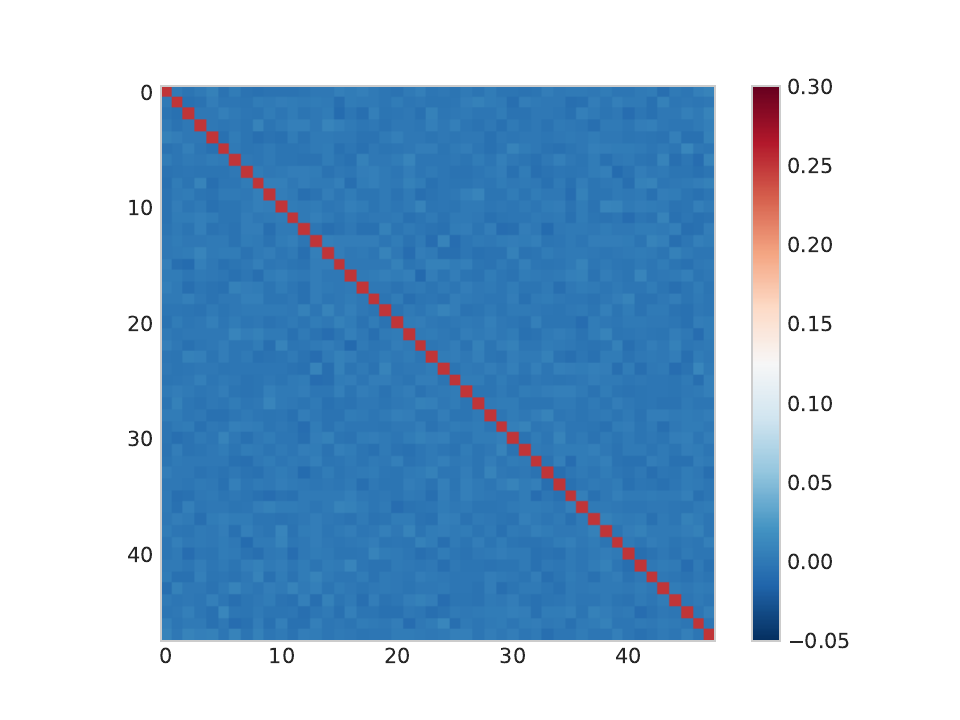} \\
        \end{tabular}
        \caption{Covariance matrices of the bits output by the watermark decoder $\mathcal{W}$ before and after whitening.}
        \label{chap3/fig:supp-assumption}
    \end{minipage}\hfill
    \begin{minipage}{0.27\textwidth}
        \centering
        \includegraphics[width=0.99\textwidth, trim=0 0 0 0, clip]{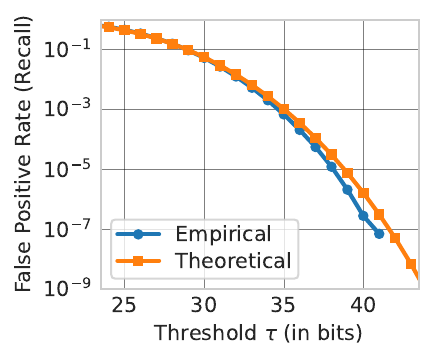}
        \caption{Empirical check of the FPR.}
        \label{chap3/fig:supp-fpr-check}
    \end{minipage}
\end{figure}

The FPR and the $p$-value \eqref{chap3/eq:p-value} are computed with the assumption that, for vanilla images (not watermarked),
the bits output by the watermark decoder $\mathcal{W}$ are independent and identically distributed (i.i.d.) Bernoulli random variables with parameter $0.5$.
This assumption is not true in practice, even when we tried using regularizing losses in the training at phase one~\citep{bardes2022vicreg, sablayrolles2018catalyser}.
This is why we whiten the output at the end of the pre-training.

\autoref{chap3/fig:supp-assumption} shows the covariance matrix of the hard bits output by $\mathcal{W}$ before and after whitening. 
They are computed over $5$k vanilla images, generated with our LDM at resolution $512\times512$ (as a reminder the whitening is performed on $1$k vanilla images from COCO at $256\times256$).
We compare them to the covariance matrix of a Bernoulli simulation, where we simulate $5k$ random messages of $48$ Bernoulli variables.
We observe the strong influence of the whitening on the covariance matrix, although it still differs a little from the Bernoulli simulation.
We also compute the bit-wise mean and observe that for un-whitened output bits, some bits are very biased.  
For instance, before whitening, one bit had an average value of $0.95$ (meaning that it almost always outputs $1$).
After whitening, the maximum average value of a bit is $0.68$.
For the sake of comparison, the maximum average value of a bit in the Bernoulli simulation was $0.52$.
It seems to indicate that the distribution of the generated images are different than the one of vanilla images, and that it impacts the output bits.
Therefore, the bits are not perfectly i.i.d. Bernoulli random variables. 
We however found they are close enough for the theoretical FPR computation to match the empirical one (see next section) -- which was what we wanted to achieve.

\subsection{Empirical check of the FPR}\label{chap3/app:fpr-check}
In \autoref{chap3/fig:tpr-fpr}, we plotted the TPR against a theoretical value for the FPR, with the i.i.d. Bernoulli assumption.
The FPR was computed theoretically with Eq.~\eqref{chap3/eq:p-value}.
Here, we empirically check on smaller values of the FPR (up to $10^{-7}$) that the empirical FPR matches the theoretical one (higher values would be too computationally costly).
To do so, we use the $1.4$ million vanilla images from the training set of ImageNet resized and cropped to $512\times512$, and perform the watermark extraction with $\mathcal{W}$.
We then fix $10$ random $48$-bits key $m^{(1)},\cdots, m^{(10)}$, and, for each image, we compute the number of matching bits $d(m', m^{(i)})$ between the extracted message $m'$ and the key $m^{(i)}$, and flag the image if $d(m', m^{(i)})\geq \tau$.

\autoref{chap3/fig:supp-fpr-check} plots the FPR averaged over the $10$ keys, as a function of the threshold $\tau$.
We compare it to the theoretical one obtained with Eq.~\eqref{chap3/eq:p-value}.
As it can be seen, they match almost perfectly for high FPR values. 
For lower ones ($<10^{-6}$), the theoretical FPR is slightly higher than the empirical one.
This is a good thing since it means that if we fixed the FPR at a certain value, we would observe a lower one in practice.

\subsection{Scaling factor at pre-training}

\begin{table}[t!]
    \centering
    \caption{
        Influence of the (discarded) watermark embedder perceptual quality. 
        We show PSNR and robustness results when using either the watermark embedder trained in the first phase, or the LDM fine-tuned in the second phase.
        The watermark embedder does not need to be very good for the second phase the watermark to be robust at the end of phase 2.
    }\label{chap3/tab:encoder-quality}
    \footnotesize
    \begin{tabular}{l *{5}{l}}
        \toprule
        Scaling factor $\alpha$ & $0.8$ & $0.4$ & $0.2$ & $0.1$ & $0.05$ \\ 
        \midrule
        Watermark embedder - PSNR $\uparrow$  & $16.1$ & $21.8$ & $27.2$ & $33.5$ & $39.3$ \\
        Fine-tuned LDM - PSNR $\uparrow$  & $27.9$ & $30.5$ & \textbf{30.8} & $28.8$ & $27.8$ \\
        \midrule
        Watermark embedder - Bit acc. $\uparrow$ on `none'& $1.00$ & $1.00$ & $0.86$ & $0.72$ & $0.62$ \\
        Fine-tuned LDM - Bit acc. $\uparrow$ on `none'& $0.98$ & \textbf{0.98} & $0.91$ & $0.90$ & $0.96$ \\
        Fine-tuned LDM - Bit acc.  $\uparrow$ on `comb.'& \textbf{0.86} & $0.73$ & $0.82$ & $0.81$ & $0.69$ \\
        \bottomrule 
    \end{tabular}
\end{table}

The watermark encoder does not need to be perceptually good and it is beneficial to degrade image quality during pre-training.
In the following, ablations are conducted on a shorter schedule of $50$ epochs, on $128\times 128$ images and $16$-bits messages.
In \autoref{chap3/tab:encoder-quality}, we train watermark encoders/extractors for different scaling factor $\alpha$ (see Sec.~\ref{chap3/subsec:pre-training}), and observe that $\alpha$ strongly affects the bit accuracy of the method.
When it is too high, the LDM needs to generate low quality images for the same performance because the distortions seen at pre-training by the extractor are too strong.
When it is too low, they are not strong enough for the watermarks to be robust: the LDM will learn how to generate watermarked images, but the extractor will not be able to extract them on edited images.

\subsection{Attack simulation layer}\label{chap3/subsec:message-decoder}

\begin{table}[t!]
    \centering
    \caption{
        Role of the attack simulation layer at pre-training.
        Certain augmentations are needed when pre-training the watermark embedder and extractor, while others are handled naturally.
    }
    \label{chap3/tab:asl}
    \footnotesize
    \begin{tabular}{c *{5}{c}}
        \toprule
        \multirow{2}{*}{ \shortstack{ Seen at  \\ $\mathcal{W}$ training \vspace*{-4pt}} } & \multicolumn{5}{c}{Bit accuracy $\uparrow$ at test time:} \\ \cmidrule{2-6}
            & Crop $0.1$ & Rot. $90$ &JPEG $50$ & Bright. $2.0$ & Res. $0.7$  \\
        \midrule
        \xmark     & 1.00 & 0.56 & 0.50 & 0.99 & 0.48 \\
        \cmark     & 1.00 & 0.99 & 0.90 & 0.99 & 0.91 \\
        \bottomrule
    \end{tabular} 
\end{table}

Watermark robustness against image transformations depends solely on the watermark extractor.
here, we pre-train them with or without specific transformations in the simulation layer, on a shorter schedule of $50$ epochs, with $128\times 128$ images and $16$-bits messages.
From there, we plug them in the LDM fine-tuning stage and we generate $1$k images from text prompts.
We report the bit accuracy of the extracted watermarks in \autoref{chap3/tab:asl}.
The extractor is naturally robust to some transformations, such as crops or brightness, without being trained with them, while others, like rotations or JPEG, require simulation during training for the watermark to be recovered at test time.
Empirically we observed that adding a transformation improves results for the latter, but makes training more challenging.

%% file: chapter-3/sections/8-conclu.tex
\section{Conclusion}\label{chap3/sec:conclusion}

By a quick fine-tuning of the decoder of Latent Diffusion Models, we can embed watermarks in all the images they generate.
This does not alter the diffusion process, making it compatible with most of LDM-based generative models.
These watermarks are robust, invisible to the human eye and can be employed to \emph{detect} generated images and \emph{identify} the user that generated it, with very high performance.
The public release of image generative models has an important societal impact.
With this work, we put to light the usefulness of using watermarking instead of relying on passive detection methods.
We hope it encourages researchers and practitioners to employ similar approaches before making their models publicly available.

The biggest drawback from the method is its intrusive nature.
It requires substantial effort in re-evaluating model image quality and increases implementation cost (because it needs to modify multiple production models instead of maintaining a single encoding model as production use-cases grow).
Embedding the watermark task within the generation process is also not conducive to industry standardization and may limit certain ``data sharing'' opportunities.
Therefore, in cases where the model is not expected to be publicly released, or where the model is expected to be used in a wide range of applications, the watermarking method may not be suitable. 
A post-hoc watermarking method may be preferred, as it is more flexible and can be applied to any model without any modification.
This is the approach we take in Chap.~\ref{chapter:audioseal} for audio watermarking.

%% file: chapter-4/main.tex
\chapter{Proactive Detection of Voice Cloning with Localized Watermarking}\label{chapter:audioseal}

This chapter is based on the paper \fullcite{san2024proactive}.

In the rapidly evolving field of speech generative models, there is a pressing need to ensure audio authenticity against the risks of voice cloning. 
This chapter presents AudioSeal, an audio watermarking technique designed specifically for localized detection of AI-generated speech. 
It is the counterpart to the image watermarking method presented in Chap.~\ref{chapter:stable-signature} in two respects: it is a post-hoc watermarking method (as opposed to watermarking during generation), and it is designed for audio signals (which present different challenges).
AudioSeal employs a generator / detector architecture trained jointly with a localization loss to enable localized watermark detection up to the sample level, and a novel perceptual loss inspired by auditory masking, that enables better imperceptibility. 
AudioSeal achieves state-of-the-art performance in terms of robustness to real life audio manipulations and imperceptibility based on automatic and human evaluation metrics. 
Additionally, AudioSeal is designed with a fast, single-pass detector, that significantly surpasses existing models in speed, achieving detection up to two orders of magnitude faster, making it ideal for large-scale and real-time applications.
Code is available at \url{github.com/facebookresearch/audioseal}.

\newpage
\input{chapter-4/sections/0-intro}

\input{chapter-4/sections/1-related}

\input{chapter-4/sections/2-method}

\input{chapter-4/sections/3-quality}
\input{chapter-4/sections/4-expes}

\input{chapter-4/sections/7-attacks}

\input{chapter-4/sections/5-ablations}
\input{chapter-4/sections/8-conclu}

%% file: chapter-4/sections/0-intro.tex
\section{Introduction}

Generative speech models are now capable of synthesizing voices that are indistinguishable from real ones~\citep{arik2018neural, kim2021conditional, casanova2022yourtts, wang2023neural}.
Though speech generation and voice cloning are not novel concepts, their recent advancements in quality and accessibility have raised new security concerns. 
A notable incident occurred where a deepfake audio misleadingly urged US voters to abstain, showcasing the potential for misusing these technologies to spread false information~\citep{murphy2024biden}.
Regulators and governments are implementing measures for AI content transparency and traceability, including forensics and watermarking -- see \citet{ChineseAIGovernance, EuropeanAIAct, USAIAnnouncement}, and Chap.~\ref{chapter:introduction} for a broader overview.

\begin{figure}[t]
    \centering
    \includegraphics[width=0.99\linewidth, clip, trim={0in 0in 0.5in 0in}]{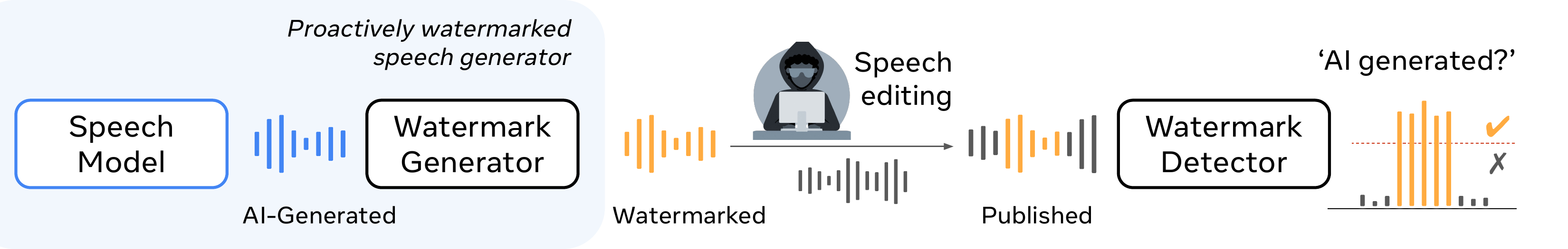}
    \caption{
        Overview of the proactive detection of AI-generated speech.
        We embed an imperceptible watermark in the audio, which can be used to detect if a speech is AI-generated and identify the model that generated it.
        It can also precisely pinpoint AI-generated segments in a longer audio with a sample level resolution (1/16k seconds).
        }
    \label{chap4/fig:fig1}
\end{figure}

The main forensics approach to detect synthesized audio is to train binary classifiers to discriminate between natural and synthesized audios, a technique highlighted in studies by~\citet{Borsos2022AudioLMAL, Kharitonov2023SpeakRA, le2023voicebox}.
We refer to this technique as \textit{passive detection} since it does not alter of the audio source. 
Albeit being a straightforward mitigation, 
it is prone to fail as generative models advance and the difference between synthesized and authentic content diminishes. 

Watermarking emerges as a strong alternative.
It embeds a signal in the generated audio, imperceptible to the ear but robustly detectable by specific algorithms.
As explained in the conclusion of Chap.~\ref{chapter:stable-signature}, post-hoc watermarking methods are more flexible and can be applied to any model without any modification, which makes them the go-to solution for most applications.
However, current post-hoc audio watermarking methods have strong limitations.
First, \emph{they are not adapted for detection}.
Most deep learning based audio watermarking methods~\citep{pavlovic2022robust, DEAR_Liu0FMZY23, chen2023wavmark} are multi-bit.
They train a generator to output the watermarked audio from a sample and a message, and an extractor retrieving the hidden message.
Our method aligns more closely with the concurrent work by \citet{juvela2023collaborative}, which trains a detector, rather than a decoder.
Moreover, the initial applications assumed any sound sample under scrutiny was watermarked (\eg, IP protection).
As a result, the decoders were never trained on non-watermarked samples.
This discrepancy between the training of the models and their practical use leads to poor or overestimated detection rates, depending on the embedded message (see Sec.~\ref{chap4/app:fpr}).

Second, they \emph{are not localized} and consider the entire audio, making it difficult to identify small segments of AI-generated speech within longer audio clips. 
This is particularly important for speech, where editing a single word can be enough to mislead the listener.
The concurrent WavMark's approach~\citep{chen2023wavmark} addresses this by repeating at 1-second intervals a synchronization pattern followed by the actual binary payload. 
This has several drawbacks. 
It cannot be used on spans less than 1 second and is susceptible to temporal edits. 
The synchronization bits also reduce the payload for the encoded message, accounting for 31\% of the total payload. 
Most importantly, the brute force detection algorithm for decoding the synchronization bits is prohibitively slow especially on non-watermarked content, as we show in Sec.~\ref{chap4/sec:speed}.
This makes it unsuitable for real-time and large-scale traceability of AI-generated content on social media platforms, where most content is not watermarked.

To address these limitations, we introduce a method for localized speech watermarking, \emph{AudioSeal}, that can detect AI-generated speech at the time-step level.
It jointly trains two networks: a \emph{generator} that predicts an additive watermark waveform from an audio input, and a \emph{detector} that outputs the probability of the presence of a watermark at each sample of the input audio. 
The detector is trained to precisely and robustly detect synthesized speech embedded in longer audio clips by masking the watermark in random sections of the signal.
The training objective is to maximize the detector's accuracy while minimizing the perceptual difference between the original and watermarked audio. 
We also extend AudioSeal to multi-bit watermarking, so that an audio can be attributed to a specific model or version without affecting the detection signal.

We evaluate the performance of AudioSeal to detect and localize AI-generated speech.
AudioSeal achieves state-of-the-art results on robustness of the detection, far surpassing passive detection with near perfect detection rates over a wide range of audio edits.
It also performs sample-level detection (at resolution of 1/16k second), outperforming WavMark in both speed and performance.
In terms of efficiency, our detector is run once and yields detection logits at every time-step, allowing for real-time detection of watermarks in audio streams. 
This represents a major improvement compared to earlier watermarking methods, which require synchronizing the watermark within the detector, thereby substantially increasing computation time.
Finally, in conjunction with binary messages, AudioSeal almost perfectly attributes an audio to one model among $1000$, even in the presence of strong audio edits.

Our overall contributions are:
\begin{itemize}
    \item We introduce AudioSeal, the first audio watermarking technique designed for localized detection of AI-generated speech up to the sample-level;
    \item A novel perceptual loss inspired by auditory masking, that enables AudioSeal to achieve better imperceptibility of the watermark signal;  
    \item AudioSeal achieves the state-of-the-art robustness to a wide range of real life audio manipulations (Sec.~\ref{chap4/sec:exps});
    \item AudioSeal significantly outperforms the state-of-the-art models in computation speed, achieving up to two orders of magnitude faster detection (Sec.~\ref{chap4/sec:speed});
    \item Insights on the security and integrity of audio watermarking techniques when open-sourcing (Sec.~\ref{chap4/sec:attacks}).
\end{itemize}

%% file: chapter-4/sections/1-related.tex
\section{Related work}

In this section we give an overview of the detection and watermarking methods for audio data. 
A complementary description of prior works can be found in Chap.~\ref{chapter:related-work}.

\paragraph{Synthetic speech detection.}
The approach of most audio generation and text-to-speech (\Gls*{TTS}) papers~\citep{Borsos2022AudioLMAL, Kharitonov2023SpeakRA, borsos2023soundstorm, le2023voicebox} is to train end-to-end deep learning classifiers on what their models generate, similarly as \citet{zhang2017investigation}.
Most of the time, these approaches rely on faint artifacts left by the vocoder that generate the final audio. 
Therefore, accuracy when comparing synthetic to real is usually good, although not performing well on out of distribution audios (compressed, noised, slowed, etc.).
For the same reason, the detection can be fooled by using a different vocoder (to generate synthetic audio that are not flagged) or by using the same vocoder on non-synthetic audios (to generate audios that spoof the detection).
This is highlighted in Sec.~\ref{chap4/sec:active-passive}, where we show that the detection fails on re-synthesized audio.

\paragraph{Invisible watermarking.} 
Traditional audio watermarking methods embed watermarks in the time or frequency domains, often using domain-specific features. 
Deep learning methods focus on multi-bit watermarking and follow a generator/decoder framework, as explained in Sec.~\ref{chap0/sec:watermarking}.
Few works have explored zero-bit watermarking~\citep{wu2023adversarial, juvela2023collaborative}, which is better adapted for detection of AI-generated content.
Our rationale is that robustness increases as the message payload is reduced to the bare minimum~\citep{furon2007constructive}.

In this study, we compare our work with the state-of-the-art watermarking method, WavMark~\citep{chen2023wavmark}, which outperforms previous ones. 
It uses invertible networks to hide 32 bits in 1-second audio segments.
Detection is done by sliding along the audio in 0.05s steps and decoding the message for each window.
If the 10 first decoded bits match a synchronization pattern the rest of the payload is saved (22 bits), and the window can directly slide 1s (instead of the 0.05).
This brute force detection algorithm is prohibitively slow especially when the watermark is absent, since the algorithm will have to attempt and fail to decode a watermark for each sliding window in the input audio (due to the absence of watermark).

%% file: chapter-4/sections/2-method.tex
\section{Method}
\label{chap4/sec:method}
The method jointly trains two models, in a similar fashion to Sec.~\ref{chap0/sec:deep learning-watermarking}.
The generator creates a watermark signal that is added to the input audio. 
The detector outputs local detection logits.
The training optimizes two concurrent classes of objectives: minimizing the perceptual distortion between original and watermarked audios and maximizing the watermark detection. 
To improve robustness to modifications of the signal and localization, we include a collection of train time augmentations.
At inference time, the logits precisely localize watermarked segments allowing for detection of AI-generated content.
Optionally, short binary identifiers may be added on top of the detection to attribute a watermarked audio to a version of the model while keeping a single detector.

\begin{figure}[b]
    \centering
    \includegraphics[width=0.75\linewidth, clip, trim={0.2 1.8in 1.2in 0}]{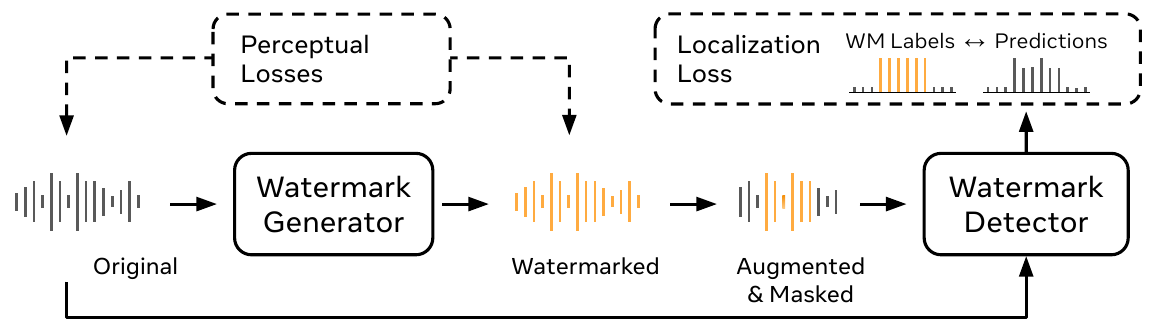}
    \caption{Generator-detector training pipeline.}
    \label{chap4/fig:method}
\end{figure}

\begin{figure}[b!]
    \centering
    \includegraphics[width=0.65\linewidth, clip, trim={0 0 0 0}]{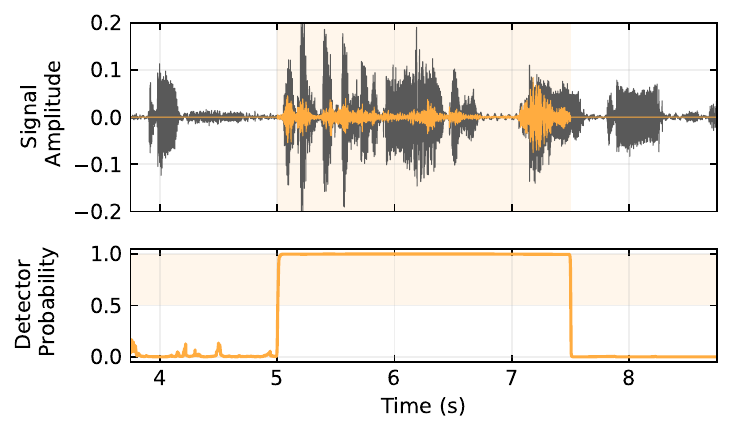}
    \caption{
    (Top) A speech signal ({\color{darkgray} gray}) where the watermark is present between 5 and 7.5 seconds ({\color{orange} orange}, magnified by 5).
    (Bottom) The output of the detector for every time step. 
    An {\color{orange!50} orange} background color indicates the presence of the watermark.
    }
    \label{chap4/fig:loc_quali}
\end{figure}

\subsection{Training pipeline} 
\autoref{chap4/fig:method} illustrates the joint training of the generator and the detector.
It consists of four critical stages:
\begin{enumerate}[label=(\roman*), itemsep=2pt, topsep=2pt, leftmargin=20pt]
    \item The watermark generator takes as input a waveform $s \in \mathbb{R}^T$ and outputs a watermark waveform $\wm \in \mathbb{R}^T$ of the same dimensionality, where $T$ is the number of samples in the  signal. The watermarked audio is then $s_w = s+\wm$.
    \item To enable sample-level localization, we adopt an augmentation strategy focused on watermark masking with silences and other original audios. This is achieved by 
    randomly selecting $k$ starting points and altering the next $T/2k$ samples from $s_w$ in one of 4 ways: revert to the original audio (\ie, $s_w(t)=s(t)$) with probability 0.4; replacing with zeros (\ie, $s_w(t)=0$) with probability 0.2; or substituting with a different audio signal from the same batch (\ie, $s_w(t) = s'(t)$) with probability 0.2, or not modifying the sample at all with probability 0.2.
    \item The second class of augmentation ensures the robustness against audio editing. 
    One of the following signal alterations is applied: 
    bandpass filter, boost audio, duck audio, echo, highpass filter, lowpass filter, pink noise, gaussian noise, slower, smooth, resample (full details in Sec.~\ref{chap4/sec:training-details}). 
    The parameters of those augmentations are fixed to aggressive values to enforce maximal robustness and the
    probability of sampling a given augmentation is proportional to the inverse of its evaluation detection accuracy.
    We implemented these augmentations in a differentiable way when possible, and otherwise (\eg, MP3 compression) with the straight-through estimator~\citep{yin2019understanding} that allows the gradients to back-propagate to the generator.
    \item 
    Detector $D$ processes the original and the watermarked signals, outputting for each a soft decision at every time step, meaning $D(s) \in [0, 1]^T$.
    \autoref{chap4/fig:loc_quali}
    illustrates that the detector's outputs
    are at one only when the watermark is present.
\end{enumerate}

\subsection{Losses} 

Our setup includes multiple perceptual losses and a localization loss. 
We balance them during training by scaling their gradients as done by~\citet{defossez2022high}.

\subsubsection*{Perceptual losses} 

Perceptual losses enforce the watermark imperceptibility to the human ear.
These include an $\ell_1$ loss on the watermark signal, a multi-scale Mel spectrogram loss~\citep{gritsenko2020spectral}, and discriminative losses based on adversarial networks that operate on multi-scale short-term-Fourier-transform spectrograms. 
\citet{defossez2022high} use this combination of losses for training the EnCodec model for audio compression. 

\paragraph*{\ploss.}\label{chap4/app:loudness}
We notably introduce a loss specific for watermarking -- called time-frequency loudness loss -- which operates entirely in the waveform domain. 
It accounts for auditory masking, a psycho-acoustic property of the auditory system exploited in the early days of watermarking~\citep{blockrep1-Kirovski2003}: the ear fails at perceiving sounds occurring at the same time and at the same frequency range~\citep{book:audio}.

\ploss\ is calculated as follows. 
First, the input signal $s$ is divided into $B$ signals based on non-overlapping frequency bands $s_0, \dots, s_{B-1}$. 
Every signal is then segmented using a window of size $W$, with an overlap amount denoted by $r$.
This procedure is applied to both the original audio signal \(s\) and the embedded watermark \(\wm\). 
As a result, we obtain segments of the signal and watermark in time-frequency dimensions, denoted as \(s_b^{w}\) and \(\wm_b^{w}\) respectively.
For every time-frequency window we compute the loudness difference:
\begin{equation}\label{chap4/eq:loudness_diff}
    l_b^w = \mathrm{Loudness}(\wm_b^w) - \mathrm{Loudness}(s_b^w).
\end{equation}
The loudness is estimated using ITU-R BS.1770-4 recommendations~\citep{loudness}\footnote{
    Our loudness implementation is a simplified version of the torchaudio~\citep{yang2021torchaudio} one.
}.
$l_b^w$ quantifies the discrepancy in loudness between the watermark and the original signal within a specific time window $w$, and a particular frequency band $b$.
The final loss is a weighted sum of the loudness differences using softmax function:
\begin{equation}\label{chap4/eq:perceptual_loss}
    \mathcal{L}_{loud} = \sum_{b, w} \left(\mathrm{softmax}(l)_b^w * l_b^w\right).
\end{equation}
The softmax prevents the model from targeting excessively low loudness where the watermark is already inaudible.

\subsubsection*{Masked sample-level detection loss}

A localization loss ensures that the detection of watermarked audio is done at the level of individual samples. 
For each time step $t$, we compute the binary cross entropy (BCE) between the detector's output $D(s)_t$ and the ground truth label (0 for non-watermarked, 1 for watermarked).
There are two possibilities for the aggregation of sample-level losses.
When no localization augmentation is applied, the loss is averaged over both the original and watermarked audio.
Otherwise, $s_w$ contains both types of segments so the loss is computed only on $s_w$.
This avoids creating an imbalance between watermarked and original data. 
Overall, this reads:
\begin{equation}\label{chap4/eq:loc_loss}
    \mathcal{L}_{loc}
    = \frac{1}{T} \sum_{t=1}^{T} \mathrm{BCE}(D(s')_t, y_t),
\end{equation}
where $s'$ might be $s$ or $s_w$, and where time step labels $y_t$ are set to 1 if they are watermarked, and 0 otherwise.

\subsection{Networks architectures}

\begin{figure}[b!]
    \centering
    \includegraphics[width=1.0\linewidth, clip, trim={0 1.6in 0.8in 0}]{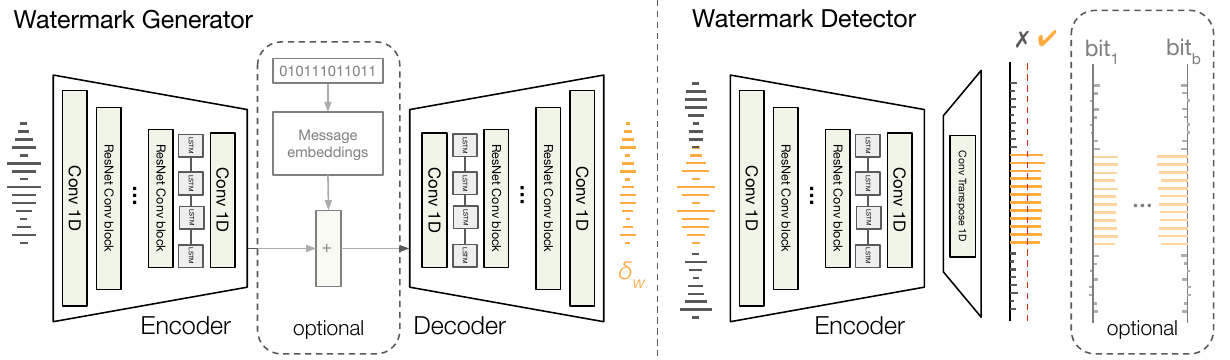}
    \caption{
        Architectures. 
        The \emph{generator} is made of an encoder and a decoder both derived from EnCodec, with optional message embeddings. 
        The encoder includes convolutional blocks and an LSTM, while the decoder mirrors this structure with transposed convolutions.
        The \emph{detector} is made of an encoder and a transpose convolution, followed by a linear layer that returns sample-wise logits. 
        Optionally, multiple linear layers can return k-bit messages.
    }
    \label{chap4/fig:archs}
\end{figure}

The watermark generator is composed of an encoder and a decoder, both incorporating elements from EnCodec~\citep{defossez2022high}. 
The encoder applies a 1D convolution with 32 channels and a kernel size of 7, followed by four convolutional blocks. Each of these blocks includes a residual unit and down-sampling layer, which uses convolution with stride $S$ and kernel size $K = 2S$.
The residual unit has two kernel-3 convolutions with a skip-connection, doubling channels during down-sampling. The encoder concludes with a two-layer LSTM and a final 1D convolution with a kernel size of 7 and 128 channels. 
Strides $S$ values are (2, 4, 5, 8) and the nonlinear activation in residual units is the Exponential Linear Unit (ELU). 
The decoder mirrors the encoder but uses transposed convolutions instead, with strides in reverse order.

The detector comprises an encoder, a transposed convolution and a linear layer. 
The encoder shares the generator's architecture (but with different weights). 
The transposed convolution has $h$ output channels and upsamples the activation map to the original audio resolution (resulting in an activation map of shape $(t, h)$). 
The linear layer reduces the $h$ dimensions to two, and a softmax function finally gives sample-wise logits.

\subsection{Multi-bit watermarking}

We extend the method to support multi-bit watermarking, which allows for attribution of audio to a specific model version.
\emph{At generation}, we add a message processing layer in the middle of the generator.
It takes the activation map in $\R ^{h, t'}$ and a binary message $m\in \{0,1\} ^{b}$ and outputs a new activation map to be added to the original one.
We embed $m$ into $ e = \sum_{i=0..b-1}{E_{2i + m_i} \in \R^h}$, where $E\in \R^{2b, h}$ is a learnable embedding layer.
$e$ is then repeated $t$ times along the temporal axis to match the activation map size ($t,h$).
\emph{At detection}, we add $b$ linear layers at the very end of the detector. 
Each of them outputs a soft value for each bit of the message at the sample-level.
Therefore, the detector outputs a tensor of shape $\R^{t, 1+b}$ (1 for the detection, $b$ for the message).

\emph{At training}, we add a decoding loss $\mathcal{L}_{dec}$ to the localization loss $\mathcal{L}_{loc}$.
This loss $\mathcal{L}_{dec}$ averages the BCE between the original message and the detector's outputs over all parts where the watermark is present.

\subsection{Detection, localization and attribution}
At inference, we may use the generator and detector for:
\begin{itemize}
\item \emph{Detection}: To determine if the audio is watermarked or not. 
To achieve this, we use the average detector's output over the entire audio and flag it if the score exceeds a threshold (default: 0.5).
\item \emph{Localization}: To precisely identify where the watermark is present. We utilize the sample-wise detector's output and mark a time step as watermarked if the score surpasses a threshold (default: 0.5).
\item \emph{Attribution}: To identify the model version that produced the audio, enabling differentiation between users or APIs with a single detector. 
The detector's first output gives the detection score and the remaining $k$ outputs are used for attribution. 
This is done by computing the average message over detected samples and returning the identifier with the smallest Hamming distance.
\end{itemize}

\subsection{Training details}\label{chap4/sec:training-details}

Our watermark generator and detector are trained on a 4.5K hours subset from the VoxPopuli~\citep{voxpopuli-WangRLWTHWPD20} dataset.
We use a sampling rate of 16~kHz and one-second samples, so $T=16000$ in our training.
A full training requires 600k steps, with Adam, a learning rate of $10^{-4}$, and a batch size of 32.
For the drop augmentation, we use $k=5$ windows of $0.1$ sec. $h$ is set to 32, and the number of additional bits $b$ to 16 (note that $h$ needs to be higher than $b$, for example $h=8$ is enough in the zero-bit case). 
For the audio 
The perceptual losses are balanced and weighted as follows: $\lambda_{\ell_1} = 0.1$, $\lambda_{msspec} = 2.0$, $\lambda_{adv} = 4.0$, $\lambda_{loud} = 10.0$. 
The localization and watermarking losses are weighted by $\lambda_{loc} = 10.0$ and $\lambda_{dec} = 1.0$ respectively.

Here are the details of each audio editing augmentation; 
(T) refers to training time and (E) to evaluation time, \ie, parameters used for Tab.~\ref{chap4/tab:wm_robustness}:
\begin{itemize}[leftmargin=2em]
    \item \emph{Bandpass Filter:} Combines highpass and lowpass filtering to allow a specific frequency band to pass through.
    (T) fixed between 300Hz and 8000Hz; (E) fixed between 500Hz and 5000Hz.
    \item \emph{Highpass Filter:} Uses a highpass filter on the input audio to cut frequencies below a certain threshold.
    (T) fixed at 500Hz; (E) fixed at 1500Hz.
    \item \emph{Lowpass Filter:} Applies a lowpass filter to the input audio, cutting frequencies above a cutoff frequency.
    (T) fixed at 5000Hz; (E) fixed at 500Hz.
    \item \emph{Speed:} Changes the speed of the audio by a factor close to 1. 
    (T) random between 0.9 and 1.1; (E) fixed at 1.25.
    \item \emph{Resample:} Upsamples to intermediate sample rate and then downsamples the audio back to its original rate without changing its shape.
    (T) and (E) 32kHz.
    \item \emph{Boost Audio:} Amplifies the audio by multiplying by a factor.
    (T) factor fixed at 1.2; (E) fixed at 10.
    \item \emph{Duck Audio:} Reduces the volume of the audio by a multiplying factor.
    (T) factor fixed at 0.8; (E) fixed at 0.1.
    \item \emph{Echo:} Applies an echo effect to the audio, adding a delay and less loud copy of the original.
    (T) random delay between 0.1 and 0.5 seconds, random volume between 0.1 and 0.5; (E) fixed delay of 0.5 seconds, fixed volume of 0.5.
    \item \emph{Pink Noise:} Adds pink noise for a background noise effect.
    (T) standard deviation fixed at 0.01; (E) fixed at 0.1.
    \item \emph{White Noise:} Adds gaussian noise to the waveform.
    (T) standard deviation fixed at 0.001; (E) fixed at 0.05.
    \item \emph{Smooth:} Smooths the audio signal using a moving average filter with a variable window size.
    (T) window size random between 2 and 10; (E) fixed at 40.
    \item \emph{AAC:} Encodes the audio in AAC format.
    (T) bitrate of 128kbps; (E) bitrate of 64kbps.
    \item \emph{MP3:} Encodes the audio in MP3 format.
    (T) bitrate of 128kbps; (E) bitrate of 32kbps.
    \item \emph{EnCodec:} Resamples at 24kHz, encodes the audio with EnCodec with $nq=16$ (16 streams of tokens), and resamples it back to 16kHz.
\end{itemize}
Implementation is done with the \texttt{julius} python library. 
For all edits except EnCodec compression, evaluation with parameters in the training range would be almost perfect.

%% file: chapter-4/sections/4-expes.tex
\section{Main experiments}
\label{chap4/sec:exps}

\subsection{Audio/speech quality}
\label{chap4/sec:quality}

We first evaluate the quality of the watermarked audio using:
Scale Invariant Signal to Noise Ratio (SI-SNR): 
$\textrm{SI-SNR}(s, s_w) = 10 \log_{10} \left( \| \alpha s \|_2^2 / \| \alpha s - s_w \|_2^2 \right)$,
with $s$ the clean audio and $s_w$ the watermarked one,
where $\alpha = \langle s, s_w \rangle / \| s \|_2^2$;
as well as Perceptual Evaluation of Speech Quality (PESQ)~\citep{rix2001perceptual}, 
Virtual Speech Quality Objective Listener (ViSQOL)~\citep{hines2012visqol} and
Short Term Objective Intelligibility (STOI)~\citep{taal2010short} which are objective perceptual metrics measuring the quality of speech signals.

\autoref{chap4/tab:audio_quality} report these metrics.
AudioSeal behaves differently than watermarking methods like WavMark~\citep{chen2023wavmark} that try to minimize the SI-SNR.
In practice, high SI-SNR is indeed not necessarily correlated with good perceptual quality.
AudioSeal is not optimized for SI-SNR but rather for perceptual quality of speech (similarly as DeAR~\citep{DEAR_Liu0FMZY23} which reports 25.96 as SI-SNR).
This is better captured by the other metrics (PESQ, STOI, ViSQOL), where AudioSeal consistently achieves better performance.
Put differently, our goal is to hide as much watermark power as possible while keeping it perceptually indistinguishable from the original.
\autoref{chap4/fig:loc_quali} also visualizes how the watermark signal follows the shape of the speech waveform.

The metric used for our subjective evaluations is the MUSHRA score~\citep{mushra}. 
It is a crowdsourced test in which participants rate the quality of various samples on a scale of 0 to 100. 
The ground truth is provided for reference. 
We utilized 100 speech samples, each lasting 10 seconds. 
Each sample was evaluated by at least 20 participants.
As part of the study, we included a low anchor, which is a very lossy compression at 1.5kbps, encoded using EnCodec. 
Participants who failed to assign the lowest score to the low anchor for at least 80\% of their assignments were excluded from the study.
In this study our samples got superior ratings than WavMark, with an average score of 77.07, 5 points higher than WavMark.
For comparison, the ground truth samples received an average score of 80.49, while the low anchor's average score was 53.21.

\begin{table}[t!]
    \centering
    \caption{
        Audio quality metrics. 
        Compared to traditional watermarking methods that minimize the SNR like WavMark, AudioSeal achieves same or better perceptual quality.
    }\label{chap4/tab:audio_quality}
    \footnotesize
        \begin{tabular}{lccccc}
            \toprule
            \textbf{Methods} & \textbf{SI-SNR} & \textbf{PESQ} & \textbf{STOI} & \textbf{ViSQOL} & \textbf{MUSHRA}  \\
            \midrule
            WavMark & \textbf{38.25} & 4.302 & 0.997 & 4.730 & 71.52 $\pm$ 7.18\\
            AudioSeal & 26.00 & \textbf{4.470} & 0.997 & \textbf{4.829} &  \textbf{77.07} $\pm$ 6.35 \\
            \bottomrule
        \end{tabular}
\end{table}

\subsection{Comparison with passive classifier}\label{chap4/sec:active-passive}

\begin{table}[t!]
    \centering
    \caption{
        Comparison with Voicebox binary classifier. 
        Percentage refers to the fraction of masked input frames.
        Detection with AudioSeal is perfect for all samples, while Voicebox classifier fails on re-synthesized audio.
    }
    \label{chap4/tab:voicebox}
    \footnotesize
        \begin{tabular}{r *{3}{c}  *{3}{c}  *{3}{c} }
            \toprule
            & \multicolumn{3}{c}{\textbf{AudioSeal (Ours)}} & \multicolumn{3}{c}{\textbf{Voicebox Classif.}} \\
            \cmidrule(rr){2-4} \cmidrule(rr){5-7}
            \textbf{\% Mask} & Acc. & TPR & FPR & Acc. & TPR & FPR \\
            \midrule
            \multicolumn{7}{l}{\emph{Original audio vs AI-generated audio}} \\
            30\% & 1.0 & 1.0 & 0.0 & 1.0 & 1.0 & 0.0 \\
            50\% & 1.0 & 1.0 & 0.0 & 1.0 & 1.0 & 0.0 \\
            90\% & 1.0 & 1.0 & 0.0 & 1.0 & 1.0 & 0.0 \\
            \midrule
            \multicolumn{7}{l}{\emph{Re-synthesized audio vs AI-generated audio}} \\
            30\% & \textbf{1.0} & \textbf{1.0} & \textbf{0.0} & 0.704 & 0.680 & 0.194 \\
            50\% & \textbf{1.0} & \textbf{1.0} & \textbf{0.0} & 0.809 & 0.831 & 0.170 \\
            90\% & \textbf{1.0} & \textbf{1.0} & \textbf{0.0} & 0.907 & 0.942 & 0.112 \\
            \bottomrule
        \end{tabular}
\end{table}

We first compare detection results on samples generated with Voicebox~\citep{le2023voicebox}.
We compare to the passive setup where a classifier is trained to discriminate between Voicebox-generated and real audios (this is done with a detector that shares the same architecture as AudioSeal's, trained until reaching 100\% accuracy on Voicebox samples).
Following the approach in the Voicebox study, we evaluate 2,000 approximately 5-second samples from LibriSpeech, these samples have masked frames (90\%, 50\%, and 30\% of the phonemes) pre-Voicebox generation.
We evaluate on the same tasks, \ie, distinguishing between original and generated, or between original and re-synthesized (created by extracting the Mel spectrogram from original audio and then vocoding it with the HiFi-GAN vocoder).

We use the True Positive Rate (\Gls*{TPR}) and the False Positive Rate (\Gls*{FPR}) as key metrics.
TPR measures correct identification of AI-generated samples, while FPR indicates the rate of genuine audio clips falsely flagged.
In practical scenarios, minimizing FPR is crucial. 
For example, on a platform processing 1 billion samples daily, an FPR of $10^{-3}$ and a TPR of $0.5$ means that 1 million samples require manual review each day, yet only half of the watermarked samples are detected.
The \Gls*{ROC} AUC (Area Under the Curve of the Receiver Operating Characteristics) gives a global measure of performance over all threshold levels, and captures the TPR/FPR trade-off.

Both active and passive setups achieve perfect classification in the case when trained to distinguish between natural and Voicebox.
Conversely, the second part of Tab.~\ref{chap4/tab:voicebox} highlights a significant drop in performance when the classifier is trained to differentiate between Voicebox and re-synthesized.
It suggests that the classifier is detecting vocoder artifacts, since the re-synthesized samples are sometimes wrongly flagged.
The classification performance quickly decreases as the quality of the AI-generated sample increases (when the input is less masked).
On the other hand, our proactive detection does not rely on model-specific artifacts but on the watermark presence. %
This allows for perfect detection over all the audio clips. %

\subsection{Robustness and comparison with watermarking}

\begin{table}[t!]
    \centering
    \caption{
        Detection results for different edits applied before detection. 
        Acc. ({\aux{TPR/FPR}}) is the accuracy (and TPR/FPR) obtained for the threshold that gives best accuracy on a balanced set of augmented samples.
        AUC is the area under the ROC curve.
    }
    \label{chap4/tab:wm_robustness}
    \footnotesize
        \begin{tabular}{l *{2}{l}  *{2}{l}}
        \toprule
        & \multicolumn{2}{l}{\textbf{AudioSeal (Ours)}} & \multicolumn{2}{l}{\textbf{WavMark}} \\
        \cmidrule(rr){2-3} \cmidrule(rr){4-5}
        \multicolumn{1}{c}{Edit} & Acc. \aux{TPR/FPR} & AUC & Acc. \aux{TPR/FPR}  & AUC \\
        \cmidrule(rr){1-1} \cmidrule(rr){2-3} \cmidrule(rr){4-5}
        None & 1.00 \aux{1.00/0.00} & 1.00 & 1.00 \aux{1.00/0.00} & 1.00 \\
        Bandpass & 1.00 \aux{1.00/0.00} & 1.00 & 1.00 \aux{1.00/0.00} & 1.00 \\
        Highpass  &  0.61 \aux{0.82/0.60} & 0.61 & \bf 1.00 \aux{1.00/0.00} & \bf 1.00 \\
        Lowpass & \bf 0.99 \aux{0.99/0.00} & \bf 0.99 & 0.50 \aux{1.00/1.00} & 0.50 \\
        Boost & 1.00 \aux{1.00/0.00} & 1.00 & 1.00 \aux{1.00/0.00} & 1.00 \\
        Duck & 1.00 \aux{1.00/0.00} &  1.00 & 1.00 \aux{1.00/0.00} & 1.00 \\
        Echo & \bf 1.00 \aux{1.00/0.00} & \bf 1.00 & 0.93 \aux{0.89/0.03} & 0.98 \\
        Pink & \bf 1.00 \aux{1.00/0.00} & \bf 1.00 & 0.88 \aux{0.81/0.05} & 0.93 \\
        White & \bf 0.91 \aux{0.86/0.04} & \bf 0.95 & 0.50 \aux{0.54/0.54} & 0.50 \\
        Fast (1.25x) & \bf 0.99 \aux{0.99/0.00} & \bf 1.00 & 0.50 \aux{0.01/0.00} & 0.15 \\
        Smooth & \bf 0.99 \aux{0.99/0.00} &  1.00   & 0.94 \aux{0.93/0.04} & 0.98 \\
        Resample & 1.00 \aux{1.00/0.00} &  1.00 & 1.00 \aux{1.00/0.00} & 1.00 \\
        AAC & 1.00 \aux{1.00/0.00} &  1.00 & 1.00 \aux{1.00/0.00} & 1.00 \\
        MP3 & \bf 1.00 \aux{1.00/0.00} & \bf 1.00 & 1.00 \aux{0.99/0.00} & 0.99 \\
        EnCodec & \bf  0.98 \aux{0.98/0.01} & \bf 1.00 & 0.51 \aux{0.52/0.50} & 0.50 \\
        \midrule
        Average & \bf 0.96 \aux{0.98/0.04} & \bf 0.97 & 0.85 \aux{0.85/0.14} & 0.84 \\
        \bottomrule
        \end{tabular}
\end{table}

\paragraph*{Audio editing.}
We evaluate the robustness of the detection on a wide range of audio editing operations: 
time modification (faster, resample), 
filtering (bandpass, highpass, lowpass), 
audio effects (echo, boost audio, duck audio), 
noise (pink noise, random noise),
and compression (MP3, AAC, EnCodec).
In order to show generalization, we chose stronger parameter to the attacks than those used during training (see Sec.~\ref{chap4/sec:training-details}).

\paragraph*{Robustness of the detection.}
Detection is done on 10k ten-seconds audios from our VoxPopuli validation set.
For each edit, we first build a balanced dataset made of the 10k watermarked/ 10k non-watermarked edited audio clips.
We quantify the performance by adjusting the threshold of the detection score, selecting the value that maximizes accuracy (we provide corresponding TPR and FPR at this threshold).
To adapt data-hiding methods (\eg, WavMark) for proactive detection, we embed a binary message (chosen randomly beforehand) in the generated speech before release. The detection score is then computed as the Hamming distance between the original message and the one extracted from the scrutinized audio. 

We observe in Tab.~\ref{chap4/tab:wm_robustness} that AudioSeal is overall more robust, with an average AUC of 0.97 vs. 0.84 for WavMark.
The performance for lowpass and highpass filters indicates that AudioSeal embeds watermarks neither in the low nor in the high frequencies (WavMark focuses on high frequencies).
We give results on more augmentations in Sec.~\ref{chap4/app:robustness}.

\subsection{Localization}

\begin{figure}[b!]
    \centering
    \includegraphics[width=0.48\linewidth, clip, trim={0 1.8in 0 0}, valign=t]{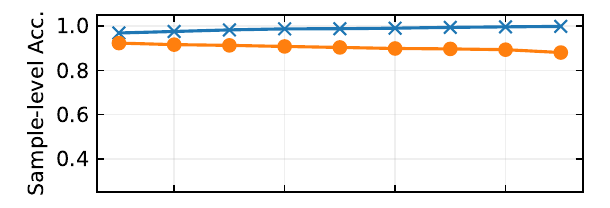}\hfill
    \includegraphics[width=0.48\linewidth, clip, trim={0 0 0 1.3in}, valign=t]{chapter-4/figs/loc_quantitative.pdf}
    \caption{\textbf{Localization results} across different durations of watermarked audio signals in terms of Sample-Level Accuracy and Intersection Over Union (IoU) metrics ($\uparrow$ is better).}
    \label{chap4/fig:loc_quantitative}
\end{figure}

We evaluate localization with the sample-level detection accuracy, \ie, the proportion of correctly labeled samples, and the Intersection over Union (IoU).
The latter is defined as the intersection between the predicted and the ground truth detection masks (1 when watermarked, 0 otherwise), divided by their union.
IoU is a more relevant evaluation of the localization of short watermarks in a longer audio.

This evaluation is carried out on the same audio clips as for detection.
For each one of them, we watermark a randomly placed segment of varying length.
Localization with WavMark is a brute-force detection: a window of 1s slides over the 10s of speech with the default shift value of 0.05s.
The Hammning distance between the 16 pattern bits is used as the detection score.
Whenever a window triggers a positive, we label its 16k samples as watermarked in the detection mask in $\{0,1\}^t$.

\autoref{chap4/fig:loc_quantitative} plots the sample-level accuracy and IoU for different proportions of watermarked speech in the audio clip.
AudioSeal achieves an IoU of 0.99 when just one second of speech is AI-manipulated, compared to WavMark's 0.35.
Moreover, AudioSeal allows for precise detection of minor audio alterations: it can pinpoint AI-generated segments in audio down to the sample level (usually 1/16k sec), while the concurrent WavMark only provides one-second resolution.
This is the reason why it lags behind in terms of IoU more than accuracy.
It is especially relevant for speech samples, where a simple word modification may greatly change meaning.

\subsection{Attribution}

\begin{table}[t!]
    \centering
    \caption{
        Attribution results.
        We report the accuracy of the attribution (Acc.) and false attribution rate (FAR). 
        Detection is done at FPR=$10^{-3}$ and attribution matches the decoded message to one of $N$ versions.
        We report averaged results over the edits of Tab.~\ref{chap4/tab:wm_robustness}.
    }\label{chap4/tab:attribution}
    \footnotesize
        \begin{tabular}{cr *{5}{c}}
            \toprule
            & N & $1$ & $10$ & $10^2$ & $10^3$ & $10^4$ \\ \midrule
    \multirow{2}{*}{FAR (\%) $\downarrow$} & WavMark      & 0.0 & \textbf{0.20} & \textbf{0.98} & \textbf{1.87} & \textbf{4.02} \\
            & AudioSeal   & 0.0 & 2.52 & 6.83 & 8.96 & 11.84 \\ \midrule
    \multirow{2}{*}{\shortstack{Acc. (\%) $\uparrow$}} & WavMark      & 58.4 & 58.2 & 57.4 & 56.6 & 54.4 \\
            & AudioSeal  & \textbf{68.2} & \textbf{65.4} & \textbf{61.4} & \textbf{59.3} & \textbf{56.4} \\ 
            \bottomrule
        \end{tabular}
\end{table}

Given an audio clip, the objective is now to find if any of $N$ versions of our model generated it (detection), and if so, which one (identification). 
For evaluation, we create $N'=100$ random 16-bits messages and use them to watermark 1k audio clips, each consisting of 5 seconds of speech (not 10s to reduce compute needs). 
This results in a total of 100k audios. 
For WavMark, the first 16 bits (/32) are fixed and the detection score is the number of well decoded pattern bits, while the second half of the payload hides the model version.
An audio clip is flagged if the average output of the detector exceeds a threshold, corresponding to FPR=$10^{-3}$.
Next, we calculate the Hamming distance between the decoded watermark and all $N$ original messages. 
The message with the smallest Hamming distance is selected.
It is worth noting that we can simulate $N>N'$ models by adding extra messages. 
This may represent versions that have not generated any sample.

False Attribution Rate (FAR) is the fraction of wrong attribution \emph{among the detected audios} while the attribution accuracy is the proportion of detections followed by a correct attributions \emph{over all audios}. 
AudioSeal has a higher FAR but overall gives a better accuracy, which is what ultimately matters.
First, we observe that the false attribution rate -- which we define as the proportion of audios that are wrongly attributed among the detected ones -- is higher in our case.
On the other hand \autoref{chap4/tab:attribution} highlights that AudioSeal gives better attribution accuracy.
It is defined as the proportion of watermarked samples that are both flagged and correctly attributed and is what ultimately matters.
In summary, decoupling detection and attribution achieves better detection rate and makes the global accuracy better, at the cost of occasional false attributions.

\subsection{Efficiency analysis}
\label{chap4/sec:speed}

\begin{figure}[b!]
    \centering
    \includegraphics[width=0.65\linewidth, clip, trim={0 0 0 0}]{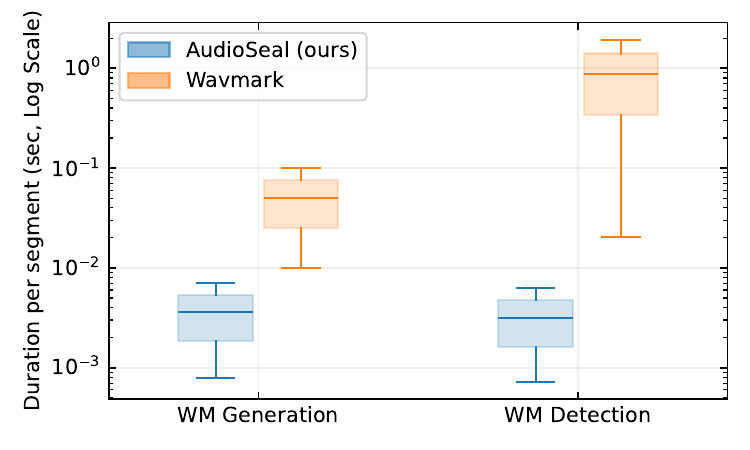}
    \caption{Mean runtime ($\downarrow$ is better) of AudioSeal versus WavMark. 
    AudioSeal is one order of magnitude faster for watermark generation and two orders of magnitude faster for watermark detection for the same audio input.
    }
    \label{chap4/fig:efficiency}
\end{figure}

To highlight the efficiency of AudioSeal, we conduct a performance analysis and compare it with WavMark. 
We apply the watermark generator and detector of both models on a dataset of 500 audio segments ranging in length from 1 to 10 seconds, using a single Nvidia Quadro GP100 GPU. 
The results are displayed in Fig.~\ref{chap4/fig:efficiency} and Tab.~\ref{chap4/tab:speed}.
In terms of generation, AudioSeal is 14$\times$ faster than WavMark. 
For detection, AudioSeal outperforms WavMark with two orders of magnitude faster performance on average, notably 485$\times$ faster in scenarios where there is no watermark (Tab.~\ref{chap4/tab:speed}). 
The speed difference in the case of WavMark between watermarked and non-watermarked audios is explained by the fact that whenever the detector flags a 1-second span as watermarked, it will directly skip to the next 1-second span, while for non-watermarked audios, it will slide the window by 0.05s.
AudioSeal's speed is due to the model's localized watermark design, which bypasses the need for watermark synchronization (recall that WavMark relies on 20 pass forwards for a one-second snippet).
AudioSeal's detector provides detection logits for each input sample directly with only one pass to the detector, significantly enhancing the detection's computational efficiency.
This makes our system highly suitable for real-time and large-scale applications.

\begin{table*}[t!]
    \centering
    \caption{
        Average runtime (ms) per sample of AudioSeal model against WavMark~\citep{chen2023wavmark} method. 
        Our experiments were conducted on a dataset of audio segments spanning 1 second to 10 seconds, using a single Nvidia Quadro GP100 GPU. 
        The results demonstrate significant speed improvements for both watermark generation and detection with and without the presence of a watermark. 
        Notably, for watermark detection, AudioSeal is 485$\times$ faster than WavMark when there is no watermark, because the latter relies on more forward passes when trying to synchronize the watermark. 
    }
    \footnotesize
    \label{chap4/tab:speed}
    \begin{tabular}{llll}
    \toprule
               Model & Watermarked &     \textbf{Detection ms (speedup)} &   \textbf{Generation ms (speedup)} \\
    \midrule
             WavMark &          \multirow{2}{*}{\xmarkg}      & 1710.70 $\pm$ 1314.02 &    -- \\
             AudioSeal (ours) &           &       \textbf{3.25 $\pm$ 1.99} \;\; (\textbf{485$\times$}) &    -- \\
    \midrule
             WavMark &         \multirow{2}{*}{\cmarkg} &    106.21 $\pm$ 66.95 & 104.58 $\pm$ 65.66 \\
    AudioSeal (ours) &          &       \textbf{3.30} $\pm$ \textbf{2.03} \;\; (\textbf{35$\times$}) &    \textbf{7.41} $\pm$ \textbf{4.52} \;\; (\textbf{14} $\times$) \\
    
    \bottomrule 
    \end{tabular}
\end{table*}

%% file: chapter-4/sections/7-attacks.tex
\section{Adversarial watermark removal}
\label{chap4/sec:attacks}

So far, we considered robustness against regular editing that may happen naturally.
We now examine more damaging deliberate attacks, where attackers might either ``forge'' the watermark by adding it to authentic samples (to overwhelm detection systems) or ``remove'' it to avoid detection. 
Our findings suggest that in order to maintain the effectiveness of watermarking against such adversaries, the code for training watermarking models and the awareness that published audios are watermarked can be made public. 
However, the detector's weights should be kept confidential.

We focus on watermark-removal attacks and consider three types of attacks depending on the adversary's knowledge:
\begin{itemize}
    \item \textit{White-box}: 
    the adversary has access to the detector (\eg, because of a leak), and performs a gradient-based adversarial attack against it.
    The optimization objective is to minimize the detector's output.
    \item \textit{Semi black-box}: 
    the adversary does not have access to any weights, but is able to re-train generator/detector pairs with the same architectures on the same dataset.
    They perform the same gradient-based attack as before, but using the new detector as proxy for the original one.
    \item \textit{Black-box}: 
    the adversary does not have any knowledge on the watermarking algorithm being used, but has access to an API that produces watermarked samples, and to negative speech samples from any public dataset.
    They first collect samples and train a classifier to discriminate between watermarked and not-watermarked.
    They attack this classifier as if it were the true detector.
\end{itemize}

\begin{figure}[b!]
    \centering
    \includegraphics[width=0.55\linewidth, clip, trim={0 0.3cm 0 0}]{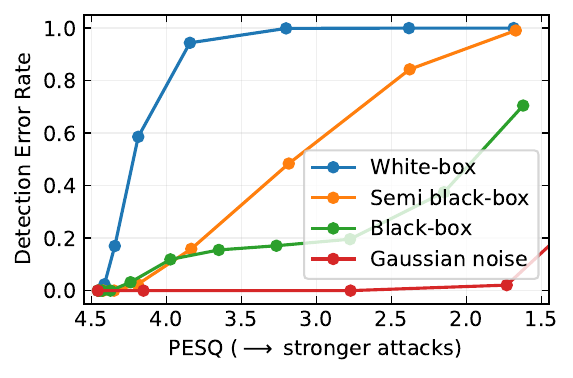}
    \caption{
    \textbf{Watermark-removal attacks.} 
    PESQ is measured between attacked audios and genuine ones (PESQ $<4$ strongly degrades the audio quality).
    The Gaussian noise is used as a reference, but better ``no-box'' attacks are possible (\eg, perceptual autoencoders).
    The more knowledge the attacker has over the watermarking algorithm, the better the attack is.
    }
    \label{chap4/fig:attacks}
\end{figure}

\paragraph{Adversarial attack against a detector.}
Given a watermarked sample $x$ and a detector $D$, we want to find $x' \sim x$ such that $D(x') = 0$.
To that end, we use a gradient-based attack.
It starts by initializing a distortion $\delta_{adv}$ with random gaussian noise.
The algorithm iteratively updates the distortion for a number of steps $n$. 
For each step, the distortion is added to the original audio via $ x' = x + \alpha . \mathrm{tanh} (\delta_{adv})$, passed through the model to get predictions. 
A cross-entropy loss is computed with label 0 (for removal, 1 would be for forging which is not explored here) and back-propagated through the detector to update the distortion, using the Adam optimizer.
At the end of the process, the adversarial audio is $x + \alpha . \mathrm{tanh} (\delta_{adv})$.
In our attack, we use a scaling factor $\alpha=10^{-3}$, a number of steps $n=100$, and a learning rate of $10^{-1}$. 
The $\mathrm{tanh}$ function is used to ensure that the distortion remains small, and gives an upper bound on the SNR of the adversarial audio\footnote{
    This approach is similar to the image optimizations of chapters~\ref{chapter:ssl-watermarking} and~\ref{chapter:active-indexing}; and to the adversarial attacks of Chap.~\ref{chapter:stable-signature}.
}.

\paragraph{Training of the malicious detector.}
For the black-box attack, we want to train and attack a surrogate detector that can distinguish between watermarked and non-watermarked samples, when access to many samples of both types is available.
To train the classifier, we use a dataset made of more than 80k samples of 8 seconds speech from Voicebox~\citep{le2023voicebox} watermarked using our proposed method and a similar amount of genuine (un-watermarked) speech samples. 
The classifier shares the same architecture as AudioSeal's detector. 
The classifier is trained for 200k updates with batches of 64 one-second samples. 
It achieves perfect classification of the samples. 
This is coherent with the findings of Voicebox~\citep{le2023voicebox}.

\paragraph*{Results.}
For every scenario, we watermark 1k samples of 5 seconds, then attack them.
\autoref{chap4/fig:attacks} contrasts various attacks at different intensities, using Gaussian noise as a reference.
The white-box attack is by far the most effective one, increasing the detection error by around 80\%, while maintaining high audio quality (PESQ $>4$).
Other attacks are less effective, requiring significant audio quality degradation to achieve $50\%$ increase the detection error, though they are still more effective than random noise addition.
In summary, the more is disclosed about the watermarking algorithm, the more vulnerable it is. 
The effectiveness of these attacks is limited as long as the detector remains confidential.

%% file: chapter-4/sections/5-ablations.tex
\section{Ablation studies and additional results}\label{chap4/sec:ablations}

\subsection{False positive rates for WavMark}\label{chap4/app:fpr}

\paragraph{Theoretical study.}

When doing detection with multi-bit watermarking, previous works usually extract the message $m'(x)$ from the content $x$ and compare it to the original binary signature $m\in \{ 0,1 \}^{k}$ embedded in the speech sample, as done in Chap.~\ref{chapter:stable-signature}, Sec.~\ref{chap3/subsec:statistical-test}.
The detection test relies on the number of matching bits $M(m,m')$:
\begin{equation} 
    \text{if } M\left(m,m'\right) \geq \tau \,\,\textrm{ where }\,\, \tau\in 
\{0,\ldots,k\},
\end{equation}
then the audio is flagged.
This provides theoretical guarantees over the false positive rates.

Formally, the null hypothesis $\H_0$ is: ``The audio signal is not watermarked'', against the alternative $\H_1$: ``The audio signal is watermarked''.
Under $\H_0$ (\ie, for unmarked audio), if the bits $m'_1, \ldots, m'_k$ are independent and identically distributed Bernoulli random variables with parameter $0.5$, then  $M(m, m')$ follows a binomial distribution with parameters ($k$, $0.5$).
The False Positive Rate (FPR) is defined as the probability that $M(m, m')$ exceeds a given threshold $\tau$. 
A closed-form expression can be given using the regularized incomplete beta function $I_x(a;b)$ (linked to the c.d.f. of the binomial distribution):
\begin{align}\label{chap4/eq:p-value}
    \text{FPR}(\tau) & = \mathbb{P}\left(M \geq \tau \mid \H_0\right) = I_{1/2}(\tau, k - \tau +1).
\end{align}

\begin{figure}[b!]
    \centering
    \includegraphics[width=0.7\linewidth, clip, trim={0.1in 0 0.1in 0}]{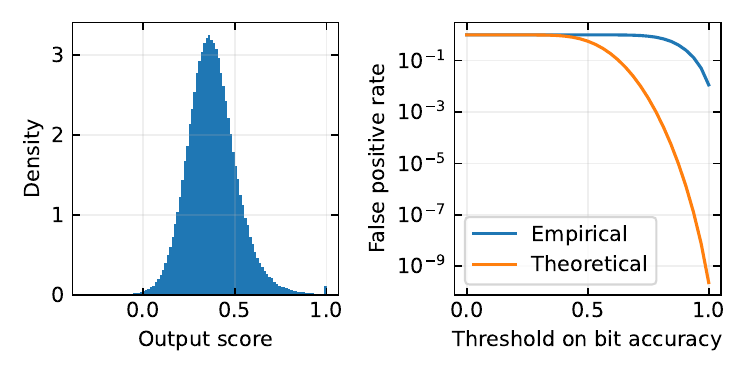}
    \caption{
        (Left) Histogram of scores output by WavMark's extractor on 10k genuine samples. 
        (Right) Empirical and theoretical FPR when the chosen hidden message is all 0.
    }
    \label{chap4/fig:app_fpr_wavmark}
\end{figure}

\paragraph{Empirical study.}
We empirically study the FPR of WavMark-based detection on our validation dataset.
We use the same parameters as in the original paper, \ie, $k=32$-bits are extracted from 1s speech samples.
We first extract the soft bits (before thresholding) from 10k genuine samples and plot the histogram of the scores in Fig.~\ref{chap4/fig:app_fpr_wavmark} (left).
We should observe a Gaussian distribution with mean $0.5$, while empirically the scores are centered around $0.38$. 
This makes the decision heavily biased towards bit 0 on genuine samples.
It is therefore impossible to theoretically set the FPR since this would largely underestimate the actual one.
For instance, \autoref{chap4/fig:app_fpr_wavmark} (right) shows the theoretical and empirical FPR for different values of $\tau$ when the chosen hidden message is full 0.
Put differently, the argument that says that hiding bits allows for theoretical guarantees over the detection rates is not valid in practice.\footnote{
    In Chap.~\ref{chapter:ssl-watermarking} and~\ref{chapter:stable-signature}, we overcome this issue by whitening the outputs of the watermark extractor (see Fig.~\ref{chap1/fig:whitening}).
    This is not possible in this case, since the watermark embedder and extractor operate jointly. 
    It would therefore require at least to retrain or regularize the model to avoid this bias.
}

\subsection{Another architecture}
\label{chap4/app:other-arch}

Our architecture relies on the SOTA compression method EnCodec. 
However, to further validate our approach, we conduct an ablation study using a different architecture DPRNN~\citep{luo2020dual}. 
The results are presented in Tab.~\ref{chap4/tab:dprnn}.
They show that the performance of AudioSeal is consistent across different architectures. 
This indicates that model capacity is not a limiting factor for AudioSeal.

\begin{table}[t!]
    \centering
    \caption{
        Results for different architectures of the generator and detector.
        The IoU is computed for 1s of watermark in 10s audios (corresponding to the leftmost point in Fig.~\ref{chap4/fig:loc_quantitative}).
    }\label{chap4/tab:dprnn}
    \footnotesize
    \begin{tabular}{lccccc}
        \toprule
        Method  & SI-SNR & STOI & PESQ & Acc. & IoU \\
        \midrule
        EnCodec & 26.00 & 0.997 & 4.470 & 1.00 & 0.802 \\
        DPRNN   & 26.7 & 0.996 & 4.421 & 1.00 & 0.796 \\
        \bottomrule
    \end{tabular}
    \end{table}

\subsection{Audio mixing}

We hereby evaluate the scenario where two signals (e.g., vocal and instrumental) are mixed together. 
We use a non-vocal music dataset for the instrumental part, and we normalize and sum the loudness of the watermarked speech and the music segments. 
\autoref{chap4/tab:mixed_signals} shows that the watermark is still detectable in the mixed signal, even when a non-watermarked background music is present, with a slight decrease in performance.

\begin{table}[t!]
    \centering
    \caption{
        Detection results for watermarked speech and music mixed signals.
        \cmarkg\ and \xmarkg\ indicate the presence or absence of the watermark.
    }
    \label{chap4/tab:mixed_signals}
    \footnotesize
    \begin{tabular}{cclll}
    \toprule
    Speech & BG Music & Acc. & \aux{FPR / TPR} & AUC \\
    \midrule
    \cmarkg & \cmarkg & 1.000 & \aux{$3\times 10^{-4}$ / 1.000} & 1.000 \\
    \cmarkg & \xmarkg & 0.979 & \aux{$3.1\times 10^{-2}$ / 0.988} & 0.996 \\
    \bottomrule
    \end{tabular}
\end{table}

\subsection{Out of domain evaluations}\label{chap4/app:ood}

\paragraph{Synthesized speech and audio.} 
\label{chap4/sec:generalization}
We first evaluate how AudioSeal generalizes on AI-generated speech/audio of various domains and languages. 
Specifically, we use the datasets ASVspoof~\citep{liu2023asvspoof} and FakeAVCeleb \citep{khalid2021fakeavceleb}. 
Additionally, we translate speech samples from a subset of the Expresso dataset~\citep{nguyen2023expresso} (studio-quality recordings) using the SeamlessExpressive translation model~\citep{seamless2023}.
We select four target languages: Mandarin Chinese (CMN), French (FR), Italian (IT), and Spanish (SP). 
We also evaluate on non-speech AI-generated audios: music from MusicGen~\citep{copet2023simple} and environmental sounds from AudioGen~\citep{kreuk2023audiogen}.

\begin{table*}[t]
    \caption{
    Evaluation of AudioSeal Generalization across domains and languages. Namely, translations of speech samples from the Expresso dataset~\citep{nguyen2023expresso} to four target languages: Mandarin Chinese (CMN), French (FR), Italian (IT), and Spanish (SP), using the SeamlessExpressive model~\citep{seamless2023}. Music from MusicGen~\citep{copet2023simple} and environmental sounds from AudioGen~\citep{kreuk2023audiogen}. 
    }
    \label{chap4/tab:ood_data}
    \centering
    \footnotesize
        \begin{tabular}{l| *{6}{p{1.0cm}} *{2}{p{1.0cm}} }
        \toprule
        Aug & \rotatebox[origin=c]{45}{Seamless (Cmn)} & \rotatebox[origin=c]{45}{Seamless (Spa)} & \rotatebox[origin=c]{45}{Seamless (Fra)} & \rotatebox[origin=c]{45}{Seamless(Ita)} & \rotatebox[origin=c]{45}{Seamless (Deu)} & \rotatebox[origin=c]{45}{Voicebox (Eng)} & \rotatebox[origin=c]{45}{AudioGen} & \rotatebox[origin=c]{45}{MusicGen}  \\
        \midrule
        None         & 1.00           & 1.00            & 1.00            & 1.00           & 1.00            &   1.00   &   1.00   &   1.00\\
        \midrule
        Bandpass   & 1.00 &   1.00  &   1.00 & 1.00 &   1.00  &  1.00 &   1.00   &   1.00   \\
            Highpass  & 0.71 &  0.68  &   0.70  & 0.70 &  0.70  &  0.64   &  0.52 &  0.52     \\
        Lowpass    & 1.00 &  0.99 &  1.00 & 1.00 & 1.00 &  1.00  &  1.00  &  1.00  \\
            Boost     & 1.00  &  1.00  &  1.00 & 1.00  &  1.00 &  1.00 &     1.00     &      1.00   \\
            Duck      & 1.00  &  1.00  &  1.00 &1.00 &  1.00 &   1.00  &   1.00   &     1.00    \\
            Echo   & 1.00  &  1.00 &  1.00 &1.00 & 1.00 &  1.00  &      1.00    &     1.00     \\
            Pink   & 0.99 &   1.00  & 0.99 & 1.00 &   0.99 &  1.00 &      1.00    &    1.00  \\
            White  & 1.00 &    1.00  & 1.00 & 1.00 &  1.00 & 1.00   &  1.00   &  1.00   \\
        Fast (x1.25)  & 0.97 & 0.98 & 0.99  & 0.98 & 0.99 & 0.98 & 0.87 & 0.87 \\
            Smooth    &  0.96  &  0.99  &   0.99  &    0.99    &      0.99          & 0.99 &  0.98  &    0.98   \\
            Resample  & 1.00 &  1.00 &  1.00 & 1.00 &    1.00  & 1.00 &   1.00    &   1.00   \\
                AAC & 0.99 &  0.99  &  0.99  & 0.99 &  0.99  &   0.97  &  0.99   &     0.98    \\
                MP3 & 0.99 &  0.99   &  0.99 & 0.99 & 0.99  &    0.97  &  0.99    & 1.00  \\
            Encodec   & 0.97 &  0.98   &  0.99 & 0.99 & 0.98 &   0.96     &  0.95    & 0.95   \\
            \midrule
            Average   &  0.97 &  0.97  & 0.98  & 0.98 & 0.98 & 0.97 & 0.95 & 0.95  \\
            \bottomrule
        \end{tabular}
\end{table*}

\begin{table}[t!]
    \centering
    \caption{
        Audio quality and intelligibility evaluations on AI-generated speech from various models and languages.
    }
    \label{chap4/tab:ood_metrics}
    \footnotesize
    \renewcommand{\arraystretch}{1.2}
    \begin{tabular}{cccccc}
        \toprule
        Model & Dataset & SI-SNR & PESQ & STOI & ViSQOL \\
        \midrule
        \multirow{3}{*}{\scriptsize \rotatebox[origin=c]{90}{AudioSeal}} & Seam. (Deu)       & 23.35 & 4.244 & 0.999 & 4.688 \\
        & Seam. (Fr)        & 24.02 & 4.199 & 0.998 & 4.669 \\
        & Voicebox             & 25.23 & 4.449 & 0.998 & 4.800 \\
        \midrule
        \multirow{3}{*}{\scriptsize \rotatebox[origin=c]{90}{WavMark}} & Seam. (Deu)    & 38.93 & 3.982 & 0.999 & 4.515 \\
        & Seam. (Fr)     & 39.06 & 3.959 & 0.999 & 4.506 \\
        & Voicebox          & 39.63 & 4.211 & 0.998 & 4.695 \\
        \bottomrule
    \end{tabular}
\end{table}

We employ the same set of augmentations and observe very similar detection results, as demonstrated in Tab.~\ref{chap4/tab:ood_data}.
Interestingly, even though we did not train our model on AI-generated speech, we sometimes notice a slight improvement in performance compared to our test data. 
No sample is misclassified among the 10k samples that comprise each of our out-of-distribution (OOD) datasets.
We also provide perceptual metrics results on some OOD data in Tab.~\ref{chap4/tab:ood_metrics}.
We observe that AudioSeal performs similarly on these datasets, with a slight decrease in performance compared to our original dataset.
We explain the decrease by the \ploss\ which makes the watermark hidden in the same frequency bands as English speech, which might not be the case for other languages or audio types.

\paragraph{Fake vs. real datasets.}
We also evaluate AudioSeal on three additional datasets containing real human speech: AudioSet~\citep{gemmeke2017audio}, ASVspoof~\citep{liu2023asvspoof}, and FakeAVCeleb~\citep{khalid2021fakeavceleb}, and observe similar performance, see Tab.~\ref{chap4/tab:other_datasets}.

\begin{table}[t!]
    \centering
    \caption{Evaluation of the detection performances on different datasets. AudioSet is an environmental sounds dataset while ASVspoof~\citep{liu2023asvspoof} and FakeAVCeleb~\citep{khalid2021fakeavceleb} are deep-fake detection datasets.}
    \label{chap4/tab:other_datasets}
    \footnotesize
    \begin{tabular}{l *{2}{l}}
        \toprule
        Dataset & Acc. \aux{TPR/FPR} & AUC \\
        \midrule
        Audioset & 0.9992 \aux{0.9996/0.0011} & 1.0 \\
        ASVspoof & 1.0 \aux{1.0/0.0} & 1.0 \\
        FakeAVCeleb & 1.0 \aux{1.0/0.0} & 1.0 \\
        \bottomrule
    \end{tabular}
\end{table}

\begin{figure}[b!]
    \centering
    \includegraphics[width=1.0\textwidth]{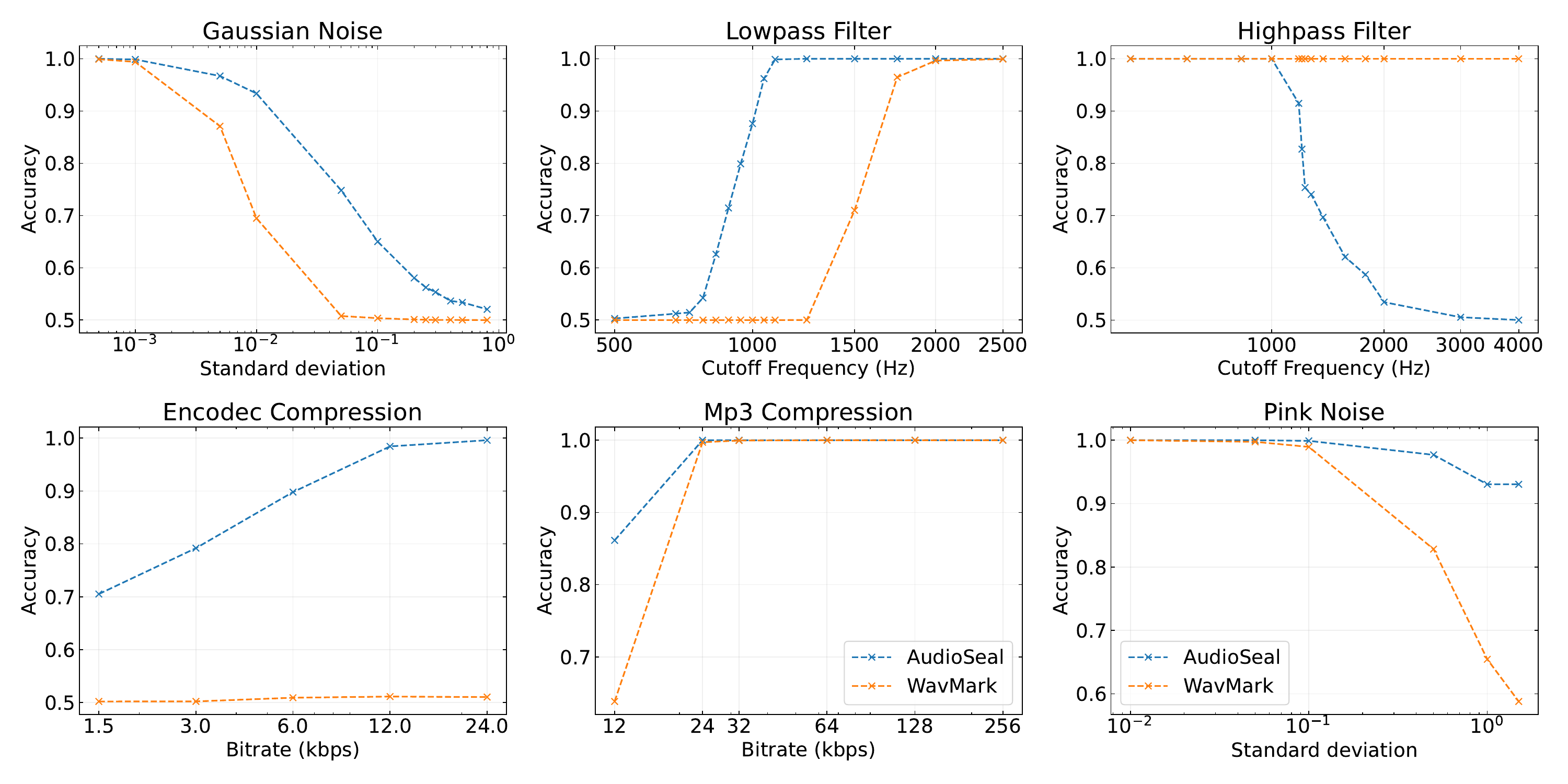}
    \caption{
        Accuracy of the detector on augmented samples with respect to the strength of the augmentation.
    }
    \label{chap4/fig:app_augmentation_curves}
\end{figure}

\subsection{More robustness results}\label{chap4/app:robustness}

We plot the detection accuracy against the strength of multiple augmentations in Fig.~\ref{chap4/fig:app_augmentation_curves}. 
AudioSeal outperforms WavMark for most augmentations at the same strength.
However, for highpass filters above our training range (500Hz) WavMark has a much better detection accuracy.
Our system's TF-loudness loss embeds the watermark where human speech carries the most energy, typically lower frequencies, due to auditory masking. 
This contrasts with WavMark, which places the watermark in higher frequency bands.
Embedding the watermark in lower frequencies is advantageous. 
For example, speech remains audible with a lowpass filter at 1500 Hz, but not with a highpass filter at the same frequency. 
This difference is measurable with PESQ in relation to the original audio, making it more beneficial to be robust against a lowpass filter at a 1500 Hz cut-off than a highpass filter at the same cut-off:

\begin{center}
    \footnotesize
    \begin{tabular}{cccc}
        Filter Type & PESQ & AudioSeal & WavMark \\
        \midrule
        Highpass 1500Hz & 1.85 \xmarkg & 0.7 & 1.0 \\
        Lowpass 1500Hz & 2.93 \cmarkg & 1.0 & 0.7 \\
    \end{tabular}
\end{center}

%% file: chapter-4/sections/8-conclu.tex
\section{Conclusion}

In this chapter we introduce a proactive method for the detection, localization, and attribution of AI-generated speech. 
AudioSeal redesigns audio watermarking to be specific to localized detection rather than data hiding. 
This removes the dependency on slow brute force algorithms, traditionally used to encode and decode audio watermarks.

A key advantage of AudioSeal is its practicality. 
Compared to the previous~\autoref{chapter:stable-signature}, it is post-hoc, meaning it can be applied to any existing audio content without the need to retrain the generator.
It stands as a ready-to-deploy solution for watermarking in voice synthesis APIs (or to authenticate real speech).
This is important for large-scale content provenance on social media and for detecting and eliminating incidents, enabling swift action on instances like the US voters' deepfake case~\citep{murphy2024biden}, long before they spread.
Note that \citet{san2024latent} introduce the in-model counterpart of this chapter.
The authors show how to use AudioSeal to watermark the weights of an audio autoregressive model~\citep{copet2023simple} through a specific watermarking of its training data, although it is less practical and less robust than post-hoc watermarking.

Two key limitations are worth mentioning.
First, the \pval\ of the watermark detection is not mathematically grounded, which makes it hard to scale to extremely low FPRs -- as done in previous and next chapters -- and to provide a theoretical guarantee of the watermark's presence.
Second, the watermark is still brittle to adversarial attacks (\eg, using new neural audio compression algorithms).
Therefore, AudioSeal should be seen as a filtering tool to flag suspicious content, rather than a definitive proof of fakeness  or authenticity.

%% file: chapter-5/main.tex
\chapter{Three Bricks to Consolidate Watermarks for Large Language Models}\label{chapter:three-bricks}

This chapter is based on the paper \fullcite{fernandez2023three}.

Discerning between generated and natural texts is increasingly challenging due to the rapid progress of large language models (LLMs).
In this context, watermarking again emerges as a promising technique for ascribing text to a specific generative model. 
It alters the sampling generation process to leave an invisible trace in the output, facilitating later detection.
This research consolidates watermarks for large language models based on three theoretical and empirical considerations. 
First, we introduce new statistical tests that offer robust theoretical guarantees which remain valid even at low false-positive rates (less than 10$^{\text{-6}}$). 
Second, we compare the effectiveness of watermarks using classical benchmarks in the field of natural language processing, gaining insights into their real-world applicability.
Third, we develop advanced detection schemes for scenarios where access to the LLM is available, as well as multi-bit watermarking. 
Code is available at \url{github.com/facebookresearch/three_bricks}.

\newpage
\input{chapter-5/sections/1-intro.tex}

\input{chapter-5/sections/2-background.tex}

\input{chapter-5/sections/3-stats.tex}

\input{chapter-5/sections/4-expes.tex}

\input{chapter-5/sections/5-extension.tex}
\input{chapter-5/sections/8-conclusion.tex}

%% file: chapter-5/sections/1-intro.tex
\section{Introduction}

The misuse of Large Language Models (LLMs) like ChatGPT~\citep{chatgpt2022}, Clau\-de~\citep{claude}, or the open-sourced Llama~\citep{touvron2023llama} may become a threat as their availability and capabilities expand~\citep{weidinger2022taxonomy, crothers2022machine, cardenuto2023age}.
LLMs might help generate fake news by reducing costs to spread disinformation at scale~\citep{kertysova2018artificial, kreps2022all}, with a potential impact on public opinion and democratic outcomes~\citep{kuvsen2018politics}.
They could help impersonate people, facilitate scams~\citep{ryan2023junk}, or make student assessments impossible.
In this context, it is crucial to enforce fair and responsible usage through regulations and technical means.

On the one hand, similarly to image or speech generation, detecting outputs of LLMs with passive forensics is difficult because generated texts are hardly distinguishable from real ones, be it for humans or algorithms~\citep{ippolito2019automatic, mitchell2023detectgpt}.
On the other hand, watermarking text is more challenging than continuous data like images or audios due to the discrete nature of words/tokens, and no end-to-end solution like the ones presented in Sec.~\ref{chap0/sec:deep learning-watermarking} or in Chap.~\ref{chapter:audioseal} provides convincing results.
Therefore, watermarking for text has been less explored, until the seminal works of~\citet{kirchenbauer2023watermark} and \citet{aaronson2023watermarking} which change the generation process to ascribe text to a specific LLM~\citep{aaronson2023watermarking,kirchenbauer2023watermark,kirchenbauer2023reliability,christ2023undetectable}.
In this case, watermarking either alters the sample generation process~\citep{aaronson2023watermarking,christ2023undetectable} or changes the probability distribution of the generated tokens~\citep{kirchenbauer2023watermark, zhao2023provable}, to leave an imperceptible trace in the generated text.
These works then describe a detection mechanism analyzing the generated tokens to see if their distribution follows the one induced by the watermark. 

We introduce three contributions to consolidate this young literature, one for each of the following paragraphs and sections.
Each part can be read independently:
\begin{itemize}
    \item 
    False positives can have serious consequences in contexts where the integrity of results are essential, such as falsely accusing a student of cheating in an exam.
    However, current approaches focus on the True Positive Rate (TPR) rather than on False Positive Rate (FPR).
    The FPR has never been empirically checked at interesting scales (with more than 1k negative examples).
    Our large-scale experiments reveal that hypotheses of previous works do not hold and that their detection thresholds largely underestimate the false positives at low FPR.
    This work provides grounded statistical tests that theoretically guarantee FPR and accurate \pval\ in real-world regimes.
    We validate them empirically and show that they provide a close-to-perfect control of the FPR, even at low values ($<10^{-6}$).
    \item
    We compare the watermarking methods on traditional Natural Language Processing (NLP) benchmarks. 
    Indeed, current watermark evaluation mainly considers the deviation from the original LLM distribution, for example using perplexity or similarity scores.
    This is in contrast with the LLM literature, where models are rather evaluated on their effective usefulness, \eg, free-form completion tasks such as question answering.
    Such evaluations are much more informative on the actual abilities of the model when used on downstream tasks.
    \item
    We expand these algorithms to advanced detection schemes.
    When access to the LLM is possible at detection time, we provide optimal statistical tests.
    We also investigate multi-bit watermarking (hiding binary messages as watermarks) when current approaches only tackle zero-bit watermarking.
    This allows not only to determine whether the text was generated by the watermarked LLM, but also to identify which version of the model generated it.
\end{itemize}

%% file: chapter-5/sections/2-background.tex
\section{Technical background and notations}\label{chap5/sec:background}

We hereunder re-introduce the concepts of the LLM watermarking techniques used later in the chapter and explained in details in \autoref{chap0/sec:llm_wm}.
More specifically, we focus on the two most popular methods~\citep{aaronson2023watermarking, kirchenbauer2023watermark} that modify the LLM decoding by hashing a watermark window.
See \autoref{chapter:related-work} for a broader overview of the detection and watermarking methods for text.

We consider a decoder-only LLM that takes as input a context $\left( x^{(-C)}, ..., x^{(-1)} \right)\in \V^C$, $\V$ being the vocabulary of the model, and outputs a vector of logits $\logit\in \R^{|\V|}$.
$\logit$ is transformed into the probability distribution $\mathbf{p}=\text{softmax}(\logit / \temperature ) \in [0,1]^{|\V|}$, where $\temperature$ is a temperature.
The text is generated by sampling the next token $x^{(0)}$ from this distribution $\Prob \left( . \mid x^{(-C)},\dots, x^{(-1)} \right)$, then appending it to the context, and repeating the process.

The \emph{watermark embedding} alters the logit vector $\logit$ or the sampling procedure depending on a secret-key $\sk$.
The output of a secret-key cryptographic function hashes $k$ previous tokens $\left(x^{(-k)},\dots, x^{(-1)} \right)$ (the watermark window) and the secret-key $\sk$.
It serves as a seed for a random number generator (RNG), that influences the choice of the next token $x^{(0)}$.
\begin{itemize}
    \item \citet{kirchenbauer2023watermark} use the RNG to partition the vocabulary $\V$ into a greenlist $\G$ and a redlist $\bar{\G}$, where $\G$ contains a proportion $\gamma$ of the vocabulary.
    The logit of each token in the greenlist is incremented by a value $\delta>0$, and the sampling proceeds as usual.
    \item \citet{aaronson2023watermarking} use the RNG to sample a vector $\vec{r}\in[0,1]^{|\V|}$.
    The next token is chosen as $x^{(0)} = \arg \max_{v \in \V } \vec{r}_v^{1/\vec{p}_v}$.
\end{itemize}
For both methods we can trade off generation quality against robustness by varying the watermarking strength.
In~\citep{kirchenbauer2023watermark}, increasing the $\delta$ parameter increases the generation of green tokens at the risk of including unlikely tokens.
In~\citep{aaronson2023watermarking}, increasing the temperature $\temperature$ has the same effect, since it flattens the probability vector, thus diminishing the relative importance of $\vec{p}_v$ over $\vec{r}_v$.

The \emph{watermark scoring} tokenizes the text and, for a text of $T$ tokens, computes a score $s_T$  based on the frequency or characteristics of certain tokens.
\begin{itemize}
    \item For \citet{kirchenbauer2023watermark}, the score is the number of greenlist tokens:
    \begin{equation}
        s_T = \sum_{t=k+1}^T \mathds{1} \left(x^{(t)}\in\G^{(t)}\right),
    \end{equation}\label{chap5/eq:score-kirchenbauer}
    where $x^{(t)}$ and $\G^{(t)}$ are the $t^{\textrm{th}}$ token and its associated partition.
    \item For \citet{aaronson2023watermarking}, the score is:
    \begin{equation}
        s_T=-\sum_{t=k+1}^T \ln \left(1-\vec{r}^{(t)}_{x^{(t)}}\right),
    \end{equation}\label{chap5/eq:score-aaronson}
    where $\vec{r}^{(t)}_{x^{(t)}}$ is the value of the secret vector corresponding to the token $x^{(t)}$.
\end{itemize}
Note that $\G^{(t)}$ and $\vec{r}^{(t)}$ depend on the $k$ tokens that precede $x^{(t)}$ and the secret-key $\sk$.

The \emph{statistical hypothesis test} distinguishes between the hypothesis $\H_0$: ``the text is natural'' and $\H_1$: ``the text has been generated with watermark''.
It is based on a $Z$-test, which compares the observed score $s_T$ against its expected value under the hypothesis $\H_0$ (no watermark).
The $Z$-test is typically used for large sample sizes assuming a normal distribution under the null hypothesis thanks to the central limit theorem.  
The $Z$ statistic is computed as:
\begin{equation}
    Z = \frac{{s_T/T - \mu_0}}{{\sigma_0 / \sqrt{T}}},
    \label{chap5/eq:Z}
\end{equation}
where $\mu_0$ and $\sigma_0$ are the expected mean and standard deviation of the score per token under $\H_0$.
A \pval\ is then calculated to determine the likelihood of observing a score as extreme as $s_T$ under $\H_0$:
\begin{equation}
    \text{p-value}(z) = \Prob(Z \geq z \mid \H_0) = 1 - \Phi(z),
    \label{chap5/eq:pvalue}
\end{equation}
where $\Phi$ is the cumulative distribution function of the normal distribution. 
Texts are flagged as watermarked if the \pval\ is less than a fixed false positive rate (FPR).

%% file: chapter-5/sections/3-stats.tex
\begin{figure}[b!]
    \centering
    \begin{subfigure}{0.8\textwidth}
        \includegraphics[width=\textwidth, trim=0 0 0 0, clip]{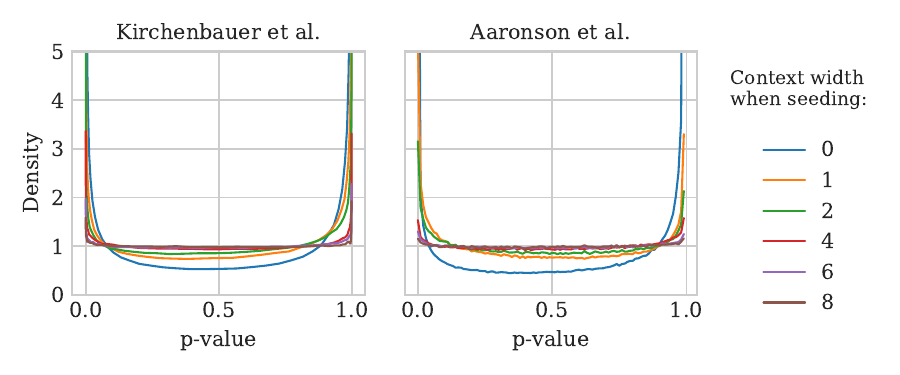}
        \vspace{-0.6cm}
        \caption{Empirical \pval\ distribution without any correction.}
        \label{chap5/fig:pvalue}
    \end{subfigure}
    \\[0.5cm]
    \centering
    \begin{subfigure}{1.0\textwidth}
        \includegraphics[height=1.8in, trim=0.2cm 0 0.3cm 0, clip]{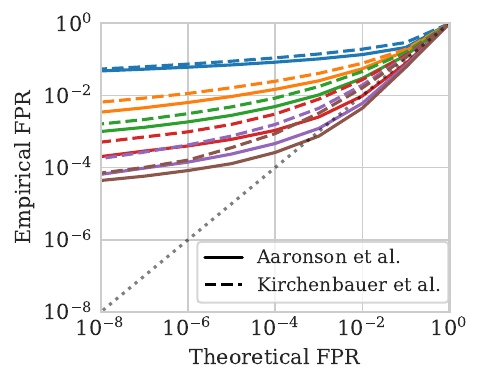}
        \includegraphics[height=1.8in, trim=0.6cm 0 0.3cm 0, clip]{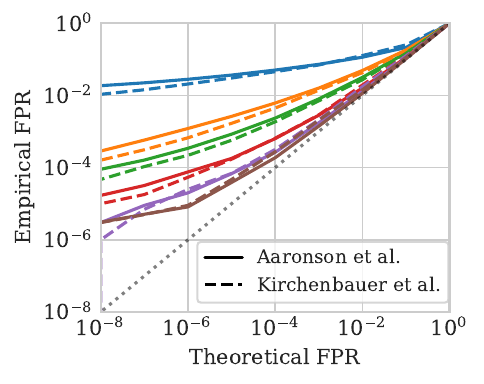}
        \includegraphics[height=1.8in, trim=0.6cm 0 0.3cm 0, clip]{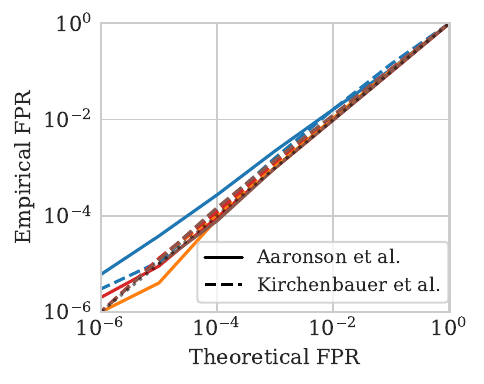}
        \caption{
            Empirical FPRs against theoretical ones.
            (\emph{Left}) using $Z$-tests;
            (\emph{Middle}) using new statistical tests presented in~\ref{chap5/sec:new-stats};
            (\emph{Right}) using the new statistical tests with the rectified scoring strategy of~\ref{chap5/sec:rect}.
        }\label{chap5/fig:fpr}
    \end{subfigure} \\
    \caption{
        Empirical checks of \pval\ and false positive rates for different watermarks and values of the context width $k$.
        Results are computed over $10$ secret keys $\times$ 100k sequences of $256$ tokens sampled from Wikipedia.
        Theoretical values do not hold in practice for $Z$-tests even for high values of $k$: \pval s are not uniformly distributed and empirical FPRs do not match theoretical ones.
        This is solved by basing detection on grounded statistical tests and analytic \pval, as well as by revising the scoring strategy with deduplication.
    }
\end{figure}

\section{Ensuring reliable \pval\ and FPR}\label{chap5/sec:stats}

In this section, large-scale evaluations of the FPR show a gap between theory and practice. 
It is closed with new statistical tests and by rectifying the scoring method.

\subsection{Empirical validation of FPR with Z-scores}\label{chap5/sec:zscore}

So far, the FPR has been checked on only around $500$ negative samples~\citep{kirchenbauer2023watermark,kirchenbauer2023reliability, zhao2023provable}.
We scale this further and select $100$k texts from multilingual Wikipedia to cover the distribution of natural text.
We tokenize with Llama's tokenizer, and take $T=256$ tokens/text.
We run detection tests with varying window length $k$ when seeding the RNG. 
We repeat this with $10$ different secret keys, which makes $1$M detection results under $\H_0$ for each method and $k$ value.
For the detection of the greenlist watermark, we use $\gamma=0.25$.

\autoref{chap5/fig:pvalue} first shows that the distribution of \pval s is not uniform.
his should be the case for a $Z$-test under $\H_0$, which calls into question whether this type of test can be used here.
To further emphasize this point, \autoref{chap5/fig:fpr} compares empirical and theoretical FPRs.
Theoretical guarantees do not hold in practice and empirical FPRs are much higher than the theoretical ones.
Besides, the larger the watermarking context window $k$, the closer we get to theoretical guarantees. 
In pratice, one would need $k>>8$ to get reliable \pval, but this makes the watermarking method less robust to attacks on generated text because it hurts synchronization.

\subsection{New non-asymptotical statistical tests}\label{chap5/sec:new-stats}

The Gaussian assumption of $Z$-tests breaks down for short or repetitive texts.
Here are non-asymptotical tests for both methods that reduce the gap between empirical and theoretical FPR, especially at low FPR values as shown in Fig.~\ref{chap5/fig:fpr}.

\paragraph*{\texorpdfstring{\citep{kirchenbauer2023watermark}}{}} 
Under $\H_0$, we assume that the event $x^{(t)}\in\G^{(t)}$ occurs with probability $\gamma$, and that these events are i.i.d.
Therefore, $S_T$~\eqref{chap5/eq:score-kirchenbauer} is distributed as a binomial of parameters $T$ and $\gamma$. Consider a text under scrutiny whose score equals $s$.
The \pval\ is defined as the probability of obtaining a score higher than $s$ under $\H_0$: %
\begin{equation}
    \text{p-value}(s) = \Prob(S_T \geq s \mid \H_0) = I_{\gamma}(s,T-s+1).
\end{equation}
This comes from the fact that $S\sim\mathcal{B}(T,\gamma)$ whose c.d.f. is expressed by $I_x(a,b)$ the regularized incomplete Beta function.

\paragraph*{\texorpdfstring{\citep{aaronson2023watermarking}}{}} 
Under $\H_0$, we assume that the text under scrutiny and the secret vector are independent, so that $\vec{r}_{x^{(t)}} \overset{i.i.d.}{\sim} \mathcal{U}(0,1)$. 
Therefore, $S_T$~\eqref{chap5/eq:score-aaronson} follows a $\Gamma(T,1)$ distribution.
The \pval\ associated to a score $s$ reads:
\begin{equation}
    \text{p-value}(s) = \Prob(S_T \geq s \mid \H_0) = \frac{\Gamma(T,s)}{\Gamma(T)},
\end{equation}
where $\Gamma$ is the upper incomplete gamma function.
Under $\H_1$, the score is expected to be higher as proven in App.~\ref{chap5/app:aaronson_score}, so the \pval\ is likely to be small.

\subsection{Rectifying the detection scores}\label{chap5/sec:rect}

\begin{figure}[b!]
    \centering
    \begin{tcolorbox}[width=0.7\linewidth, colframe=metablue, colback=white]
        \includegraphics[width=0.99\linewidth, trim=0 110 0 18, clip]{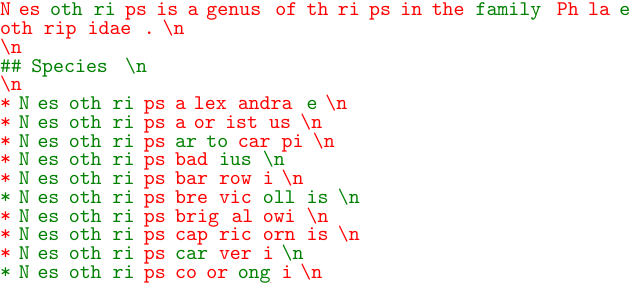}
    \end{tcolorbox}
    \centering
    \caption{
    Typical example of a non-watermarked text with low \pval\ because of repeated tokens (we only show half of the text).
    Here \pval$=10^{-21}$ on $256$ tokens.
    }
    \label{chap5/fig:low-pval-text}
\end{figure}

Even with grounded statistical tests, empirical FPRs are still higher than theoretical ones.
In fact, \citet{kirchenbauer2023watermark} mention that random variables are only pseudo-random since repeated windows generate the same secret. 
This can happen even in a short text and especially in formatted data.
For instance in a bullet list, the sequence of tokens \texttt{$\backslash$n$\backslash$n*\_} repeats a lot as shown in Fig.~\ref{chap5/fig:low-pval-text}.
In this text of $256$ tokens, the \pval\ is $10^{-21}$ for the scheme of~\citet{kirchenbauer2023watermark} and with $\gamma=0.25$ and $k=2$.
Repetition pulls down the assumption of independence necessary for computing the \pval.

We experiment with two simple heuristics mitigating this issue.
The first one takes into account a token only if the watermark context window has not already been seen during the detection.
The second scores the tokens for which the $k+1$-tuple formed by \{watermark context + current token\} has not already been seen.
Note, that the latter is present in~\citep{kirchenbauer2023watermark}, although without ablation and without being used in further experiments.
Of the two, the second one is better since it counts more ngrams, and thus has better TPR. 
It can also deal with the specific case of $k=0$.
\autoref{chap5/fig:fpr} reports empirical and theoretical FPRs when choosing not to score already seen $k+1$-tuples.
They now match perfectly, except for $k=0$ where the FPR is still slightly underestimated.
\emph{In short, we guarantee FPR thanks to new statistical tests and by scoring only tokens for which \{watermark context + current token\} has not been scored.}

%% file: chapter-5/sections/4-expes.tex
\section{Better watermarking evaluations}

\autoref{chap5/sec:robustness-analysis} evaluates the detection with the revised statistical tests and smaller FPRs than the previous literature. 
\autoref{chap5/sec:free-form} shows the impact on NLP benchmarks.

\subsection{Robustness analysis}
\label{chap5/sec:robustness-analysis}

\begin{table}[t!]
    \centering
    \caption{
    Robustness analysis of the watermarks, with rectified statistical tests.
    We report the TPR@FPR=$10^{-5}$ and the S-BERT scores over $10\times 1$k completions, for different hyperparameters controlling the strength of the watermark 
    ($\delta$ in \citep{kirchenbauer2023watermark} and $\temperature$ in \citep{aaronson2023watermarking} - see Sec.~\ref{chap5/sec:background}).
    The `TPR aug.' is the TPR when texts are attacked before detection by randomly replacing tokens with probability 0.3.
    }
    \label{chap5/tab:robustness}
    \footnotesize
    \begin{tabular}{rl *{4}{p{1cm}} @{\hspace{0.5cm}} *{4}{p{1cm}}}
        \toprule
        & & \multicolumn{4}{c}{\citep{aaronson2023watermarking}} &  \multicolumn{4}{c}{\citep{kirchenbauer2023watermark}}  \\
        $k$ & Metric & $\temperature:$ 0.8 & 0.9 & 1.0 & 1.1 & $\delta:$ 1.0 & 2.0 & 3.0 & 4.0 \\
        \cmidrule(rr){3-6} \cmidrule(rr){7-10}
        \multirow{3}{*}{$0$} 
            & S-BERT    & 0.60 & 0.56 & 0.52 & 0.44 & 0.63 & 0.61 & 0.57 & 0.50 \\
            & TPR       & 0.20 & 0.31 & 0.42 & 0.51 & 0.00 & 0.16 & 0.58 & 0.70 \\
            & TPR aug.  & 0.04 & 0.06 & 0.09 & 0.10 & 0.00 & 0.02 & 0.20 & 0.39 \\[4pt]
        \multirow{3}{*}{$1$} 
            & S-BERT    & 0.62 & 0.61 & 0.59 & 0.55 & 0.63 & 0.62 & 0.60 & 0.56 \\
            & TPR       & 0.35 & 0.51 & 0.66 & 0.77 & 0.02 & 0.41 & 0.77 & 0.88 \\
            & TPR aug.  & 0.04 & 0.10 & 0.20 & 0.36 & 0.00 & 0.05 & 0.30 & 0.58 \\[4pt]
        \multirow{3}{*}{$4$} 
            & S-BERT    & 0.62 & 0.62 & 0.61 & 0.59 & 0.62 & 0.62 & 0.60 & 0.57 \\
            & TPR       & 0.43 & 0.59 & 0.71 & 0.80 & 0.02 & 0.44 & 0.76 & 0.88 \\
            & TPR aug.  & 0.01 & 0.02 & 0.06 & 0.18 & 0.00 & 0.00 & 0.03 & 0.14 \\
        \bottomrule
    \end{tabular}
\end{table}

We now compare watermarking methods by analyzing the TPR when detecting watermarked texts.
For detection, we employ the previous statistical tests and scoring strategy.
They enable precise control over the FPR and therefore to operate at operating-points not yet seen in the literature.
More specifically we flag a text as watermarked if its \pval\ is lower than $10^{-5}$ ensuring an FPR=$10^{-5}$.
We prompt Guanaco-7-B~\citep{dettmers2023qlora}, an instruction fine-tuned version of Llama, with the first $1$k prompts from the Alpaca dataset~\citep{alpaca}.
For generation, we use top-$p$ sampling with $p=0.95$, and in the case of \citep{kirchenbauer2023watermark} a temperature $\theta =0.8$ and $\gamma=1/4$.
We simulate synonym attacks by randomly replacing tokens with probability $0.3$ (other attacks are studied in related work~\citep{kirchenbauer2023reliability}).

\autoref{chap5/tab:robustness} reports the TPR for different strength of the watermark (see Sec.~\ref{chap5/sec:background}), and the S-BERT~\citep{reimers2019sentence} similarity score between the generated texts with and without watermarking to measure the semantic distortion induced by the watermark. 
Results reveals different behaviors.
For instance, \citep{kirchenbauer2023watermark} has a finer control over the trade-off between watermark strength and quality.
Its TPR values ranges from 0.0 to 0.9, while \citep{aaronson2023watermarking} is more consistent but fails to achieve TPR higher than 0.8 even when the S-BERT score is degraded a lot.

The watermark context width also has a big influence. 
When $k$ is low, we observe that repetitions happen more often because the generation is easily biased towards certain repetitions of tokens.
It leads to average S-BERT scores below 0.5 and unusable completions.
On the other hand, low $k$ also makes the watermark more robust, especially for \citep{kirchenbauer2023watermark}.
It is also important to note that $k$ has an influence on the number of analyzed tokens since we only score tokens for which the $k+1$-tuple has not been seen before (see Sec.~\ref{chap5/sec:rect}).
If $k$ is high, almost all these tuples are new, while if $k$ is low, the chance of repeated tuples increases.
For instance in our case, the average number of scored tokens is around 100 for $k=0$, and 150 for $k=1$ and $k=4$.

\subsection{Impact on free-form generation tasks}
\label{chap5/sec:free-form}
Previous studies measure the impact on quality using distortion metrics such as perplexity or similarity score as done in Tab.~\ref{chap5/tab:robustness}.
However, such metrics are not informative of the utility of the model for downstream tasks~\citep{holtzman2019curious}, where the real interest of LLMs lies. 
Indeed, watermarking LLMs could be harmful for tasks that require very precise answers, like code or maths.
This section rather quantifies the impact on typical NLP benchmarks, in order to assess the practicality of watermarking.

LLMs are typically evaluated by comparing samples of plain generation to target references (free-form generation) or by comparing the likelihood of predefined options in a multiple choice question fashion. 
The latter makes little sense in the case of watermarking, which only affects sampling.
We therefore limit our evaluations to free-form generation tasks.
We use the evaluation setup of Llama:
1) Closed-book Question Answering (Natural Questions~\citep{kwiatkowski2019natural}, TriviaQA~\citep{joshi2017triviaqa}): we report the $5$-shot exact match performance;
2) Mathematical reasoning (MathQA~\citep{hendrycks2021measuring}, GSM8k~\citep{cobbe2021training}), we report exact match performance without majority voting;
3) Code generation (HumanEval~\citep{chen2021Evaluating}, MBPP~\citep{austin2021program}), we report the pass@1 scores.
For \citep{kirchenbauer2023watermark}, we shift logits with $\delta=1.0$ before greedy decoding.
For \citep{aaronson2023watermarking}, we use $\theta = 0.8$, apply top-p at $0.95$ to the probability vector, then apply the watermarked sampling.

\autoref{chap5/tab:bench-full} reports the performance of Llama models on the aforementioned benchmarks, with and without the watermark and for different window size $k$. 
The performance of the LLM is not significantly affected by watermarking. 
The approach of \cite{kirchenbauer2023watermark} is slightly more harmful than the one of \cite{aaronson2023watermarking}, but the difference w.r.t. the vanilla model is small.
Interestingly, this difference decreases as the size of the model increases: models with higher generation capabilities are less affected by watermarking. A possible explanation is that the global distribution of the larger models is better and thus more robust to small perturbations.
Overall, evaluating on downstream tasks points out that watermarking may introduce factual errors that are not well captured by perplexity or similarity scores.

\begin{table}[H]
    \centering
    \caption{ 
        Performances on free-form generation benchmarks when completion is done with watermarking.
        $k$ is the watermark context width. 
        We report results for methods: [AK]~\citep{aaronson2023watermarking} / [KGW]~\citep{kirchenbauer2023watermark}.
        ``-'' means no watermarking. 
    }
    \label{chap5/tab:bench-full}
    \resizebox{0.95\textwidth}{!}{
    \begin{tabular}{lllrrrrrrr}
    \toprule
     &  &  & GSM8K & Human Eval & MathQA & MBPP & NQ & TQA & Average \\
    Model & Method & $k$ &  &  &  &  &  &  &  \\
    \midrule
     \multirow[t]{15}{*}{7-B} 
     & None & - & 10.31 & 12.80 & 2.96 & 18.00 & 21.72 & 56.89 & 20.45 \\
     \cmidrule{2-10} 
     & [AK] & 0 & 10.54 & 12.80 & 3.00 & 18.00 & 21.77 & 56.88 & 20.50 \\
      &  & 1 & 10.31 & 12.80 & 2.88 & 18.20 & 21.75 & 56.87 & 20.47 \\
      &  & 2 & 10.31 & 12.80 & 2.94 & 18.00 & 21.75 & 56.86 & 20.44 \\
      &  & 4 & 10.39 & 12.80 & 2.98 & 17.80 & 21.80 & 56.88 & 20.44 \\
      &  & 8 & 10.46 & 12.80 & 2.90 & 18.20 & 21.75 & 56.85 & 20.49 \\
      \cmidrule{2-10} 
      & [KGW] & 0 & 9.63 & 12.80 & 2.20 & 16.20 & 20.06 & 55.09 & 19.33 \\
      &  & 1 & 11.14 & 9.76 & 2.82 & 16.00 & 19.50 & 55.30 & 19.09 \\
      &  & 2 & 11.07 & 6.71 & 2.62 & 16.00 & 20.44 & 55.07 & 18.65 \\
      &  & 4 & 10.77 & 9.15 & 2.76 & 16.40 & 20.17 & 55.14 & 19.06 \\
      &  & 8 & 11.37 & 11.59 & 2.90 & 16.40 & 20.66 & 55.36 & 19.71 \\
    \midrule
    \multirow[t]{15}{*}{13-B} 
    & None & - & 17.21 & 15.24 & 4.30 & 23.00 & 28.17 & 63.60 & 25.25 \\
    \cmidrule{2-10} 
    & [AK] & 0 & 17.29 & 15.24 & 4.24 & 22.80 & 28.17 & 63.60 & 25.22 \\
     &  & 1 & 17.21 & 15.24 & 4.30 & 22.80 & 28.20 & 63.61 & 25.23 \\
     &  & 2 & 17.51 & 15.24 & 4.20 & 22.80 & 28.20 & 63.59 & 25.26 \\
     &  & 4 & 17.21 & 15.24 & 4.20 & 22.60 & 28.20 & 63.63 & 25.18 \\
     &  & 8 & 17.21 & 15.24 & 4.22 & 22.80 & 28.20 & 63.62 & 25.22 \\
     \cmidrule{2-10} 
     & [KGW] & 0 & 14.33 & 14.02 & 3.04 & 20.80 & 24.32 & 62.13 & 23.11 \\
     &  & 1 & 17.29 & 14.63 & 3.62 & 21.20 & 25.12 & 62.23 & 24.02 \\
     &  & 2 & 16.45 & 11.59 & 3.54 & 20.60 & 25.54 & 62.44 & 23.36 \\
     &  & 4 & 16.76 & 15.85 & 4.08 & 21.20 & 24.49 & 62.24 & 24.10 \\
     &  & 8 & 17.29 & 14.63 & 3.68 & 21.00 & 25.46 & 62.17 & 24.04 \\
     \midrule
    \multirow[t]{14}{*}{30-B} 
     & None & - & 35.10 & 20.12 & 6.80 & 29.80 & 33.55 & 70.00 & 32.56 \\
     \cmidrule{2-10} 
     & [AK] & 0 & 35.48 & 20.12 & 6.88 & 29.80 & 33.52 & 69.98 & 32.63 \\
     & & 1 & 35.33 & 20.73 & 6.88 & 29.60 & 33.52 & 70.03 & 32.68 \\
     & & 2 & 35.33 & 20.73 & 6.94 & 30.00 & 33.49 & 70.00 & 32.75 \\
     & & 4 & 35.10 & 20.12 & 6.90 & 29.80 & 33.49 & 70.01 & 32.57 \\
     & & 8 & 35.33 & 20.73 & 6.94 & 30.00 & 33.52 & 70.01 & 32.75 \\
     \cmidrule{2-10} 
     & [KGW] & 0 & 31.84 & 21.95 & 6.88 & 28.40 & 31.66 & 69.03 & 31.63 \\
     &  & 1 & 35.56 & 20.73 & 7.54 & 28.80 & 31.58 & 68.98 & 32.20 \\
     &  & 2 & 33.21 & 17.07 & 6.48 & 27.40 & 31.83 & 69.44 & 30.91 \\
     &  & 4 & 34.12 & 22.56 & 6.96 & 28.80 & 31.55 & 68.74 & 32.12 \\
     &  & 8 & 34.95 & 20.12 & 7.42 & 27.20 & 32.08 & 69.31 & 31.85 \\
    \bottomrule
    \end{tabular}
    }
\end{table}

%% file: chapter-5/sections/5-extension.tex
\section{Advanced detection schemes}
This section introduces improvements to the detection schemes of Sec.~\ref{chap5/sec:stats}.
Namely, it develops a statistical test when access to the LLM is granted, as well as multi-bit decoding.

\subsection{Neyman-Pearson and simplified score function} 
The following is specific for the scheme of~\citet{aaronson2023watermarking} -- a similar work may be conducted with the one of~\citet{kirchenbauer2023reliability}.
Under $\H_0$, we have $\vec{r}_v\sim\mathcal{U}_{[0,1]}$, whereas $\vec{r}_v\sim Beta(1/p_v,1)$ under $\H_1$ (see Corollary~\eqref{chap5/eq:Coro} in App.~\ref{chap5/app:aaronson_prob}). 
The optimal Neyman-Pearson score function is thus:
\begin{equation*}
    s_T = \sum_{t=1}^{T} \ln\frac{f_{\H_1}(\vec{r}_{x^{(t)}})}{f_{\H_0}(\vec{r}_{x^{(t)}})} = \sum_{t=1}^T \left(\frac{1}{\vec{p}_{x^{(t)}}}-1\right)\ln(\vec{r}_{x^{(t)}})+A
\end{equation*}
where $A$ is a constant that does not depend on $\vec{r}$ and can thus be discarded. 
There are two drawbacks: (1) detection needs the LLM to compute $\vec{p}_{x^{(t)}}$, (2) there is no close-form formula for the p-value.  

This last point may be fixed by resorting to a Chernoff bound, yet without guarantee on its tightness:
$\text{p-value}(s) \leq e^{\sum_t \ln\frac{\lambda_t}{\lambda_t + c} -cs}$,
with $c$ solution of $\sum_t (c+\lambda_t)^{-1}=-s$ and $\lambda_t = p_{x^{(t)}} / (1-p_{x^{(t)}})$.
Experiments show that this detection yields extremely low \pval\ for watermarked text, but they are fragile: any attack increases them to the level of the original detection scheme~\eqref{chap5/eq:score-aaronson}, or even higher because generated logits are sensitive to the overall LLM context. 
An alternative is to remove weighting:
\begin{equation}
 s_T = \sum_{t=1}^T \ln\left(\vec{r}_{x^{(t)}}\right),
 \label{chap5/eq:Detection2}
\end{equation}
whose \pval\ is given by: $\text{p-value}(s) = \frac{\gamma(T,-s)}{\Gamma(T)}$.
In our experiments, this score function does not match the original detection presented in~\citep{aaronson2023watermarking}.

\subsection{Multi-bit watermarking}

\paragraph*{Theory.} It is rather easy to turn a zero-bit watermarking scheme into multi-bit watermarking, by associating a secret key per message. 
The decoding runs detection with every key and the decoded message is the one associated to the key giving the lowest \pval\ $p$. 
The global \pval\ becomes $1-(1-p)^M$, where $M$ is the number of possible messages.

Running detection for $M$ keys is costly, since it requires $M$ generations of the secret vector.
This is solved by imposing that the secret vectors of the messages $m\in\{0,\ldots,M-1\}$ are crafted as circular shifts of $m$ indices of $\vec{r}=\vec{r}(0)$:
\begin{align*}
\vec{r}(m) &= \mathsf{CyclicShift}(\vec{r},m) \\
    &= \left( \vec{r}_m, \vec{r}_{m+1}, ..,\vec{r}_{d}, \vec{r}_{0}, ..,  \vec{r}_{m-1}  \right).
\end{align*}
Generating $\vec{r}$ as a $d$-dimensional vector, with $d\geq|\V|$, we are able to embed $M\leq d$ different messages, by keeping only the first $|\V|$ dimensions of each circularly-shifted vector. 
Thus, the number of messages may exceed the size of the token vocabulary $|\V|$.
One way is to choose $d =  \textrm{max}(M, |\V|)$.

Besides, the scoring functions~\eqref{chap5/eq:score-kirchenbauer}~\eqref{chap5/eq:score-aaronson}
may be rewritten as:
\begin{equation}
s_T(m) = \sum_{t=1}^T f\left(\vec{r}^{(t)}(m)\right)_{x^{(t)}}  ,
\end{equation}
where $f: \R^d \mapsto \R^d$ is a component-wise function, and $x^{(t)}$ is the selected token during detection. 
This represents the selection of $f\left(\vec{r}^{(t)}(m)\right)$ at position $x^{(t)}$.
From another point of view, if we shift $f\left(\vec{r}^{(t)}\right)$ by $x^{(t)}$, the score for $m=0$ would be its first component, $m=1$ its second one, etc.
We may also write:
\begin{equation}
\vec{S}_T = \sum_{t=1}^T \mathsf{CyclicShift}\left( f\left(\vec{r}^{(t)}\right), x^{(t)} \right) ,
\label{chap5/eq:DetectionMultibit}
\end{equation}
and the first $M$ components of $\vec{S}_T$ are the scores for each $m$.
As a side note, this is a particular case of the parallel computations introduced by~\citet{JAWS}.
The simplified algorithms are given in Alg.~\ref{chap5/alg:multi-bit-gen} and Alg.~\ref{chap5/alg:multi-bit-dec}. 

\noindent
\begin{minipage}{0.48\textwidth}
    \begin{algorithm}[H] \small
    \caption{Generation (one step)}
    \label{chap5/alg:multi-bit-gen}
    \begin{algorithmic}
        \State\hspace*{-0.3cm} \textbf{Requires}: {LLM, dimension $d$, watermark window $k$, message $m\in\{0,\ldots,M-1\}$} \\
        \State logits $\vec{\boldsymbol\ell} \gets \text{LLM} \left( x^{(-C)},\dots, x^{(-1)} \right)$
        \State seed $\gets \mathsf{Hash}(x^{(-k)},\dots, x^{(-1)})$
        \State $\vec{r} \gets \mathsf{RNG_{seed}}(d)$
        \State $\vec{r}(m) \gets \mathsf{CyclicShift}(\vec{r},m)$
        \State $x^{(0)} \gets \mathsf{Sample}(\vec{\boldsymbol\ell},\vec{r}(m)_{1,\dots,|\V|})$
    \end{algorithmic}
\end{algorithm}
\end{minipage}\hfill
\begin{minipage}{0.48\textwidth}
    \begin{algorithm}[H] \small
    \caption{Decoding/identification}
    \label{chap5/alg:multi-bit-dec}
    \begin{algorithmic}
        \State $\vec{S} \gets \vec{0}_d$
        \State{\textbf{for} $t \in \{ k+1, \dots, T\}$:}
            \State \quad seed $\gets \mathsf{Hash}(x^{(t-k)},\dots, x^{(t-1)})$
            \State \quad $\vec{r}^{(t)} \gets \mathsf{RNG_{seed}}(d)$
            \State \quad $\vec{S} \gets \vec{S} +  \mathsf{CyclicShift}(f(\vec{r}^{(t)}),x^{(t)})$
        \State $\vec{p} \gets \textrm{p-value}(\vec{S}_{1,\dots,M})$
        \State $m \gets \textrm{argmin}({\vec{p}}) $
        \State $p \gets 1 - (1 - \vec{p}_m)^M$
    \end{algorithmic}
    \end{algorithm}
\end{minipage}

\paragraph*{Experiments.} 
In a tracing scenario the message is the identifier of a user or a version of the model.
The goal is to decide if any user or model generated a given text (detection) and if so, which one (identification).
There are 3 types of error: \emph{false positive}: flag a vanilla text; \emph{false negative}: miss a watermarked text; \emph{false accusation}: flag a watermarked text but select the wrong identifier.

\begin{table}[t]
    \caption{Identification accuracy for tracing users by watermarking. 
    Sequences are between $4$ and $252$ tokens long, and $149$ on average.
    }
    \label{chap5/tab:identification}
    \centering
    \footnotesize
    \begin{tabular}{cl cccc}
        \toprule
        & & \multicolumn{4}{c}{Number of users $M$} \\
        \cmidrule{3-6}
        & Method & $10$ & $10^2$ & $10^3$ & $10^4$ \\ \midrule
        \multirow{2}{*}{FPR$=10^{-3}$} & [AK] \citep{aaronson2023watermarking}      & 0.80 & 0.72 & 0.67 & 0.62 \\
        & [KGW] \citep{kirchenbauer2023watermark}  & 0.84 & 0.77 & 0.73 & 0.68 \\ \midrule
        \multirow{2}{*}{FPR$=10^{-6}$} & [AK] \citep{aaronson2023watermarking}	      & 0.61 & 0.56 & 0.51 & 0.46 \\
        & [KGW] \citep{kirchenbauer2023watermark} 	                              & 0.69 & 0.64 & 0.59 & 0.55 \\
        \bottomrule
    \end{tabular}
\end{table}

We simulate $M'$=$1000$ users that generate $100$ watermarked texts each, using the Guanaco-7B model. 
Accuracy can then be extrapolated beyond the $M'$ identifiers by adding identifiers with no associated text, for a total of $M>M'$ users.
Text generation uses nucleus sampling with top-p at $0.95$.
For~\citep{kirchenbauer2023watermark}, we use $\delta=3.0$, $\gamma=1/4$ with temperature $\theta$ at $0.8$.
For~\citep{aaronson2023watermarking}, we use $\theta = 1.0$.
For both, the context width is $k=4$.
A text is deemed watermarked if the score is above a threshold set for a given \emph{global} FPR (see~\ref{chap5/sec:stats}).
Then, the source is identified as the user with the lowest p-value.

\autoref{chap5/tab:identification} shows that watermarking performance for identification is dissuasive enough. 
For example, among $10^5$ users, we successfully identify the source of a watermarked text 50\% of the time while maintaining an FPR of $10^{-6}$ (as long as the text is not attacked).
At this scale, the false accusation rate is zero (no wrong identification once we flag a generated text) because the threshold is set high to avoid FPs, making false accusations unlikely. 
The identification accuracy decreases when $M$ increases, because the threshold required to avoid FPs gets higher.
In a nutshell, by giving the possibility to encode several messages, we trade some accuracy of detection against the ability to identify users.

%% file: chapter-5/sections/8-conclusion.tex
\section{Conclusion}

This chapter offers theoretical and empirical insights that were kept aside from the literature on watermarks for LLMs.
Namely, existing methods resort to statistical tests which are biased, delivering incorrect false positive rates.
This is fixed with grounded statistical tests and a revised scoring strategy.
We additionally introduce evaluation setups, and detection schemes to consolidate watermarks for LLMs.

Overall, LLM watermarking seems to be both reliable and practical. 
It already holds many promises as a technique for identifying and tracing LLM outputs, while being relatively new in the context of language models. 
However open questions remain, such as how to adapt watermarks for more complex sampling schemes -- \eg, beam search as in \citep{kirchenbauer2023watermark} -- since generation yields significantly better quality with these methods; how to adapt watermarking to more complex tasks like math or code generation, where the outputs are structured and verifiable; and how to make watermarking more robust to adversarial attacks, which are still a major threat to the integrity of the watermarking process; etc.

%% file: chapter-6/main.tex
\chapter{Watermarking Makes Language Models Radioactive}\label{chapter:radioactive}

This chapter is based on the paper \fullcite{sander2024watermarking}.

It is challenging for AI companies to ensure that their models -- which cost thousands if not millions -- are used in compliance with licensing agreements and prevent theft, because their complexity and lack of transparency make it difficult to track their usage. 
In this chapter, we investigate the \emph{radioactivity} of text generated by large language models (LLM), \ie,  whether it is possible to detect that such synthetic input was used as training data.
Current methods like membership inference or active IP protection either work only in confined settings (\eg, where the suspected text is known) or do not provide reliable statistical guarantees.
We discover that, on the contrary, LLM watermarking as presented in the previous Chap.~\ref{chapter:three-bricks} allows for reliable identification of whether the outputs of a watermarked LLM were used to fine-tune another language model.
Our new methods, specialized for radioactivity, detects with confidence weak residuals of the watermark signal in the fine-tuned LLM.
We link the radioactivity contamination level to the following properties: the watermark robustness, its proportion in the training set, and the fine-tuning process.
We notably demonstrate that training on watermarked synthetic instructions can be detected with high confidence (\pval\ $< 10^{-5}$) even when as little as $5\%$ of training text is watermarked.
Code is available at \url{github.com/facebookresearch/radioactive-watermark}.

\newpage

\input{chapter-6/sections/1-intro.tex}

\input{chapter-6/sections/2-background.tex}

\input{chapter-6/sections/3-radioactivity.tex}

\input{chapter-6/sections/4-experiments.tex}

\input{chapter-6/sections/5-analyzes.tex}

\input{chapter-6/sections/7-approaches.tex}
\input{chapter-6/sections/6-additional.tex}

\input{chapter-6/sections/8-conclusion.tex}

%% file: chapter-6/sections/1-intro.tex
\begin{figure*}[t]
    \centering
    \includegraphics[width=1.0\textwidth, clip, trim=0 0.5cm 0 0]{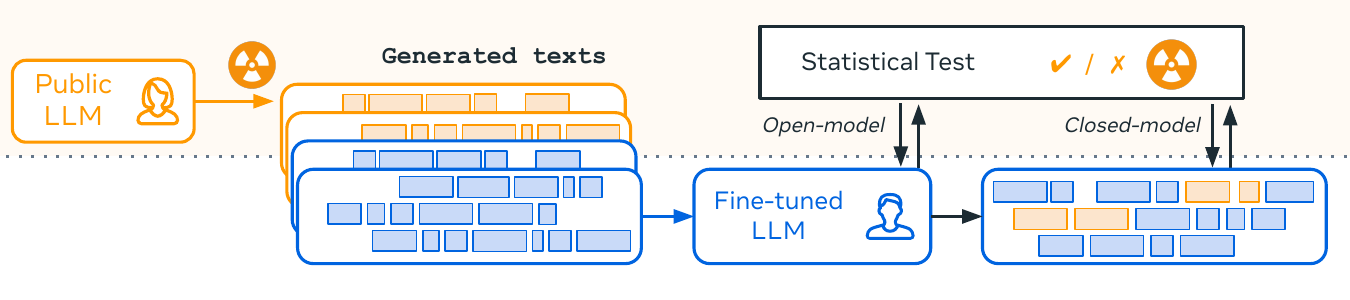}
    \captionsetup{font=small}
    \caption{
    Bob fine-tunes his LLM on data with a small proportion coming from Alice's LLM.
    This leaves traces in Bob's model that Alice can detect reliably when her texts are watermarked.
    Thus, a side effect of Alice's watermark, intended for machine-generated text detection, is to reveal what data Bob's model was fine-tuned on.
    }
    \label{chap6/fig:fig1}
\end{figure*}

\section{Introduction}

Large Language Models (LLMs) are often instruction fine-tuned to align them with human prompts and improve their performance and generalization~\citep{ouyang2022training, wei2022finetuned, chung2022scaling}.
Fine-tuning requires expert knowledge to balance diversity and quality in the instruction dataset and a costly collection of manual annotations, especially for alignment~\citep{openai2023gpt, touvron2023llama2, team2023gemini}.
To address the cost and the difficulties of fine-tuning, practitioners often train on synthetic data generated by a model that has already been instructed, such as Bard, ChatGPT, or Claude. 
For example, works by \citet{wang2022self, honovich2022unnatural, peng2023instruction} created instruction data for many of the first wave of instruction fine-tuned LLMs~\citep{alpaca, xu2023baize, gunasekar2023textbooks, mukherjee2023orca}. 
This may also be unintentional when, for example, Turkers use ChatGPT to perform their tasks~\citep{veselovsky2023artificial}.
Such imitation raises questions about whether the fine-tuned model is a derivative work of the original model~\citep{wallace2020imitation}. 
It may also be seen as theft since the training data is extracted from the model without consent. 
For instance, OpenAI, Google, Meta, and Anthropic prohibit the use of their generated output to train other AI models under their terms-of-service.
In this context, it is crucial to understand how to detect when LLM outputs are used as training data.

Meanwhile, recent AI regulations enforce the transparency of generative models.
This is increasingly important in cases where the generated content may be used for malicious purposes~\citep{weidinger2022taxonomy, crothers2022machine}.
One approach to enhance transparency is to watermark the generated content, as presented by~\citet{aaronson2023watermarking,kirchenbauer2023watermark, kirchenbauer2023reliability} and in Chap.~\ref{chapter:three-bricks}.
It embeds a secret trace in the synthetic content that can be detected to identify the generating model.
In the context of LLMs, recent techniques make detection efficient with minimal degradation of the generated text quality by altering the sampling of next tokens.

\noindent
Based on these two observations, this study addresses the following question:
\begin{center}
\textit{What occurs when watermarked texts are employed as fine-tuning data?}
\end{center}
We explore the potential ``radioactivity'' -- a term coined by~\citet{sablayrolles2020radioactive} -- of LLM watermarking, which refers to the capacity of watermarked training data to contaminate a model.

We examine a model that has been fine-tuned on a corpus that may contain watermarked texts (see Fig.~\ref{chap6/fig:fig1}).
The baseline method for detecting radioactivity executes the original watermark detection on the outputs generated by this model. 
However, this approach proves ineffective because the residual of the watermark is a weak signal hardly detectable in plain output text.
In this work, we are able to demonstrate that LLM watermarking is indeed radioactive thanks to our specific protocol designed for revealing weak contamination traces.

The contributions of the chapter are as follows:
\begin{itemize}
    \item We design radioactivity detection methods for four scenarios based on model (\textit{open} / \textit{closed}) and training data (\textit{supervised} / \textit{unsupervised}) access. 
    Notably, our open-model detection (Fig.~\ref{chap6/fig:open_model}) improves the performance by orders of magnitudes.
\item 
    We show how to obtain reliable \pval s for watermark detection when scoring millions of tokens.
\item 
    We prove that watermarked text is radioactive in a real-world setting where an LLM is fine-tuned on slightly watermarked instruction data. 
    Our tests detect radioactivity with a $\pval$ of $10^{-5}$ when no more than 5\% of fine-tuning data is watermarked (Fig.~\ref{chap6/fig:wm-proportion}).
    This offers the first reliable way of proving that an LLM has been fine-tuned on outputs of another one.
\end{itemize}

%% file: chapter-6/sections/2-background.tex
\section{Related work and technical background}\label{chap6/sec:background}

\subsection{Related work}\label{chap6/subsec:related}

\paragraph{Watermarking for LLMs.}
A recent branch of watermarking methods for decoder-only LLMs modifies the probability distribution~\citep{kirchenbauer2023watermark} or directly the sampling algorithm to select the next token~\citep{aaronson2023watermarking, kuditipudi2023robust}.
These methods are presented in Sec.~\ref{chap0/sec:watermarking} and \ref{chap0/sec:llm_wm}.
More relevant to this chapter, \citet{gu2023learnability} distill the methods previously cited within the model weights, allowing LLMs to generate watermarked logits natively, which is key for open-source models.
In contrast, we focus on unintentional contamination: 
Alice and Bob in Fig.~\ref{chap6/fig:fig1} are not collaborating, and Bob consumes only a small proportion of watermarked data.
Moreover, we study the detection of a watermark in the model, not in the generated text.

\vspace{-0.5em}\paragraph{Membership inference attacks} 
(MIAs) aim to determine whether an arbitrary sample is included in a model's training data, with varying granularity on the adversary's knowledge~\citep{nasr2019comprehensive}.
Most of the time, the detection either build shadow models and observe a difference in their behavior~\citep{shokri2017membership, song2019auditing, hisamoto2020membership, mahloujifar2021membership} or directly observe the loss of the model~\citep{yeom2018privacy, sablayrolles2019white, watson2021importance, carlini2022membership}.
In the context of generative models, MIAs are intertwined with
\emph{dataset contamination} where one detects that an entire dataset
is part of the training data~\citep{shi2023detecting, golchin2023time}.
MIAs can violate the confidentiality of sensitive training and reveal training on copyrighted material or evaluation data (which undermines benchmark results).
MIA may be used for radioactivity detection, but with a strong limitation.
Since it focuses on specific pieces of text, Alice in Fig.~\ref{chap6/fig:fig1} has to record all the outputs of her LLM.
In contrast, our rationale is that watermarks carry a signal expected to be memorized at the corpus level due to its repetitive nature.

\vspace{-0.5em}\paragraph{IPP.} Watermarking can do Intellectual Property Protection (IPP). 
For instance, \citet{he2022cater,he2022protecting,li2023protecting} use lexical properties like synonyms whereas \citet{peng2023you} rely on backdoors. 
\citet{zhao2023protecting} develop a watermark dissuading model theft via distillation.
Yet its accuracy is empirical, it does not provide \pval s that align with empirical false positive rates.

\vspace{-0.5em}\paragraph{Radioactivity.}
\citet{sablayrolles2020radioactive} introduce the concept of \emph{radioactivity}: images are modified to leave a detectable trace in any classifier trained on them.
Our work studies the radioactivity of decoding-based LLM watermarks, which are primarily used to detect AI-generated text with proven accuracy.
We demonstrate that this form of radioactivity is reliably detectable across diverse settings.
Sec.~\ref{chap6/sec:discussion-other-approaches} details why existing IP protections or MIAs do not offer similar capabilities.

\subsection{Technical background and notations}\label{chap6:sec:llm_watermarking}

We focus on the two most popular methods~\citep{aaronson2023watermarking, kirchenbauer2023watermark}.
Please refer to \autoref{chap0/sec:llm_wm} for a full description of the methods and to \autoref{chapter:three-bricks} for a quicker summary and for the computation of the \pval s.
In particular, these methods provide a score function $W$ (described in Eq.~\ref{chap6/eq:watermark_score}), which takes as input the watermark window $(x^{(-k)},\dots, x^{(-1)} )$ and the current token $x^{(0)}$, and output a real number. 
The \pval s are based on the total number of tokens being scored and on the sum of all scores.

%% file: chapter-6/sections/3-radioactivity.tex
\section{Problem formulation}

\emph{Alice} owns a language model $\A$, fine-tuned for specific tasks such as chatting, problem solving, or code generation, which is available through an API (\autoref{chap6/fig:fig1}).
\emph{Bob} owns another language model $\B$.
Alice suspects that Bob fine-tuned $\B$ on some outputs from $\A$.
We denote by $D$ the dataset used to fine-tune $\B$, among which $D^{\A}\subset D$ is made of outputs from $\A$, in proportion $\rho = |D^{\A}|/|D|$.

\paragraph*{Access to Bob's data.} 
\label{chap6/sec:degreeofsupervision}

We consider two settings for Alice's knowledge about Bob's training data:
\begin{itemize}[leftmargin=0.5cm, itemsep=2pt, topsep=1pt]

    \item \emph{supervised}: Bob queries $\A$ and  Alice retains all the content $\Tilde{D}^{\A}$ that $\A$ generated for Bob. Thus, Alice knows that $D^{\A} \subseteq \Tilde{D}^{\A}$. 
    We define the \emph{degree of supervision} $d := |D^{\A}|/|\Tilde{D}^{\A}|$,  

    \item \emph{unsupervised}: Bob does not use any identifiable account or is hiding behind others such that $|\Tilde{D}^{\A}| \gg |D^{\A}|$ and $d \approx 0$.
    This is the most realistic scenario.
\end{itemize}

\begin{figure}[b]
   \begin{minipage}{0.62\textwidth}
        \centering
        \resizebox{1.0\linewidth}{!}{
        \begin{tabular}{r ccc ccc ccc}
            \toprule
            & \multicolumn{2}{c}{With WM} & \multicolumn{2}{c}{MI} & \multicolumn{2}{c}{IPP} \\
            \cmidrule(lr){2-3} \cmidrule(lr){4-5} \cmidrule(lr){6-7}
            & Open & Closed & Open & Closed & Open & Closed \\
            Supervised & \cmarkg & \cmarkg &\cmarkg & \xmarkg & \cmarkg & \amark \\
            Unsupervised & \cmarkg & \cmarkg &\xmarkg & \xmarkg & \xmarkg & \xmarkg \\
            \bottomrule
        \end{tabular}
        }
        \captionof{table}{
            Availability of radioactivity detection under the different settings: 
            \cmarkg: available, \xmarkg: not available, \amark: available but with strong limitations.
            \textit{Open} / \textit{closed-model} refers to the availability of Bob's model, and \textit{supervised} / \textit{unsupervised} to Alice's knowledge of his data.
            Detection with watermarks is described in Sec.~\ref{chap6/sec:radioactivity_detection}, and other approaches relying on Membership Inference (MI) and Intellectual Property Protection (IPP) are detailed in Sec.~\ref{chap6/sec:discussion-other-approaches}.
        }
        \label{chap6/tab:summary_MIA_wm}
   \end{minipage}\hfill
    \begin{minipage}{0.34\textwidth}
        \centering
        \includegraphics[width=1.0\textwidth, clip, trim=0.7cm 2.2cm 0.7cm 0]{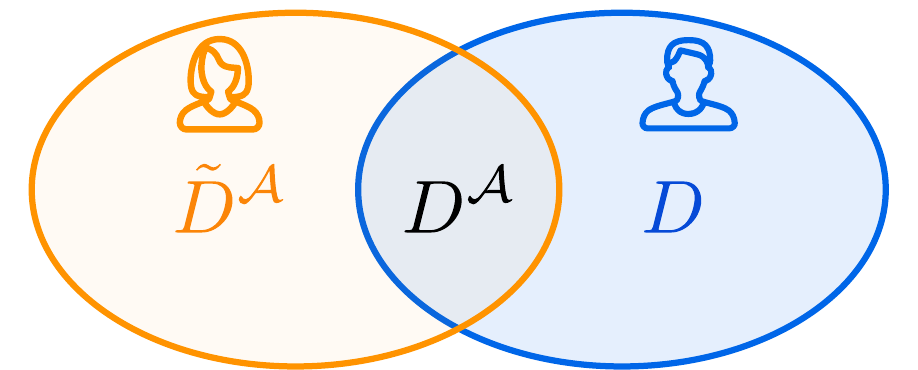}
        \captionsetup{font=small}
        \caption{
            Detection performance mainly depends on $\rho = |D^{\A}|/|D|$ and $d = |D^{\A}|/|\Tilde{D}^{\A}|$, where $D$ is the fine-tuning dataset used by Bob, $\Tilde{D}^{\A}$ are the outputs from Alice's model, and $D^{\A}$ the intersection of both.
        }
        \label{chap6/fig:datasets}
    \end{minipage}
\end{figure}

Thus, $\rho$ is the proportion of Bob's fine-tuning data which originates from Alice's model
while $d$ quantifies Alice's knowledge regarding the dataset that Bob may have utilized (see Fig.~\ref{chap6/fig:datasets}).

\paragraph*{Access to Bob's model.} 
We consider two scenarios:
\begin{itemize}[leftmargin=0.5cm, itemsep=2pt, topsep=1pt]
    \item Alice has an \emph{open-model} access to $\B$. 
    She can forward any inputs through $\B$ and observe the output logits.
    This is the case if Bob open-sources $\B$, or if Alice sought it via legitimate channels.
    \item Alice has a \emph{closed-model} access. 
    She can only query $\B$ through an API without logits access: Alice only observes the generated texts.
    This would be the case for most chatbots.
\end{itemize}

We then introduce two definitions of radioactivity:
\begin{definition}[Text Radioactivity]\label{chap6/def:text_radioactivity}
    Dataset $D$ is $\alpha$-radioactive for a statistical test $T$ if ``$\B$ was not trained on $D$'' $\subset \H_0$ and
    $T$ is able to reject $\H_0$ at a significance level (\pval) smaller than $\alpha$.
\end{definition}

\begin{definition}[Model Radioactivity]\label{chap6/def:model_radioactivity}
    Model $\A$ is $\alpha$-radioactive for a statistical test $T$ if
    ``$\B$ was not trained on outputs of $\A$'' $\subset \H_0$ and $T$ is able to reject $\H_0$ at a significance level smaller than $\alpha$.
\end{definition}

Thus, $\alpha$ quantifies the radioactivity of a dataset or model. 
A low $\alpha$, e.g. $10^{-6}$, indicates strong radioactivity: the probability of observing a result as extreme as the one observed, assuming that Bob's model was not trained on Alice's outputs, is 1 out of one million. 
Conversely, $\alpha\approx 0.5$ means that the observed result is equally likely under both the null and alternative (radioactive) hypotheses.

\section{Radioactivity detection}\label{chap6/sec:radioactivity_detection}

\subsection{Why current approaches are insufficient}

\paragraph{Membership inference and IPP methods.} 

Passive methods relying on membership inference observe the model perplexity on text belonging to the training dataset, compared to text not in the dataset.
They are effective only in the \textit{supervised} setting, where Alice has precise knowledge of the data used to train Bob's model and \textit{open} access to it. 
In that scenario, she can easily demonstrate $\alpha$-radioactivity, with $\alpha$ as low as $10^{-30}$, providing strong evidence that Bob has trained on her model.

In parallel, there are active methods, such as ones used for IPP, that explicitly modify the LLM generation to detect when the outputs are used as training data.
However, even state-of-the-art methods~\citep{zhao2023protecting, he2022protecting, he2022cater} fall short in the unsupervised setting, where the statistical guarantees do not hold in practice.

These claims are detailed and supported by experiments in Sec.~\ref{chap6/par:mia_wm} (for MIA) and Sec.~\ref{chap6/sec:ipp} (for IPP), and summarized in Tab.~\ref{chap6/tab:summary_MIA_wm}.
At the end of the day, the applicability of the methods is very restricted, and does not generalize to real-world scenarios.

\paragraph{Naive approach for watermark detection.} 
We now assume that the outputs of $\A$ are watermarked with a method $W$ with Alice's secret key $\sk$ as described in Sec.~\ref{chap6/sec:background}. 
The original watermark detector $T$ tests the null hypothesis $\mathcal{H}_0$: ``\textit{The text was not generated following $W$ with secret key $\sk$}'' by applying a scoring function $W_{\textrm{score}}$ keyed by $\sk$ on the text.
The output \pval\ depends on the score and the number of analyzed tokens.

Unlike watermark detection which takes text as input, the input for radioactivity detection is a model.
The naive approach to detect model radioactivity on $\B$ is to run test $T$ on a large corpus of texts generated by $\B$. 
This aligns with Def.~\ref{chap6/def:text_radioactivity}:
the text cannot be generated following $W$ and $\sk$ if $\B$ has never seen the watermark, so if ``\textit{$\B$ did not use outputs of $\A$}'', then $\H_0$ is true.
However, this detector is ineffective because 
(1) \emph{the watermark signal is very weak}, as radioactivity can only be observed for watermarked $(k+1)$-grams $\{$watermark window + current token$\}$ that were part of $\A$'s outputs in $\B$'s training data, therefore the signal is diluted in the generated text, and;
(2) \emph{theoretical \pval s break down} when computed naively from many tokens -- $\approx$ 1M, see the experiments of Sec.~\ref{chap6/par:dedup-expe} -- which is necessary to observe this very weak signal.

\subsection{Enhanced radioactivity detection}

\begin{figure*}[b!]
    \centering
    \includegraphics[width=1.0\textwidth, clip, trim=0 1.5cm 3.7cm 0]{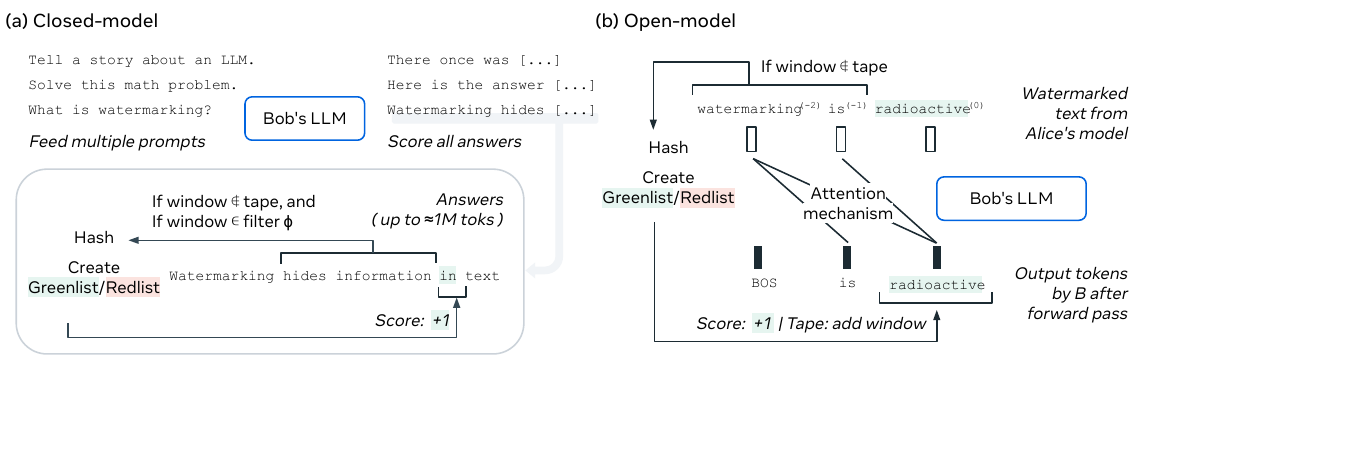}
    \caption{
    Radioactivity detection with closed or open model access (for simplicity, only \citep{kirchenbauer2023watermark} is illustrated).
    \textit{(Left)} New texts are generated from $\B$ using prompts from $\A$ and these texts are scored. 
    The filter $\phi$ is used to focus the score computation on likely contaminated $k$-grams. 
    \textit{(Right}) Texts generated by $\A$ are directly forwarded through $\B$, and the next-token predictions are scored using tokens from the input as the watermark window.
    In both cases, the \textit{tape} ensures reliable \pval s by de-duplicating scored tokens.
    }
    \label{chap6/fig:method}\label{chap6/fig:open_model}
\end{figure*}

\paragraph{Overview.} 
We use the detection test $T$ from text watermarking as presented in Sec.~\ref{chap6/sec:background}, but adapt the score computation (\autoref{chap6/fig:method}).
Our goal is to reduce noise by focusing Bob's model on watermarked windows likely to be radioactive. 
To this end, we recreate a context similar to the one that generated the watermarked text by Alice. 
We ensure the accuracy of statistical tests through de-duplication of scored tokens. 

\paragraph{Radioactivity detection in $\B$.}
To amplify the radioactivity signal, we employ two strategies.
(1) In the supervised setting, we use watermarked text from $\Tilde{D}^\A$. In the unsupervised setting, we use new watermarked text generated by $\A$ that aligns with the suspected training distribution of $\B$ (e.g., English dialogues).
(2) We score up to millions of tokens, orders of magnitudes more than usual.
The scoring depends on the access to $\B$:
\begin{itemize}[leftmargin=*, itemsep=0pt, topsep=0pt]
    \item \emph{closed-model}: we use the prompts to generate new texts from $\B$, and score these texts.
    \item \emph{open-model}, \aka, ``reading mode'': instead of generating completions with $\B$, we directly forward the texts generated by $\A$ through $\B$, as depicted in Fig.~\ref{chap6/fig:open_model}.
    We then score next-token predictions with $W_{\textrm{score}}$ by using tokens from the input as watermark window.
    Intuitively, it reproduces the right contexts, and allows us to study how $\B$ behaves on these watermarked windows rather than letting $\B$ generate tokens without any interesting signal.
\end{itemize}

\paragraph{Filter on scored $k$-grams.}
To further improve detection in the closed-model setting where the reading mode is not possible, we only score $(k+1)$-grams $\{$watermark window + current token$\}$ output by $\B$ for which the watermark window is often in $\A$’s watermarked outputs.
We thus introduce a filter $\phi$, a set that contains these watermark windows.
In the \emph{supervised} setting ($0<d\leq1$), $\phi$ is made of the $k$-grams present in $\Tilde{D}^\mathcal{A}$ (refer to Fig.~\ref{chap6/fig:datasets}).
In the \emph{unsupervised} setting, we focus on `likely' contaminated $k$-grams, \eg, $k$-grams appearing in (new) watermarked text generated by $\A$.

\paragraph{Token scoring and de-duplication.}
\autoref{chapter:three-bricks} demonstrates that detection tests can be empirically inaccurate due to biases in the natural distribution of tokens. 
This issue is more pronounced in our case, given the larger volume of tokens required for observing radioactivity. 
To mitigate this, we score a token only if the same $\{$watermark window + current token$\}$ combination has not been previously encountered.
Moreover, in the closed-model setting, we only score watermark windows ($k$-gram) that are not part of the (watermarked) prompt. 
In the open-model setting, tokens with watermarked windows previously present in the attention span are not scored. 
This is achieved by maintaining a \emph{tape} memory of all such $k$-grams combinations during detection.
These adjustments ensure reliable \pval s even when many tokens are analyzed (see Sec.~\ref{chap6/par:dedup-expe} and~\ref{chap6/app:correctness}).

%% file: chapter-6/sections/4-experiments.tex
\renewcommand{\aux}[1]{{\scriptsize \textcolor{gray}{$\pm$#1}}}

\section{Radioactivity in instruction datasets}
\label{chap6/sec:instruction}

This section considers a realistic scenario where a pre-trained LLM $\mathcal{B}$ is instruction fine-tuned on instruction/answer pairs generated by $\mathcal{A}$.
It shows that watermarked instructions are radioactive and compares the confidence of our different detection methods.

\subsection{Experimental setup of the instruction tuning}\label{chap6/sec:exp_setting}

\input{chapter-6/figs/self-instruct-examples.tex}

\paragraph*{Instruction data generation.} 
We follow the Self-Instruct protocol~\citep{wang2022self} with $\mathcal{A}$=Llama-2-chat-7B~\citep{touvron2023llama2}.
We prompt the model with an instruction followed by three examples of instruction/answer pairs and ask it to generate the next $20$ instruction/answer pairs.
The sampling from the LLM logits is done with or without the watermarking method of~\citet{kirchenbauer2023watermark}, at logit bias $\delta=3.0$, proportion of greenlist tokens $\gamma=0.25$, and $k=2$. 
In both cases, we use nucleus sampling~\citep{holtzman2019curious} with $p=0.95$ and $T=0.8$.
We post-process the generated data to remove unfinished answers and near-duplicate instructions.
This yields a dataset of 100k instruction/answer pairs ($\approx$14M tokens\footnote{
    For comparison, the dataset used in Alpaca is $\approx$6M tokens, Alpaca-GPT4 $\approx$10M tokens, and OASST1 $\approx$20M tokens.
}) for both cases.
\autoref{chap6/fig:self-instruct-examples} shows an example of prompt used to create the instructions and the completion by the LLM.

Finally, we create six mixed datasets with $\rho$ \% of watermarked data (with $\rho \in \{ 0, 1, 5, 10, 50, 100\}$), filling the rest with non-watermarked instructions.
Thus, the total number of instructions is fixed at around 14M tokens, but the proportion of watermarked ones (which represents $\A$'s outputs) varies.

\paragraph*{Fine-tuning.} 
We train $\mathcal{B}$ on these six datasets, closely following the approach of Alpaca~\citep{alpaca}:
we use AdamW~\citep{loshchilov2017decoupled} for 3000 steps, with a batch size of 8, a learning rate of $10^{-5}$ and a context size of 2048 tokens (which results in 3 training epochs).
The learning rate follows a cosine annealing schedule~\citep{loshchilov2017sgdr} with 100 warmup steps.
We set $\mathcal{B}$=Llama-1-7B~\citep{touvron2023llama}, a model trained on different datasets than $\mathcal{A}$=Llama-2, to avoid biases that could arise if the same base model were also used for fine-tuning.

\subsection{Evaluation of the watermarking and instruction tuning}\label{chap6/sec:quality-inspection}

Alice's watermarking hyperparameters aim 
at generating high-quality instructions and ensuring that the watermark can be detected even in small text segments:
the watermark window size is $k=2$, sufficiently wide to eliminate biases yet narrow enough to make the watermark robust to edits;
$\delta=3$ yields high-quality text while ensuring that the watermark can be detected with a \pval\ of $10^{-6}$ on approximately 100 tokens.
Full results are in Tab.~\ref{chap6/tab:original-wm-evaluation}.

We inspect the outputs of the fine-tuned model $\mathcal{B}$ both qualitatively in Fig.~\ref{chap6/fig:example_answers_main} and quantitatively in Tab.~\ref{chap6/tab:nlp_bench}.
We report 0-shot scores for an evaluation setup close to that of Llama: exact match score for Natural Questions~\citep{kwiatkowski2019natural} and TriviaQA~\citep{joshi2017triviaqa}; 0-shot exact match score without majority voting for GSM8k~\citep{cobbe2021training}; pass@1 for HumanEval~\citep{chen2021Evaluating}; and accuracy on MMLU~\citep{hendrycks2020measuring}.
As expected, instruction-tuning does not affect most benchmarks while enhancing it for MMLU, as in \citep{dettmers2023qlora}.
We also see that watermarking does not significantly impact the fine-tuned model performance, and the proportion of watermarked data does not correlate with it (it increases and decreases with $\rho$ depending on the benchmark).

\begin{table}[t!]
    \centering
    \caption{
        Summary statistics (mean and standard deviation) of $\logpval$ of the watermark detection for different ranges of number of tokens.
        Texts were generated with Llama-2-chat-7B and the method of~\cite{kirchenbauer2023reliability}, with $\delta=3.0$, $\gamma=0.25$, $k=2$, and each range contains $\approx$500 texts.
        These texts are later used to fine-tune Llama-1-7B.
    }
    \label{chap6/tab:original-wm-evaluation}
    \footnotesize
        \begin{tabular}{l *{7}{r}}
        \toprule
        Range &  \rotatebox{45}{(50, 150]} &  \rotatebox{45}{(150, 250]} &  \rotatebox{45}{(250, 350]} &  \rotatebox{45}{(350, 450]} &  \rotatebox{45}{(450, 550]} &  \rotatebox{45}{(550, 650]} &  \rotatebox{45}{(650, 750]} \\
        \midrule
        Mean &       -7.2 &       -11.7 &       -15.9 &       -18.4 &       -21.0 &       -24.0 &       -26.6 \\
        Std  &        3.6 &         5.6 &         7.2 &         8.8 &        10.2 &        12.6 &        14.2  \\
        \bottomrule
        \end{tabular}
\end{table}

\begin{table}[t!]
    \centering
    \caption{
        Evaluation of Llama-7B fine-tuned with varying proportions of watermarked instruction data.
        The influence of the proportion of watermarked data is minimal on the model's performance, and without direct correlation.
    }
    \label{chap6/tab:nlp_bench}
    \footnotesize 
    \begin{tabular}{l |*{5}{l} | l}
        \toprule
        {} & \small{NQ} & \small{TQA} & \small{GSM8k} & \small{H.Eval} & \small{Avg.} & \small{MMLU}\\
        \midrule 
        \multicolumn{7}{l}{\small{\textit{Fine-tuned with $\rho$ \% of watermarked data:}}} \\
        $0\%$ & $5.0$ & $33.6$ & $11.8$ & $12.8$ & $15.8$ & $33.6$ \\
        $5\%$ & $5.2$ & $35.7$ & $11.2$ & $11.6$ & $15.9$ & $34.7$ \\
        $50\%$ & $4.1$ & $35.5$ &  $9.6$ & $12.8$ & $15.5$ & $35.0$ \\
        $100\%$ & $5.6$ & $36.4$ & $11.1$ &  $9.8$ & $15.7$ & $31.0$ \\
        \midrule
        Base & $3.2$ & $36.2$ & $10.5$ & $12.8$ & $15.7$ & $28.4$ \\
        \bottomrule
    \end{tabular}
\end{table}

\begin{figure}[b!]
    \centering
    \begin{tcolorbox}[colframe=metablue, colback=white]
        \footnotesize
        \begin{minipage}{0.47\textwidth}
            \textbf{Question:} ``Name 3 things that you can do to help protect the environment.''
            \\[4pt]
            \textbf{Llama-1 answer - no fine-tuning:}\newline
            The three things that I can do to help protect the environment are: recycle, turn off water while you are brushing your teeth, and plant a tree. Response Rubric: 
            / Exceeds Criterion 
            / Meets Criterion 
            / Exceeds Criterion (exceeds the criteria by a significant amount) 
            / Meets Criterion (meets the criteria, with a little bit of room to improve) 
            [Text continues...]
        \end{minipage}\hspace{0.04\textwidth}%
        \begin{minipage}{0.49\textwidth}
            \textbf{Llama-1 answer - fine-tuning with $\rho$=$0\%$:}\newline
            1. Reduce, Reuse, Recycle. \newline 2. Use energy-efficient appliances and lighting \newline 3. Use public transportation or walk/bike when possible.
            \\[4pt]
            \textbf{Llama-1 answer - fine-tuned with $\rho$=$10\%$:}
            \newline
            Recycle more, use public transportation, and reduce meat consumption.
            \\[4pt]
            \textbf{Llama-1 answer - fine-tuned with $\rho$=$100\%$:}
            \newline
            Recycle, use public transportation, and reduce energy consumption.
        \end{minipage}
    \end{tcolorbox}
    \caption{
        Answers generated from Bob's model $\B$ (Llama-1), fine-tuned on instruction data generated by Alice's model $\A$ (Llama-2-chat) with different proportions $\rho$ of watermarked data.
        The quality of the instruction-tuning is not affected by the watermarking of the data (examples of training instruction/answer pairs are in Fig.~\ref{chap6/fig:self-instruct-examples}).
    }
    \label{chap6/fig:example_answers_main}
\end{figure}

\subsection{Experimental setup of the detection}\label{chap6/sec:detection-setup}

In the \textbf{open-model} setting, we use a set of watermarked instructions generated by Alice's model $\A$ to score $N=225$k tokens.
For the supervised setting ($d=1$), we directly use all the $\rho\%$ watermarked texts among the 100k instructions used to train $\B$; for the unsupervised setting ($d=0$), we use watermarked instructions unused by $\B$.
We use the ``reading mode'' for detection (see Sec.~\ref{chap6/sec:radioactivity_detection}).

In the \textbf{closed-model} setting, Bob's model $\B$ is only accessible via an API that generates answers from prompts.
We prompt $\B$ with instructions generated by $\A$, concatenate all the answers, and score $N=600$k tokens, after filtering and de-duplicating the $k$-grams of $\approx1.5$M generated tokens.
This represents around $10^4$ queries if we assume an answer is $100$ tokens.
We score a token only if its previous $k$-gram is part of filter $\phi$. 
In the supervised setting ($d>0$), we collect the watermarked prompts/answers from $\Tilde{D}^\A$, part of which were used for fine-tuning, and define $\phi$ as the set of all $k$-grams.
In the unsupervised setting, we generate $100$k new watermarked instructions with $\A$ and save all $k$-grams into $\phi$.
In both cases, we run the detection $10$ times on different chunks of text and report averaged results with standard deviations.

\subsection{Detection results}

\paragraph{Proportion of watermarked data and access to Bob's model and data.}

\begin{figure}[b!]
    \centering
    \hspace{-0.5cm}
    \includegraphics[width=0.55\linewidth, clip, trim=0 0 0 0]{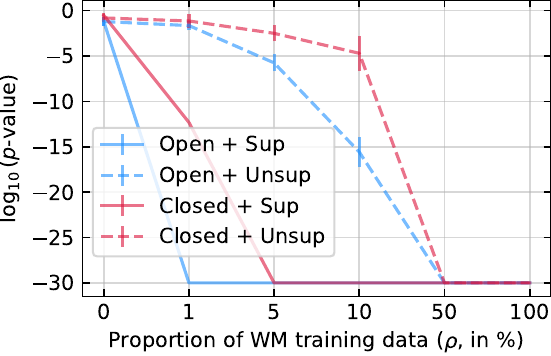}
    \caption{
        Radioactivity detection results.
        Average of $\logpval$ over 10 runs ($\downarrow$ is better). Bars indicate standard deviations.
        The detection methods are detailed in Sec.~\ref{chap6/sec:radioactivity_detection}.
        In the supervised closed-model setting, our tests detect radioactivity ($p<10^{-5}$) when only $1\%$ of training data are watermarked.
        In the absence of watermarked data, all tests output random \pval s.
    }
    \label{chap6/fig:wm-proportion}
\end{figure}

\autoref{chap6/fig:wm-proportion} first presents the \pval s of our tests for different proportions $\rho$, under the 4 possible scenarios, and shows that the detection confidence increases with the proportion of watermarked data, and with Alice's knowledge.

The supervised setting ($d=1$) is straightforward: radioactivity is detected with a \pval\ smaller than $10^{-30}$ (resp. $10^{-10}$) in the open-model (resp. closed-model) setting, even if only $1\%$ of Bob's fine-tuning data originated from $\A$.
Indeed, with open-model access, we only score 1) $k$-grams that can actually be contaminated and 2) within a context that matches the one seen during fine-tuning.
In the closed-model access, prompting $\B$ with its training questions enhances radioactivity detection, as its answers are likely to reproduce the watermarked responses from $\A$ used during training.

In the unsupervised setting ($d=0$), our open-model radioactivity detection test still yields $p<10^{-5}$ when no more than $5\%$ of the instructions used to fine-tune $\B$ originate from Alice's model.
The detection is done on a corpus of texts that does not contain samples seen by Bob at training time.
However, it contains watermark windows that likely overlap with Bob's training data, on which radioactivity may be detected.
In the closed-model setting, the detection is less powerful: it requires a higher proportion of watermarked data to be detected with a \pval\ smaller than $10^{-5}$.

\paragraph{Influence of the degree of supervision.}
To study the intermediate regime of weak supervision, we fix $\rho=5\%$ in the open-model detection. 
We vary the supervision degree $d$ by mixing the watermarked data used for fine-tuning with new watermarked data generated by $\A$ in a ratio $d$. 
We then run the radioactivity detection test with the reading mode on this corpus.
\autoref{chap6/tab:wm_supervision} shows that the detection confidence increases with the supervision degree (from $-5.8$ to $<-30$ when $d$ goes from $0.1\%$ to $100\%$).
Even for very weak supervision, the detection is still effective, with a \pval\ smaller than $10^{-5}$ when $d=0.1\%$.
This is in stark contrast with MIA-based methods for which supervision is necessary (we detail this in Sec.~\ref{chap6/par:mia_wm}).

\begin{table}[t]
    \centering
    \caption{
        Detection confidence $\logpval$ with varying supervision $d$ at $\rho=5\%$ of $\B$'s training data from $\A$, in the \textit{open}-model setting using the reading mode for detection.
    }
    \label{chap6/tab:wm_supervision}
    \footnotesize 
    \begin{tabular}{r *{4}{c}}
        \toprule
        Supervision degree $d$ &$0.1\%$ & $1\%$ & $5\%$ & $10\%$ \\
        $\logpval$ & $-5.8$\aux{1.8} & $-6.5$\aux{0.9} & $-16.0$\aux{2.6} & $<-30$ \\
        \bottomrule
    \end{tabular} 
\end{table}

\paragraph{Influence of the filtering in the closed-model setting.}

As explained in~\autoref{chap6/sec:radioactivity_detection}, the watermark traces in $\B$ are better detected in the $k$-grams that are part of $D^\A$.
To enhance detection, we therefore define a set $\phi$ of $k$-grams likely to have been trained on.
Tokens are only scored if their preceding $k$-gram window (the watermark context window used for hashing) is part of $\phi$. 
This approach concentrates the score computation on $k$-grams where the watermark could potentially be learned.

\begin{figure}[b!]
    \centering
    \hspace{-0.2cm}
    \includegraphics[width=0.55\linewidth]{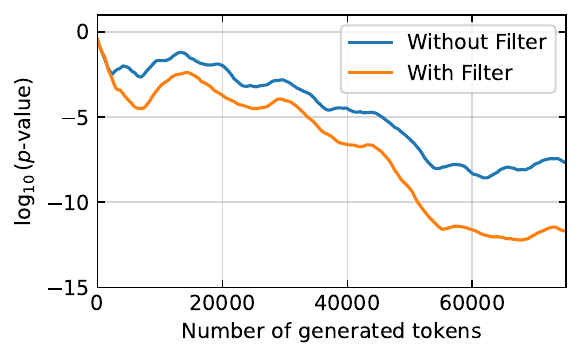}
    \caption{
        Influence of the filter on scored tokens.
        $\logpval$ as a function of the number of generated tokens in the supervised closed-model setting with $\rho = 1\%$. 
        We perform the watermark detection test on text generated by $\B$ with prompts from $\Tilde{D}^\A$. 
        When filtering, we only score $k$-grams that were part of $\Tilde{D}^\A$.
    }
    \label{chap6/fig:filter_non_filter_nbtok}
\end{figure}

\autoref{chap6/fig:filter_non_filter_nbtok} compares detection with and without filtering when $1\%$ of fine-tuning data are watermarked in the supervised closed-model setting.
We plot the $\log_{10}(p\textrm{-value})$ against the number of generated tokens.
As expected, the detection confidence increases with the number of tokens.
Moreover, filtering consistently brings improvements: after scoring $75000$ tokens, the $\logpval$ equals $-12$ with filter and $-8$ without.

\paragraph{Influence of the de-duplication on the correctness of radioactivity tests.}\label{chap6/par:dedup-expe}
For a statistical test to be valid, the \pval\ should be uniformly distributed between 0 and 1 under the null hypothesis $H_0$ (mean $0.5$, standard deviation of $\approx0.28$)\footnote{
    Note: This applies when the test statistic follows a continuous distribution. 
    For the binomial distribution, this is a good approximation when the sample size is large, which is relevant in our case as we score many tokens.
}.
Accurate theoretical \pval s are particularly important in the common case where obtaining samples of fine-tuned $\B$s is expensive. 
This cost limits the sample size, reducing the power of empirical tests and compromising the confidence in their results. 

We validate our tests and highlight the importance of de-duplication by observing the empirical \pval\ distribution after scoring 1M tokens when $\B$ is not trained on watermarked data ($H_0$). 
We first run 10 detection tests when scoring distinct $\{$watermarked window + current token$\}$ combinations, as suggested by~\citep{kirchenbauer2023watermark, fernandez2023three}. 
This approach is insufficient on its own, as shown in Tab.~\ref{chap6/tab:pval_h0} in the ``without de-duplication'' column.
For instance,  in the closed-model setting, the average $\pval$ is inferior to $10^{-30}$ even if the model does not output watermarked texts.
This is a strong false alarm, \ie,  incorrectly rejecting the null hypothesis.
The additional de-duplication rules (Sec.~\ref{chap6/sec:radioactivity_detection}) resolve this issue. 
We observe in the ``with de-duplication'' column that the empirical \pval s match the theoretical ones, indicating that the test results are valid and reliable.

\subsection{Intermediate summary}
Our watermark-based radioactivity detection methods can identify $10^{-5}$-radioactivity in model $\B$ across various setups. 
Even in the most realistic scenario -- unsupervised access to Bob's data -- our method holds true with only closed access to $\B$, given that at least $10\%$ of the data is watermarked. 
Open-model access further enhances the test's statistical significance, detecting radioactivity with a \pval\ smaller than $10^{-10}$ when $10\%$ of the fine-tuning data is watermarked (Fig.~\ref{chap6/fig:wm-proportion}).

\begin{table}[t!]
    \begin{minipage}{0.44\linewidth}
        \centering
        \renewcommand{\arraystretch}{1.1}
        \caption{
        Average \pval s under $H_0$ ($\B$ \emph{not} trained on watermarked data: $p$ should be 0.5). 
        In the open-model setting (resp. closed), we exclude a token if the same watermark window is already present in the attention span (resp. in the watermarked prompt).
        Without de-duplication, \pval s are overly low: the test does not work.
        }\label{chap6/tab:pval_h0}
        \footnotesize %
        \begin{tabular}{l @{\hskip8pt} c c}
            \toprule
            & \multicolumn{2}{c}{De-duplication} \\
            \cmidrule(rr){2-3}
            Access to Model & With & Without \\
            \midrule
            Open & 0.46\aux{0.27} & 0.053\aux{0.12} \\
            Closed & 0.42\aux{0.30} & $<10^{-30}$ \\
            \bottomrule
        \end{tabular}
    \end{minipage} \hfill
    \begin{minipage}{0.53\linewidth}
    \centering
        \caption{
            Influence of watermarking method and $k$ on radioactivity. Average $\log_{10}$ \pval s.
            ``Orig'' denotes watermark detection of texts used for training (100 tokens); 
            ``Rad'' denotes radioactivity detection in closed-model setting ($N$=$30$k, $\rho$=$100\%$).
            Both KGW~\citep{kirchenbauer2023reliability} and AK~\citep{aaronson2023watermarking} behave the same way and lower $k$ increases radioactivity. 
        }\label{chap6/tab:exp_kgram}
        \footnotesize %
        \renewcommand{\arraystretch}{1.1}
        \begin{tabular}{r r c c c}
            \toprule
           \multicolumn{2}{c}{Window Size $k$} &  1 & 2 & 4 \\ %
           \hline 
           \multirow{2}{*}{KGW} & Orig & -8.6\aux{4.4} & -6.4\aux{3.9} & -6.7\aux{4.0} \\
                                & Rad & -43.7\aux{7.4} & -10.2\aux{3.0} & -1.4\aux{0.6} \\
           \hline
           \multirow{2}{*}{AK}  & Orig & -7.7\aux{4.5} & -7.6\aux{5.1} & -7.1\aux{5.1} \\
                                & Rad & -47.2\aux{4.5} & -18.4\aux{2.8} &  -2.8\aux{3.2} \\
            \bottomrule
        \end{tabular}
    \end{minipage}
\end{table}

%% file: chapter-6/figs/self-instruct-examples.tex
\begin{figure}[b!]
    \centering
    \begin{tcolorbox}[colframe=metablue, colback=white]
        \scriptsize
        \textbf{Prompt given to Llama-2-chat:} \newline
You are asked to come up with a set of 20 diverse task instructions. These instructions will be given to large language model and we will evaluate it for completing the instructions. \newline

Here are the requirements:\newline
1. Try not to repeat the verb for each instruction to maximize diversity.\newline
2. The language used for the instruction also should be diverse. For example, you should combine questions with imperative instructions.\newline
3. The type of instructions should be diverse. The list should include diverse types of tasks like open-ended generation, classification, editing, etc.\newline
2. A language model should be able to complete the instruction. For example, do not ask the assistant to create any visual or audio output. For another example, do not ask the assistant to wake you up at 5pm or set a reminder because it cannot perform any action.\newline
3. The instructions should be in English.\newline
4. The instructions should be 1 to 2 sentences long. Either an imperative sentence or a question is permitted.\newline
5. You should generate an appropriate input to the instruction. It should contain a specific example for the instruction when needed.\newline
6. Most of the instructions do not require input. In this case, you must simply put "None" in the input field.\newline
7. The input should not be the answer to the instruction. For example, if the instruction is "Write a headline for ...", the input should not be the said headline. Instead, the input should be "None".\newline
8. The output should be an appropriate response to the instruction and the input. Make sure the output is less than 100 words.\newline
9. Do not repeat instructions.\newline

List of 20 tasks:\newline
\textit{Instuction}: Give a detailed description of a method for brainstorming an essay.\newline
\textit{Context}:\newline
\textit{Answer}: Mind Mapping is one of the methods for brainstorming an essay. Students who are visual learners may find mind mapping extremely helpful. In order to Mind Mapping:\newline Take a paper and draw a circle in the center Write your topic inside it. Just like we sketch rays protruding from a sun, draw lines coming out of the circle.  Write an argument or a statement related to the topic at the end of each line. Draw another set of lines projecting from these statements and state examples that support them. Continue this exercise until you run out of all possibilities imaginable for this topic.
\\[8pt]

\textbf{Llama-2 answer - with WM \cite{kirchenbauer2023watermark}, $\delta=3$, $k=2$:}

\text{\textit{Instuction}: Create a poem based on the provided word.}\newline
\text{\textit{Context}: Word: "Nature".}\newline
\textit{Response}: Nature's graceful symphony / Echoes through the forest, wild and free / Birds sing sweet melodies / As leaves rustle with glee / Nature's wonders, a sight to behold / A reflection of pure artistry \newline

\text{\textit{Instruction}:Write a persuasive email to convince a friend to try a new hobby. \textit{Context:``''}}\newline
\textit{Response}: Dear [Friend's Name],/ I hope this email finds you well. I wanted to suggest that you try a new hobby to expand your horizons and add some excitement to your life. [Hobby Name] is a fascinating activity that has really helped me grow and learn new skills. Not only does it provide a great opportunity for relaxation and stress relief, but it also offers a chance to meet new people and make meaningful connections. I think you would really enjoy it and I would love to see you give it a shot. Let me know if you have any questions or if you want to learn more.\newline

    \end{tcolorbox}
    \caption{
        Example of
        instruction data using Self-instruct. 
        We show the prompt which is given to Llama-2-chat-7B, and the completion with the watermarking method used in Sec.~\ref{chap6/sec:instruction}.
        We do not observe a noticeable decrease in quality compared to the original completion.
    }
    \label{chap6/fig:self-instruct-examples}
\end{figure}

%% file: chapter-6/sections/5-analyzes.tex
\section{Investigating radioactivity}\label{chap6/sec:fine-tuning-abl}

\autoref{chap6/sec:instruction} considers detection in a practical scenario.
This section further studies what influences radioactivity from different angles like fine-tuning, watermarking algorithm, and data distribution.

\subsection{Fine-tuning}

\input{chapter-6/tables/fine-tuning-abl.tex}

We first study the influence of fine-tuning on the same setup as Sec.~\ref{chap6/sec:instruction}, with regards to: 
(a) the learning rate,
(b) the fine-tuning algorithm,
\eg, with Q-LoRA~\citep{dettmers2023qlora} a widely used method for efficient fine-tuning;
(c) the number of epochs,
(d) the model size.
We fine-tune $\B$ with the same dataset of $\rho=100\%$ watermarked instructions and the same parameters.
We detect radioactivity in the \emph{open-model} / \emph{unsupervised} setting.
This is done on $N=10$k next-predictions, and where the texts that are fed to $\B$ are watermarked instructions generated with $\A$.
\autoref{chap6/tab:ft-abl} reports the results. The more the model fits the data, the easier its radioactivity is to detect.
For instance, multiplying the learning rate by $10$ almost doubles the average $\logpval$ of the test.

\subsection{Bigger teachers}\label{chap6/app:bigger-teachers}

\begin{table}[t!]
    \centering
    \caption{
        Influence of the teacher model size on radioactivity detection.
    }
    \label{chap6/table:results_Teachers}
    \footnotesize
    \begin{tabular}{ *{5}{l} }
        \toprule
        Teacher & Without & 7B & 13B & 65B \\
        \midrule
        NQ & 3.2 & 5.6 & 5.4 & 5.8 \\
        GSM8k & 10.0 & 11.1 & 10.4 & 11.0 \\
        MMLU & 28.4 & 31.0 & 32.9 & 33.8 \\
        \midrule
        $\log_{10}$ p-value & -0.3 & -32.4 & -31.1 & -31.7 \\
        \bottomrule
    \end{tabular}
\end{table}

We conduct the same experiment replacing the Llama-2-chat-7B teacher model by the 13B or 65B versions.
\autoref{chap6/table:results_Teachers} reports the results for benchmarks NQ, GSM8k, and MMLU and the average $\logpval$ of the radioactivity detection test under the same conditions as the previous paragraph.
Our observations align with the previous conclusions: watermarking does not significantly affect the benchmarks (except for MMLU where the improvement appears larger). 
Moreover, the detection of radioactivity is not significantly impacted by the teacher model size.

\subsection{Watermarking method}

\paragraph{Multi-bit scenario.} 

We adopt the same framework as in Sec.~\ref{chap6/sec:instruction}, but we use the watermarking method of \cite{yoo2023advancing}, \aka, MPAC.
It is a multi-bit method, where the watermark is a binary message of size $n$.
More precisely, we take bits 2 by 2 to generate a message $m = m_1 m_2 \ldots m_b$ with the $r$-ary $m_i = 0, 1, 2,$ or $3$, corresponding to $r=4$ and $b = n/2$ with the notations from the original paper.
The method proceeds as the one of \cite{kirchenbauer2023reliability} by altering the logits before generating a token.
However, the hash created from the watermarked window and the key is now used 
(1) to randomly partition the vocabulary into $r$ disjoint sets, 
(2) to select the position $i$ and corresponding $m_i$ that is hidden for this particular token.
A bias $\delta$ is added to the logits of the tokens belonging to the $m_i$-th set.
Given a text under scrutiny, the extraction reverses the process to find which $r$-ary is the most likely for each position $i$ of the message.

\begin{figure}[b!]
    \centering
    \includegraphics[width=0.99\linewidth,clip, trim=0 0 0 0cm]{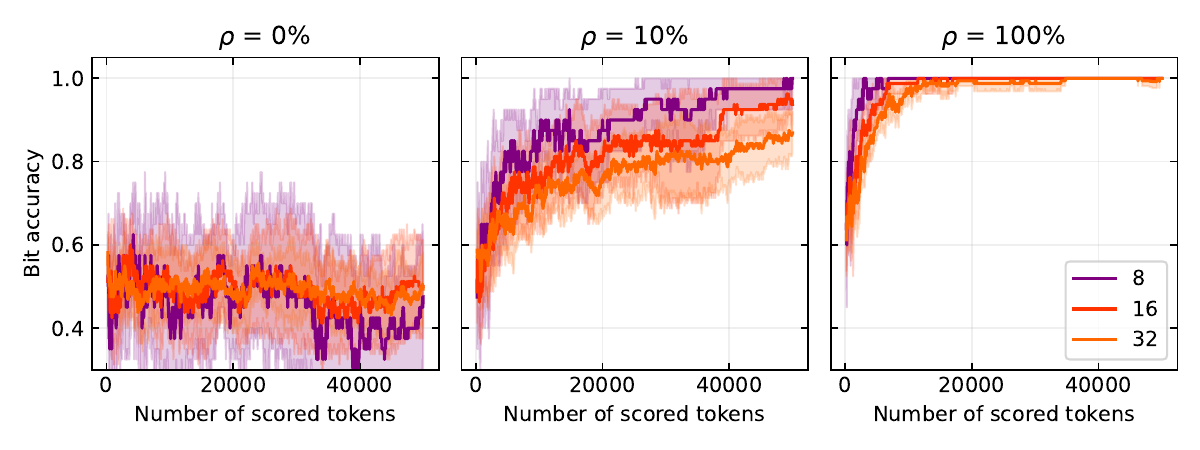}
    \caption{
        Bit accuracy when watermarked instruction data are generated with MPAC~\citep{yoo2023advancing}, against the number of scored tokens generated by the fine-tuned model.
        This is done under the supervised/closed-model setup, for various lengths ($n$=8, 16, 32) of the message.
    }
    \label{chap6/fig:bit-accuracy}
\end{figure}

We report in Fig.~\ref{chap6/fig:bit-accuracy} the extraction results in the the supervised/closed-model setup.
We filter and deduplicate the tokens as in Sec.~\ref{chap6/sec:instruction}, and plot the observed bit accuracy against the number of scored tokens -- note that since the watermark is a binary message, we now measure the bit accuracy of the extraction, instead of the \pval\ of the detection.
This is done for several lengths of the binary message. 
Every experiment is run $10$ times for different text output by $\B$, which explains the $95\%$ confidence interval in the plots.
We observe as expected that the bit accuracy significantly increases with the proportion of watermarked data in the fine-tuning data, and that the longer the message, the harder it is to extract it.
This suggests that radioactivity still holds in the multi-bit scenario, and could therefore be used to identify a specific version of the model or a specific user from which the data was generated.

\paragraph{Watermark window size.} 
To reduce the experimental requirements for generation, training and detection -- and introduce more variety to the data under study -- we now prompt $\A$=Llama-2-7B with the beginnings of Wikipedia articles in English and generate the next tokens with or without watermarking. 
We then fine-tune $\B$=Llama-1-7B on the natural prompts followed by the generated answers.
The fine-tuning is done in 1000 steps, using batches $8\times2048$ tokens (similarly to Sec.~\ref{chap6/sec:instruction}).
This section fine-tunes $\B$ on $\rho=100\%$ English watermarked texts.

\autoref{chap6/tab:exp_kgram} highlights that the confidence of the detection decreases for larger window sizes $k$ when fixing the \pval\ of the watermark detection of the training texts.
There are two explanations. 
First, for lower $k$, the chances that a $k$-gram repeats in the training data are higher, which increases its memorization.
Second, the number of possible $k$-grams is $|\V|^k$ and therefore increases with $k$, while the number of watermarked tokens is fixed. 
Thus, at detection time, the number of radioactive $k$-grams decreases with increasing $k$, diminishing the test's power. 
This experiment also demonstrates that the methods of \cite{aaronson2023watermarking} and \cite{kirchenbauer2023reliability} behave the same way.

\paragraph{Discussion on other watermarking schemes.}
Our goal is to derive a general method for detecting radioactivity in decoding-based watermarks, rather than testing all watermarking schemes or identifying the most radioactive one. 
We focus on the works of~\citet{aaronson2023watermarking, kirchenbauer2023watermark}, which are representative of the two main families of methods and provide reliable \pval s for detection. These schemes rely on key management using the hash of previous tokens, a mechanism also used by other potentially radioactive schemes~\citep{lee2023wrote, fu2024gumbelsoft}.

Some LLM watermarking schemes do not rely on hashing, such as ``semantic'' watermarks~\citep{liu2023semantic, liu2024adaptive, fu2024watermarking} and those using pre-defined key sequences~\citep{kuditipudi2023robust}. 
However, we do not evaluate the radioactivity of these methods due to the lack of \pval\ computation, a limitation also noted in other studies~\citep{piet2023mark}. 
For example, the complexity of computing \pval s for the scheme in~\citet{kuditipudi2023robust} would be prohibitively expensive, requiring $10^{16}$ times more operations than the schemes presented previously\footnote{
    \citet{kuditipudi2023robust} use the Levenshtein distance with complexity $O(mnk2)$, where $m$ is the number of tokens, $n=256$, and $k=80$ (default parameters from the authors). 
    This results in a complexity approximately $10^6$ times greater than previous schemes. 
    To evaluate \pval s of $10^{-10}$, a Monte Carlo simulation would require running the statistic $10^{10}$ times, a total $10^{16}$ more operations.
}. 
This limitation would also apply to works based on the same key management~\citep{christ2023undetectable, liu2023semantic}.

\subsection{Data distribution}

\begin{table}[t!]
        \centering
        \caption{
            Influence of the target text distribution on detection.
            $\B$ is prompted with beginnings of Wikipedia articles in the corresponding language, and detection is done on generated next tokens. 
            For each language, we score $N=250$k $k$-grams using the \textit{closed-model} setting.
        }
        \label{chap6/tab:exp_language}
        \footnotesize
        \begin{tabular}{ c *{6}{c} }
            Language &  English & French & Spanish & German & Catalan & Combined p-value (Fisher) \\
            \shline
                $\logpval$ & $<$-50 & -7.8 &  -5.7 &  -4.0 &  -2.1 & $<$-50  \\
        \end{tabular}
    \end{table}

\paragraph*{Radioactivity detection in different languages.}\label{chap6/par:distrib}
We consider an unsupervised setting where Alice has no prior knowledge about $D^\A$, the data generated with $\A$ used to fine-tune $\B$.
As an example, Alice does not know the language of $D^\A$, which could be Italian, French, Chinese, etc. 
We run the detection on text generated by $\B$, with prompts from Wikipedia in different languages.
The confidence of the test on another language -- that might share very few $k$-grams with $D^\A$ -- can be low, as shown in Tab.~\ref{chap6/tab:exp_language}.

Alice may, however, combine the \pval s of each test with Fisher's method.  
This discriminates against $\mathcal{H}_0$: ``\textit{none of the datasets are radioactive}'', under which the statement ``\textit{Bob did not use any outputs of $\A$}'' falls.
Therefore, the test aligns with our definition of model radioactivity as per definition~\ref{chap6/def:model_radioactivity}.
From Tab.~\ref{chap6/tab:exp_language}, Fisher's method gives a combined \pval\ of $<10^{-50}$. 
Thus, even if Alice is unaware of the specific data distribution generated by $\A$ that Bob may have used to train $\B$ (\eg, problem-solving scenarios), she may still detect radioactivity by combining the significance levels.

\paragraph*{Mixing instruction datasets from different sources.}

\begin{table}[t!]
    \centering
    \begin{minipage}{0.48\textwidth}
        \centering
        \caption{
            Mixing instruction datasets from different sources.
            The fine-tuning is done with the setup presented in Sec.~\ref{chap6/sec:instruction}, with $\rho$=$10\%$ of watermarked data, mixing either with human or synthetic instructions.
        }
        \label{chap6/table:data-sources}
        \footnotesize
        \begin{tabular}{c|c}
            \toprule
            Major data source
            & Average $\logpval$ \\
            \midrule
            Machine & -15 \\
            Human & -32 \\
            \bottomrule
        \end{tabular}
    \end{minipage}
    \hfill
    \begin{minipage}{0.48\textwidth}
        \centering
        \caption{
            Sequential fine-tuning to remove the watermark traces when fine-tuning.
            The first fine-tuning is done with the setup presented in Sec.~\ref{chap6/sec:instruction}, with $\rho$=$10\%$ of watermarked data, and the second on OASST1.
            }
        \label{chap6/table:results_FineTuning}
        \footnotesize
        \begin{tabular}{c|c}
        \toprule
        Second fine-tuning & Average $\logpval$ \\
        \midrule
        \xmarkg & -15 \\
        \cmarkg & -8 \\
        \bottomrule
        \end{tabular}
    \end{minipage}
\end{table}

We conduct the same experiment as in Sec.~\ref{chap6/sec:instruction}, but replace the non-watermarked synthetic instructions by human-generated ones (from the Open Assistant dataset OASST1~\citep{kopf2024openassistant}).
We report in Tab.~\ref{chap6/table:data-sources} the detection results in the open / unsupervised scenario, with $\rho=10\%$ of watermarked data.
Interestingly, with the exact same setting as the one explored in Sec.~\ref{chap6/sec:detection-setup}, the radioactivity signal is stronger in this case.
Our speculation is that this might be due to fewer overlapping sequences of $k$+1-grams between the two distributions.

\subsection{Possible defenses}

We now assume Bob is aware of watermarking radioactivity and tries to remove the watermark traces before, during or after fine-tuning.
For instance, he might attempt to rephrase the watermarked instructions, use a differentially private training, or fine-tune his model on human-generated data -- which we do in the following experiment.
The logic is overall the same as previously pointed out in the fine-tuning ablations.
If the original watermark is weaker or if the fine-tuning overfits less, then radioactivity will be weaker too.
Therefore, the radioactivity detection test will be less powerful, but given a sufficient amount of data, it may still be able to detect traces of the watermark.

\paragraph{Radioactivity ``purification''.}\label{chap6/ref:purification}
We investigate the impact of a second fine-tuning on human-generated data to remove the watermark traces, through the following experiment.
After having trained his model on a mix of watermarked and non-watermarked data, as in Sec.~\ref{chap6/sec:instruction}, Bob fine-tunes his model a second time on human-generated text (from OASST1, as in the previous paragraph), with the same fine-tuning setup.
\autoref{chap6/table:results_FineTuning} shows that the second-fine-tuning divides by $2$ the significance level of the statistical test, although it does not completely remove the watermark traces.

%% file: chapter-6/tables/fine-tuning-abl.tex
\begin{table}
    \caption{
        Influence of the model fine-tuning on the radioactivity.
        We report the $\logpval$ for $10$k scored observations (lower means more radioactive).
        {\setlength{\fboxsep}{2pt}\colorbox[HTML]{F2F2F2}{Gray}} indicates values used in Sec.~\ref{chap6/sec:instruction}.
    }
    \label{chap6/tab:ft-abl}
    \renewcommand{\arraystretch}{1.2}
    {\footnotesize
    \centering
  	\subfloat[ Learning rate.]{
            \begin{minipage}{0.23\linewidth}
                {\begin{center}
                    \begin{tabular}{ccc}
                       \colorcell  $10^{-5}$ & $5\cdot 10^{-5}$ & $10^{-4}$\\
                        \shline
                       \colorcell  -32.4 & -49.6 & -58.0\\
                    \end{tabular}
                \end{center}}
            \end{minipage} 
        } \hspace{0.03\linewidth}
  	\subfloat[ Epoch.]{
            \centering
            \begin{minipage}{0.27\linewidth}
                {\begin{center}
                    \begin{tabular}{ *{4}{c} }
                        1 & 2 & \colorcell 3 & 4 \\
                        \shline
                        -20.8 & -29.2 & \colorcell -33.2 & -34.8 \\
                    \end{tabular}
                \end{center}}
            \end{minipage} 
        } \hspace{0.03\linewidth}
        \subfloat[ Adapters.]{
            \centering
            \begin{minipage}{0.17\linewidth}
                {\begin{center}
                    \begin{tabular}{cc}
                        \colorcell Full & Q-LoRA \\
                        \shline
                        \colorcell -32.4 & -11.0 \\
                    \end{tabular}
                \end{center}}
            \end{minipage} 
        } \hspace{0.03\linewidth}
        \subfloat[ Model size.]{
            \centering
            \begin{minipage}{0.14\linewidth}
                {\begin{center}
                    \begin{tabular}{cc}
                       \colorcell  7B & 13B\\
                        \shline
                       \colorcell -32.4 & -33.2 \\
                    \end{tabular}
                \end{center}}
            \end{minipage} 
        } 
    }
 \end{table}

%% file: chapter-6/sections/7-approaches.tex
\section{Discussion on other approaches}\label{chap6/sec:discussion-other-approaches}

\subsection{Membership inference}\label{chap6/par:mia_wm}

\paragraph*{Method.}
In the open-model/supervised case, MIA evaluates the radioactivity of one sample/sentence by observing the loss (or perplexity) of $\B$ on carefully selected sets of inputs.
The perplexity is expected to be smaller on samples seen during training.
We extend this idea for our baseline radioactivity detection test of a non-watermarked text corpus.
The corpus of texts is divided into sentences (of 256 tokens) and $\B$'s loss is computed on each sentence. 
We calibrate it with the zlib entropy~\citep{roelofs2017zlib}, as done by \citet{carlini2021extracting} for sample-based MIA. 
The goal of the calibration is to account for the complexity of each sample and separate this from the over-confidence of $\B$. 

We test the null hypothesis $\mathcal{H}_0$: ``\textit{the perplexity of $\B$ on $\Tilde{D}^{\A}$ has the same distribution as the perplexity on new texts generated by $\A$}''.
Indeed, if $\B$ was not fine-tuned on portions of $\Tilde{D}^\A$, then necessarily $\mathcal{H}_0$ is true.
To compare the empirical distributions we use a two-sample Kolmogorov-Smirnov test~\citep{massey1951kolmogorov}. 
Given the two cumulative distributions $F$ and $G$ over loss values, we compute the K-S distance as $d_{\mathrm{KS}}(F,G) = \mathrm{sup}_x |F(x) -G(x)|$.
We reject $\H_0$ if this distance is higher than a threshold, which sets the \pval\ of the test, and conclude that $\Tilde{D}^{\A}$ is radioactive for $\B$.
This is inspired by \citet{sablayrolles2018d}, who perform a similar K-S test in the case of image classification. 
It significantly diverges from the approach of~\citet{shi2023detecting}, which derives an empirical test by looking at the aggregated score from one tail of the distribution. 

\paragraph*{Experimental results}

We proceed as in Sec.~\ref{chap6/sec:radioactivity_detection} for the setup where MIA is achievable:
Alice has an \emph{open-model} access to $\B$ and is aware of all data $\Tilde{D}^\A$ generated for Bob (supervised setting). 
Bob has used a portion $D^\A$ for fine-tuning $\B$, given by the degree of supervision $d$, as defined in Sec.~\ref{chap6/sec:degreeofsupervision}.
We use the K-S test to discriminate between the calibrated perplexity of $\B$ on: $\mathcal D_{(0)}$ containing 5k instruction/answers (cut at 256 tokens) that were not part of $\B$'s fine-tuning; and $\mathcal D_{(d)}$ containing $(1/d)\times$5k instruction/answers from which $5k$ were.
Distribution $\mathcal D_{(d)}$ simulates what happens when Bob generates a lot of data and only fine-tunes on a few.

\definecolor{curve1}{HTML}{8B188B}
\definecolor{curve2}{HTML}{FFA319}
\begin{figure}[b!]
    \centering
    \begin{subfigure}[b]{0.45\textwidth}
        \centering
        \includegraphics[width=\linewidth]{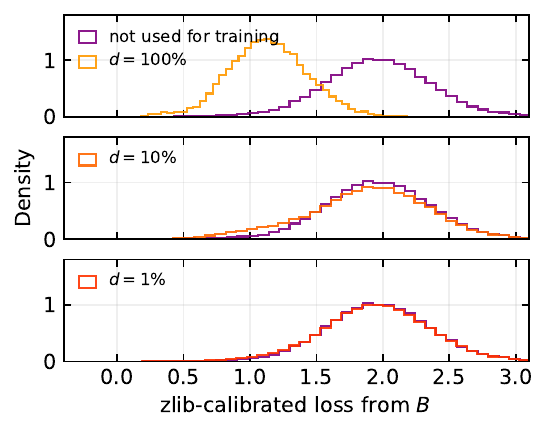}
        \caption{
            Distributions of the calibrated loss of $\B$ across two types of distributions generated by $\A$: 
            texts generated by $\A$ outside of $\B$'s fine-tuning data ({\color{curve1} purple}), texts of $\Tilde{D}^{\A}$ of which $d\%$ were used during training ({\color{curve2} orange}).
        }
        \label{chap6/fig:calibrated-loss}
    \end{subfigure}\hfill
    \begin{subfigure}[b]{0.5\textwidth}
        \centering
        \includegraphics[width=\linewidth,clip, trim=0 0 0 0cm]{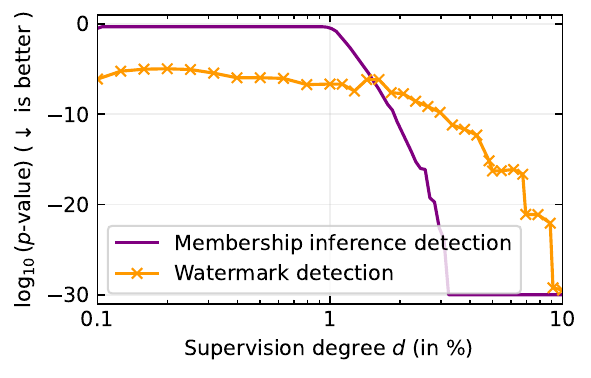}
        \caption{
            We report the \pval s of the {\color{curve1} K-S detection test} (no WM when training) and of the {\color{curve2} WM detection} ($\rho=5\%$ of WM when training) against the degree of supervision $d$ (proportion of Bob's training data known to Alice).
        }
        \label{chap6/fig:mia_vs_wm}
    \end{subfigure}
    \caption{
        Comparative analysis of membership inference and watermarking for radioactivity detection, in the open-model setup.
        (\emph{Left}) MIA aims to detect the difference between the two distributions. 
        It gets harder as $d$ decreases, since the actual fine-tuning data is mixed with texts that Bob did not use.
        (\emph{Right}) Therefore, for low degrees of supervision ($<2\%$), MIA is no longer effective, while WM detection gives \pval s lower than $10^{-5}$.
    }
    \label{chap6/fig:mia-comparative-analysis}
\end{figure}

\autoref{chap6/fig:calibrated-loss} compares the distributions for $d=0$ and $d>0$. 
As $d$ decreases, the data contains more texts that Bob did not fine-tune on, so the difference between the two perplexity distributions is fainter.
The direct consequence is that the detection becomes more challenging.
\autoref{chap6/fig:mia_vs_wm} shows that when $d>2\%$, the test rejects the null hypothesis at a strong significance level: 
$p < 10^{-5}$ implies that when radioactive contamination is detected, the probability of a false positive is $10^{-5}$.
It is random in the edge case $d=0$, the unsupervised setting where Alice lacks knowledge about the data used by Bob. 
In contrast, radioactivity detection on watermarked data succeeds in that setting.

\subsection{IP protection methods}\label{chap6/sec:ipp}

There are active methods that embed watermarking in a text specifically to detect if a model is trained on it~\citep{zhao2023protecting,he2022cater, he2022protecting}. 
Our setup is a bit different because Alice's primary goal is not to protect against model distillation but to make her outputs more identifiable. 
Consequently, ``radioactivity'' is a byproduct.
Although our focus is not on active methods, we discuss three state-of-the-art methods for intellectual property protection and their limitations in our detection setting.

\paragraph{\citet{zhao2023protecting}.}
The goal is to detect if a specific set of answers generated by Alice's model has been used in training. 
There are three main limitations.
The authors design a watermark dependent on the previous prompt (and unique to it). 
Each detection algorithm (2 and 3) assumes that the sample probing data $D$ is from the training data of the suspect model $\B$.
Therefore, the method is only demonstrated in the \textit{supervised} setting.
Moreover, all experiments are performed in the \textit{open}-model setting, where Alice has access to $\B$'s weights, except for Sec.~5.2 ``Watermark detection with text alone'' (still assuming a \textit{supervised} access), where one number is given in that setting.
Finally, it does not rely on a grounded statistical test and requires an empirically set threshold.

\paragraph{\cite{he2022protecting} and \cite{he2022cater}.} 
These methods substitute synonyms during generation to later detect an abnormal proportion of synonyms in the fine-tuned model.
However, the main hypothesis for building the statistical test -- the frequency of synonyms in a natural text is fixed -- fails when scoring a large number of tokens.
Using the \href{https://github.com/xlhex/NLG_api_watermark}{official author's code} on non-watermarked instruction-answers yields extremely low \pval s, making the test unusable at our scale, as shown in \autoref{chap6/table:pvalues_ginsew}.

\begin{table}[t!]
\centering
\caption{\pval s for non-watermarked instruction-answers}
\label{chap6/table:pvalues_ginsew}
\footnotesize
\begin{tabular}{ *{3}{l} }
    \toprule
    Number of lines & Number of characters (in thousands) & \pval \\
    \midrule
    10 & 1.5 & 0.76 \\
    100 & 30.9 & 0.16 \\
    500 & 148.3 & 2.2 $\times 10^{-7}$ \\
    1000 & 333.9 & 8.8 $\times 10^{-13}$ \\
    \bottomrule
\end{tabular}

\end{table}

%% file: chapter-6/sections/6-additional.tex
\section{Details on \pval s}\label{chap6/sec:additional}

This section provides more information on the experimental setup, reporting, and correctness experiments for the tests used in the chapter.

\subsection{Reporting}\label{chap6/app:repporting}

\begin{figure}[b!]
    \centering
    \includegraphics[width=0.5\linewidth]{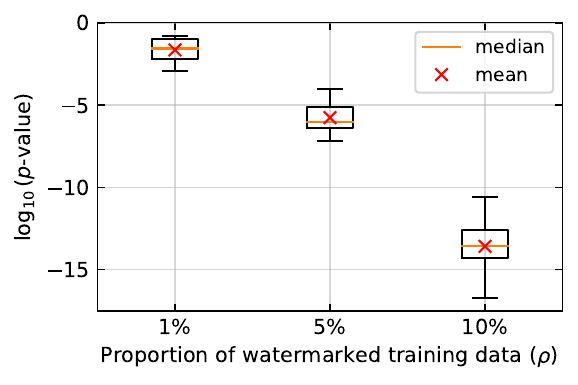}
    \captionsetup{font=small}
    \caption{Box plot for the $\logpval$ in the open/unsupervised setting with varying $\rho$, the proportion of watermarked fine-tuning data.}
    \label{chap6/fig:box_plot_open}
\end{figure}

\paragraph{Interpretation of the average $\logpval$.}
Given that \pval s often span various orders of magnitude, we report the average of the $\logpval$ over multiple runs rather than the average of the \pval s themselves. 
We interpret the average $\logpval$ as though it could be directly read as a \pval, although a direct translation to a rigorous statistical \pval\ is not possible.
Fig.~\ref{chap6/fig:box_plot_open} shows box-plots with additional statistics.

\paragraph{Average over multiple runs.} 
Due to computational constraints, standard deviations for the $\logpval$ are not calculated across multiple models trained on different instruction data for each setting. 
Instead, we generate the same volume of data (14M tokens) in addition to the data used to fine-tune the model. 
In the open-model setting, we run detection on ten distinct chunks of this additional data. 
In the closed-model setting, we prompt the model with ten different chunks of new sentences and score the responses.

\begin{figure}[b!]
    \begin{minipage}{0.48\textwidth}
        \centering
        \includegraphics[width=0.99\linewidth,clip, trim=0 0 0 0cm]{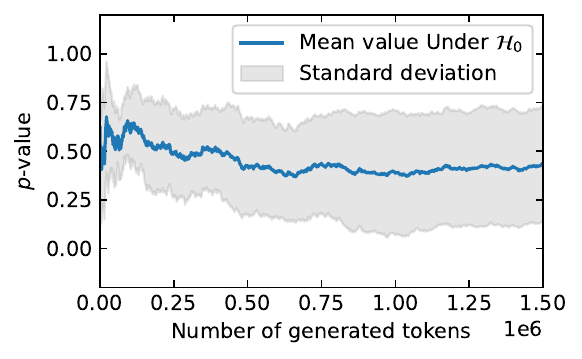}
        \caption{
            \pval\ under $\mathcal{H}_0$ with closed-model access.
            We fine-tune $\B$ on non-watermarked instructions and prompt $\B$ with watermarked instructions, scoring the distinct $(k+1)$-grams from the answers, excluding $k$-grams from the instruction. 
        }
        \label{chap6/fig:pvalu_under_H0_closed}
    \end{minipage} \hfill
    \begin{minipage}{0.48\textwidth}
        \centering
        \includegraphics[width=0.99\linewidth,clip, trim=0 0 0 0cm]{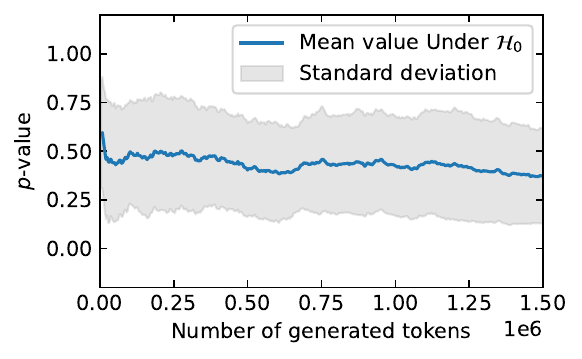}
        \caption{
            \pval\ under $\mathcal{H}_0$ with open-model access. 
            We fine-tune $\B$ on non-watermarked instructions, perform the open-model detection and score distinct $(k+1)$-grams only on $k$-grams that $\B$ did not previously attend to when generating the token. 
        }
        \label{chap6/fig:pvalu_under_H0}
    \end{minipage}
\end{figure}

\subsection{Correctness experiments}\label{chap6/app:correctness}

We validate the correctness of our statistical tests by examining the null hypothesis $\mathcal{H}_0$, which represents when the model was not fine-tuned on any watermarked data ($\rho=0$).
Our goal is to show that the \pval\ is approximately uniform under $\mathcal{H}_0$, with a mean of $0.5$ and a standard deviation of $\approx 0.28$.
Instead of fine-tuning model $\B$ with various datasets, we vary the hyper-parameters of the detection algorithm at a fixed fine-tuned model to save computation and memory. 
This approach is not sufficient on its own to establish the validity of the test, but it provides some evidence for it.
Here, $\B$ is the model fine-tuned on non-watermarked instructions as described in Sec.~\ref{chap6/sec:instruction}.

\paragraph{Closed-model.}
We prompt $\B$ with $\approx10$k watermarked instructions using the method by~\citet{kirchenbauer2023watermark}, $\delta=3$ and $k=2$, and three different seeds $\mathsf{s}$. 
We score the answers using the proposed de-duplication and repeat this $10$ times on different datasets. 
\autoref{chap6/fig:pvalu_under_H0_closed} shows that the average \pval\ is close to $0.5$ and the standard deviation is close to $0.28$, as expected under $\mathcal{H}_0$.

\paragraph{Open-model.}
We generate text with eight distinct watermarking methods (methods of~\citet{aaronson2023watermarking} and ~\citet{kirchenbauer2023watermark} for $k\in\{1,2,3,4\}$).
We divide each dataset into three and apply the radioactivity detection test on these 24 segments, each containing over 1.5 million tokens. 
We score the distinct $(k+1)$-grams from the answers, excluding $k$-grams from the instruction.
\autoref{chap6/fig:pvalu_under_H0} shows again that the average \pval\ is close to $0.5$ and the standard deviation is close to $0.28$.

%% file: chapter-6/sections/8-conclusion.tex
\section{Conclusion}

This chapter formalizes the concept of ``radioactivity'' in language models.
It introduces methods to detect traces that LLM-generated texts leave when used as training data.
This makes it possible to identify with high confidence if outputs from a watermarked model have been used to fine-tune another one.

There are (at least) two limitations that are worth mentioning. 
First, we do not extensively discuss how attempts to remove the original watermark may impact radioactivity, except in Sec.~\ref{chap6/ref:purification} where Bob re-fine-tunes his model. 
A skilled attacker may be able to eliminate these traces. 
Second, the computational efficiency of the algorithms is linear in the number of tokens to analyze and depends on the suspected LLM. 
In the closed-model setting, about 1000 queries to the model are needed to confidently detect radioactivity, which could be a limitation if the suspect model is only accessible through an API.

%% file: chapter-7/main.tex
\chapter{Functional Invariants to Watermark Large Transformers}\label{chapter:invariants}

This chapter is based on the paper \fullcite{fernandez2024functional}.

The rapid growth of transformer-based models increases the concerns about their integrity and ownership insurance. 
In particular, models are vulnerable to leaks and unauthorized redistribution, which can result in significant financial losses and reputational damage for companies. 
Model watermarking addresses this issue by embedding a unique identifier into the model weights, while preserving its performance. 
It may then be extracted to identify the model's owner.
However, most existing approaches require to optimize the weights to imprint the watermark signal, which is not suitable at scale due to the computational cost.
This chapter explores watermarks with virtually no computational cost, applicable to a non-blind white-box setting (assuming access to both the original and watermarked networks). 
They generate functionally equivalent copies by leveraging the invariance of models, via operations like dimension permutations or scaling/unscaling. 
This enables to watermark models without any change in their outputs and remains stealthy.
Experiments demonstrate the effectiveness of the approach  and its robustness against various model transformations (fine-tuning, quantization, pruning), making it a practical solution to protect the integrity of large models.

\newpage
\input{chapter-7/sections/1-introduction}

\input{chapter-7/sections/2-related}

\input{chapter-7/sections/3-method}

\input{chapter-7/sections/4-experiments}

\input{chapter-7/sections/8-conclusion}

%% file: chapter-7/sections/1-introduction.tex
\section{Introduction}

Large-scale transformer models are a leap forward in the field of machine learning, with large language models like GPT-4~\citep{openai2023gpt} and Llama~\citep{touvron2023llama}, or vision ones like ViT-22b~\citep{dehghani2023scaling} and DINOv2~\citep{oquab2023dinov2}.
As these models grow in size, it becomes important to be able to trace weights to safeguard it from unauthorized usage and distribution.
Note that, differently to the previous part~\ref{part:genai-tracing}, we are interested in tracing the model itself, not content it generates.
Watermarking deep neural networks (DNN)~\citep{uchida2017embedding, adi2018turning} aims to address this issue by embedding an identifier into the model, which can be extracted to identify the model's owner in case of unauthorized redistribution.

However, watermarking large transformer models poses new challenges. 
Current watermarking methods optimize the weights to infuse the watermark, either during pre-training or by fine-tuning the weights with additional losses.
While these techniques have shown success for smaller models, they become computationally infeasible for large-scale models and for the burgeoning number of potential users and applications.

To address these challenges, we introduce a new approach to watermarking large transformers, when access to both the original and watermarked model is granted, \ie, in a non-blind white-box setting (as defined in Sec.~\ref{chap0/sec:other-considerations}).
Our method capitalizes on the inherent invariance of transformers. 
For a given model, it generates equivalent copies that serve as carriers for arbitrary signatures, by employing operations such as dimension permutation and coupled matrix multiplications.
We therefore create model replicas without changing the model's outputs and without any training or fine-tuning.
We conduct experiments on Llama models ranging from 7 to 70 billion parameters to evaluate the applicability of our approach and its robustness against model processing (\eg, fine-tuning, pruning, quantization, etc.).
We also discuss the main drawbacks of this setting. 

\begin{figure}[t]
    \centering
    \includegraphics[width=0.99\linewidth]{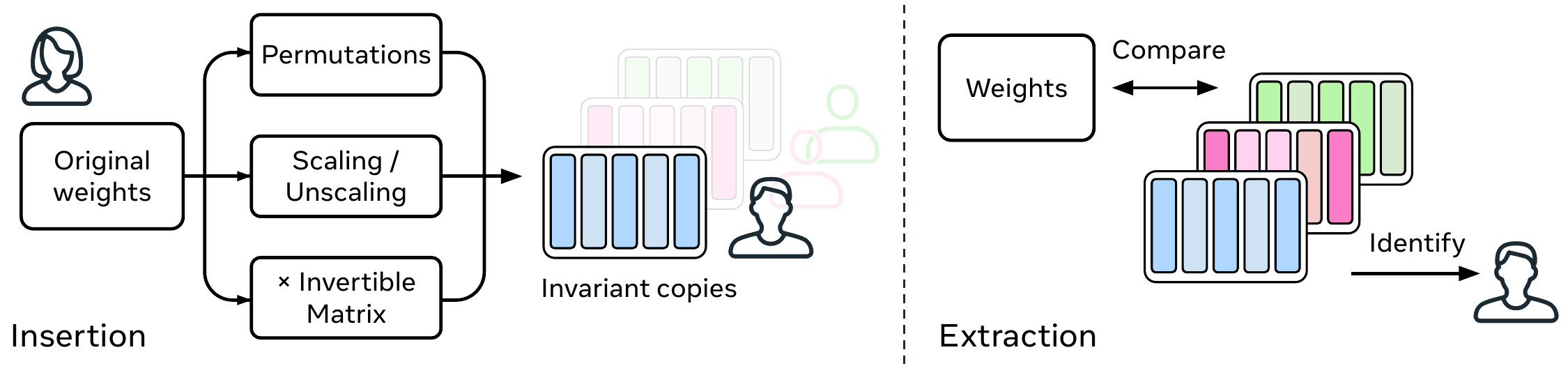}
    \caption{
        Overview of watermarking through invariance. 
        We identify each model by applying invariance operations to the original weights.
        The resulting model is functionally equivalent to the original, but carries an identifier that can be extracted to identify the model's owner.
    }
    \label{chap7/fig:overview}
\end{figure}

The chapter is organized as follows:
\begin{itemize}
    \item 
    \autoref{chap7/sec:related} provides an overview of related works on DNN watermarking and background on the transformer architecture;
    \item
    \autoref{chap7/sec:method} details three of the transformer's invariants and how to exploit them for watermarking;
    \item
    \autoref{chap7/sec:experiments} presents experimental results on the Llama LLMs, and discusses the robustness of the watermarking against various model transformations.
\end{itemize}

\paragraph*{Problem statement.}
A provider \emph{Alice}, distributes her model to various users \emph{Bob} (either individuals or organizations).
She aims to trace the model back to a specific user, in case of unauthorized distribution or leaks. 
As a precautionary measure, Alice embeds a unique signature in the model's weights for each user. 
In a white-box setting, Alice has access to the models' weights and extracts the signature from it to identify Bob.
Besides, Bob may evade detection intentionally (trying to remove the watermark) or unintentionally (fine-tuning, quantization, etc.).

This setting is quite common. 
Indeed few entities (``Alices'') have the necessary computation resources, data and expertise to generate the base model. 
For example, the training of the 70-B Llama model took around $1$B GPU-hours.
Therefore, there are few variants of such large models in the world. 
Besides, when Bob gains access to the base model, it is common that he transforms it and that it re-emerges in a public forum or through another channel, so that Alice can analyze it. 
This can be either because Bob re-distributed it or because Alice sought the model through legal channels, as suggested by~\citet{fan2021deepip}.
For example, many variants of the Llama models have been fine-tuned on instruction datasets and been made available online.

%% file: chapter-7/sections/2-related.tex
\newcommand{\att}[1]{\mathrm{Att}^{#1}}
\newcommand{\ffn}[1]{\mathrm{Ffn}^{#1}}
\newcommand{\lnatt}[1]{\mathrm{Ln}_{\mathrm{att}}^{#1}}
\newcommand{\lnffn}[1]{\mathrm{Ln}_{\mathrm{ffn}}^{#1}}
\newcommand{\lnout}{\mathrm{Ln}_{\mathrm{out}}}

\section{Related work and technical background}\label{chap7/sec:related}

\subsection{Deep Neural Network (DNN) Watermarking}\label{chap7/par:watermarking} 

DNN watermarking robustly embeds a unique identifier into the model without affecting its performance, in order to later verify the model's identity.
\textit{robustness}: the watermark should be detected even after the model has been modified.
Modifications may be unintentional -- models are fine-tuned, pruned and quantized
-- or intentional -- adversaries may try to remove the watermark or embed their own~\citep{fan2019rethinking, zhang2020passport, kallas2022rose}.
For instance, some adversarial transforms employ invariance operations in neurons and ReLU layers to evade detection~\citep{yan2023rethinking}, in a similar fashion as the techniques of this work.

We distinguish between white-box and black-box settings, depending on whether the model weights are accessible at verification time, or only through a remote API.
In white-box, the pioneering work~\citep{uchida2017embedding} embeds watermarks into the DNN's weights. 
A regularization loss term during training constrains the weights to carry a specific signal, while minimizing the impact on the model's performance. 
The watermark is then retrieved directly by analyzing the model's weights.
The  Deep·Signs·Marks~\citep{darvish2019deepsigns,chen2019deepmarks} extends this to target black-box settings and propose building collusion-resistant watermarks, 
RIGA~\citep{wang2021riga} improves its covertness and robustness, 
and greedy residuals~\citep{liu2021watermarking} improves the selection of the weights to modify.

Another line of work, called trigger-set based methods, embeds the watermark in the behavior of the model with regards to certain inputs. 
A recurrent idea is to use \emph{backdoors}, \ie, memorize certain sequences of input-output pairs~\citep{adi2018turning, zhang2018protecting}. 
Watermarking generative models is also an active field of research, either by employing triggers~\citep{lim2022protect, ong2021protecting}, or by watermarking their outputs~\citep{fernandez2023stable, kim2023wouaf}.

The literature on watermarking large models is scarce, and none of the current papers operate at our scale. 
The most recent works~\citep{liu2023trapdoor, jiang2023ipcert, tondi2024robust} focus on ResNet/AlexNet ($\approx$20M parameters). 
PLMmark~\citep{li2023plmmark} also needs training and evaluates at most on BERT-large, around 300M parameters. 
This is 100 times smaller than the models we consider in our work (\eg, Llama-30B, Llama-70B). 
Previous methods could be adapted in the context of LLMs, but all of them would require training or at least fine-tuning. 
Their impact on the quality of the text generation and the robustness of the watermark is also not demonstrated. 
Thus, the feasibility of existing watermarking methods to these models remains an open question.

\subsection{Transformers}

\begin{figure}[b!]
    \centering
    \includegraphics[width=0.99\linewidth, clip, trim={1cm 0 1cm 0.5cm}]{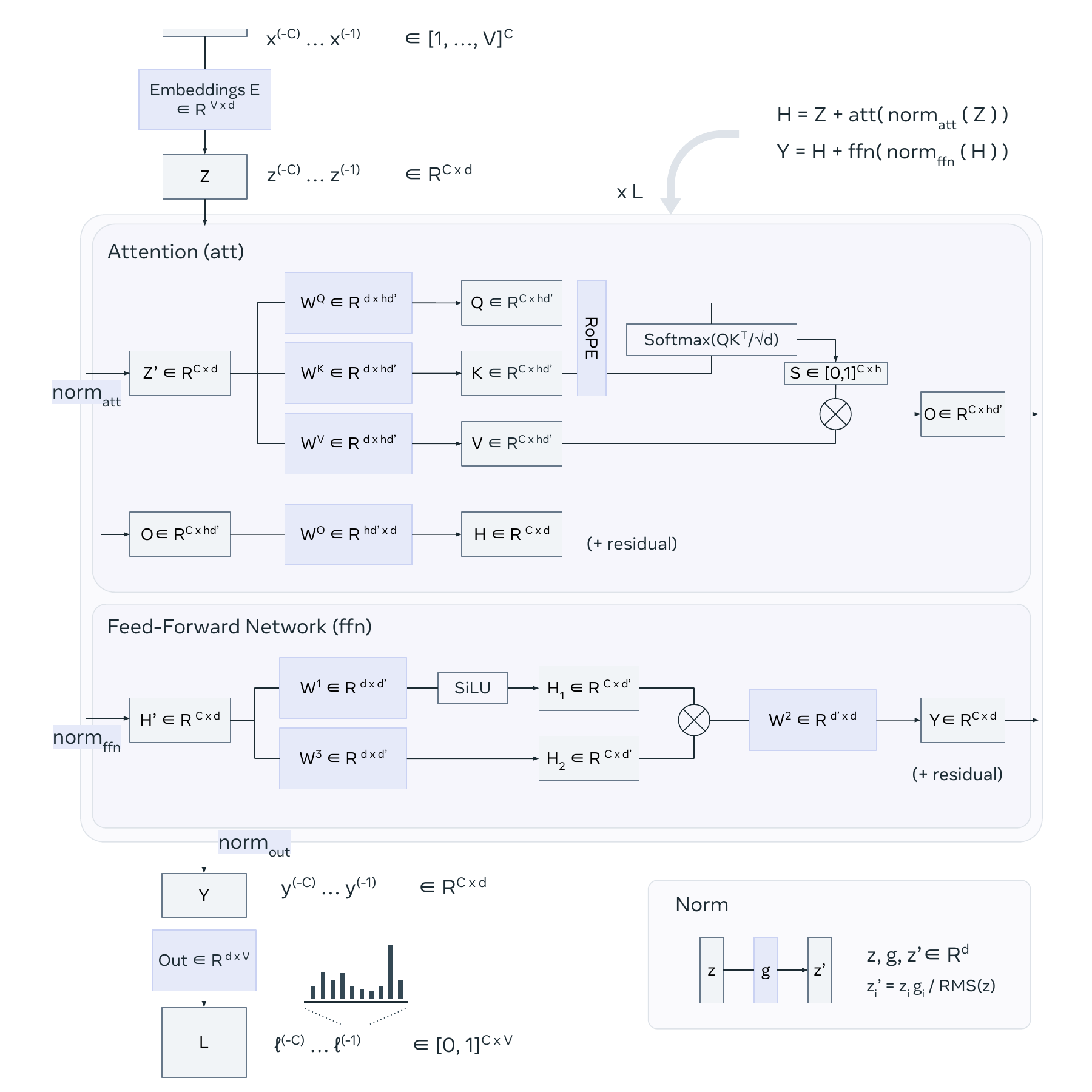}
    \caption{
    Detailed structure of the operations that take place in a transformer.
    $L$: number of layers, $d$: embedding dimension, $h$: number of heads, $d_\mathrm{k}$ and $d_\mathrm{v}$: dimensions of the keys and values, $d_\mathrm{ff}$: dimension of the feed-forward network, $|\mathcal V|$: size of the vocabulary.
    }\label{chap7/fig:transformer}
\end{figure}

The transformer~\citep{vaswani2017attention} has become the standard architecture for many applications in the last few years.
It can be trained efficiently on GPUs and scale well to large datasets and models, in both NLP~\citep{raffel2020exploring, kaplan2020scaling} and computer vision~\citep{dehghani2023scaling, oquab2023dinov2}.
For the sake of completeness and to help better understand the invariant properties of the model, we describe in details the architecture from GPT-2~\citep{radford2019language} (see Fig.~\ref{chap7/fig:transformer}).

\paragraph*{Tokenization.} The input string is first tokenized into a sequence of integers $(x_1, \dots, x_n) \in \V^n$.
An embedding layer $E \in \R^{|\V|\times d}$ maps each token $x_i$ to a continuous vector $z_i^0 = E_{x_i} \in \R^d $, where $d$ is the embedding dimension.

\paragraph*{Attention layers.} 
The self-attention mechanism enables long-range dependencies between sequence elements.
A self-attention transforms an input sequence $\vec{z} \in \R^{n \times d}$ into queries $Q$, keys $K$, and values $V$:
\begin{equation}
    Q = \vec{z}W^\mathrm{Q} \in \R^{n \times d_\mathrm{k}};\;
    K = \vec{z}W^\mathrm{K} \in \R^{n \times d_\mathrm{k}};\;
    V = \vec{z}W^\mathrm{V} \in \R^{n \times d_\mathrm{v}}.
\end{equation}
It then computes attention weights by taking a scaled dot product between the queries and keys:
\begin{align}\label{chap7/eq:attention}
    \mathrm{Attention}(Q,K,V) = \mathrm{softmax} \left( \frac{QK^\top}{\sqrt{d_\mathrm{k}}} \right)V.
\end{align}
Where the $\mathrm{Softmax}$ operator is applied column-wise.

This attention operator is applied $h$ times in parallel, yielding $h$ output \emph{heads}.
The results are concatenated and projected back to the original dimension:
\vspace{-0.1cm}
\begin{align}\label{chap7/eq:multihead}
\mathrm{MultiHead}(Q,K,V) = \mathrm{Concat}(\mathrm{head}_1,., \mathrm{head}_h)W^\mathrm{O},
\end{align}
where $\mathrm{head}_i = \mathrm{Attention}(QW_i^\mathrm{Q}, KW_i^\mathrm{K}, VW_i^\mathrm{V})$. 
The projections $W_i^\mathrm{Q}, W_i^\mathrm{K} \in \R^{d \times d_\mathrm{k}}$, $W_i^\mathrm{V} \in \R^{d \times d_\mathrm{v}}$ and $W^\mathrm{O} \in \R^{hd_\mathrm{v} \times d}$ are learned.

\paragraph*{The feed-forward network.}
The output is fed to a feed-forward network (FFN), \eg, two linear layers with a ReLU activation:
\vspace{-0.1cm}
\begin{align}
\mathrm{FFN}(\vec{h}) = \mathrm{ReLU}(\vec{h}W_1 + b_1)W_2 + b_2,
\end{align}
where $W_1 \in \R^{d \times d_\mathrm{ff}}$, $b_1 \in \R^{d_\mathrm{ff}}$, $W_2 \in \R^{d_\mathrm{ff} \times d}$ and $b_2 \in \R^{d}$ are learned parameters (other variants like SwiGLU~\citep{shazeer2020glu} frequently replace ReLU).

\paragraph*{A stack of residual connections.}
Instead of directly feeding $\vec{z}$ and $\vec{h}$ to the attention and FFN layers, residual connections are applied and the inputs are normalized using layer normalization~\citep{ba2016layer} (or variants like RMSnorm~\citep{zhang2019root}):
$\label{chap7/eq:layernorm}
\mathrm{LayerNorm}(\vec{z}) = \frac{\vec{z} - \mu}{\sigma} \odot g + b,
$
where $\mu$ and $\sigma$ are the mean and standard deviation of $\vec{z}$ along its second dimension, and $g\in \R^d$ and $b\in \R^d$ are learned parameters.
This is repeated for each layer $l\in \{1, ..., L\}$ of the transformer:
\vspace{-0.1cm}
\begin{align}
\vec{h}^{l} &= \att{l} \left( \lnatt{l} \big( \vec{z}^{l} \big) \right) + \vec{z}^{l} \\
\vec{z}^{l+1} &= \ffn{l} \left( \lnffn{l} \big( \vec{h}^{l} \big) \right) + \vec{h}^{l}.
\end{align}
The output is fed to a normalization layer $\lnout$ and a linear layer $W_\mathrm{out} \in \R^{d \times |\mathcal V|}$ to generate logits, and a softmax outputs the probability distribution of the next token.

\paragraph*{Positional embeddings.} 
For many tasks, it is useful to encode the position of tokens in the input sequence.
Positional embeddings are what allows to encode this information.
They were originally sinusoidal functions of the position~\citep{vaswani2017attention} added to the input embeddings. 
There are now several variants~\citep{raffel2020exploring, su2021roformer, press2021train, kazemnejad2023impact}, that may change Eq.~\eqref{chap7/eq:attention}.
For instance, rotary embeddings~\citep{su2021roformer} multiply queries and keys depending on their relative position in the sequence.
If $m$ is the position of the query ($Q_m= \vec{z}_m W^\mathrm{Q}$) and $n$ the position of the key, then it rewrites the product of~\eqref{chap7/eq:attention} as:
\vspace{-0.1cm}
\begin{align}\label{chap7/eq:rotary}
    \qquad Q_m K^\top_n = z_m W^\mathrm{Q} R_{\Theta, n-m} (z_n W^\mathrm{K})^\top.
\end{align}
$R_{\Theta,n}$ is a block diagonal matrix with $2\times2$ rotation matrix entries:
\vspace{-0.1cm}
\begin{align*}
   \left( R_{\Theta,n} \right) _i  =
   \begin{pmatrix}
   \cos n \theta_i & -\sin n \theta_i \\
   \sin n \theta_i & \cos n \theta_i
   \end{pmatrix},
\end{align*}
with rotation frequencies chosen as $\theta_i = 10,000^{-2i/d}$.

%% file: chapter-7/sections/3-method.tex
\section{Watermarking through Invariance}\label{chap7/sec:method}

\subsection{Invariants in the weights of transformers}

\paragraph*{Definition.}
We define an invariant as a series of operation applied on the model's weights $\theta \rightarrow \theta ' $ such that for any input $x$, the output $f_{\theta'}(x)$ is the same as before the application of the invariant.

\paragraph*{Permutation invariance} appears in (at least) four levels of the transformer.
We note $\Pi^d$ the set of permutations of $\{1,..., d\}$. 
For a matrix $M\in\R^{d\times d}$ and $\pi \in \Pi^d$, we denote by $M_{:,\pi}$ (resp. $M_{\pi,:}$) the matrix where columns (resp. rows) are permuted according to $\pi$.

\noindent
\emph{Embedding dimension.} 
The embedding matrix $E$ can be permuted along its second dimension without changing the output of the model, as long as the permutation is propagated to other matrices in the model. 
More formally, for $\pi \in \Pi^d$, if $E' = E_{:,\pi}$, then matrices $\{ W^\mathrm{Q}, W^\mathrm{K}, W^\mathrm{V}, W_1, W_{out}, \lnatt{}, \lnffn{}, b_2 \} \subset \theta$ need to be permuted along their first dimension by $\pi$ and all matrices $\{ W^\mathrm{O}, W_2 \}$ along their second one.
Formally, it means 
$(W^\mathrm{Q})' = W^\mathrm{Q}_{\pi, :}$, $(W^\mathrm{O})' = W^\mathrm{O}_{:, \pi}$, etc.

\noindent
\emph{FFN layer dimension.}
All neurons making up matrices $W_1$ and $W_2$ of feed-forward networks can be permuted:
for $\pi \in \Pi^{d_\mathrm{ff}}$, if $W_1' = (W_1)_{:,\pi}$ and $W_2' = (W_2)_{\pi,:}$, then $f_{\theta'}(\cdot) = f_{\theta}(\cdot)$.

\noindent
\emph{Attention heads.}
Heads are interchangeable in~\eqref{chap7/eq:multihead} provided that $W^\mathrm{O}$ is permuted in blocks of $d_\mathrm{v}$ according to its first dimension.

\noindent
\emph{Inside the head.}
Depending on the type of positional embeddings, the previous permutations can be extended.
For instance if they do not impact~\eqref{chap7/eq:attention} (this is not the case for rotary embeddings) then $W^\mathrm{Q}$ and $W^\mathrm{K}$ can be permuted along their second dimension.

\paragraph*{Scaling/Unscaling.}\label{chap7/sec:scaling}
Whenever layer norms or variants are directly followed (or preceded) by linear layers, \eg, at every attention or FFN block, we can rescale component-wise the parameters $g$, $b$ of $\mathrm{LayerNorm}(\vec{z})$
by a vector $\alpha\in \R^d$.
Invariance is obtained by dividing the rows of the following (or preceding) linear layers by the same vector.

\paragraph*{Invertible matrices in QK products.}\label{chap7/sec:invertible}
We hereby assume the positional embeddings do not impact~\eqref{chap7/eq:attention}.
If $P \in \R ^{d_\mathrm{k}\times d_\mathrm{k}}$ is invertible, then choosing $(W^\mathrm{Q})' = W^\mathrm{Q} P$ and $(W^\mathrm{K})' = W^\mathrm{K} (P^\top)^{-1}$ is invariant in~\eqref{chap7/eq:attention}.
This also applies to the case of rotary embeddings by restricting $P$ to be block diagonal of $2\times 2$ matrices that apply a 2D rotations and scaling by a factor $\lambda$ (thanks to the commutativity of 2D rotations).

\paragraph*{Combining invariants.}
All previous parameter transformations may be seen as invertible right or left matrix multiplications applied to the model parameters. 
They do not interfere and may be combined in a sequence of arbitrary order, yielding $\theta \rightarrow \theta' \rightarrow \theta'' \rightarrow \cdots$.
Combining transformations at all levels improves robustness to removal attacks and to collusion (\ie, when several Bobs share their weights to evade detection).
Indeed, if Bob tries to remove the watermark by re-applying one invariant, it will still be present in the other invariants.
In the same way, if several Bobs compare their models, it will be hard for them to identify which operations were applied to their models, since the order in which they were applied is unknown, and since the weights will differ a lot between them.

We presented some invariant operations that can be applied to the weights of a transformer model similar as GPT-2. 
Note that depending on the model's specificities, invariants may appear or disappear. 
For instance, rotary embeddings~\citep{su2021roformer} and QK normalization layers~\citep{dehghani2023scaling} impact the way to apply invertible matrices in QK products (see below); the order in which normalization is applied (post-norm or pre-norm) may impact the scaling/unscaling invariance; etc.

\subsection{From invariants to watermarks}\label{chap7/sec:invariantsaswatermarks}

\paragraph*{Insertion.}
Before starting the watermark process, for each invariant and each level of the network, we restrict the set of transformations to $2^k$.
For example, we randomly sample $2^k$ possible permutations in $\Pi^d$ for the Embedding dimension (out of the total $d!$). 

Therefore, we can encode $k$ bits for each combination of an invariant and a level. 
We encode a model's identifier as the concatenation of $m$ chunks of $k$ bits ($2^{mk}$ possibilities).
For instance, let $k=4$ and the model have $32$ layers. 
We choose to embed two permutations per layer, one for the attention block and one for the FFN block. 
The total number of bits is $2\times 32\times 4=256$, representing $10^{77}$ possible models (approximately the number of atoms in the universe, an upper bound of the number of Bobs). 

\paragraph*{Extraction.}
To extract the $k$-bits message from a weight matrix, we re-apply all $2^k$ possible invariants to the original matrix.
We then compute the Frobenius norm of the difference (MSE) between the observed weight and the possible watermarked ones.
We choose the one with lowest MSE among the $2^k$ and this choice is encoded as a $k$-bit integer.
Doing that on every blocks of the network and every invariant, we end up with a full message made of $m$ chunks of $k$ bits.
In the case of intertwined invariants, we do the same in the order in which we inserted the invariants, reverting them as we go.

To speed up extraction, we may select a subset of the matrices' rows before extraction.
This speeds up the extraction (in the order of 100$\times$), but makes the detection slightly less robust. 
For instance, in the case of scaling/unscaling we may select the first $100$ components of $\alpha$ from $\R^{d}$ to $\R^{100}$ and $W$ from $\R^{d\times d'}$ to $\R^{100\times 100}$.  

\paragraph*{Matching.}
To match an extracted message (made of $m$ chunks of $k$ bits) with a model's identifier, we compute the number $s$ of chunk-wise errors with respect to all possible identifiers.
We return a match if $s$ is bellow a fixed threshold $\tau$ to ensure resilience to corruption and to provide a confidence score.
A theoretical p-value, \ie, the probability of obtaining a number of errors lower than $s$ for a random model, is given by the regularized incomplete beta function $\mathcal{I}$:
\begin{equation}\label{chap7/eq:pvalue}
    \textrm{p-value}(s) = 1- \left(1-\mathcal{I}_{ 1/2^{k} } ( m-s, s+1) \right)^N,
\end{equation}
where $N$ is the number of distributed models.

\begin{figure}[b!]
    \centering
    \includegraphics[width=1.0\linewidth]{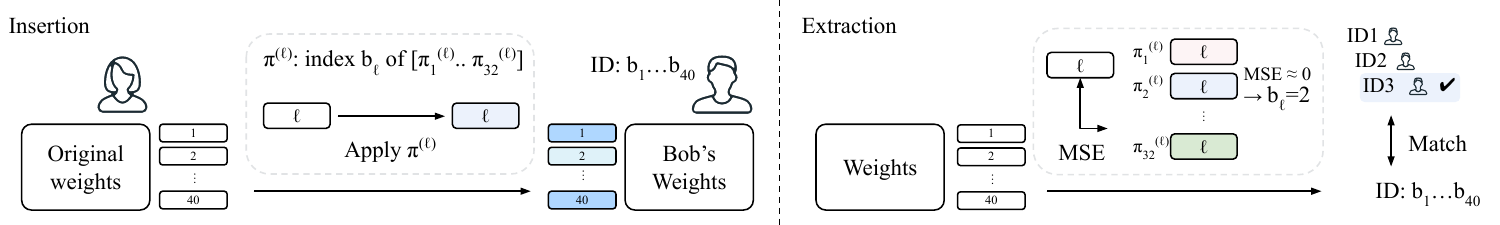}
    \caption{
    Detailed illustration of watermark insertion and extraction, with the example of permutation on $L$=40 blocks. 
    A user ID is a list $b_1...b_L$ of $L$ bytes, that are used to select the permutation to apply for each block $\ell$.
    For each $\ell$, the extraction computes the MSE between the observed weights and all original permuted weights. 
    It then selects the one with minimum MSE, which in turn gives $b_\ell$.
    }
    \label{chap7/fig:method}
\end{figure}

\paragraph*{Robustness and security.}
Watermarking models through invariance is stealthy, because it does not change their outputs.
However, a distortion-free watermark is also a weakness: Alice can hide the watermark without impacting the model's utility, but on the other hand an adversarial Bob may do the same at no cost.
In short, most of these watermarks are very robust against classical model manipulations (fine-tuning, quantization, etc.) but not against a malicious user who knows the method.
In this case we would only know that the model is an unauthorized copy, without knowing the leaker.

%% file: chapter-7/sections/4-experiments.tex
\newcommand{\valtab}[1]{{\color{gray}\footnotesize ({#1})}}

\section{Experiments}\label{chap7/sec:experiments}

The purpose of the experiments is to evaluate the effectiveness and the robustness of the watermarks to transformations on large language models.

\subsection{Setup}

\paragraph*{Model.}
We use Llama~\citep{touvron2023llama} models as main benchmark. 
The architectural differences with regards to the original transformer architecture are pre-normali\-zation~\citep{radford2019language} with RMSnorm~\citep{zhang2019root}, SwiGLU activation~\citep{shazeer2020glu} and rotary embeddings~\citep{su2021roformer}.
To evaluate that the utility of the model is not degraded, we show results on a next-token prediction task.
This is done on random sequences of text taken from Wikipedia, then tokenized using the default tokenizer of Llama.
Unless stated otherwise, we use the $7$B-parameter model.

\paragraph*{Attacks.}
We consider the following attacks.
\emph{Fine-tuning.} 
We fine-tune the model in a supervised manner with the same settings as~\citep{alpaca}, on 3 epochs with learning-rate of $2\times 10^{-5}$.
\emph{Noise.} 
We add zero-mean Gaussian noise with standard deviation $\sigma$ to the model weights.
\emph{Quantization.}
We quantize the model weights into $b$ bits. 
To allow flexible rates and ease the experiments, this is done by uniformly quantizing the weights between their minimum and maximum values.
\emph{Pruning.}
We prune the model weights by zeroing the ones with smallest L1 norms, with sparsity given in percentage of zero weights.

\paragraph*{Watermark settings.}
We apply the encoding process of Sect.~\ref{chap7/sec:invariantsaswatermarks}.
For permutation invariance, we permute attention heads and FFN layers.
For scaling, we alter the layers' RMSnorms and following matrices.
The scaling vector $\alpha$ is such that $\log_{10}(\alpha) \sim \mathcal{U}(-1, 1)$.
For QK products, as mentioned in Sect.~\ref{chap7/sec:invertible}, the invertible matrix has to be block diagonal of $2$ by $2$ rotation matrices, so we randomly sample $d/2$ rotation angles.
We fix the number of possible choices at $k$=$8$, \ie, each choice is encoded with a byte.
Therefore, we encode $2$ bytes at every layer, except in QK products where we encode $1$.
When combining all invariants together, we proceed the same way for all blocks: we start with permutation, then apply invertible matrices in QK products, then scale/unscale the layer norms and matrices.

For instance, the $7$B model has $L$=$32$ layers so the watermark is $64$ bytes long except for the QK products invariant where it is $32$.
In the case of combined invariants, the total number of bytes is $160$.

\subsection{Results.}\label{chap7/sec:results}

\begin{table}
    \centering
    \caption{
    Distortion induced on generation 
    and robustness of watermark extraction under various processes. 
    Each line stands for a different invariant.
    We present results of the sped-up extraction, the ones for no speed-up are given as \valtab{acc}.
    }
    \label{chap7/tab:robustness}
    \footnotesize
        \begin{tabular}{lc *{4}{l}}  
            \toprule
            \multirow{2}{*}{Method} & \multirow{2}{*}{Distortion} & \multicolumn{4}{c}{Byte accuracy ($\%$) on:} \\
            \cmidrule(lr){3-6}
                        &           & Noise $1.0$    & Quantization $3$b & Pruning $50\%$ & Fine-tune  \\
            \midrule
            Perm. & 0.20$\%$ & 51.4  \valtab{99.6} & 72.0 \valtab{100.0}   & 100.0     & 100.0 \\
            QK          & 0.18$\%$ & 100.0          & 100.0         & 100.0     & 100.0 \\
            Scaling     & 0.24$\%$ & 100.0          & 98.1 \valtab{100.0}       & 100.0     & 100.0 \\
            All        & 1.77$\%$ & 60.8  \valtab{99.8} & 70.0   \valtab{99.4} & 100.0 & 100.0 \\
            \bottomrule
        \end{tabular}
\end{table}

\paragraph*{Robustness.}
We evaluate the robustness of the watermark using the \emph{byte accuracy}, \ie, the percentage of bytes correctly recovered.
Results are averaged over $N$=$100$ watermarked models except for fine-tuning where we only fine-tune one model.
We speed-up the extraction by selecting a subset of $100$ rows of the matrices (see Sect.~\ref{chap7/sec:invariantsaswatermarks}); 
time needed for extraction is around 20 minutes when using the full matrix instead.

\autoref{chap7/tab:robustness} reports the byte accuracy for different processing applied before extraction.
We observe that the watermark is robust to all attacks with byte accuracy >$50\%$.
Errors mainly come from the speed-up of the extraction process.
We also consider the \emph{\pval} of the associated statistical test~\eqref{chap7/eq:pvalue}.
A byte accuracy of 50\% on 64-bytes messages is more than enough to reliably identify a model: 
the \pval s are always bellow $10^{-60}$, due to the very low probability of simultaneously observing a match between tens of pairs of random bytes.
As an illustration, 8 matching bytes on 64-bytes messages already gives a \pval\ of $10^{-8}$.

\paragraph*{Model's utility.}
In fact, previous invariants are not perfect because of quantization (weights are stored as 16bits floating point numbers).
Thus, we quantitatively compare watermarked and original models.
We feed to both of them $1$k sequences of $256$ tokens.
Predicted next tokens are greedily chosen as the argmax of the $256$k observed logits.

\autoref{chap7/tab:robustness} reports the distortion as the proportion of predicted tokens that differ between watermarked and original models.
As expected this proportion is very low (<$1.8\%$) and higher for the scaling invariant since it further affects quantization.

Besides, the distortion increases when the token is far in the sequence \eg, for sequences of length 1024 tokens, the average distortion at the last token rises to $2.5\%$ for the scaling invariant.
This is still very low and does not affect the utility of the model since predicted tokens are still very likely.

\paragraph*{Computational efficiency.}
Larger models have more layers and parameters, which increases the computational cost of insertion and extraction.
In \autoref{chap7/tab:modelsize}, we report results for different model sizes.
Insertion and extraction times are averaged over $100$ runs and measured on 2 Intel(R) Xeon(R) 6230 @ 2.10GHz cores and a total of 480GB of RAM.
The low computational costs and requirements (no GPU needed) makes it possible to scale to very large models.

\begin{table}
    \centering
    \caption{
        Computational cost of watermark insertion and extraction for different model sizes and the different invariants. 
    }
    \label{chap7/tab:modelsize}
    \footnotesize
        \begin{tabular}{rcc *{3}{r} *{3}{r}}
            \toprule
            \multirow{2}{*}{Model} & \multirow{2}{*}{$L$} & \multirow{2}{*}{$d$} &  \multicolumn{3}{c}{Insertion (s)} & \multicolumn{3}{c}{Extraction (s)} \\
            & & & Perm. & Scaling & QK   & Perm. & Scaling & QK  \\
             \cmidrule(rr{2pt}){4-6} \cmidrule(rr){7-9}
            7-B & 32 &  4096  & 3.5  & 2.7 &    7.4    &  9.2 &  31.7 &  6.0    \\
            13-B & 40 & 5120  & 7.0  & 4.9 &    15.8   & 14.1 &  30.3 &  7.7    \\
            30-B & 60 & 6656  & 19.3 & 8.7 &    47.3   & 31.7 &  54.7 & 13.5    \\
            70-B & 80 & 8192  & 37.1 & 17.5 &   106.0 &   56.3 & 110.0 & 21.5   \\
            \bottomrule
        \end{tabular}
\end{table}

%% file: chapter-7/sections/8-conclusion.tex
\section{Conclusion}

Our work presents a lightweight approach for watermarking large transformers. 
We leverage invariance properties to generate equivalent copies for watermark embedding. 
It ensures that the model's outputs are preserved while providing close-to-perfect robustness against processes like fine-tuning or quantization. 

Yet, this approach has limitations. 
Namely, it is limited to white-box scenarios. 
Additionally, if a sophisticated attacker identifies all invariants, they may remove the watermark by applying the same transformation techniques. 
Although the model could still be identified as an unauthorized copy, the unique binary signature associated with the original watermark would be lost.
Despite these challenges, this work is a starting point to exploit invariance properties that stem from the extreme redundancy of parameters of large networks, for watermarking applications.

%% file: 0-conclusion/conclusion.tex
\chapter{Conclusion}\label{chapter:conclusion}
\chaptermark{Conclusion}

\section{Summary of the contributions}

This thesis first presents a comprehensive overview of watermarking techniques in \autoref{chapter:related-work} and \autoref{chapter:technical-background}.
It then introduces and evaluates new watermarking techniques for a more and more digital and AI-driven internet.
The contributions are structured around three main axes: content moderation, tracing AI-generated content, and monitoring AI models.
Each part addresses current challenges and leverages watermarking to enhance traceability of content and models.

Content moderation is a critical issue for online platforms, which need to manage the spread of harmful or illegal content.
\autoref{part:content-moderation} introduces watermarking techniques that significantly enhance the robustness of content moderation systems. 
By imperceptibly modifying images as they enter a platform, these techniques improve origin tracing and verification and limits the spread of already-flagged material.
\autoref{chapter:ssl-watermarking} hides information in the latent representations of self-supervised neural networks through an iterative image optimization, enabling watermarking images with varying resolutions, adjustable payload, and a customizable tradeoff between robustness and quality. 
\autoref{chapter:active-indexing} combines copy detection and watermarking to greatly improve the robustness of copy detection systems, with a similar optimization scheme.

In parallel, the need for robust detection mechanisms becomes critical as AI-generated content becomes increasingly indistinguishable from human-generated content, and used for misinformation or fraud. 
\autoref{part:genai-tracing} presents watermarking methods that embed unique identifiers into content generated by AI models, allowing for easy tracing of the content's origin. 
Regulatory frameworks are increasingly recognizing the importance of such techniques, and this thesis provides practical solutions that align with these emerging standards.
\autoref{chapter:stable-signature} introduces a method that fine-tunes latent generative models such that all images they produce hide an invisible signature, which can be used to detect and track the origin of synthesized images, even when models are openly shared.
\autoref{chapter:audioseal} introduces a watermarking solution for the detection of AI-generated speech, which proactively watermarks the speech signal and predicts for each time step if the watermark is present or not.
\autoref{chapter:three-bricks} brings three improvements to state-of-the-art watermarking methods for large language models (LLM), notably theoretically grounded and empirically validated statistical tests that guarantee false positive rates.

The economic and intellectual property aspects of AI models are also increasingly important, since organizations invest heavily in training and deploying these models.
\autoref{part:model-monitoring} introduces techniques that allow organizations to protect their intellectual property and ensure compliance with licensing agreements.
\autoref{chapter:radioactive} examines whether it is possible to detect when an LLM has been fine-tuned on the output of another model. 
More precisely it shows that LLM watermarks can in some cases be identified in fine-tuned models even when only a small proportion of the fine-tuning data is watermarked.
\autoref{chapter:invariants} introduces a training-free watermark for the weights of large transformers that allows for traitor tracing.
It leverages the model's invariance through operations like dimension permutations to generate different functionally equivalent copies with different weights.

\section{Perspectives on generative AI watermarking}

Watermarking is the most robust and efficient solution to trace AI-generated content, but governance and interoperability challenges remain.

\subsection{Robustness and security}

Because watermarking works by actively modifying content, there is a common belief that these traces can be easily removed. 
However, this thesis has outlined the technical superiority of watermarking, particularly in terms of detection confidence and robustness, compared to other content tracing methods.
Additionally, when viewed as part of a broader ecosystem that includes detection algorithms, legal frameworks, and social norms, watermarking may arguably be robust enough for its intended purpose.

\paragraph*{Invisible watermarking surpasses other methods.}
Watermarking presents undeniable advantages.
First, it intentionally injects traces into content, whence the greater robustness than that of passive methods like forensics or fingerprinting. 
For instance, \autoref{chapter:active-indexing} shows that watermarking achieves $\approx\times$8 better recall than fingerprinting after crops that keep $50\%$ of the original images;
and \autoref{chapter:stable-signature} compares watermarking to forensics and shows that it achieves the same true positive rate (probability of correctly flagging a watermarked piece of content) for a $10$ million times smaller false positive rate (probability of wrongly flagging a non-watermarked one), on images that are cropped, resized and compressed.
Attacking the watermark is always possible but this always damages the quality, contrary to visible watermark or metadata erasure -- this is also true for forensics~\citep{barni2018adversarial} and fingerprinting methods~\citep{tolias2019targeted}.
Second, a sound watermarking design has a low false positive rate.
Most importantly, it is provably low (see Chap.~\ref{chapter:three-bricks}), unlike with forensics and fingerprinting. 
Data provenance is expected to be tested on millions of pieces of content, as for content moderation, therefore requiring extremely low false positive rates. 
This is beyond reach of an empirical validation.  

\paragraph*{Are attacks really a limitation?}\label{conclusion:attacks}
Robustness and security are close yet different concepts in watermarking~\citep{cayre2005watermarking}.
While watermarking is built to be robust, it is not foolproof to intentional attacks~\citep{cox1998some, bas2011break}, be it for image and audio~\citep[\autoref{chapter:stable-signature}, \autoref{chapter:audioseal}]{jiang2023evading, zhao2023invisible, saberi2023robustness}, or LLM-generated texts~\citep{sadasivan2023can, krishna2024paraphrasing, hu2024stable, jovanovic2024watermark, pang2024attacking, chang2024watermark}.
It is subject to attacks that are roughly categorized based on the attacker's knowledge, as done in Chap.~\ref{chapter:stable-signature} and \ref{chapter:audioseal}.
\textit{White-box} attacks have full access to the watermarking algorithm and its parameters (\eg, model weights);
\textit{black-box} attacks only have access to inputs and outputs, for instance through an API;
and \textit{no-box} attacks do not have any knowledge of the system.
The effectiveness and ease of an attack generally increases with the attacker's level of knowledge about the watermarking system, the hardest ones being no-box attacks, where there is not even the feedback on if an attack was successful or not.

Most attacks are removal attacks, where the goal is to eliminate the watermark from the content. 
Watermark forging, where the attacker creates a counterfeit watermark, may pose a more significant problem. 
Currently, without white-box access to the embedder or extractor, forging a watermark is considerably difficult. 
There is always a trade-off to consider: an attack may succeed in removing or forging a watermark, but at the cost of degrading the quality of the content itself and of making the attack more detectable.

\emph{Watermarking keeps honest people honest.}
Most mafia organizations or belligerent countries now have the expertise and resources to train their own generative models. 
They will include neither watermarking nor metadata, and forensic methods are also doomed to fail due to the lack of such data to train a detector. 
Watermarking's goal is not to protect against these cases.
Rather, it aims to dissuade 99\% of the population, by making the removal of the watermark complex enough and voluntary -- or even criminal by law, as what happened with DRM systems~\citep{enwiki:1221667351}.
This aligns with the motto ``keep honest people honest,'' which Hollywood popularized in the 2000s about DRMs.

\subsection{The real challenges are governance and interoperability}

The technical aspects of watermarking, like its robustness to adversarial attacks, are far from being the only considerations to take into account. 
It is essential to address overlooked challenges that concern governance, control, and, maybe naively, how to even use the detection outcomes. 

\paragraph*{Who controls watermark detection?}

While everybody is a priori willing to know when they are interacting with generated content, making watermark detectors publicly available introduces security risks.
Open-source detectors can lead to white-box attacks, and API access can facilitate black-box attacks.
Consequently, no record of watermark detection by anyone other than the generative model's owner currently exists.
This situation, where the model provider is both judge and jury, is problematic. 
It would be more trustworthy if watermarking and detection were managed by trusted, unbiased entities. 
This raises questions about who these entities should be and how they are governed. 

\paragraph*{Open-source generative models?}
They present a unique challenge since they are freely available and usable without post-hoc watermarks (applied after generation).
The case in point is Stable Diffusion~\citep{rombach2022high}, which was open-sourced in late 2022. 
Removing the watermark in its source code was as simple as commenting out a single line.
Ideally, models should be trained or fine-tuned to generate watermarked content natively as in Chap.~\ref{chapter:stable-signature} or related work~\citep{yu2020responsible, kim2023wouaf, juvela2023collaborative, gu2023learnability}.
Determining responsibility in this context is complex: should it be the responsibility of the individual who uploads a model to a platform, or should hosting platforms like GitHub or Hugging Face enforce in-model watermarking? 
This issue needs clear regulatory guidance and possibly new technological solutions to ensure compliance.

\paragraph*{What to do with detection?}
This question is not clearly addressed by current regulations.
The use of watermarks for labeling authentic or fake content on social networks and search engines, as suggested by current texts like 22949.90.3.(a) of \cite{ca_ab3211_2024}, may lead to a rebound effect. 
It may conversely exacerbate misinformation by placing undue emphasis on content that is either not detected, generated by unknown models, or authentic but used out of context.
Moreover, detection of watermarks extends beyond individual pieces of content, often involving the aggregation of evidence from multiple submissions linked to a single account. 
\citet{kirchenbauer2024on} notably showed that watermarked text may be detected even under strong paraphrasing after observing enough words.
Finally, current regulations lead different entities to quickly develop their own watermarking methods.
This results in a fragmented ecosystem where nobody is responsible for detection.
For instance, the music generation startup \citet{sunov3} watermarks their outputs, but no platforms (Facebook, X, Spotify, Youtube, etc.) actually detect them.
Collaborative efforts are needed to establish standards that ensure watermarks are robust, but, most importantly, recognizable across platforms.
It should involve regulators, model providers and content hosting platforms.

\section{Perspectives and open questions}

Watermarking to trace AI-generated content, is far from being (1) a solved problem, (2) the only use-case of watermarking.

\subsection{Security}
A conceptual problem of watermarking is that allowing third-parties to detect the watermark discloses information which can be used to remove the watermark or forge fake ones, \eg, the watermark extractor/detector (see Sec.~\ref{conclusion:attacks}).
By default, it does not follow Kerckhoffs's principle~\citep{Kerckhoffs1883}, which states that the security of a system should assume that the attacker knows the system, except for the secret key.

Two key concepts from cryptography deal with this issue in the context of watermarking.
In \emph{asymmetric watermarking} (or public watermarking), the embedder and the detector are different entities.
Anyone with access to the detector may detect a watermark, but only the entity with the private key may embed or remove it.
It is similar to public-key cryptography, where anyone can encrypt a message, but only the recipient can decrypt it.
It has been applied to image or video watermarking~\citep{furon1999asymmetric, furon2003asymmetric, hartung1997fast} and recently to watermarking for LLMs~\citep{fairoze2023publicly}.
In the context of \emph{zero-knowledge watermark detection}, the prover should convince the verifier of the presence of a watermark without revealing information that can be used to remove the watermark, similar to zero-knowledge proofs (ZKP), where a prover convinces a verifier that a statement is true without revealing any information about the statement itself~\citep{fiege1987zero}.
This is done by replacing the watermark detection process with a cryptographic protocol~\citep{craver2000zero, adelsbach2001zero, adelsbach2003watermark}.
The aforementioned methods are not yet widely adopted because too computationally expensive and very mathy (and for some a bit rusty).
Watermarking research could benefit from more research in these directions, with the ultimate goal of creating a watermark that anyone can detect, but only authorized parties can embed.

\subsection{Semantic watermarking in generative models}

Most watermarking methods modify pixel values or audio samples, which can be made robust to common editing operations by incorporating them into the training process (see Sec.~\ref{chap0/sec:deep learning-watermarking}).
However, not all attacks can be covered, \eg, for images, noising and denoising with a diffusion model~\citep{nie2022diffusion}.
A promising approach is to perform watermarking at a higher level, in the semantic content of the generation. 
While image noising/denoising or compression alters the pixel values, most image/audio editing methods and attacks leave the image-level semantics unchanged. 
This also holds true for text: paraphrasing alters the ``surface'' tokens, but the sentence-level semantics remain unchanged.
Modifying the content semantics is undesirable in traditional watermarking, as it would alter the content itself.
However, for generated content, this is not an issue since the primary concern is that the generated content follows the given instructions, regardless of what would have been generated without the watermark.
Recent works have explored this direction, such as Tree-Ring~\citep{wen2023tree}, which is the first watermark that does not rely on minor modifications of generated images, but instead alters the image generation process. 
Similar approaches have been proposed for text watermarking~\citep{liu2023semantic,hou2023semstamp}.
Despite these advancements, there are still gaps in this research area. 
Notably, no semantic watermarking method has been proposed for audio or video, and no existing methods work ``in-model'', or post-hoc for either images or text (as defined in Sec.~\ref{chap0/sec:generation-watermarking}).
This is a highly promising research direction, as it offers the potential for more robust watermarking methods that are less sensitive to common editing operations and no-box attacks.

\subsection{Tracing training data}
Data attribution, \ie, the ability to trace model's outputs back to its training data, is an active field of research. 
It has applications in model interpretability, fairness, and privacy, and would have huge economical downfalls in the context of copyrights of training data~\citep{deng2024economic}.
Current methods for data attribution, are based on old concepts in machine learning, like influence functions~\citep{cook1977detection} and Shapley values~\citep{shapley1953value}.
They often fall short when applied to large-scale models due to their computational complexity and the necessity for model retraining~\citep{koh2017understanding, feldman2020neural, pruthi2020estimating, park2023trak} or the absence of causation between training data and generated outputs~\citep{wang2023evaluating}.

The introduction of active watermarking techniques could provide a scalable and efficient solution to the data attribution problem.
By embedding watermarks directly into the data used for training AI models, it becomes possible to trace the influence of specific training sets on the model's behavior at test time, with better efficiency and guarantees than passive methods.
This is the approach followed by \citet{sablayrolles2020radioactive}, which watermark training set of image classifiers to detect if it was used to train a given model and of \autoref{chapter:radioactive} for synthesized data from large language models.
It is further explored by \citet{asnani2024promark}, who attribute synthetically generated images to specific training data concepts through watermarking.
This could be applied at scale on different types of generative models and for various use-cases, to properly credit the data sources when generating new content.

%% file: 0-conclusion/proofs.tex
\section{Demonstration of the hypercone volume}
\label{app/ssl-watermarking}

\paragraph*{Proposition.}
Let a secret key $a\in \mathbb R ^d$ s.t. $\|a\|=1$, $\theta \in (0, \pi/2)$ and the dual hypercone:
\begin{equation*}
     \mathcal{D} := \left\{  x \in \R^d :  \abs{ x\T  a} > \norm{ x} \cos (\theta) \right\}.
\end{equation*}
We have:
\begin{equation*}
   \mathbb{P}\left(\phi(I)\in \mathcal D \mid \text{``key } a \text{ uniformly distributed''}\right)
    = 1-I_{\cos^2(\theta)} \left(\frac{1}{2}, \frac{d-1}{2} \right)
\end{equation*}
where $I_\tau (\alpha, \beta)$ is the regularized Beta incomplete function.

\paragraph*{Proof.}
Let $ U $ be a random vector uniformly distributed on the unit sphere of $ \mathbb{R}^d $ ($\|U\| = 1$) and $ v $ a fixed vector of the same hypersphere. 
We are interested in the distribution of the cosine similarity, which is in this case $U^\top v$.

One way to sample $U$ is the following: First sample a white Gaussian vector $G \sim \mathcal{N}(0,I)$ in $\mathbb R ^d$ and then normalize: $U=G/\norm{G}$.
Without loss of generality, we can assume that $v = (1,0,...,0)$ (up to a change of axes). Therefore, $$U^T v = \frac{G_1}{\sqrt{\sum_{i=1}^d G_i^2}}$$
Let $\tau \in [0,1]$. We have 
\begin{align*}
    \mathbb{P}\left((U^T v)^{2} \geq \tau^{2}\right) &=\mathbb{P}\left(\frac{G_{1}^{2}}{\sum_{i=1}^{d} G_{i}^{2}} > \tau^{2}\right)
    =\mathbb{P}\left(\frac{G_{1}^{2}}{\sum_{i=2}^{d} G_{i}^{2}} > \frac{\tau^{2}}{1-\tau^{2}}\right)\\
    &=\mathbb{P}\left((d-1)\frac{G_{1}^{2}}{\sum_{i=2}^{d} G_{i}^{2}} > (d-1)\frac{\tau^{2}}{1-\tau^{2}}\right)
\end{align*}
Note that $Y := (d-1)\frac{G_{1}^{2}}{\sum_{i=2}^{d} G_{i}^{2}}$ is the ratio of two independent chi-squares random variables of degree 1 and $d-1$. By definition, $Y$ follows a Fisher distribution $F(1,d-1)$. Its cumulative density function is: $F(x; 1, d-1) = I_{x/(x+d-1)}(1/2, (d-1)/2)$. 

It follows, using $x=(d-1)\frac{\tau^2}{1-\tau^2} = -(d-1) + (d-1)\frac{1}{1-\tau^2}$ and setting $\tau = \cos\theta$, that:
$$\mathbb P (\abs{U^T v} > \cos \theta) = \mathbb P (Y> \cos^2 \theta) = 1 - I_{\cos^2 \theta}(1/2, (d-1)/2),$$
where $I_x(a,b)$ is the regularized incomplete beta function.
(It also equals $ I_{\sin^2 \theta}( (d-1)/2,1/2)) $ by symmetry of the beta function.)

Finally, since 
$$\mathbb{P}\left(\phi(I)\in \mathcal D \mid \text{``key } a \text{ uniformly distributed''}\right) = \mathbb{P}\left(\abs{U^T \phi(I)} > \cos \theta\right),$$
we have the result.

\section{Demonstrations for \texorpdfstring{\cite*{aaronson2023watermarking}}{}}\label{app/three-bricks}

We here provide the proofs of the LLM watermarking results presented by~\cite{aaronson2023watermarking}, which are absent from the literature.

\subsection{Sampling probability}
\label{chap5/app:aaronson_prob}

\paragraph*{Proposition.}
Consider a discrete distribution $\vec{p}=(p_1,\ldots,p_V)$
and $V=|\V |$ random variables $\vec{R} = (R_1,\ldots,R_V)$ s.t. $R_v\overset{iid}{\sim}\mathcal{U}_{[0,1]}$. 
Let $V^\star = \arg \max_v R_v^{1/p_v}$.
Then: $$\Prob(V^\star=v) = p_v.$$

\paragraph*{Proof.}
For any $v \in \V$, $R_v\overset{iid}{\sim}\mathcal{U}_{[0,1]}$ so, $- \ln(R_v)$ follows an exponential distribution $\mathcal{E}(1)$.
Let $Z_v := -\frac{1}{p_v} \ln(R_v)$. By construction, $Z_v\sim\mathcal{E}(p_v)$, with density $f_{Z_v}(z) = p_v e^{-p_v.z}$.
We now have:
\begin{equation}
V^\star = \arg \max_v R_v^{\frac{1}{p_v}} = \arg \min_v Z_v.
\end{equation}
A well known result about exponential laws is that (see \href{https://francisbach.com/the-gumbel-trick/}{the-gumbel-trick} for following lines):
\begin{eqnarray}
\underline{Z}  &=& \min_v Z_v \sim \mathcal{E}\left(\sum_v p_v\right)=\mathcal{E}\left(1\right),\\ \label{chap5/eq:sampling}
\Prob(V^\star=v) &=& \frac{p_v}{\sum_j p_j}  = p_v.
\end{eqnarray}

This shows that for a given secret vector $\vec{r}$, the watermarking chooses a word which may be unlikely (low probability $p_{V^\star}$). 
Yet, on expectation over the secret keys, \ie, over r.v. $\vec{R} = (R_1, \ldots, R_V)$, the distribution of the chosen token follows the distribution given by the LLM.

\paragraph*{Corollary.} $R_{V^\star} \sim Beta(1/p_{V^\star}, 1)$.

\paragraph*{Proof.}
\begin{equation}
\underline{Z}  = Z_{V^\star} = -\frac{1}{p_{V^\star}} \ln(R_{V^\star}) \sim \mathcal{E}(1),
\label{chap5/eq:Coro}
\end{equation}
which translates to $R_{V^\star} = e^{-p_{V^\star} E}$ with $E\sim\mathcal{E}(1)$, with p.d.f. $f_{R_{V^\star}}(r) = \frac{r^{\frac{1}{p_{V^\star}}-1}}{p_{V^\star}}$. 
Therefore, $R_{V^\star} \sim Beta(1/p_{V^\star}, 1)$.

\subsection{Detection}
\label{chap5/app:aaronson_score}

We denote by $x^{(1)}, \ldots, x^{(T)}$ the sequence of tokens in the text, 
by $\vec{p}^{(t)}$ the probability vector output by the LLM and by $\vec{R}^{(t)} \in [0,1]^{|\mathcal{V}|}$ the key random vector at time-step $t$.
We define $R_t := R^{(t)}_{x^{(t)}}$ and $p_t := p^{(t)}_{x^{(t)}}$ at time-step $t$.
The score is $S_T=-\sum_{t=1}^{T} \ln (1-R_t)$.

\paragraph*{Proposition ($p$-value under $\H_0$).}
The $p$-value associated to a score $s$ is defined as:
\begin{equation}
\text{$p$-value}(s) = \Prob(S_T>s \mid \H_0) = \frac{\Gamma(T,s)}{\Gamma(T)},
\end{equation}
where $\Gamma(T,s)$ is the \emph{upper} incomplete gamma function.

\paragraph*{Proof.}
Under $\H_0$, the assumption is s.t. $R_t\overset{iid}{\sim}\mathcal{U}_{[0,1]}$. 
Then, $- \ln(1-R_t)$ follows an exponential distribution $\mathcal{E}(1)$.
Therefore $S\sim\Gamma(T,1)$ (see \href{https://en.wikipedia.org/wiki/Gamma_distribution#Summation}{sum of Gamma distributions}). Therefore the $p$-value associated to a score $s$ is 
\begin{equation}
    \text{$p$-value}(s) = 1 - \frac{\gamma(T,s)}{\Gamma(T)} = \frac{\Gamma(T, s)}{\Gamma(T)} ,
\end{equation}
where $\Gamma(T,s)$ is the \emph{upper} incomplete gamma function, $\gamma(T,s)$ is the \emph{lower} \href{https://en.wikipedia.org/wiki/Incomplete_gamma_function}{incomplete gamma function}. 

\paragraph*{Corollary.} Per token, 
\begin{equation}
\mu_0 = \E(S_T/T|\H_0) = 1,\quad \sigma_0^2 = \mathbb{V}(S_T/T|\H_0) = 1/T.
\end{equation}

\paragraph*{Proposition (Bound on expected score under $\H_1$).}
Under $\H_1$, 
$\displaystyle \mathbb{E}(S_T) \geq T +  \left( \frac{\pi^2}{6} -1 \right) H_T $, 
where $H_T = - \sum_{t=1}^T p_t\ln(p_t)$ is the entropy of the completion.

\paragraph*{Proof.}
From~\eqref{chap5/eq:Coro}, $R_t=\exp(-p_t E)$ with $E\sim \mathcal{E}(1)$, so:
\begin{align*}
    \mathbb{E}(S) &= - \mathbb{E} \left[ \sum_{t=1}^T \ln (1-\exp(-p_t E)) \right] \\
    &= - \sum_{t=1}^T \int_0^\infty \ln (1-e^{-p_t x}) e^{-x} dx \\
    &= - \sum_{t=1}^T \int_0^1 \frac{1}{p_t} r^{1/p_t-1} (-\ln ( 1 - r)) dr  \\ 
    & \text{ \qquad (by change of variable $x = -1/p_t \ln (r) $ )} 
\end{align*}
Then, using integration by parts with $u = 1 - r^{1/p_t}$ and $v = \ln(1-r)$, the integral becomes:
\begin{align*}
    -\int_0^1 \frac{1}{p_t} r^{1/p_t-1} \ln ( 1 - r) dr &= \int_0^1 \frac{1-r^{1/p_t}}{1-r} dr = \H_{1/p_t}
\end{align*}
where $\H_{z}$ is the $z$-th \href{https://en.wikipedia.org/wiki/Harmonic_number}{harmonic number}
also defined as $\H_{z} = \sum_{n=1}^\infty \frac{1}{n} - \frac{1}{n+z}$.
Therefore, we have:
\begin{align*}
    -\int_0^1 \frac{1}{p_t} r^{1/p_t-1} \ln ( 1 - r) dr &= 
        \sum_{n=1}^\infty \frac{1}{n} - \frac{1}{n+1/p_t} \\
    &= 1 + \sum_{n=1}^\infty \frac{1}{n+1} - \frac{1}{n+1/p_t}.
\end{align*}
Now, $\forall n\in \mathbb{N^\star}$, we have:
\begin{align*}
   (n+1)^2 \left(\frac{1}{n+1} - \frac{1}{n+1/p_t}\right) &= \frac{(n+1)(n+1/p_t) - (n+1)^2}{n + 1/p_t} \\
    &=  \frac{1+n}{1/p_t + n} \left( 1/p_t -1\right) \\
    &\geq -  \frac{1+n}{1/p_t + n} \ln(p_t) \\
    &\geq -  \, p_t \ln(p_t).
\end{align*}
Therefore, by summing over all $t\in [1,T]$,
\begin{align*}
    \mathbb{E}(S) &\geq T +  \left(\sum_{n=1}^\infty \frac{1}{(n+1)^2}\right)\left(\sum_{t=1}^T- p_t\ln(p_t) \right) \\
    &=T +  \left( \frac{\pi^2}{6} -1 \right) H_T.
\end{align*}

\paragraph*{Proposition (Variance of score under $\H_1$).}
$\displaystyle\mathbb{V}(S_T)\leq T\frac{\pi^2}{6}$.

\paragraph*{Proof.}
For $R_{t}\sim Beta(1/p_t, 1)$:
\begin{equation}
\mathbb{V}(\ln(1-R_t)) = \psi_1(1) - \psi_1(1+1/p_t)
\end{equation}
where $\psi_1$ is the trigamma function, which can be expressed as the following serie $\psi_1(z) = \sum_{n=0}^{\infty} 1/(n+z)^2$. 
Then $\psi_1(1) = \pi^2/6$ and $\psi_1(1+1/p_t)>0$, so that $\mathbb{V}(\ln(1-R_t)) \leq \pi^2/6$.
The results comes because the sampled tokens are independent.

%% file: 0-conclusion/publications.tex
\noindent
The main material of this thesis appeared in the following publications:

{\footnotesize
\setstretch{1.05}
\begin{itemize}
    \item \fullcite{fernandez2022sslwatermarking}
    \item \fullcite{fernandez2022active}
    \item \fullcite{fernandez2023stable}
    \item \fullcite{fernandez2023three}
    \item \fullcite{fernandez2024functional}
    \item \fullcite{san2024proactive}
    \item \fullcite{fernandez2024what}
    \item \fullcite{sander2024watermarking}
\end{itemize}}
\vspace*{1em}

\noindent
Some works were adapted to French venues for the national community: 

{\footnotesize
\setstretch{1.05}
\begin{itemize}
    \item \fullcite{fernandez2022tatouage}
    \item \fullcite{fernandez2023tatouage}
\end{itemize}}
\vspace*{1em}

\noindent
Contributions were also made to the following publications, not included in this thesis, but relevant for the field:

{\footnotesize
\setstretch{1.05}
\begin{itemize}
    \item \fullcite{sander2025watermark} 
    \begin{spacing}{1.0}
        \scriptsize Redefines watermarking as a segmentation task, introducing a method to embed and extract watermarks localized to specific areas of an image. \\
        \scriptsize Contributions: co-first author with Tom Sander.
    \end{spacing}
    \item \fullcite{san2024latent} 
    \begin{spacing}{1.0}
        \scriptsize Introduces a method to watermark audio generative models at the latent space level, through watermarking training data in a way that the watermark is preserved by the audio tokenizer. \\
        \scriptsize Contributions: advised on the watermarking part, and wrote the paper with Robin San Roman.
    \end{spacing}
    \item \fullcite{kinakh2024evaluation} 
    \begin{spacing}{1.0}
        \scriptsize Introduces a copy attack on the scheme presented in Chap.~\ref{chapter:ssl-watermarking}. \\
        \scriptsize Contributions: advised at the beginning of the project, and contributed to the writing.
    \end{spacing}
    \item \fullcite{labiad2024log} 
    \begin{spacing}{1.0}
        \scriptsize Introduces new black-box attacks against fake detection systems, and evaluates if they can be detected. \\
        \scriptsize Contributions: regularly advised Ismail on the watermarking and fake detection parts, and contributed to the writing.
    \end{spacing}
\end{itemize}
}

\noindent
Contributions were also made to the bigger projects at FAIR:

{\footnotesize
\setstretch{1.05}
\begin{itemize}
    \item \fullcite{oquab2023dinov2}
    \begin{spacing}{1.0}
        \scriptsize Contributions: first experiments with the Kozachenko-Leonenko regularizer, classification fine-tuning on ImageNet, zero-shot and fine-tuning evaluations on video benchmarks.
    \end{spacing}
    \item \fullcite{balestriero2023cookbook}
    \begin{spacing}{1.0}
        \scriptsize Contributions: wrote parts on vision transformers and video.
    \end{spacing}
    \item \fullcite{seamless2023} 
    \begin{spacing}{1.0}
        \scriptsize Contributions: wrote part on watermarking, jointly with Hady Elsahar and Robin San Roman.
    \end{spacing}
\end{itemize}
}